\documentclass[epj,nopacs]{svjour}
\usepackage{graphics}
\usepackage{epsfig}
\usepackage{caption}
\usepackage{graphicx}
\usepackage{color}
\usepackage{bm}
\def\inbar{\,\vrule height1.5ex width.4pt depth0pt}
\def\IR{\relax{\rm I\kern-.18em R}}
\def\IC{\relax\hbox{$\inbar\kern-.3em{\rm C}$}}

\def\ifm#1{\relax\ifmmode#1\else$#1$\fi}
\def\DAF{DA\char8NE}    
\def\epem{e$^{+}$e$^{-}$}
\def\x{\ifm{\times}}   \def\pt#1,#2,{\ifm{#1\x10^{#2}}}
\def\up#1{\ifm{^{#1}}}   \def\dn#1{\ifm{_{#1}}}

\def\kl{\ifm{K_L}}   \def\ks{\ifm{K_S}}

\def\rmk{\rm\kern.5mm }

\def\figb#1;#2;{\parbox{#2cm}{\epsfig{file=#1.eps,width=#2cm}}}
\def\toP{\ifm{\rightarrow}}
\newcommand{\bea}{\begin{eqnarray}}
\newcommand{\eea}{\end{eqnarray}}
\newcommand{\be}{\begin{equation}}
\newcommand{\ee}{\end{equation}}
\newcommand{\beq}{\begin{equation}}
\newcommand{\eeq}{\end{equation}}
\newcommand{\ba}{\begin{array}}
\newcommand{\ea}{\end{array}}
\newcommand{\beqa}{\begin{eqnarray}}
\newcommand{\eeqa}{\end{eqnarray}}

\newcommand{\nn}{\nonumber}
\newcommand{\dis}{\displaystyle}

\def\Re{{\rm Re}}
\def\Im{{\rm Im}}

\newcommand{\BR}{\ensuremath{\mathcal{B}}}

\newcommand{\ket}[1]{\ensuremath{\left|#1\right>}}

\renewcommand{\vec}[1]{\ensuremath{\mathbf{#1}}}

\newcommand{\Vud}{\ensuremath{|V_{ud}|}}
\newcommand{\Vus}{\ensuremath{|V_{us}|}}

\newcommand{\text}{\rm}

\newcommand{\gammo}[1]{\ensuremath{\Gamma(#1)}}

\newcommand{\Pem}{\ensuremath{e^-}}
\newcommand{\Pep}{\ensuremath{e^+}}

\newcommand{\Pnu}{\ensuremath{\nu}}
\newcommand{\Pnubar}{\ensuremath{\bar{\nu}}}

\newcommand{\Ppim}{\ensuremath{\pi^-}}
\newcommand{\Ppip}{\ensuremath{\pi^+}}

\newcommand{\DKSeIII}{\ensuremath{K_S\rightarrow\pi e \nu}}

\newcommand{\DKLeIII}{\ensuremath{K_L\rightarrow\pi e \nu}}

\def\H{H\hskip-8.5pt/\hskip2pt}
\def\coeff#1#2{{\textstyle{#1\over #2}}}

\newcommand{\kn}{\ensuremath{\mathrm{K^0}}}
\newcommand{\knb}{\ensuremath{\mathrm{\bar{K}^0}}}

\begin{document} %
%
%
%
%
%
%
\hugehead
\title{Physics with the KLOE-2 experiment at the upgraded \DAF\ }
\author{ G.~Amelino-Camelia\inst{1,2} \and F.~Archilli\inst{3,4} 
\and D.~Babusci\inst{5} 
\and D.~Badoni\inst{4} 
\and G.~Bencivenni\inst{5} \and J. Bernabeu\inst{6} 
\and R.A.~Bertlmann\inst{7} 
\and D.R.~Boito\inst{8} 
\and C.~Bini\inst{1,2} 
\and C.~Bloise\inst{5} \and V.~Bocci\inst{2} 
\and F.~Bossi\inst{5} \and P.~Branchini\inst{9} \and A.~Budano\inst{9} 
\and S.A.~Bulychjev\inst{10}
\and P.~Campana\inst{5} \and G.~Capon\inst{5} \and F.~Ceradini\inst{9,11} 
\and P.~Ciambrone\inst{5} 
\and E.~Czerwinski\inst{5} \and H.~Czyz\inst{12} \and G.~D'Ambrosio\inst{13}
\and E.~Dan\'e \inst{5} \and E.~De~Lucia\inst{5} \and G.~De~Robertis\inst{14}
\and A.~De~Santis\inst{1,2} 
\and P.~De~Simone\inst{5} \and G.~De~Zorzi\inst{1,2} 
\and A.~Di~Domenico\inst{1,2}
\and C.~Di~Donato\inst{13} 
\and B.~Di~Micco\inst{9,11}\thanks{\emph{Present address:} CERN, Geneve, Switzerland} 
\and D.~Domenici\inst{5} \and S.I.~Eidelman\inst{15} 
\and O.~Erriquez\inst{14,16} 
\and R.~Escribano\inst{8} \and R.~Essig\inst{17} 
\and G.V.~Fedotovich\inst{15} \and G.~Felici\inst{5} 
\and S.~Fiore\inst{1,2} \and P.~Franzini\inst{1,2}
\and P.~Gauzzi\inst{1,2} \and F.~Giacosa\inst{18} 
\and S.~Giovannella\inst{5} 
\and F.~Gonnella\inst{3,4} \and E.~Graziani\inst{9} 
\and F.~Happacher\inst{5} \and B.C.~Hiesmayr\inst{7,19} 
\and B.~H\"oistad\inst{20} \and E.~Iarocci\inst{5,21} 
\and S.~Ivashyn\inst{12,22}
\and M.~Jacewicz\inst{5} \and F.~Jegerlehner\inst{23} 
\and T.~Johansson\inst{20} 
\and J.~ Lee-Franzini\inst{5} \and W.~Kluge\inst{24}
\and V.V.~Kulikov\inst{10} \and A.~Kupsc\inst{20}
\and R.~Lehnert\inst{25} 
\and F.~Loddo\inst{14} \and P.~Lukin\inst{15} 
\and M.A.~Martemianov\inst{10} 
\and M.~Martini\inst{5,21} \and  M.A.~Matsyuk\inst{10} 
\and N.E.~Mavromatos\inst{26} \and F.~Mescia\inst{27}
\and R.~Messi\inst{3,4} 
\and S.~Miscetti\inst{5}  \and G.~Morello\inst{28,29} 
\and D.~Moricciani\inst{4}  
\and P.~Moskal\inst{30} 
\and S.~M\"uller\inst{5}\thanks{\emph{Present address:} Institut f\"ur Kernphysik, Gutenberg Universit\"at, Mainz, Germany}  
\and F.~Nguyen\inst{9}  
\and E.~Passemar\inst{6,31} \and M.~Passera\inst{32} \and A.~Passeri\inst{9}
\and V.~Patera\inst{5,21}  \and M.R.~Pennington\inst{33} 
\and J.~Prades\inst{34}
\and L.~Quintieri\inst{5} 
\and A.~Ranieri\inst{14} \and M.~Reece\inst{35} 
\and P.~Santangelo\inst{5} \and S.~Sarkar\inst{26} 
\and I.~Sarra\inst{5} \and M.~Schioppa\inst{28,29} 
\and P.C.~Schuster\inst{17} \and B.~Sciascia\inst{5} 
\and A.~Sciubba\inst{5,21} \and M.~Silarski\inst{30} 
\and C.~Taccini\inst{9,11} \and N.~Toro\inst{36} \and L.~Tortora\inst{9}
\and G.~Venanzoni\inst{5} 
\and R.~Versaci\inst{5}\footnotemark[1] 
\and L.-T.~Wang\inst{37}
\and W.~Wislicki\inst{38} \and M.~Wolke\inst{20}
\and J.~Zdebik\inst{30}
}                     
%
%
\institute{ Dipartimento di Fisica, Universit\`a ``Sapienza", Roma, Italy
\and INFN, Sezione di Roma, Roma, Italy
\and Dipartimento di Fisica, Universit\`a ``Tor Vergata", Roma, Italy
\and INFN, Sezione Roma 2, Roma, Italy
\and INFN, Laboratori Nazionali di Frascati, Frascati, Italy
\and Departamento de F\`isica Te\`orica and IFIC, Universidad de
Valencia-CSIC, Valencia, Spain
\and University of Vienna, Vienna, Austria 
\and Grup de F\`isica Te\`orica and IFAE, Universitat Aut\`onoma de Barcelona, Spain  
\and INFN, Sezione Roma 3, Roma, Italy
\and Institute for Theoretical and Experimental Physics, Moscow, Russia
\and Dipartimento di Fisica, Universit\`a ``Roma Tre", Roma, Italy
\and Institute of Physics, University of Silesia, Katowice, Poland
\and INFN, Sezione di Napoli, Napoli, Italy  
\and INFN, Sezione di Bari, Bari, Italy 
\and Budker Institute of Nuclear Physics, Novosibirsk, Russia
\and Dipartimento di Fisica, Universit\`a di Bari, Bari, Italy 
\and SLAC National Accelerator Laboratory, Menlo Park, U.S.A.
\and Institute for Theoretical Physics, Goethe University, Frankfurt, Germany
\and University of Sofia, Sofia, Bulgaria
\and University of Uppsala, Uppsala, Sweden
\and Dipartimento di Energetica, Universit\`a ``Sapienza", Roma, Italy 
\and Institute for Theoretical Physics, NSC ``Kharkov Institute of Physics and Technology'', Kharkov, Ukraine 
\and Humboldt-Universit\"at zu Berlin, Institut f\"ur Physik, Berlin, Germany
\and Institut f\"ur Experimentelle Kernphysik, Universit\"at Karlsruhe, Karlsruhe, Germany
\and Instituto de Ciencias Nucleares, Universidad 
Nacional Aut\`onoma de M\'exico, M\'exico D.F., M\'exico 
\and Department of Physics, King's College, London, England
\and Universitat de Barcelona, Departament ECM and ICC, Barcelona, Spain 
\and Universit\`a della Calabria, Cosenza, Italy 
\and INFN, Gruppo collegato di Cosenza, Cosenza, Italy 
\and Jagiellonian University, Cracow, Poland 
\and Albert Einstein Center for Fundamental Physics, University of Bern, Bern, Switzerland 
\and INFN, Sezione di Padova, Padova, Italy
\and Institute for Particle Physics Phenomenology, Durham University, Durham, U.K.
\and CAFPE and Departamento de F\'isica Te\'orica y del Cosmos, Universidad
de Granada, Granada, Spain
\and Princeton Center for Theoretical Science, Princeton University, Princeton, U.S.A.
\and Stanford Institute for Theoretical Physics, Stanford University, Stanford, U.S.A.
\and Department of Physics, Princeton University, Princeton, U.S.A. 
\and A. Soltan Institute for Nuclear Studies, Warsaw, Poland 
}
\date{Received: date / Revised version: date}
\abstract{ Investigation at a $\phi$--factory can shed light on 
several debated issues in particle physics. We discuss: i) recent 
theoretical development and experimental progress in kaon physics relevant 
for the Standard Model tests in the flavor sector, ii) the sensitivity 
we can reach in probing CPT and Quantum Mechanics from time evolution 
of entangled kaon states, iii) the interest for improving on the present 
measurements of non-leptonic and radiative decays of kaons 
and $\eta$/$\eta\prime$ mesons, iv) the contribution to understand 
the nature of light scalar mesons, and 
v) the opportunity to search for narrow di-lepton 
resonances suggested by recent models proposing a hidden dark-matter sector.
We also report on the $e^+ e^-$ physics in the continuum with the 
measurements of (multi)hadronic cross sections and the study of 
$\gamma \gamma$ processes. 
%
}
\maketitle
\setcounter{tocdepth}{3}
\tableofcontents

%
%

%
\section{Introduction}
\label{intro}
This report 
results from discussion started at the workshop held at the 
Frascati Laboratory of INFN 
to review the major topics of interest for investigation at  
the upgraded $\phi$--factory \DAF\ .
The scientific program with a high-performance detector such as KLOE covers 
several fields in particle physics: from measurements of interest for 
the development of the Effective Field Theory (EFT) in 
quark-confinement regime 
to fundamental tests of Quantum Mechanics (QM) and CPT invariance. 
It includes  
precision measurements to probe lepton universality, CKM unitarity 
and settle the hadronic vacuum polarization contribution to the  
anomalous magnetic moment of the muon and to 
the fine--struc\-ture constant at the $M_Z$ scale. 

During year 2008 the Accelerator Division of the 
Fra\-sca\-ti Laboratory  
has tested a new interaction
 scheme on the \DAF\ $\phi$-factory collider, with the goal
 of reaching a peak luminosity of 5$\times$10$^{32}$ cm$^{-2}$s$^{-1}$,
 a factor of three larger than what previously obtained.
 The test has been successful and presently \DAF\ is
 delivering up to 15 pb$^{-1}$/day, with solid hopes to reach
 soon 
 20 pb$^{-1}$/day \cite{Milardi:2009zz,Zobov:2009zz}.
Following these achievements, the data-taking campaign
of the KLOE detector on the improved machine that was proposed in
 2006 \cite{KLOE2EoI}, will start 
 in 2010.
 
 KLOE is a multipurpose detector,
 mainly consisting of a large cylindrical drift chamber with an internal
 radius of 
 25 cm and an external one of 2 m, surrounded by
 a lead-scintillating fibers electromagnetic calorimeter.
 Both are immersed in the 0.52 T field of a superconducting solenoid.
 Peculiar to KLOE is the spherical, 10 cm radius, beam pipe which
 allows all of the $K^{0}_{S}$ mesons produced in $\phi$ decays to move in
 vacuum before decaying. Details
 of the detector can be found in Refs. \cite{Adinolfi:2002uk,Adinolfi:2002zx,Adinolfi:2002hs,Adinolfi:2002me,Aloisio:2004ig}.
 From 2000 to 2006, KLOE has acquired 2.5 fb$^{-1}$ of data
 at the $\phi$(1020) peak, plus additional 250 pb$^{-1}$ at
 energies slightly higher or lower than that. 
 The $\phi$ meson predominantly decays into charged and neutral kaons,  
thus allowing KLOE to make precision studies in the fields of 
flavor physics and low energy QCD. The latter can also 
be addressed using $\phi$ radiative decays into scalar or
pseudoscalar particles. Test of discrete symmetries conservation can 
 be performed using several different methods. Most notably, 
CPT conservation can be tested via quantum interferometry measurements
with neutral kaons, a technique which exploits the quantum correlation
between the $K^{0}_{L}$ and the $K^{0}_{S}$ produced in $\phi$ decays.  
A collection of the main physics results of KLOE can be found in
 Ref. \cite{Bossi:2008aa}. \par
 For the forthcoming run \cite{KLOE2LoI}, 
 upgrades have also been proposed for the detector.
 In a first phase, two different devices (LET and HET) 
 will be installed 
along the beam line to detect the scattered electrons/positrons
 from $\gamma\gamma$ interactions.
 In a second phase, a 
light--material internal tracker (IT) will be installed
 in the region 
 between the beam pipe and the drift chamber to improve charged vertex
 reconstruction and to increase the acceptance for low p$_{T}$ 
 tracks \cite{Archilli:2010xb}.
 Crystal calorimeters (CCALT) will cover the low $\theta$ region, aiming at
 increasing acceptance for very forward electrons/photons down to 
 8$^\circ$. A new tile calorimeter (QCALT) will be used to instrument 
 the \DAF\ focusing system for the detection of photons coming from \kl\  
 decays in the drift chamber.  
 Implementation of the second phase is planned for late 2011.
 The integrated luminosity 
for the two phases, from here on dubbed
 as  step-0 and  step-1, will be 5 fb$^{-1}$ and 20 fb$^{-1}$, 
 respectively. \par
 \DAF\ can run 
in a range of $\pm$20 MeV 
from the $\phi$ peak 
without loss of luminosity, 
with the same magnetic configuration. 
Minor modifications, i.e., a new   
final particle focusing system,     
are needed to extend the range 
to $\pm$100 MeV
while 
a major 
upgrade of the machine is required to  
extend it above this limit.  
The improved KLOE detector is perfectly suited for taking data also 
at energies away 
from the $\phi$ mass. 
Therefore a proposal to perform the challenging and needed   
precision measurements of (multi)hadronic 
and $\gamma \gamma$ cross sections   
at energies up to 2.5 GeV has also
been put forward.  \par
We will refer to the entire plan of run 
as the KLOE-2 project. \par
The present paper 
is organised into sections, 
devoted to 
the main physics topics of the experiment,  
where we 
briefly review the present--day situation for the theoretical
 and experimental achievements 
and the improvements expected for the next few years, 
paying particular attention   
to the 
discussion of 
the
KLOE-2 contribution in the field. 
\par \noindent 
We begin in Sect. \ref{sec:SMtest} 
with precision measurements of CKM unitarity and 
lepton universality.  Section \ref{sec:symm} is devoted to the tests of 
QM and CPT invariance and particurarly to the study of the 
entangled kaon states. Issues on low-energy QCD of interest for Chiral 
Perturbation Theory (ChPT) and for the modelization of underlying 
quark dynamics are addressed in Sect. \ref{sec:leqcd}, while the 
two following sections, Sects. \ref{sec:hadcs}--\ref{sec:gg}, discuss the
physics in the continuum, with the measurements of (multi)hadronic 
cross sections and $\gamma \gamma$ processes. 
Section \ref{sec:DMquest} presents the program for searches in the 
hidden particle sector proposed by recent papers on the Dark Matter (DM) 
quest. 
Finally, in Sect. \ref{sec:fine} we summarize our proposal 
of physics reach at the upgraded \DAF\ . 
%
\section{CKM Unitarity and Lepton Universality}
\label{sec:SMtest}
%
\newcommand{\abs}[1]{\ensuremath{\left|#1\right|}}
\newcommand{\kpm} {\ensuremath{K^{\pm}}}
\newcommand{\kp} {\ensuremath{K^{+}}}
\newcommand{\km} {\ensuremath{K^{-}}}
\newcommand{\pipm} {\ensuremath{\pi^{\pm}}}
\newcommand{\vub} {\ensuremath{|V_\mathrm{ub}|}}
\newcommand{\fVus}{\ensuremath{|V_{us}|\times f_+(0)}}
\newcommand{\fzero}  {\ensuremath{f_+(0)}}
\newcommand{\fzerokpi}  {\ensuremath{f_+^{K^0\pi^-}(0)}}
\newcommand{\kzero} {\ensuremath{K^{0}}}
\newcommand{\kltre}  {\ensuremath{K_{l3}}}
\newcommand{\kletre}  {\ensuremath{K_{L}e3}}
\newcommand{\ksetre}  {\ensuremath{K_{S}e3}}
\newcommand{\klmutre}  {\ensuremath{K_{L}\mu3}}
\newcommand{\kpmetre}   {\ensuremath{K^{\pm} {e3}}}
\newcommand{\kpmmutre}  {\ensuremath{K^{\pm} {\mu3}}}
\newcommand{\pippimpio}  {\ensuremath{\pi^+\pi^-\pi^0}}
\newcommand{\pippim}  {\ensuremath{\pi^+\pi^-}}
\newcommand{\duepio}  {\ensuremath{2\pi^0}}
\newcommand{\trepio}  {\ensuremath{3\pi^0}}
\newcommand{\mudue}  {\ensuremath{\mu^{\pm}\nu}}
\newcommand{\taus}  {\ensuremath{ \tau_S}}
\newcommand{\taul}  {\ensuremath{\tau_L}}
\newcommand{\taupm}  {\ensuremath{\tau_{\pm}}}
\newcommand{\kppipigall}{\ensuremath{K^{+} \rightarrow \pi^{+}\pi^0\,(\gamma)}}
\let\kpppg=\kppipigall

\def\ff{$\phi-$factory}  \def\DAF{DA\-\char8\-NE}
\def\up#1{$^{#1}$}  \def\dn#1{$_{#1}$}
\def\pt#1,#2,{\ifm{#1\x10^{#2}}}
\newcommand{\ka}[1]{\ensuremath{K^{{#1}}}}
\newcommand{\lam}[2]{\ensuremath{\lambda^{{#1}}_{{#2}}}}
\newcommand{\eV}{{e\kern-.07em V}}
\newcommand{\MeV}{{\rm \,M\eV}}
\hyphenation{Flavia-Net}
\newcommand{\mtaud} {m^2_{\tau}}
\newcommand{\mkd} {m^2_K}
\newcommand{\mkq} {m^4_K}
\newcommand{\mpid} {m^2_\pi}
\def\ni {\noindent}
\def\nn {\nonumber}
\def\rd{\mbox{d}}
\def\IK{I_{K_{e3}^0}}
\def\eqalign#1{\null\,\vcenter{\openup\jot\m@th
  \ialign{\strut\hfil$\displaystyle{##}$&$\displaystyle{{}##}$\hfil
      \crcr#1\crcr}}\,}
\newcommand{\szero}{{\it step-0}}
\newcommand{\suno}{{\it step-1}}

Purely leptonic and semileptonic  decays of K mesons ($K  \to \ell \nu, K \to   \pi \ell \nu$, $\ell = e, \mu$)  
are mediated in the Standard Model (SM) by tree-level W-boson exchange. 
Gauge coupling universality and three-generation  quark mixing imply 
that 
semileptonic processes such as $d^i \to  u^j \ell \nu$ are governed by 
the effective Fermi constant $G_{ij} = G_\mu \, V_{ij}$, 
where:  (i) $G_\mu$ is the  muon decay constant; 
(ii) $V_{ij}$ are the elements of the unitary  Cabibbo--Kobayashi 
Maskawa (CKM) matrix. This fact has simple but deep consequences, 
that go under the name of universality relations:
\begin{itemize}
\item In the SM  the effective semileptonic 
constant $G_{ij}$ does not depend on the lepton flavor 
({\it lepton universality}).
\item If one extracts $V_{ij}$ from different semileptonic transitions assuming 
quark-lepton gauge universality (i.e. normalizing the decay rates with $G_\mu$), the CKM unitarity 
condition $\sum_j  |V_{ij}|^2 = 1$ should be verified. 
\end{itemize}
Beyond the SM,  these universality relations can be violated 
by new  contributions 
to the low-energy  $V$-$A$  four fermion operators,  
as well as new non $V$-$A$ structures.  
Therefore,  precision tests of the universality relations 
probe physics beyond the SM and are sensitive to several SM extensions~\cite{Marciano:1987ja,Hagiwara:1995fx,Kurylov:2001zx,Cirigliano:2009wk}.
A simple dimensional analysis argument reveals that deviations from the universality relations 
scale as  $\Delta \sim M_W^2/\Lambda^2$, where $\Lambda$ is the scale 
associated with new physics (NP). 
Therefore, testing the universality relations at the 0.1\% level  allows us to put non-trivial 
constraints on physics scenarios at the TeV scale that will be directly 
probed at the LHC. 
Kaon physics currently plays a prominent role in testing both 
quark-lepton universality, through the  $V_{us}$ entry in the 
CKM unitarity relation 
$|V_{ud}|^2 +  |V_{us}|^2 +  |V_{ub}|^2 = 1$, and
lepton universality, through the measurement of \fVus\  
and the helicity 
suppressed ratio $\Gamma(K \to e \nu (\gamma))/ \Gamma(K \to \mu \nu (\gamma))$. In the case of 
the latter, theoretical prediction, at the $0.04$ \% level, 
 is extremely clean  \cite{Cirigliano:2007xi,Cirigliano:2007ga} 
(it is only affected by hadronic structure dependence to 
two loops in the chiral effective theory).

The experimental precision was at the 6\% level \cite{Amsler:2008zzb}
before the recently-published KLOE 
measurement \cite{Ambrosino:2009rv} which has 
improved the sensitivity to  
1\%. New results from the NA62 collaboration at CERN are 
expected to reduce 
in near-term the error  
to the $0.5$\% level.  
This helicity suppressed ratio is very  promising in terms of 
uncovering physics beyond 
the SM \cite{Antonelli:2008jg}, including tests of Higgs-induced 
lepton flavor violation in
supersymmetry \cite{Masiero:2005wr}.  
Improvements on the experimental side are highly  motivated. 

Large amount of
data  has been collected on the semileptonic 
modes $K \to \pi \ell \nu$ by several experiments,  
BNL-E865, KLOE, KTeV, ISTRA+, and NA48   
in the last few years.
These data 
have stimulated a sub\-stan\-ti\-al pro\-gress on the theoretical 
inputs, so that 
most of the theory-do\-mi\-na\-ted errors associated to radiative 
corrections \cite{Cirigliano:2004pv,Cirigliano:2001mk,Andre:2004tk,Cirigliano:2008wn}  and hadronic form 
factors \cite{Antonelli:2008jg} have been reduced below $1\%$.
Presently,  the unitarity test
\begin{equation}
|V_{ud}|^2 + |V_{us}|^2 +|V_{ub}|^2 = 1 + \Delta_{\rm CKM}
\label{eq:unitarity}
\end{equation}
implies that $\Delta_{\rm CKM}$ is consistent with zero 
at the level of $6 \times 10^{-4}$.  
$V_{us}$ from $K \to \pi \ell \nu$ decays contributes 
about half of this uncertainty, 
mostly coming from the hadronic matrix element. 
Both experimental and the\-o\-re\-ti\-cal pro\-gress in $K_{\ell 3}$  decays 
will be needed in order to improve the accuracy 
on $\Delta_{\rm CKM}$ in the future.  

The kaon semileptonic decay rate is given by:  
$$ 
\Gamma(\kltre) = \frac{C_K^2 G_F^2 M_K^5}{192 \pi^3} S_{EW} \Vus^2
|\fzero|^2 \times  
$$
\begin{equation}
I_{K,l}(\lambda)(1+2\Delta_K^{SU(2)}+2\Delta_{K,l}^{EM})
\label{eq:gammakl3}
\end{equation}
where $K = \kzero, \kpm$, $l = e, \mu$ and $C_K$ is a Clebsch-Gordan
coefficient, equal to $1/2$ and $1$ for \kpm\  and \kzero,
respectively. The decay width $\Gamma(\kltre)$ is experimentally
determined  by measuring the kaon lifetime and the semileptonic 
branching fractions (BRs)
totally inclusive of radiation. The theoretical inputs are: the universal
short-distance electroweak correction $S_{EW} = 1.0232$, the
$SU(2)$-breaking $\Delta_K^{SU(2)}$ and the long-distance electromagnetic
corrections $\Delta_{K,l}^{EM}$ which depend on the kaon charge and on
the lepton flavor, and the form factor $\fzero\equiv\fzerokpi$
parametrizing the hadronic matrix element of  the $K \rightarrow \pi$
transition, evaluated at zero momentum transfer and for neutral kaons.
The form factor dependence on the momentum transfer is needed 
for the calculation of the phase space integral $I_{K,l}(\lambda)$. 
It can be described by
one or more slope parameters $\lambda$ measured from the decay 
spectra. 

Complementary to $K_{\ell 3}$ decays, 
the kaon (pion) leptonic radiation-inclusive decays $K(\pi) \to
\mu \bar{\nu}_\mu (\gamma)$ 
provide a precise determination of $\Vus/\Vud$.
The ratio of these decay rates is:
\begin{equation}
{\Gamma(K^+_{\mu2})\over\Gamma(\pi^+_{\mu2})}=
{\:m_K\left(1-{m_\mu^2\over m_K^2}\right)^2
\over \:m_\pi\left(1-{m_\mu^2\over m_\pi^2}\right)^2}\:
 \frac{f_K^2\Vus^2} {f_\pi^2\Vud^2}
\:\frac{\:1 + \frac{\alpha}{\pi} C_K\:}
             {\:1 + \frac{\alpha}{\pi} C_\pi\:}
\label{eq:gammakmu2pimu2}
\end{equation}
where $f_K$ and $f_\pi$ are, respectively, the kaon and the pion decay
constants; $C_\pi$ and $C_K$ parametrize the radiation-inclusive electroweak
corrections accounting for bre\-ms\-stra\-h\-lung emission of real 
photons and 
virtual-photon loop contributions. 
\par 
It has been shown that presently semileptonic processes and the 
related universality tests provide
constraints on NP that cannot be obtained from other electroweak
precision tests and/or direct measurements at the 
colliders~\cite{Cirigliano:2009wk}.

\par
In the following subsections, an overview of the present scenario and possible
improvements on results from kaon physics will be 
given on both theoretical and experimental sides. 

\subsection{Determination of $f_+(0)$ and $f_K/f_\pi$}
\label{sec:Theo_CKM}

The vector form factor at zero-momentum transfer, $f_+(0)$, is the key hadronic 
quantity required for the extraction of the CKM matrix element $|V_{us}|$ from 
semileptonic $K_{\ell 3}$ decays as in Eq.(\ref{eq:gammakl3}).
Within SU(3) ChPT one can perform a systematic expansion of $f_+(0)$ of the type 
 \be
    f_+(0) = 1 + f_2 + f_4 + ... ~ , 
    \label{eq:chiralPT}
 \end{equation}
where $f_n = {\cal O}[M_{K, \pi}^{n} / (4 \pi f_\pi)^n]$ and the first term is 
equal to one due to the vector current conservation in the SU(3) limit. 
Because of the Ademollo-Gatto (AG) theorem \cite{Ademollo:1964sr}, 
the first non-trivial term 
$f_2$ does not receive contributions from the local operators of the effective 
theory and can be unambiguously computed in terms of the kaon and pion masses 
($m_K$ and $m_\pi$) and the pion decay constant $f_\pi$. 
At the physical point the value is $f_2 = -0.023$ \cite{Leutwyler:1984je}.
Thus the problem is reduced to the evaluation of 
$\Delta f$, defined as
 \begin{equation}
    \Delta f \equiv f_4 + f_6 + ... = f_+(0) - (1 + f_2) ~ ,
    \label{eq:deltaf}
 \end{equation}
which depends on the low-energy constants (LECs) of the effective theory and 
cannot be deduced from other processes.
The original estimate made by Leutwyler and Roos (LR) 
\cite{Leutwyler:1984je} was based on the quark model yielding 
$\Delta f = -0.016(8)$.
More recently other analytical approaches have been devised to obtain  
the next-to-next-to-leading (NNLO) 
order term $f_4$ by writing it as:  
 \begin{equation}
    f_4 = L_4(\mu) + f_4^{loc}(\mu),
    \label{eq:f4}
 \end{equation}
where $\mu$ is the renormalization scale, $L_4(\mu)$ is the loop contribution computed 
in Ref. \cite{Bijnens:2003uy} and $f_4^{loc}(\mu)$ is the ${\cal O}(p^6)$ local contribution.
For the latter, various models have been adopted  
obtaining values compatible with zero within uncertainty 
leading to $f_+(0)$  values significantly larger than the LR estimate 
(cf. Tab. \ref{tab:fplus}) and  
to the prediction of smaller SU(3)\--bre\-aking 
effects. 
Even if in principle the NNLO term $f_4$ could
be obtained from the measurement 
of the slope ($\lambda^{\prime}_{0}$) and the curvature
($\lambda^{\prime\prime}_{0}$) of the scalar form factor $f_0(q^2)$,
this is experimentally impossible. 
Even in the case of a perfect knowledge of \lam{'}{+} and \lam{''}{+},
with one million of $K_{\mu 3}$ events, the error $\delta$\lam{''}{0} would
be about four times the expected value of \lam{''}{0}. 
Thus, one has to rely on dispersion relations which allow the calculation 
of the curvature from the
measurement of only one parameter 
and then to perform a matching with ChPT \cite{Bernard:2007tk}.
At present, the precision reached is not sufficient  
for an accurate determination of $f_+(0)$. 

A precise evaluation of $f_+(0)$, or equivalently $\Delta f$, requires the use of 
non-perturbative methods based on the fundamental theory of the 
strong interaction, 
such as lattice QCD simulations.
Such precision determinations began recently with the quenched 
simulations of Ref. \cite{Becirevic:2004ya}, 
where it was shown that $f_+(0)$ can be determined at the physical point with $\simeq 1 \%$ accuracy.
The findings of Ref. \cite{Becirevic:2004ya} triggered various unquenched calculations of 
$f_+(0)$, with both $N_f = 2$ and $N_f = 2 + 1$ dynamical flavors. 
The error associated with the chiral
extrapolation was significantly reduced thanks to the lighter pion masses
( 
the minimal value of the simulated pion mass is $260$ \MeV).
The results for $f_+(0)$ 
are summarized in Tab. \ref{tab:fplus}.
It can be seen that all lattice results are in agreement with the LR estimate, 
while the analytical approaches of Refs.
\cite{Bijnens:2003uy,Jamin:2004re,Cirigliano:2005xn} are systematically higher.
\begin{table}[!htb]
\begin{center}
\caption{Summary of model and lattice results for $f_+(0)$.
The lattice errors include both statistical and systematic uncertainties, 
unless otherwise indicated. 
\label{tab:fplus}}
\vskip 2mm 
\begin{tabular}{||c|c|c||}
\hline
 $\mbox{Ref.}$            & $Model/Lattice$ & $f_+(0)$  \\ \hline \hline
 \cite{Leutwyler:1984je}  & $LR$            & $0.961~(~8)$ \\ \hline
 \cite{Bijnens:2003uy}    & $ChPT+LR$       & $0.978~(10)$ \\ \hline
 \cite{Jamin:2004re}      & $ChPT+disp.$    & $0.974~(11)$ \\ \hline 
 \cite{Cirigliano:2005xn} & $ChPT+1/N_c$    & $0.984~(12)$ \\ \hline 
 \cite{Kastner:2008ch} & $ChPT+1/N_c$    & $0.986~(~7)$ \\ \hline \hline
 \cite{Becirevic:2004ya}  & $SPQ_{cd}R$     & $0.960~(~9)$ \\ \hline \hline
 \cite{Tsutsui:2005cj}    & $JLQCD$         & $0.967~(~6)$ \\ \hline
 \cite{Dawson:2006qc}     & $RBC$           & $0.968~(12)$ \\ \hline
 \cite{Brommel:2007wn}    & $QCDSF$         & $0.965~(2_{\small{stat}})$ \\ \hline
 \cite{Lubicz:2009ht}     & $ETMC$          & $0.957~(~8)$ \\ \hline \hline
 \cite{Boyle:2007qe}      & $RBC+UKQCD$     & $0.964~(~5)$ \\ \hline \hline
\end{tabular}
\end{center}
\end{table}
Since simulations of lattice QCD are carried out in a finite volume, the 
momentum transfer $q^2$ for the conventionally used periodic fermion
boundary conditions takes values corresponding to the Fourier modes of the 
kaon or pion. Using a phenomenological ansatz for the 
$q^2$-\-de\-pen\-den\-ce of the form factor one interpolates to $q^2=0$ where
$f_+(0)$ is extracted, thereby introducing a major systematic uncertainty.  
A new method based on the use of partially twisted boundary conditions
has been developed \cite{Boyle:2007wg} which allows 
this uncertainty to be entirely removed by simulating directly at
 the desired kinematical point $q^2=0$. 
A systematic study of the scaling behavior of 
$f_+(0)$, using partially twisted boundary conditions 
 and the extension of the simulations to lighter pion masses in order 
to improve the chiral extrapolation 
are priorities for the upcoming lattice studies of $K_{\ell 3}$ decays.

In contrast to $f_+(0)$, the pseudoscalar 
decay constants $f_K$ and $f_\pi$ are not protected by 
the AG theorem \cite{Ademollo:1964sr} against 
linear corrections in the SU(3) breaking.
Moreover the first non-trivial term in the chiral 
expansion of $f_K / f_\pi$, of order ${\cal O}(p^4)$, 
depends on the LECs and therefore it cannot be 
unambiguously predicted within ChPT.
This is the reason why the most precise determinations of $f_K / f_\pi$ 
 come from lattice QCD simulations.

In recent years various collaborations have provided new results for 
$f_K / f_\pi$ using unquenched gauge configurations with both 
$N_f=2$ and $N_f=2+1$ 
dynamical flavors. They are summarized in Tab. \ref{tab:fKfPi}.
At the current level of precision, the comparison of 
$N_f=2$ and $N_f=2+1$ results indicate a rather 
small contribution of the strange sea quarks to the ratio 
of the decay constants.
\begin{table}[!htb]
\begin{center}
\caption{Summary of lattice results for $f_K / f_\pi$.
The errors include both statistical and systematic uncertainties.
\label{tab:fKfPi}}
\vskip 2mm
\begin{tabular}{||c|c|c||}
\hline
 $\mbox{Ref.}$                     & $Collaboration$ & $f_K / f_\pi$         \\ \hline \hline
 \cite{Aubin:2004fs,Bazavov:2009bb}& $MILC$          & $1.197~_{-13}^{+~7}~$ \\ \hline
 \cite{Follana:2007uv}             & $HPQCD$         & $1.189~(~7)$          \\ \hline
\cite{Durr:2010hr}                 & $BMW$           & $1.192~(~9)$          \\ \hline 
 \cite{Blossier:2009bx}            & $ETMC$          & $1.210~(18)$          \\ \hline 
 \cite{Aubin:2008ie}               & $Aubin\, \emph{et al.}$   & $1.191(23)$ \\ \hline\hline
 \cite{Beane:2006kx}               & $NPLQCD$        & $1.218~_{-24}^{+11}~$ \\ \hline
 \cite{Allton:2008pn}              & $RBC/UKQCD$     & $1.205~(65)$          \\ \hline
 \cite{Aoki:2008sm}                & $PACS-CS$       & $1.189~(20)$          \\ \hline \hline
\end{tabular}
\end{center}
\end{table}
Table \ref{tab:fKfPi} deserves few comments:
i) the convergence of the SU(3) chiral expansion for $f_K / f_\pi$ 
is quite questionable, mainly because large next-to-leading order (NLO) 
corrections are already required to 
account for the large difference between the experimental value of $f_\pi$ 
and the value of this decay constant in the massless SU(3) limit.
Instead, the convergence of the SU(2) chiral expansion 
is much better and, thanks 
to the light pion masses reached in recent lattice calculations, the 
uncertainty related to chiral extrapolation to the physical point is kept 
at the percent level \cite{Allton:2008pn};
ii) little is known about the details of the chiral and continuum
extrapolation in Ref. \cite{Follana:2007uv} (HPQCD) which is currently
the most precise lattice prediction for $f_K/f_\pi$;
iii) there is also some concern about the staggered fermion formulation used 
by MILC \cite{Aubin:2004fs,Bazavov:2009bb}, HPQCD \cite{Follana:2007uv}, 
Aubin \emph{et al.} \cite{Aubin:2008ie}  and NPLQCD  \cite{Beane:2006kx}. 
These results would have 
to be confirmed by conceptually clean fermion formulations.

Therefore it is not obvious to decide which is the value 
to be used for $f_+(0)$ and 
$f_K / f_\pi$. A dedicated working group, the FLAVIAnet 
lattice averaging group (FLAG), has just started  
to compile and publish 
lattice QCD results for SM observables and parameters. 
They have identified
those published results 
demonstrating good control over systematic uncertainties. 
A first status report is given in Ref. \cite{FLAG:Colangelo}.
\subsection{Experimental results from kaon decays}
\label{sec:Exp_CKM}
In the last years, many efforts have been dedicated to the correct 
averaging of the rich
harvest of recent results in kaon physics. 
The FLAVIAnet kaon working group 
has published a comprehensive review~\cite{Antonelli:2008jg} 
where a detailed description of
the averaging procedure can be found.
In this paper we will focus on the contribution 
from KLOE,  
with both the present data set (2.5 fb$^{-1}$) 
and the 5 fb$^{-1}$ 
planned for KLOE-2/step-0 (cf. Sect. \ref{intro}). 

After four years of data analysis, KLOE  
has produced the most comprehensive set of results
from a single experiment, measuring the main 
BRs of $K_L$~\cite{Ambrosino:2005ec}, 
$K^{\pm}$~\cite{Ambrosino:2007xm,Ambrosino:2005fw,Ambrosino:2008pz}    
and $K_S$~\cite{Ambrosino:2006si,Ambrosino:2006sh} (unique to KLOE), including
semileptonic and two-body decays;
 lifetime measurements for $K_L$ \cite{Ambrosino:2005vx} 
and $K^\pm$ \cite{Ambrosino:2007xz};
form factor slopes from the analysis of $K_{L}e3$ \cite{Ambrosino:2006gn} and
$K_{L} \mu3$ \cite{Ambrosino:2007yza,Testa:2008xz}. 
The value of \fVus\ 
has been obtained \cite{Ambrosino:2008ct} 
using the $K_S$ lifetime
from PDG \cite{Amsler:2008zzb} as the only non--KLOE input.
These data 
together with the value of $\Vus/\Vud$ from the measurement 
of the  $\kpm
\rightarrow \mudue (\gamma)$ branching ratio~\cite{Ambrosino:2005fw}
 and the extraction of $\Vud$ from superallowed nuclear $\beta$ decays, 
provide the basis for testing
the unitarity of the quark-flavor mixing matrix.
A test of lepton universality 
and improvements of the bounds on NP scenarios have also obtained from 
the measurements of $\Vus$ from leptonic and 
semileptonic decays.
\subsubsection{Branching ratios}
KLOE has measured all of the main branching ratios 
of \kl, \ks, and \kpm\ decays
using for the \kl\ the whole data set and only one fifth of the 
entire sample 
for \ks\ and \kpm. A
summary of the results on \kl, \ks, and \kpm\ 
is shown in Tab. \ref{tab:brs}.  
The measurement of the absolute BR of the \kpm\ semileptonic decays 
on the basis of the entire data sample of 2.5 fb$^{-1}$ has just 
started.
The analysis of the subdominant \ksetre\ modes will be improved at KLOE-2 also 
with the upgrade of the tracking system through the installation 
of the inner chamber   
 (cf. Sect. \ref{intro}). 
Preliminary studies have shown that the upgrade  
will increase i) the acceptance for low-momentum tracks and 
ii) the vertex resolution for decays close 
to the beam interaction point by a factor of three. 
%
\begin{table}[htb]
\caption{Summary of KLOE results on \kl, \ks, and \kpm\ branching ratios.}
\label{tab:brs}
\vskip 2mm
{\parbox{0.4\textwidth}
{\centering
\begin{tabular}{ll}
        \hline
 {\em $K_L \to \pi e \nu$}   & 0.4008 $\pm$ 0.0015~\cite{Ambrosino:2005ec} \\
 {\em $K_L \to \pi\mu\nu$}   & 0.2699 $\pm$ 0.0014~\cite{Ambrosino:2005ec} \\
 {\em $K_L \to 3\pi^0$}   & 0.1996 $\pm$ 0.0020~\cite{Ambrosino:2005ec} \\
 {\em $K_L \to \pi^{+}\pi^{-}\pi^{0}$}   & 0.1261 $\pm$ 0.0011~\cite{Ambrosino:2005ec} \\
 {\em $K_L \to \pi^{+}\pi^{-}$} & (1.963 $\pm$ 0.21)$\times$10$^{-3}$~\cite{Ambrosino:2006up} \\
 {\em $K_L \to \gamma\gamma$}   & (5.569 $\pm$0.077)$\times$10$^{-4}$~\cite{Adinolfi:2003ca} \\

 {\em $K_S \to \pi^{+}\pi^{-}$}   & 0.60196 $\pm$ 0.00051~\cite{Ambrosino:2006sh}\\
 {\em $K_S \to \pi^{0}\pi^{0}$}   & 0.30687 $\pm$ 0.00051~\cite{Ambrosino:2006sh}     \\
 {\em $K_S \to \pi e \nu$}   & (7.05 $\pm$ 0.09)$\times$10$^{-4}$~\cite{Ambrosino:2006si}  \\
 {\em $K_S \to \gamma\gamma$}   & (2.26 $\pm$ 0.13)$\times$10$^{-6}$~\cite{Ambrosino:2007mm} \\
 {\em $K_S \to 3\pi^{0}$} & $<$1.2$\times$10$^{-7}$ at 90$\%$ C.L.~\cite{Ambrosino:2005iw} \\
 {\em $K_S \to e^{+}e^{-}(\gamma)$} & $<$9 $\times$10$^{-9}$ at 90$\%$ C.L.~\cite{Ambrosino:2008zi} \\

 {\em $K^{+} \to \mu^{+}\nu(\gamma)$}   & 0.6366 $\pm$ 0.0017~\cite{Ambrosino:2005fw}     \\
 {\em $K^{+} \to \pi^{+}\pi^{0}(\gamma)$}   & 0.2067 $\pm$ 0.0012~\cite{Ambrosino:2008pz}  \\
 {\em $K^{+} \to \pi^{0}e^{+}\nu(\gamma)$}   & 0.04972 $\pm$ 0.00053~\cite{Ambrosino:2007xm}     \\
 {\em $K^{+} \to \pi^{0}\mu^{+}\nu(\gamma)$} & 0.03237 $\pm$ 0.00039~\cite{Ambrosino:2007xm}     \\
 {\em $K^{+} \to \pi^{+}\pi^{0}\pi^{0}$}   & 0.01763 $\pm$ 0.00034~\cite{Aloisio:2003jn}  \\
\hline
 \end{tabular}}}
\vspace{-0.5cm}
 \end{table}
\subsubsection{Kaon lifetimes}
KLOE has measured the $K_L$ 
\cite{Ambrosino:2005vx,Ambrosino:2005ec} and the $K^{\pm}$ 
lifetimes \cite{Ambrosino:2007xz} using
approximately one fifth of the collected data set.
Two independent measurements of the $K_L$ lifetime have been
obtained, yielding an overall fractional accuracy of 1\% : 
\begin{equation}
\begin{array}{ll}
{\mathrm {(a)}} & \tau_L = (50.92\pm 0.17\pm 0.25)\;\mathrm{ns} \\
{\mathrm {(b)}} & \tau_L = (50.72\pm 0.11\pm 0.35)\;\mathrm{ns} \\
\end{array}
\label{eq:taul}
\end{equation}
The first measurement uses the fit to the proper-time distribution of the
$K_L\rightarrow\pi^0\pi^0\pi^0$ decay channel and is currently being 
repeated with the complete data set. 
From the preliminary result, $\tau_L = (50.56\pm 0.14\pm
0.21)\;\mathrm{ns}$  \cite{bib:tauL_prelim}, a fractional error of 0.38\%
is expected on the final measurement. 
Adding the 5 fb$^{-1}$ from the first year of data-taking with KLOE-2, 
the fractional accuracy is reduced to 0.27\%. 
Furtermore, the insertion of new quadrupole 
instrumentation~\cite{bib:nim_qcalt}
as planned in 2011 will improve the photon reconstruction allowing better   
control of the systematics and leading to an error 
 $\Delta\tau_L /\tau_L < 0.2\%$. 
\par    
The second $\tau_L$ measurement was obtained from the simultaneous 
measurements of the main $K_L$ branching ratios, imposing the constraint 
on the their sum, $\sum BR_i=1$. 
This technique yielded a result of remarkable precision, but
severely dominated by systematics. 
\par 
With the available sample at KLOE-2 and
the improved tracking performance from the installation of the IT, 
it is possible to exploit 
the measurement of the 
proper-time distribution 
of all of the charged modes to further improve on $\tau_L$. 
Critical point of this approach is the control of 
small variations in the tag efficiency of different decay channels. 
Succeeding in the latter, the result combined with the measurement 
obtained from the
neutral channel would reduce the relative error on $\tau_L$ to 0.1\%.
\par
The $K^{\pm}$ lifetime has been determined by KLOE with two 
independent tecnhiques \cite{Ambrosino:2007xz}: 
the measurement of the decay length by reconstructing the $K^{\pm}$ track,
and 
of the decay time with a time--of--flight technique using modes with
one $\pi^0$ in the final state ($K^{\pm}\to\pi^0X$).
Their combination yields a relative error of 0.25\%.
\begin{equation}
\begin{array}{ll}
{\mathrm {decay\,  length}} & \tau_{\pm} = (12.364\pm 0.031\pm 0.031)\;\mathrm{ns} \\
{\mathrm {decay\,  time}} & \tau_{\pm} = (12.337\pm 0.030\pm 0.020)\;\mathrm{ns} \\
\end{array}
\label{eq:taupm}
\end{equation}
For both methods there is still room for improving both
statistical and systematic errors by increasing the analyzed sample.
Making use of the KLOE and 5 fb$^{-1}$ of the KLOE-2
data sample, the $\tau_{\pm}$ relative error can be reduced to the 
0.1\% level. Moreover, the decay length technique would become much more
accurate with the inner tracker allowing the detection of  
$K^{\pm}$ tracks closer to the interaction point. This is the step-1 
scenario, aiming to improve by a factor of two  
$\Delta\tau_{\pm} /\tau_{\pm}$.
\par
Finally, 
a new promising technique is currently being developed at KLOE to measure
the $K_S$ lifetime. 
This 
consists in an event-by-event determination of the $\phi$ and  
the $K_S$ decay-point positions by a 
kinematic fit. With the full KLOE statistics the preliminary result $\tau_S
=  (89.56\pm 0.03\pm 0.07)\;\mathrm{ps}$~\cite{bib:tauL_prelim}
has been obtained, with the aim of reaching $\sim$0.03 ps final
systematic uncertainty. 
A relative error of 0.03\% on $\tau_S$ is expected scaling this result to the 
 KLOE-2/step-0 data sample.
\subsubsection{Kaon form factors}

To compute the phase space integrals appearing in 
the SM photon-inclusive $K_{\ell3}$ decay rates as in Eq.~(\ref{eq:gammakl3}),
we need experimental or theoretical inputs about the  $t$-\-de\-pen\-den\-ce
of $f_{+,0}(t)$. In principle,  ChPT 
and Lattice QCD are useful to set theoretical constraints.
In practice, at present the  $t$-dependence of the form factors (FFs)  
is better determined by measurements and by combining measurements 
with dispersion relations. 

A complete description of the experimental situation on FFs can be found in 
Ref.~\cite{Antonelli:2008jg}. 
KLOE-2 can contribute to improve present knowledge of FF.
Besides, precise measurements of the FF properties impose stringent test
on quantities which can be obtained only from theory.
In particular, the theoretical determination of 
$f_{+}(0)$
(from Lattice and ChPT),
and the ratio of the kaon and pion decay constants
defined in the $m_u = m_d$ and $\alpha_{\rm em} \to 0$ limit, $f_K/f_\pi$ 
(from Lattice only). 

In the physical region, $m_\ell^2<t<(m_K-m_\pi)^2$,
a very good approximation of the FFs is given by a Taylor expansion
up to $t^2$ terms
\begin{equation}
  \tilde{f}_{+,0}(t) \equiv\frac{{f}_{+,\,0}(t)}{{f}_{+}(0)} = 1 + 
\lambda'_{+,0}~\frac{t}{m_\pi^2}+\frac{1}{2}\;\lambda''_{+,0}\,\left(\frac{t}{m_\pi^2}\right)^2\,
 + \ \dots .
  \label{eq:Taylor}
\end{equation}
For the vector form factor $\tilde{f}_{+}$, the experimental information 
from both 
$K_{e3}$ and $K_{\mu3}$ data is quite accurate and so far superior 
to theoretical predictions.
For the scalar form factor, $\tilde{f}_0(t)$, the situation is 
more complicated: 
i) 
first of all $\tilde{f}_0(t)$ is accessible 
from  $K_{\mu3}$ data only;
ii) the correlation between \lam{'}{+,0} and \lam{''}{+,0} is close to $-1$ 
so that a fit will 
trade \lam{'}{+,0} for 
\lam{''}{+,0} and consequently enlarge the errors, 
on the other hand, ignoring the $t^2$ term would increase the value of
\lam{'}{+,0} by $\sim$3.5 \lam{''}{+,0}; 
iii) it is not fully justified to fit with two parameters 
related by the simple relation \lam{''}{+,0}$=2$\lam{'}{+,0}$^2$ 
from the pole
parametrisation. Indeed, if for the vector FF the K$^\star$(892) resonance
dominates, for the scalar FF there is not such an obvious pole dominance.
Instead, a suitable single--parameter function could be introduced 
for $\tilde{f}_{+,0}(t)$ using 
a dispersive approach 
as emphasized in Refs. \cite{Antonelli:2008jg,Bernard:2006gy,Bernard:2009zm}.
A dispersive 
pa\-ra\-me\-tri\-za\-tion for $\tilde{f}_+$ has been built in
 Ref. \cite{Bernard:2009zm}, 
with good analytical and unitarity
properties and a correct threshold behaviour. 
For the scalar FF, particularly appealing is an improved dispersion 
relation proposed in Ref. \cite{Bernard:2009zm,Bernard:2007cf} 
where two subtractions are performed, at $t=0$
(where by definition, $\tilde f_0(0)\equiv 1$), and at
the so-called Callan-Treiman point $t_{CT} \equiv (m_K^2-m_\pi^2)$.
At this second point, the Callan-Treiman 
theorem \cite{Callan:1966hu,Dashen:1969bh} implies 
\begin{equation}
\tilde f_0(t_{CT})=\frac{f_K}{f_\pi}\frac{1}{f_+(0)}+ \Delta_{CT},
\label{eq:CTrel}
\end{equation}
where $\Delta_{CT} \sim  {\cal O} (m_{u,d}/4 \pi F_{\pi})$ is a small
quantity which can be estimated from theory
( ChPT at NLO in the isospin limit estimates 
$\Delta_{CT}=(-3.5\pm 8)\times 10^{-3}$ ) \cite{Gasser:1984ux}.

Using $\sim$500 pb$^{-1}$ of integrated 
luminosity 
KLOE has obtained the vector FF in $K_{L}e3$ 
decays \cite{Ambrosino:2006gn} and 
the vector and scalar FF in $K_L\mu3$ decays \cite{Ambrosino:2007yza}.
For scalar and vector FFs measured in $K_L\mu3$ events, KLOE has 
also presented the preliminary result based on an integrated 
luminosity of about 
1.5 fb$^{-1}$ \cite{Testa:2008xz}.
The statistical error on FF
should improve by a factor of three with respect to the KLOE published 
measurements with the analysis of KLOE-2/step-0 data. 
Since in KLOE the systematic errors are partially statistical in nature 
(efficiencies are measured with selected control samples and  
downscaled samples are commonly used)
also these contributions to the total uncertainty decrease approximately  
by a comparable amount.

For \ka{\pm} semileptonic FF the situation is different: 
the analyses of the
KLOE data sample has just started. The main difference with
respect to the $K_L$ analysis is
the absence of the ambiguity on the lepton charge assignment due to the 
unique charge of the \kpm\ decay
product. Moreover, due to the
$\pi^0$ presence, for $K^\pm\mu3$ there is not the $\pi-\mu$ ambiguity  
which is 
a limiting systematic factor on the $K_L\mu3$ FF measurement.
All new FF measurements will use dispersive ap\-pro\-a\-ches to fit the data.

The Callan-Treiman relation fixes the
value of scalar form factor at $t_{CT}$
to be equal to  
the $(f_K/f_\pi)/f_+(0)$ ratio as in Eq.(\ref{eq:CTrel}). 
The dispersive parametrization for the scalar
form factor proposed in Ref. \cite{Bernard:2007cf} allows 
a precise determination of $f_K/f_\pi/f_+(0)$ from the 
available measurements 
of the scalar form factor. 
This experimental evaluation is completely independent from lattice estimates. 
\begin{figure}[t]
\centering
\includegraphics[width=0.4\textwidth]{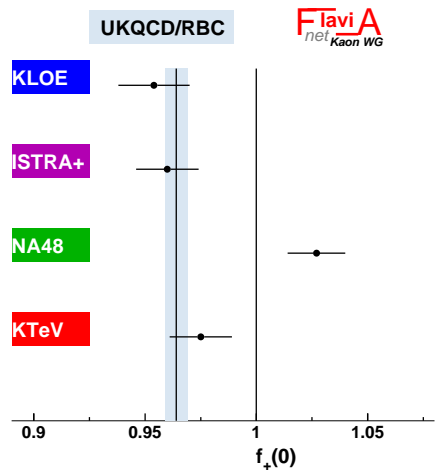}
\vspace{-3mm}
\caption{Values of $f_+(0)$ determined from 
experimental measurement of the scalar FF slope 
compared with the lattice calculations from the UKQCD/RBC collaboration.
\label{fig:CTtest} }
\end{figure}
Figure~\ref{fig:CTtest} shows the 
experimental results on $f_+(0)$ 
from the scalar FF dispersive
measurements, using Callan-Treiman relation and
$f_K/f_\pi=1.189(7)$. 
The value of $f_+(0)=0.964(5)$ from lattice UKQCD/RBC is also shown 
for comparison.
The NA48 result is difficult to accommodate  
and it violates the theoretical bound $f_+(0)<1$ 
pointing to the
presence of 
right-handed currents from physics beyond the SM.
The fit probability of the world $\tilde{f}_0(t)$ data (excluding NA48) 
is 39\% 
and the average is in good agreement with lattice calculations. 
Improvements on the experimental determination of the ratio  
$f_K/f_\pi/f_+(0)$ 
can be very effective in constraining right-handed currents 
~\cite{Bernard:2007cf} and also for the validation of more precise 
lattice calculations.
\subsubsection{Kaon form factors from $\tau$ decays}
\label{sec:tauVus}

To complete the discussion on the determination of kaon semileptonic 
FFs we present the status of the analysis 
of $\tau\rightarrow K\pi\nu_\tau$ decays.
The recent precise results 
by BaBar \cite{Aubert:2007jh} and Belle \cite{Epifanov:2007rf} 
provide complementary, higher-energy measurements of the 
kaon form factors.
Indeed, the differential decay distribution,
$$
 \frac{\rd\Gamma_{K\pi}}{\rd\sqrt{s}} \,=\, \frac{G_F^2|V_{us}|^2 m_\tau^3}
{32\pi^3s}\,S_{\mbox{\tiny EW}}\left(1-\frac{s}{m_\tau^2}\right)^{\!2}
\times
$$
\begin{equation}
\left[
\left( 1+2\,\frac{s}{m_\tau^2}\right) q_{K\pi}^3\,|f_+(s)|^2 +
\frac{3\Delta_{K\pi}^2}{4s}\,q_{K\pi}|f_0(s)|^2 \right], 
\label{dGamma}
\end{equation}
is expressed in terms of $f_+(s)$, with $s  = (p_K + p_\pi )^2$, 
the $K \pi$ vector form factor, and $f_0(s)$, the scalar FF. 
$S_{\mbox{\tiny EW}}$ is an 
electroweak correction factor, $\Delta_{K\pi}=m_K^2-m_\pi^2$,
and $q_{K\pi}$ is the absolute value of the momentum in
the $K \pi $ center of mass reference frame,
$$
q_{K\pi}(s) \,=\, \frac{1}{2\sqrt{s}}\sqrt{\Big(s-(m_K+m_\pi)^2\Big)
\Big(s-(m_K-m_\pi)^2\Big)}
$$
\begin{equation}
 \times \theta\Big(s-(m_K+m_\pi)^2\Big)\ .
\label{kinvff}
\end{equation}

A fit to the measured decay distribution is performed to determine the FFs.
Several FF parametrizations have been proposed
\cite{Moussallam:2007qc,Jamin:2006tk,Jamin:2008qg,Boito:2008fq}. 
The form factors are obtained in the framework of 
chiral theory with resonances imposing additional constraints 
from dispersion relations 
\cite{Jamin:2006tk,Jamin:2008qg,Boito:2008fq}, 
whereas in Ref. \cite{Moussallam:2007qc} a coupled channel analysis 
has been performed taking into
account, through analyticity requirements, 
the experimental information on elastic and inelastic
$K \pi$ scattering from the LASS collaboration \cite{Aston:1987ir}. 

The parametrizations relying 
on dispersion relations 
guarantee that the form factors fulfil the 
analyticity and unitarity properties. 
In this framework,  
a vector FF description 
as model independent as possible 
through a three-times-subtracted dispersive parametrization 
has been proposed in Ref. \cite{Boito:2008fq},
\begin{equation}
\tilde{f}_+(s)=\exp\left[a_1\frac{s}{m_{\pi^-}^2}+ a_2\left(\frac{s}{m_{\pi^-}^2}\right)^2
+a_3(s) \right]
\label{Dispfp}
\end{equation}
with
$$
a_1 = \lambda_+',\quad  
 a_2 = \frac{1}{2}\left(\lambda_+''-\lambda_+'^2\right),
$$
and
$$
a_3(s)= \frac{s^3}{\pi} \int_{(m_K + m_\pi)^2}^{\infty} ds'\frac{\delta_1^{K\pi}(s')}{(s')^3(s'-s-i\epsilon)}.
$$
Slope and curvature of the form factor, $\lambda_+'$ and $\lambda_+''$,  
are obtained 
from a fit. The phase, $\delta_1^{K\pi}$, considered elastic 
and therefore equal to the $P$-wave elastic $I = 1/2$ $K\pi$ 
scattering phase shift, 
is determined by a model relying on the inclusion 
of two resonances, K*(892) and K*(1410) whose 
parameters are also fitted from data. 
Due to the presence of the K*(892) resonance which 
dominates the decay distribution between 0.8 and 1.2 GeV, 
the vector FF 
dominates this decay except for the 
threshold region 
where the scalar form factor plays an important role. 
With the present accuracy the scalar FF cannot be 
determined from data and is
thus taken as an input to the analysis.   
The values of $\lambda'_+$ and $\lambda''_+$ in Tab. \ref{resultsVFF} 
have been obtained in  Ref. \cite{Boito:2008fq} from a fit to the 
Belle data.
These are compared with the results from 
others 
pa\-ra\-me\-tri\-za\-tions \cite{Moussallam:2007qc,Jamin:2006tk,Jamin:2008qg} 
and the measurements from $K_{e3}$ 
analyses \cite{Antonelli:2008jg,Abouzaid:2009ry}. 
\begin{table}[h!!]
\caption{Recent results on the slope $\lambda_+^{'}$ and the
curvature $\lambda_+^{''}$ of the vector form factor in units
of $10^{-3}$ from the analysis of $\tau \rightarrow K \pi \nu_\tau$  
\cite{Boito:2008fq,Moussallam:2007qc,Jamin:2006tk,Jamin:2008qg} 
and $K_{e3}$ decays \cite{Antonelli:2008jg,Abouzaid:2009ry}.
\label{resultsVFF}}
\begin{center}
\vskip 2mm
\begin{tabular}{ccc}
\hline \hline
Ref.                & $\lambda_+^{'}\,[10^{-3}]$    & $\lambda_+^{''}\,[10^{-3}]$ \\ \hline
\cite{Boito:2008fq}   & $\qquad 24.66 \pm 0.77\qquad$ & $\quad 1.199 \pm 0.020\quad$ \\
\cite{Moussallam:2007qc}     & $26.05^{+0.21}_{-0.58}$       & $1.29^{+0.01}_{-0.04}$ \\
\cite{Jamin:2006tk,Jamin:2008qg}   & $25.20 \pm 0.33$              & $1.285 \pm 0.031$ \\
\cite{Antonelli:2008jg}& $25.2 \pm 0.9$                & $1.6 \pm 0.4$ \\
\cite{Abouzaid:2009ry}     & $25.17 \pm 0.58$              & $ 1.22 \pm 0.10$ \\
\hline
\hline
\end{tabular}
\end{center}
\end{table}
The different precision results on  $\lambda_+^{'}$  
are in good agreement. 
The curvature  $\lambda_+^{''}$ 
from the average of recent $K_{e 3}$ 
measurements \cite{Antonelli:2008jg},
using a quadratic Taylor pa\-ra\-me\-tri\-za\-tion 
has 25\% uncertainty 
reduced to 8\% 
by the dispersive pa\-ra\-me\-tri\-za\-tion of the FF~\cite{Abouzaid:2009ry}. 

The knowledge of the t-dependence of vector FF and thus 
the $K_{e 3}$ phase space integral in Eq.(\ref{eq:gammakl3})
can be improved by combining  $\tau$ and $K_{e 3}$ results. 
The potentiality of such a combined analysis 
has been presented in Ref. \cite{Boito:2009pv}
leading to the following results: 
\begin{equation}
\lambda_+'=(25.10\pm 0.44)\times 10^{-3}\mbox{,}\;
\lambda_+''=(1.213\pm 0.021)\times 10^{-3}\mbox{.}
\end{equation}
The uncertainty on $\lambda_+'$ is 
mainly driven 
by statistics thus offering good prospects for more accurate results and  
a better determination of the phase space integral.
The correlation between slope and curvature imposed by 
the dispersion relation 
 \cite{Bernard:2009zm} 
used in the $K_{\ell 3}$ dispersive analyses 
~\cite{Abouzaid:2009ry,Lai:2007dx,Ambrosino:2007yza} 
will also be tested.

Moreover, the combination of $\tau$, $K_{e 3}$ and $K_{\mu 3}$ 
measurements using the dispersive representation 
as in Eq.(\ref{Dispfp})
and a 
double-subtracted dispersion relation for the normalized scalar form 
factor \cite{Bernard:2006gy,Bernard:2009zm},
could also lead to an improvement in the scalar form factor description.
In particular, a more precise determination of the free parameter $ln C$, 
where $C=\bar{f}_0(\Delta_{K\pi})$, is of interest 
for testing the SM 
and the presence of right-handed quark or scalar currents 
\cite{Antonelli:2008jg,Bernard:2006gy,Bernard:2009zm}. 
\subsubsection{Unitarity and universality}
\label{sec:unituniv}

To  extract \fVus\ we use Eq.(\ref{eq:gammakl3}) 
together with the $SU(2)$-breaking 
\cite{Cirigliano:2004pv,Cirigliano:2001mk,Andre:2004tk,Cirigliano:2008wn} and 
long-distance $EM$ corrections 
to the radiation-inclusive decay rate 
\cite{Cirigliano:2004pv,Cirigliano:2001mk,Andre:2004tk,Cirigliano:2008wn,DescotesGenon:2005pw}.
The measured values of \fVus\ from KLOE are shown in Tab. \ref{tab:vusf}. 
\begin{table}[!h]
\begin{center}
\caption{\label{tab:vusf}Values of \fVus\  extracted from \kltre
\ decay rates.}
\vskip 3mm
\begin{tabular}{c c c c c} 
\hline
\hspace{-1mm}$K_Le3$&\hspace{-1mm}$K_L\mu3$&\hspace{-1mm}$K_Se3$&\hspace{-1mm}$K^{\pm}e3$&\hspace{-1mm}$K^{\pm}\mu3$\\ \hline
\hspace{-1mm}0.2155(7)&\hspace{-1mm}0.2167(9)&\hspace{-1mm}0.2153(14)&\hspace{-1mm}0.2152(13)&\hspace{-1mm}0.2132(15) \\ \hline
\end{tabular}
\end{center}
\vspace{-0.4cm}
\end{table}
The five decay modes agree well within the errors and average 
to \fVus\ = 0.2157$\pm0.0006$, with 
$\chi^2/ndf =7.0/4$ (Prob=13\%).
The 0.28\% accuracy of this result has to be compared with the 0.23\% 
of the world average 
\fVus\  = $0.2166 \pm 0.0005$ \cite{Ambrosino:2008ct}.
Significant lepton-universality tests are provided by the comparison of 
the results from different leptonic channels. 
Defining the ratio $r_{\mu e}=\fVus_{\mu3}^2/\fVus_{e3}^2$
and using Eq.(\ref{eq:gammakl3}), we have $r_{\mu e} = g_{\mu}^2/ g_{e}^2$,
with $g_{\ell}$ the coupling strength at the $W \to \ell \nu$ vertex.
Lepton universality can be then tested comparing the measured value 
of $r_{\mu e}$ with the SM prediction
$r_{\mu e}^{SM}=1$. Averaging charged- and neutral-kaon modes, 
we obtain $r_{\mu e} =1.000(8)$, to be compared with the results  
from leptonic pion decays,
$(r_{\mu e})_{\pi} =1.0042(33)$ \cite{RamseyMusolf:2007yb}, and from 
leptonic $\tau$ decays 
$(r_{\mu e})_{\tau} = 1.0005(41)$ \cite{Davier:2005xq}. Using the world average
of all kaon measurements, $r_{\mu e} = 1.004(4)$ \cite{Antonelli:2008jg} 
has the same
precision than $\tau$ decays. 
\par
Using the determination of \fVus\  from \kltre\ decays and 
the value $\fzero = 0.964(5)$ from \cite{Boyle:2007qe}, 
we get $\Vus = 0.2237(13)$.
 
Furthermore, a measurement of $\Vus/\Vud$ can be obtained from the  
comparison of the radiation-inclusive decay 
rates of $\kpm \rightarrow \mudue (\gamma)$ 
and $\pipm \rightarrow \mudue (\gamma)$, combined 
with lattice calculation of $f_K/f_{\pi}$ \cite{Marciano:2004uf}.
Using BR$(\kpm \rightarrow \mudue) = 0.6366(17)$ 
from KLOE \cite{Ambrosino:2005fw}
 and the lattice result $f_K/f_\pi = 1.189(7)$ 
from the HP/UKQCD '07 \cite{Follana:2007uv},
we get $\Vus/\Vud = 0.2323(15)$. 
This value can be used in a fit together with the measurements of
\Vus\ from \kltre\ decays and $\Vud = 0.97418(26)$ \cite{Towner:2007np} 
from superallowed nuclear $\beta$ decays.
The result of this fit is $\Vud = 0.97417(26)$ and $\Vus = 0.2249(10)$, with 
$\chi^2/ndf =2.34/1$ (Prob$ = 13\%$), from which we get $1-(\Vud^2 +
\Vus^2+\vub^2)=4(7)\times10^{-4}$ compatible with unitarity 
at 0.6--$\sigma$ level. 
Using these results, we evaluate 
$G_{\rm CKM}=G_{\rm F} (\Vud^2 + \Vus^2+\vub^2)^{1/2}=(1.16614\pm0.00040)\times 10^{-5}$ GeV$^{-2}$ 
which is in perfect agreement with the measurement from the muon
lifetime 
$G_{\rm F}=(1.166371\pm0.000006)\times 10^{-5}$ GeV$^{-2}$ 
\cite{Marciano:1999ih}.
At present, the sensitivity of the quark--lepton 
universality test 
through the $G_{\rm CKM}$ measurement is competitive and even 
better than  
the measurements from $\tau$ decays and the electroweak precision 
tests \cite{Marciano:2007zz}.  
Thus unitarity can also be interpreted as a test of 
the universality of lepton and quark weak couplings to the $W$ boson,  
allowing bounds to be set on extensions of the SM leading to some kind of 
universality breaking.
For instance, the existence of additional $Z^\prime$ gauge bosons, 
giving different loop-contributions to muon and 
semileptonic decays, can break gauge universality
\cite{Marciano:1987ja}. 
The measurement of $G_{\rm CKM}$ can set constraints on 
the $Z^\prime$ mass which
are competitive with direct search at the colliders. 
When considering supersymmetric extensions, differences between muon 
and semileptonic decays can arise in the loop 
contributions from SUSY 
particles~\cite{Hagiwara:1995fx,Kurylov:2001zx}.  
The slepton-squark mass difference could be investigated improving 
present accuracy on the unitarity relation by a factor of $\sim$2-3. 
\subsubsection {$K_{\ell 2}$  and $K_{\ell 3}$  beyond the SM}
\label{sec:Ke2Kl3}
\par
The ratio of the \Vus\ values
obtained from helicity-suppressed and helicity-allowed kaon modes,
\begin{equation}
R_{\ell23} = \frac{\Vus(K_{\ell2})}{\Vus(K_{\ell3})} \times 
\frac{V_{ud}(0^+\to0^+)}{V_{ud}(\pi_{\mu2})}
\label{eq:busratio}
\end{equation}
is equal to unity in the SM.
The presence of a  scalar current due to a charged Higgs $H^+$ exchange 
is expected to lower
the value of $R_{\ell23}$\cite{Isidori:2006pk}:
\begin{equation}
R_{\ell23} = \left| 1 - \frac{m^2_{K^+}}{m^2_{H^+}}\left( 1 -  \frac{m^2_{\pi^+}}{m^2_{K^+}}\right)
\frac{\tan^2\beta}{1+0.01 \tan\beta} \right|,
\label{eq:rl23}
\end{equation}
with $\tan \beta$ the ratio of the two Higgs vacuum expectation values 
in the Minimal Supersymmetric Standar Model, MSSM.
\begin{figure}[!htb]
\centering
\includegraphics[width=0.4\textwidth]{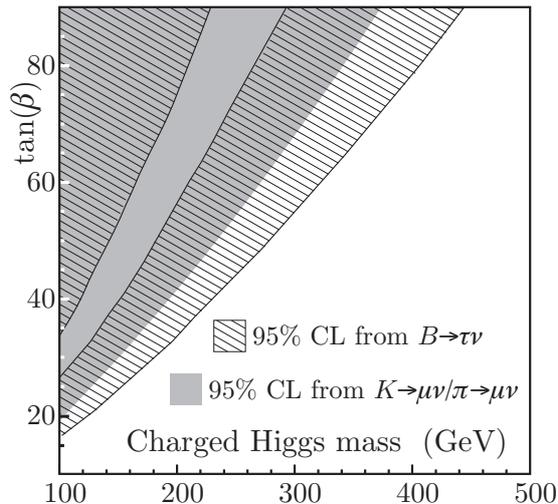}
\caption{\label{fig:fig1} Excluded region in the $m_{H^\pm} - \tan\beta$
  plane from the $R_{\ell23}$ as measured by KLOE. The region excluded 
by $B \rightarrow \tau \nu$ is indicated with cross-hatching.}
\end{figure} 
The experimental data on $K_{\mu2}$ and $K_{\ell3}$ decays are fit 
in order to evaluate $R_{\ell23}$, using as external inputs the most 
recent lattice determinations of \fzero\ \cite{Boyle:2007qe} 
and $f_K/f_\pi$ \cite{Follana:2007uv},
the value of \Vud\ from \cite{Towner:2007np}, 
and $\Vud^2 + {\ensuremath{|V_{us}(K_{l3})|^2}}= 1$ as a constraint.
The result obtained with KLOE data is $R_{\ell23} = 1.008 \pm
0.008$ \cite{Ambrosino:2008ct}.
From this measurement 
bounds can be set on the values of 
the charged Higgs mass and $\tan \beta$ 
in a region not reachable
with the present measurements of the  
BR$(B\rightarrow \tau \nu)$ \cite{Ikado:2006un}. 
Figure \ref{fig:fig1} shows the region excluded at 95\% CL in the
charged Higgs mass $m_{H^\pm}$ and $\tan\beta$ plane together with
the bounds from BR$(B \rightarrow \tau \nu)$.
The measurement of $R_{\ell23}$ compatible with the SM prediction gives
stringent bounds on the presence of charged right-handed currents
appearing for instance in some Higgsless low-energy effective 
 theories \cite{Bernard:2009zm}.
\subsubsection {Test of lepton-flavor conservation: $R_K$}
\label{sect:lfv}
\begin{figure}[!htb]
\centering
\includegraphics[width=0.4\textwidth]{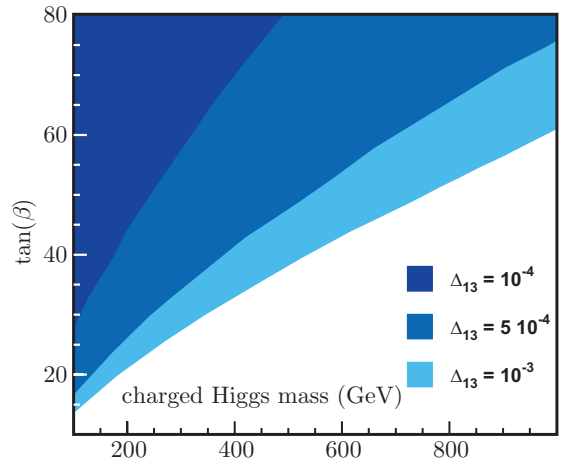}
\caption{\label{fig:fig1a} Excluded region at 95\% C.L. in the 
$m_{H^\pm} - \tan\beta$
  plane from the $R_{\rm K}$ as measured by KLOE,  
for $\Delta_{13}$=$10^{-3}\text{, }~  0.5\times10^{-3}\text{, }~  
 10^{-4}$.}
\end{figure}

KLOE performed a comprehensive study of the process $\Gamma(K\to e\nu(\gamma))$
based on the complete data set \cite{Ambrosino:2009rv}. 
The ratio of $\Gamma(K\to e\nu(\gamma))$ and
$\Gamma(K\to \mu\nu(\gamma))$ decay widths has been measured for photon
energies smaller than 10 \MeV, without photon detection 
requirement,  
$R_{10}$=$(2.333\pm0.024_\mathrm{stat}\pm0.019_\mathrm{stat})\times 10^{-5}$.
The systematic fractional error of $\sim0.8\%$, to be compared
with the statistical accuracy of $1\%$, is dominated 
by the statistics of the 
control-sample used (0.6\% contribution)  
thus making possible the improvement of the results with larger 
data samples.
The radiation-inclusive ratio  
R$_K$=(2.493$\pm$0.025$_\mathrm{stat}\pm0.019_\mathrm{syst})\times 10^{-5}$ 
has been derived,
in excellent agreement with the SM prediction.
This result improves the accuracy on $R_K$ by a factor of five 
with respect to the
previous world average, putting severe constraints 
on new physics contributions from MSSM 
with lepton-flavor violating couplings.
The excluded region in the $\tan\beta$\--$m_{H^\pm}$ plane   
as a function of the 1-3 slepton-mass matrix element $\Delta_{13}$ 
is shown in
Fig. \ref{fig:fig1a}.
Recently, the NA62 collaboration presented a preliminary 
result \cite{Goudzovski:2009me}, 
$R_K=(2.500\pm0.012_\mathrm{stat}\pm0.011_\mathrm{syst})\times 10^{-5}$, 
obtained with 40\%
of the sample collected with two dedicated runs at CERN in 2007 and 2008.
\par
The new world average is 
$R_K=(2.498\pm0.014)\times 10^{-5}$, with an accuracy of 0.6\%.
\par
KLOE-2 with 25 fb$^{-1}$ of integrated luminosity  
can reach 0.4\% accuracy on R$_K$, a noticeable step forward for 
the lepton-flavor conservation test. 
\par
In the mid--term, the best sensitivity on  R$_K$   
would be met 
by the NA62 experiment with the analysis   
and a severe control over the systematics    
of the run dedicated to the measurement of the   
BR($K^\pm \to \pi^\pm \nu \bar{\nu}$). 

\subsection{KLOE-2 prospects on \fVus}
\label{sec:future}
KLOE-2 can significantly improve 
the accuracy on the measurement of \kl, \kpm\ lifetimes and \ksetre\
branching ratio with respect to present world average 
\cite{Antonelli:2008jg} 
with data from the first year 
of data taking, at KLOE-2/step-0.
\begin{table}[!htb]
\begin{center}
\caption{\label{tab:vusf_future}KLOE-2/step-0 prospects on \fVus\
  extracted from \kltre\ decay rates.}
\vskip 3mm
\begin{tabular}{|c|c|c|c|c|c|c|} 
\hline
Mode & $\delta\fVus\:$(\%) & $\BR$ & $\tau$ & $\delta$ & I$_{Kl}$ \\ \hline
$K_Le3$      & 0.21 & 0.09 & 0.13 & 0.11 & 0.09 \\ \hline
$K_L\mu3$    & 0.25 & 0.10 & 0.13 & 0.11 & 0.15 \\ \hline
$K_Se3$      & 0.33 & 0.30 & 0.03 & 0.11 & 0.09 \\ \hline
$K^{\pm}e3$  & 0.37 & 0.25 & 0.05 & 0.25 & 0.09 \\ \hline
$K^{\pm}\mu3$& 0.40 & 0.27 & 0.05 & 0.25 & 0.15 \\ \hline
\end{tabular}
\end{center}
\vspace{-0.4cm}
\end{table}
The present 0.23\% fractional uncertainty on \fVus\ 
can be reduced to 0.14\% 
using KLOE present data set together with the KLOE-2/step-0 statistics. 
The world-average uncertainties on phase space integrals 
and \kl\ semileptonic BRs \cite{Antonelli:2008jg} have been used  
in Tab. \ref{tab:vusf_future} to summarize  
the expected accuracy on \fVus\ 
for each decay mode and with the contributions 
from branching ratio, lifetime, SU(2)-breaking
and long--distance $EM$ corrections, and phase space integral.
Statistical uncertainties on the measurement of BRs 
and lifetimes have been obtained
scaling to the total sample of 7.5 fb$^{-1}$ of integrated luminosity 
available at the completion of KLOE-2/step-0. 
The estimate of systematic errors is rather conservative, being based   
on KLOE published analyses  
without 
including 
any improvement from the detector upgrade.
\par \noindent
Lattice QCD simulations have made tremendous progress in the last few years
and in near-term the accuracy on \fzero\
could reach the 0.1\% level thus allowing \Vus\ to be determined 
from Tab. \ref{tab:vusf_future} with 0.17\%
relative error. At this level of precision the accuracy on \Vud\, 
presently at 0.026\% \cite{Antonelli:2009ws} and dominated by the
uncertainty on radiative corrections, becomes the limiting factor in
testing CKM unitarity.
With 0.02\% accuracy on \Vud\ and with \Vus\ measured at 0.17\%, 
the sensitivity on
the unitarity relation 
would improve by a factor of two reaching the level of a few 10$^{-4}$:   
a significant opportunity  
to investigate beyond--SM models with gauge universality 
breaking 
 \cite{Marciano:1987ja,Hagiwara:1995fx,Kurylov:2001zx,Cirigliano:2009wk}.

\par

\section{CPT Symmetry and Quantum Mechanics}
\label{sec:symm}
%
\label{sec:neutk}
The $CPT$ theorem  
\cite{Lueders:1992dq,Pauli:1988xn,Bell:1955sd,Jost:1957zz}
proves the 
invariance of physics under 
combination of the discrete 
trasformations,  
$C$ (charge conjugation), 
$P$ (parity), and $T$ (time reversal).
The theorem 
holds for any quantum field theory formulated on flat space-time
 assuming i) Lorentz invariance, ii) locality, and iii) unitarity  
(conservation of probability).

Experimental tests of $CPT$ invariance therefore probe
the most fundamental
assumptions of our present understanding of particles 
and their interactions.

The sensitivity of the neutral kaon system to 
a variety of $CPT$-violating effects makes it one of the best candidates
for probing $CPT$ invariance. 

A unique feature of the $\phi$-factory is the production of neutral kaon 
pairs in a pure quantum state so that we can 
study quantum interference effects and  
tag pure monochromatic 
\ks\ and \kl\ beams. 
The neutral kaon doublet is one of the most intriguing systems in nature.
During its time evolution a neutral kaon oscillates 
to--and--fro 
between particle and 
antiparticle states with a beat frequency 
$\Delta m 
\approx 5.3 \times 10^9 
\hbox{ s}^{-1}$, where $\Delta m$ 
is the mass difference between 
\kl\ and \ks\ .  
%
%
%
The observation of a rich variety 
  of interference phenomena in the time evolution and decay of neutral kaons 
  is then possible thanks to the special circumstance of  
  a $\Delta m$ value that is half of the \ks\ decay width. 
\par
The $CPT$ violation 
could manifest for instance 
in conjunction with tiny 
modifications of the initial correlation of the kaon doublet, with  
decoherence effects 
or Lorentz symmetry violations, which in turn might be 
justified in the context of a quantum theory of gravity.
 
At KLOE-2  
the sensitivity to some observables can meet the
level of the Planck scale,
$\mathcal {O} (m^2_K/M_{Planck}) \sim 2 \times 10^{-20} \,\mbox{GeV}$,
unreachable in other similar systems,  
such as B mesons.
As a figure of merit, 
the fractional mass 
difference $\left({m_{\kn}-m_{\knb}}\right)/{m_{\kn}}$ can be 
measured 
to $\mathcal{O}(10^{-18})$ 
while in the neutral B system experimental accuracy is of the order 
of 
$10^{-14}$, 
and the proton--antiproton mass difference provide a limit at  
the $10^{-9}$ level 
\cite{Amsler:2008zzb}. 

Besides measurements of most of the kaon parameters, the entangled 
neutral kaons at the $\phi$-factory are also suitable to 
perform several tests of the foundations of QM, 
such as state coherence over macroscopic distances, 
Bohr's complementarity principle, 
and quantum-erasure and quantum-marking concepts, 
as explained in 
Sects. \ref{sec:DSQMdeco}, \ref{sec:DSQMeraser}\--\ref{kaonicQE}.
 
%
%
\newtheorem{thm}{Theorem}
\newtheorem{lem}{Lemma}
\subsection{The neutral kaon system at the $\phi$ factory}
\label{sec:DSQMexp}
%
%
A pure two-kaon state is produced from 
$\phi$ decays:
\begin{eqnarray}|\psi^-(t)\rangle&=&\frac{1}{\sqrt{2}}\lbrace
|K^0\rangle\otimes|\bar K^0\rangle-|\bar K^0\rangle\otimes|K^0\rangle\rbrace\nonumber\\
&=&\frac{N_{SL}}{\sqrt{2}}\lbrace
|K_S\rangle\otimes|K_L\rangle-|K_S\rangle\otimes|K_L\rangle\;,
\label{eq:state}\end{eqnarray}
with  
$N_{SL}={\sqrt{(1+|\epsilon_S|^2)(1+|\epsilon_L|^2)}}/{(1-\epsilon_S\epsilon_L)} \simeq 1$ and $\epsilon_S=\epsilon+\delta$,  $\epsilon_L=\epsilon-\delta$, 
where $\epsilon$ and $\delta$ denote the $CP$- and $CPT$-violation 
parameters in the mixing. 
The state of Eq.(\ref{eq:state}) is antisymmetric with respect to permutations 
of kaons, and exhibits maximal entanglement. 
decay mechanism into a Lindblad operator of a master equation.
The maximal entanglement of kaon pairs as in 
Eq.(\ref{eq:state}) has been  
observed in the 
$\phi\rightarrow \ks\kl \rightarrow 
\pi^+\pi^- \pi^+\pi^-$\cite{Ambrosino:2006vr} by the KLOE collaboration in year 2005.
 Since then, more data (a total of 1.7 fb$^{-1}$) and improvements 
  in the analysis procedure have brought to the results on several  
  decoherence and $CPT$-violating parameters 
\cite{DiDomenico:2009xw} that will be presented in this section.

\begin{figure}[!h]
\begin{center}
\resizebox{0.85\columnwidth}{!}{
\includegraphics{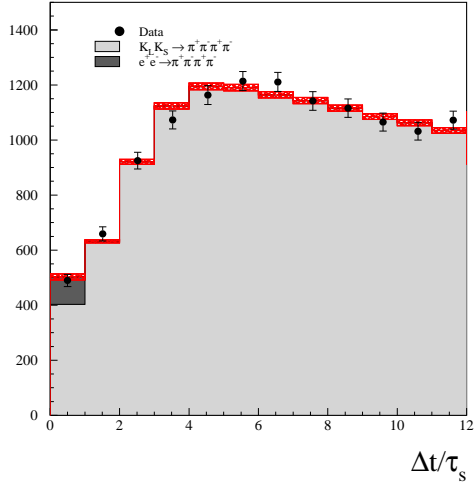}
}
\caption{Number of events as a fuction of the difference in the decay time 
of the two $\pi^+ \pi^-$ vertices, in $\tau_S$ units. Bin width is 1. 
The measured $I(\pi^+\pi^-,\pi^+\pi^-; \Delta t)$ distribution is 
fitted: 
black points with errors are data and 
the fit result is shown by the histogram.
}
\label{fig:fit_zsl}
\end{center}
\end{figure}

The selection of the signal requires two vertices,
each with two opposite-curvature tracks inside the drift chamber, 
with an invariant mass and total momentum
compatible with the two neutral kaon decays.
The resolution on $\Delta t$, 
the absolute value of the time
difference of the two $\pi^+\pi^-$ decays, 
benefits of the 
precision momentum measurements   
and of the kinematic closure of the 
events. 
The  $\Delta t$ distribution
can be fitted with the function:
\begin{eqnarray}
\label{eq:deckloe}
I(\pi^+\pi^-,\pi^+\pi^-;\Delta t)\propto 
e^{-\Gamma_L \Delta t}+
e^{-\Gamma_S \Delta t}
\nonumber \\
-2 
e^{-{{(\Gamma_S+\Gamma_L)}\over{2}}\Delta t}\cos( \Delta m \Delta t)~,
\end{eqnarray}
evaluated from QM, vanishing at $\Delta t =0$ \cite{Bertlmann:2002wv} 
for the complete antisymmetry of the state of Eq.(\ref{eq:state}).
The fit of the $\Delta t$  distribution is performed taking into account  
  resolution and detection efficiency, the background from coherent 
  and incoherent \ks\--regeneration on the beam pipe wall, and  
  the small contamination from the non-resonant 
$e^+e^-\rightarrow\pi^+\pi^-\pi^+\pi^-$ channel.
  The result obtained fixing $\Delta m$, $\Gamma_S$ and $\Gamma_L$
to the Particle Data Group (PDG) values \cite{Amsler:2008zzb},
  is shown in Fig. \ref{fig:fit_zsl}.
%
%
%


%
\par
Different hypotheses on decoherence and $CPT$-violating phenomena are 
  expressed by different modifications of the function of 
Eq.(\ref{eq:deckloe}).  
  The modified expressions have been then used to obtain the best 
values of the QM- and $CPT$-violating parameters 
  ($\zeta_{K_S K_L}$, $\zeta_{K^0 \bar K^0}$,
$\gamma$, $\Re\omega$, $\Im\omega$, $\Delta a_X$, $\Delta a_Y$, 
$\Delta a_Z$) presented in the following subsections.  

\subsection{Test of quantum coherence}
\label{sec:DSQMdeco}
The kaon states 
evolve exponentially in time according to the Wigner--Weisskopf approximation
with an effective Hamiltonian consisting of the Hermitean and anti-Hermitean 
components.
Time evolution can be also described using a Hermitean Hamiltonian and 
the master equation in the 
Kossakowski-Lindblad form of Ref. \cite{Bertlmann:2006fn}.

Due to conservation of the quantum numbers, it is safe to assume
that the initial state is indeed the maximally entangled state given
by Eq.(\ref{eq:state}). 
However, it is by no means proven that the assumed quantum mechanical time
evolution (the Wigner--Weisskopf approximation) is the correct one. 
It may well be that the kaon pair 
undergoes decoherence effects, 
i.e., the initially pure state would become mixed 
for $t>0$. 


Decoherence 
may result from fundamental modifications of QM and may be traced
back to the influence of quantum gravity~
\cite{Penrose:1998dg,Mavromatos:2007xb}
---quantum fluctuations in the space--time structure on the Planck
mass scale--- or to dynamical state reductions
\cite{Ghirardi:1985mt,Ghirardi:1987nr}. 

Thus 
from the experimental study of the decoherence  
we can answer to:
\begin{itemize}
\item How accurately is the QM 
interference term verified by the experimental data?
\item Is the purity of the two--kaon state maintained for times $t>0$ ?
\end{itemize}

The general formalism of open quantum systems, 
that means a master equation of type Lindblad \cite{Lindblad:1975ef} 
and Gorini--Kossakowski--Sudarshan 
\cite{Gorini:1975nb}, derived under rather general assumptions, 
allows the study of different scenarios of  
decoherence without 
modeling the environment explicitly. 
Such scenarios have been proposed and, interestingly, 
a direct relation between
the decoherence parameter and the loss of entanglement 
 has been established
~\cite{Bertlmann:2002wv}. The decoherence parameter depends on 
the basis used to describe time evolution of the kaon doublet.

A simpler but very effective approach 
to treat decoherence is based on an idea of
Schr\"odinger~\cite{Schrodinger:1935zz} and 
Furry~\cite{Furry:1936xx}, proposed already in 1935.
They raised the question whether an initially entangled state can
spontaneously factorize into product states, 
with a basis--de\-pen\-dent spontaneous factorization.
We can
quantify the effect by multiplying the QM 
interference term by a factor $(1-\zeta)$, where $\zeta$ represents the 
decoherence parameter \cite{Bertlmann:1996at,Bertlmann:1999np} 
and varies between 0$\leq\zeta\leq$1, the limits of QM and 
spontaneous factorization.

KLOE has provided the first experimental evidence for  
quantum interference in the $CP$--suppressed process 
$\phi\longrightarrow K_S K_L\longrightarrow\pi^+\pi^-\pi^+\pi^-$ 
obtaining for
the decoherence parameter $\zeta$ in $K_S K_L$ and $K^0 \bar{K^0}$ bases  
\cite{DiDomenico:2009xw,Ambrosino:2006vr}:
\begin{eqnarray}
\zeta_{K_S K_L}&=&0.003\pm 0.018_{stat}\pm 0.006_{sys}\\
\zeta_{K^0 \bar {K^0}}&=&(1.4\pm 9.5_{stat}\pm 3.8_{sys})\cdot
10^{-7}\;.
\end{eqnarray}
Since decoherence in the $K^0 \bar{K^{0}}$ basis would result in the
$CP$--allowed $K_S K_S \rightarrow \pi^+ \pi^- \pi^+ \pi^-$ decays,
the sensitivity for the parameter $\zeta_{K^0 \bar{K^0}}$ is
naturally much larger than for $\zeta_{K_S K_L}$.
The KLOE results represent a large improvement with respect to  
the analysis published in year 1998 \cite{Bertlmann:1999np}, based  
on the results of the 
CPLEAR experiment \cite{Apostolakis:1997td}:
$\zeta_{K_S K_L} = 0.13\pm 0.16$, 
$\zeta_{K^0 \bar{K^0}} = 0.4\pm 0.7$.

The upper limits 
can be definitely 
improved with KLOE-2,  
meeting   
the scale of quantum gravity and 
dynamical state reduction. 
The decoherence parameter $\zeta$ is a time--integrated value and 
in the framework of open quantum systems we have the 
following relation~\cite{Bertlmann:2002wv}:
\begin{eqnarray}
\zeta(\min\{t_l,t_r\})&=&1-e^{-\lambda \min\{t_l,t_r\}}\,,
\end{eqnarray}
where $\lambda$ is a 
constant 
representing the strength of the interaction with the environment. 
Since we are interested in the decoherence (or entanglement loss) 
of the two--kaon system, we consider time evolution of the 
total state up to the 
minimum of the two times $t_l,t_r$, when the measurements are performed 
on the left-- and right--hand side. After the first measurement,  
the two--kaon state collapses and the remaining one--kaon state 
evolves according to QM until it is measured too.

The 
parameter $\lambda$ is considered to be 
the fundamental one. A comparison with other meson systems is  
of great interest to understand how universal $\lambda$ is, how does it 
depend on the kind of environment and how does it typically scale 
with the mass of the system. 

The Belle collaboration, at the KEK 
B--factory in Japan, has analyzed 
data from the two--B-meson system 
obtaining, in the flavour basis 
\cite{Go:2007ww}, 
$\zeta_{B^0,\bar B^0}$ = $0.029\pm0.057$.

A precision measurement of the 
decoherence parameter can provide stringent upper bounds that in turn 
can 
rule out several sources of decoherence, using data from 
neutral mesons and also from photon, neutron, and molecular  
systems.

%
%
%
\subsection{$CPT$ violation from  quantization of space--time}
\label{sec:DSQMpheno}
%
%
%
The study of $CPT$ symmetry at KLOE-2 can contribute to the research 
  program of Quantum Gravity Phenomenology 
~\cite{AmelinoCamelia:2008qg}. 

The $CPT$ and Lorentz symmetry are to be placed under
scrutiny whenever 
quantization of the 
space-time is adopted. The simplest line of reasoning that
supports this thesis takes as starting point the duality
relation between the (smooth, classical, Riemannian) light-cone
structure of Minkowski space-time and the (classical, Lie-algebra)
structure of Poincare' symmetries. This duality between
space and its symmetry algebra is rather rigid and very sensitive
to the introduction of new elements in the structure of the space.
Galilei space-time and the Galilei algebra are linked by
an analogous duality relation, and  in going from Galilei  
to Minkowski space-time, thus adding the light-cone structure,
one must replace the Galilei symmetries with the Poincare' ones.
Similarly, we found that the additional space-time 
  elements introduced by different 
  proposals of the space-time quantization typically affect also 
the description 
  of the symmetries.  
And indeed over the last
decade or so 
the scenarios with broken 
  Poincare's symmetries 
~\cite{AmelinoCamelia:1997gz,Gambini:1998it,Aloisio:2000cm,AmelinoCamelia:2000zs} 
or those alternatively introducing 
  the novel concept of deformed Poincare's symmetries 
~\cite{AmelinoCamelia:2000mn,Magueijo:2002am,AmelinoCamelia:2003xp}, 
  have experienced a fast-growing popularity in the 
  quantum-gravity studies.



 These studies on Poincare's transformations have often  
  implications on the $CPT$ symmetry, producing scenarios with broken 
  $CPT$ invariance~\cite{AmelinoCamelia:1997em} or alternatively 
with a deformed concept of $CPT$  
  at the Planck-scale~\cite{AmelinoCamelia:1999pm,AmelinoCamelia:2007zzb}. 

Several tests of quantum-spacetime-inspired scenarios with broken 
  or deformed Poincare's symmetries have been proposed and 
are being performed by 
  astrophysics experiments 
~\cite{AmelinoCamelia:1997gz,Aloisio:2000cm,AmelinoCamelia:2000zs,AmelinoCamelia:1998ax,Jacobson:2002ye,Jacob:2006gn}.

The sensitivity reach of these studies is 
 very 
  promising although the control on possible competitive phenomena 
  is typically rather limited. As a result of our lack of knowledge,  
  the experimental bounds on the 
  parameters of interest are often conditional ~\cite{AmelinoCamelia:2008qg}
and rely on some 
  assumptions on the properties of the astrophysical systems.      
  For this reason the quantum gravity community still looks with 
  great interest at the results from controlled laboratory experiments 
  such as KLOE-2.



A key characteristic of the effects expected from spacetime quantization at the Planck scale
is the peculiar dependence of the effects on energy. For example, the most
studied effects of violation of classical Poincare' symmetry 
are propagation effects,
often codified with modifications of the dispersion relation: 
\begin{equation}
 m^2 \simeq E^2 - \vec{p}^2
+  \eta \vec{p}^2 {E^n \over E^n_{p}} + \dots
~,
\label{displeadbisNEW}
\end{equation}
where $E_p$ denotes the Planck scale, 
the parameter $\eta$ sets the magnitude of
the effect, and the integer $n$ (usually $n=1$ or $n=2$) fixes 
the dependence on energy
(and the Planck-scale) of the leading Planck-scale effect. 

The possibility to formulate these effects within effective field theories
is not always viable. 
Some aspects of the space-time noncommutativity introduce non-analytic
features that must be handled with care 
when attempting to reformulate quantum space-time
properties effectively as novel properties of an ordinary 
field theory in a classical
space-time. 
In cases where an effective-field-theory formulation appears to be viable
these effects
 end up requiring the introduction of 
non-renormalizable terms in the Lagrangian density,
such as dimension-5 and dimension-6 
operators (in the case of theories formulated in 4 space-time
dimensions). 

Since the whole quantum gravity problem is plagued by
objective difficulties at the level of perturbative renormalizability,
the quantum-gravity motivation for this type of studies
of space-time quantization conceptually renders 
formalization  
in terms of nonrenormalizable field theories
rather natural. 

On the other hand, this
nonrenormalizability still represents a serious
challenge for the efforts on the phenomenology side.
A study of the implications for KLOE-2 of these specific 
quantum-spacetime Poincare'-violation effects
is presently under way~\cite{AmelinoCamelia:2009inprep}.

The prospects to investigate with KLOE-2 some 
scenarios for $CPT$ symmetry emerging from quantum space-time
studies are even more interesting. 
While experimental tests of the Poincare' or Lorentz symmetry in astrophysics 
are rather powerful, the controlled, laboratory experiments 
are essential 
for the $CPT$ tests. 


 Since space-time quantization is expected to imply non-trivial effects 
  on the coherence of multiparticle states,   
  the tests of $CPT$ invariance at KLOE-2 based on the 
  coherent neutral kaon system from $\phi$ decay, are 
  particularly significant 
  from a quantum space-time perspective. 


In our current (pre-quantum-gravity) theories one obtains multiparticle states
from single-particle states by a standard use of the trivial tensor product
of Hilbert spaces, but there is theoretical evidence that this
recipe might not be applicable in a quantum-spacetime environment.

A valuable phenomenological formulation of this intuition has been 
 recently proposed in Ref. \cite{Bernabeu:2003ym}
and is presented 
in Sect. \ref{sec:DSQMCPT}.

  A robust derivation of the properties of multiparticle states within 
  specific quantum space-time proposals
  is still obscured by the mathematical complexity of the theoretical 
  framework. Nevertheless, 
recent preliminary results \cite{AmelinoCamelia:2007zzb} on theories 
  formulated in noncommutative space-time 
  are promising for providing in not-too-far future a 
valuable theory of the   
  multiparticle systems to be tested with KLOE-2.


%
\subsection{$CPT$ violation and decoherence}
\label{sec:DSQMCPT}
%
In this and following subsections we shall concentrate  
on the possible breakdown of the $CPT$ symmetry and 
in testing it 
with the entangled states of neutral kaons at the $\phi$-factory. 
In general, there are two ways by which $CPT$ breakdown is encountered 
in a quantum gravity model. 

The first is through the non--commutativity of a well--defined 
quantum mechanical $CPT$ operator 
with the Hamiltonian of the system.
This is 
the breakdown of $CPT$ symmetry dealt with in standard Lorentz-violating 
Extensions of the Standard 
Model (SME) \cite{Kostelecky:2008zz,Greenberg:2002uu,Greenberg:2003nv}.
 
In the second way,  
the $CPT$ operator is 
ill--defined as a quantum mechanical operator, but in a 
perturbative sense to be described below. 
This ill--defined operator is a consequence of the foamy 
structure~\cite{Wheeler:1998vs} of 
space-time, whereby the quantum metric fluctuations at Planck scales 
induce 
quantun decoherence of matter. 
The particle field theoretical system is 
an open quantum mechanical system interacting with the environment of 
quantum gravity. 
The reason for ill-defined $CPT$ operator in such cases 
is of more fundamental nature than the mere non--commutativity of 
this operator with the local effective Hamiltonian 
of the matter system in Lorentz-symmetry violating SME models. 
Quantum-gravity induced decoherence is in operation and the very concept 
of a local effective Lagrangian may itself break down. 
R.~Wald~\cite{Wald:1980nm} has elegantly argued, based on elementary 
quantum mechanical analysis of open systems,
that the $CPT$ operator 
is ill--defined 
for systems which exhibit quantum decoherence, 
that is they are characterised by an evolution of initially pure 
quantum mechanical states to mixed ones.
This was interpreted as a microscopic time arrow in quantum 
gravitational media which induce such decoherence. 
Hence such open 
systems are characterised by 
intrinsic $CPT$ violation, a terminology we shall use from now 
in order to describe this particular type of $CPT$ symmetry breakdown.

{\it Lindblad Parametrizations of Decoherent Evolution in Quantum Gravity:} 
in the past~\cite{Ellis:1983jz,Ellis:1995xd}, quantum gravity decoherence 
has been paramaterised in the neutral kaon system
by means of a Lindblad-type evolution of the reduced density matrix for 
kaons, 
 which satisfies 
energy conservation 
on average,  positivity of the density matrix 
and probability conservation. 
In its simplest form, taking into account
additional constraints relevant for neutral kaon physics 
(such as $\Delta S = \Delta Q$ rule), 
the relevant evolution operator for the density matrices 
of a single neutral kaon state is 
given by~\cite{Ellis:1983jz,Ellis:1995xd}:
$$\partial_t \rho = i[\rho, H] + \delta\H \rho~,$$
where {\small $H_{\alpha\beta}=\left( \begin{array}{cccc}  - \Gamma
& -\coeff{1}{2}\delta \Gamma
& -{\rm Im} \Gamma _{12} & -{\rm Re}\Gamma _{12} \\
 - \coeff{1}{2}\delta \Gamma
  & -\Gamma & - 2{\rm Re}M_{12}&  -2{\rm Im} M_{12} \\
 - {\rm Im} \Gamma_{12} &  2{\rm Re}M_{12} & -\Gamma & -\delta M    \\
 -{\rm Re}\Gamma _{12} & -2{\rm Im} M_{12} & \delta M   & -\Gamma
\end{array}\right)~$},
and {\small ${\delta\H}_{\alpha\beta} =\left( \begin{array}{cccc}
 0  &  0 & 0 & 0 \\
 0  &  0 & 0 & 0 \\
 0  &  0 & -2\alpha  & -2\beta \\
 0  &  0 & -2\beta & -2\gamma \end{array}\right)~.$}
Positivity of $\rho$ requires: $\alpha, \gamma  > 0,\quad
\alpha\gamma>\beta^2$. Notice that $\alpha,\beta,\gamma$ violate
both $CPT$, due to their decohering nature~\cite{Wald:1980nm}, and $CP$
symmetry.

However, as pointed out in
Ref. \cite{Benatti:1997rv}, although the above parametrization 
guarantees a positive--definite density matrix for single kaon states, 
this is 
not true for
entangled kaon states, 
as those produced  
at the $\phi$-factory.
For entangled states, the requirement of complete positivity 
implies different parametrizations for the foam effects.
For instance, if one chooses the above-mentioned  parametrization 
of Refs. \cite{Ellis:1983jz,Ellis:1995xd}, for consistency with the 
requirement of universal action of quantum gravity on both single-particle 
and entangled states, then in the latter case,
complete positivity is guaranteed only if
the further conditions
$\alpha = \gamma~ {\rm and} ~\beta = 0$
are imposed. The latest measurement of $\gamma$ 
from KLOE 
yields the 
result \cite{DiDomenico:2009xw,Ambrosino:2006vr}: 
\begin{eqnarray}
  \label{eq:kloegamma}
  \gamma &=& \left( 0.7 \pm 1.2_{\mbox{stat}} \pm 0.3_{\mbox{syst}} \right) \times 10^{-21}\, \mbox{GeV}~,
\end{eqnarray}
consistent with zero. Nevertheless, the experiment can measure 
all three decoherence parameters $\alpha,\beta, \gamma$ 
and hence one can continue using the full set~\cite{Huet:1994kr} 
when discussing neutral kaon effects at the $\phi$-factory. 
In this way one can provide a formalism for
testing experimentally
the assumption of complete positivity, which notably may not be a 
property of quantum gravity.

{\it Decoherence, Ill-defined CPT Operator and the $\omega$-effect in 
Entangled States:} in addition to the decoherence 
evolution parameters $\alpha,\beta,\gamma$, in entangled states 
one may have~\cite{Bernabeu:2003ym,Bernabeu:2005pm} some other effects 
associated with the ill-defined nature of the $CPT$ operator in 
decoherent models of quantum gravity.
As a result of the weak nature of 
quantum gravitational interactions, this ill-defined nature of $CPT$ operator 
is 
perturbative in the sense that the anti-particle state still exists, 
but its properties, 
which under normal circumstances would be connected by the action of 
this operator, are modified \cite{Bernabeu:2003ym,Bernabeu:2005pm}. 
The modifications can be perceived \cite{Bernabeu:2006av,Mavromatos:2008bz} 
as a result of the dressing of the (anti-)particle states by 
perturbative interactions expressing the effects of the medium.
\par \noindent
As argued in Refs. \cite{Bernabeu:2003ym,Bernabeu:2005pm}, 
the perturbatively ill-defined nature of the $CPT$ operator implies 
modified Einstein--Podolsky--Rosen (EPR) correlation among the 
entangled states in meson factories, which are uniquely associated 
with this effect, 
termed as  
$\omega$-effect.  

\begin{figure}[h]
\centering
\resizebox{0.8\columnwidth}{!}{
  \includegraphics{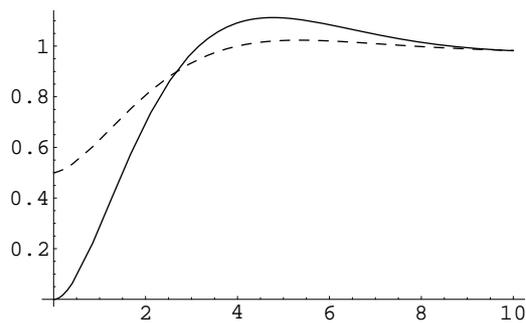}
}
\caption{The intensity
$I(\Delta t)$ in arbitrary units, 
with $\omega = |\omega |e^{i\Omega} = 0$ (solid line)  
and $\omega = |\eta_{+-}|e^{i(\phi_{+-} - 0.16\pi)}$ 
(dashed line)  
~\cite{Bernabeu:2003ym,Bernabeu:2005pm}.}
\label{intensomega}
\end{figure}

{\it Experimental Signatures of the $\omega$-effect in kaons at the $\phi$-  
factory:}  if $CPT$ is 
intrinsically violated,
in the sense of being not well--defined due to decoherence \cite{Wald:1980nm},
the neutral mesons $K^0$ and ${\overline K}^0$ should 
no longer be treated as    
particles-antiparticles states.
As a
consequence \cite{Bernabeu:2003ym,Bernabeu:2005pm}, the initial 
entangled state at the $\phi$- factory
assumes the form:
\begin{eqnarray}|\psi\rangle&=&{\cal N} \lbrace\lbrace
|K^0\rangle_{{\vec k}}\otimes|\bar K^0\rangle_{-{\vec k}} -
|\bar K^0\rangle_{-{\vec k}}\otimes|K^0\rangle_{{\vec k}}\rbrace + \nonumber\\
&\omega &  \lbrace |K^0\rangle_{{\vec k}}\otimes|\bar 
K^0\rangle_{-{\vec k}} +
|\bar K^0\rangle_{-{\vec k}}\otimes|K^0\rangle_{{\vec k}}\rbrace\rbrace ,
\label{eq:omega}
\end{eqnarray}
where $\omega = |\omega |e^{i\Omega}$ is a complex parameter,
related to the intrinsic $CPT$--violating modifications of the EPR
correlations.
The $\omega$ parameter controls 
perturbatively the amount of contamination of the right-symmetry state 
by the wrong-symmetry state.
The appropriate observable (Fig. \ref{intensomega})
is the intensity \cite{Bernabeu:2003ym,Bernabeu:2005pm}: $$I(\Delta t)
= \int_{\Delta t \equiv |t_1 - t_2|}^\infty
|A(X,Y)|^2~,$$ with $A(X,Y)$ the decay
amplitude of the state $\psi$,
where one kaon decays to
the  final state $X$ at $t_1$ and the other to $Y$
at time $t_2$. 

At the $\phi$-factory 
there is a particularly good channel,
with 
$\pi^+\pi^-$ as final decay products for both kaons,
which 
enhances the sensitivity 
to the $\omega$-effect by three orders of magnitude.
This is due to the fact that
this channel is $CP$-forbidden for the right symmetry, whereas is 
$CP$-allowed for the wrong symmetry. Then, the relevant 
terms~\cite{Bernabeu:2003ym,Bernabeu:2005pm} in the 
intensity $I(\Delta t)$ (Fig. \ref{intensomega})
contain the combination $\omega/|\eta_{+-}|$, where $\eta_{+-}$ 
is the $CP$-violating
amplitude for the $\pi^+\pi^-$ states, which is of the order $10^{-3}$.
The KLOE experiment has obtained the first measurement of the
$\omega$ parameter~\cite{DiDomenico:2009xw,Ambrosino:2006vr}: 
\begin{eqnarray}
\label{eq:resomega}
\Re(\omega) &=&\left( {-1.6^{+3.0}_{-2.1}}_{\mbox{stat}} \pm 0.4_{\mbox{syst}} \right)\times{10^{-4}}\\
\Im(\omega) &=&\left( {-1.7^{+3.3}_{-3.0}}_{\mbox{stat}}\pm 1.2_{\mbox{syst}} \right)\times{10^{-4}} ~,
\end{eqnarray}

that is $|\omega | < 1.0 \times 10^{-3}$ at 95\% C.L.. 
At least 
one 
order of magnitude 
improvement is expected
with  
KLOE-2 at the upgraded
\DAF  ~\cite{DiDomenico:2009xw,Ambrosino:2006vr}.

This sensitivity is not far from
certain models of space time foam leading to
$\omega$-like effects~\cite{Bernabeu:2006av,Mavromatos:2008bz}. 
Indeed, in such models, inspired by string theory,
the $\omega$-effect is the result of  local distortions of 
space-time 
in the neighborhood of space-time
defects, which interact --via topologically non-trivial 
interactions (string capture / splitting)--
only with electrically neutral matter string states, due to electric 
charge conservation.
The recoil of the Planck-mass defect results in metric deformations 
along the direction of motion of the string state, 
$g_{0i} \sim \Delta k^i /M_P = \zeta k^i /M_P$, 
where $\Delta k^i = \zeta k^i$ denotes the momentum transfer 
of the matter state. On average, $\langle \zeta k^i \rangle = 0$, 
so Lorentz invariance holds macroscopically, but one has non trivial 
quantum fluctuations $\langle \zeta^2 k_i k_j \rangle \propto \delta_{ij} 
\overline{\zeta}^2 |\vec k|^2 $. 
It can be shown~\cite{Bernabeu:2006av,Mavromatos:2008bz} that as a result 
of such stochastic 
effects of 
the space time foam, 
neutral entangled states -- such as the ones in meson factories -- 
exhibit $\omega$-like effects, with the order of magnitude 
estimate: $|\omega |^2 \sim \frac{|\vec k|^4\overline{\zeta}^2}
{M_P^2 (m_1 - m_2)^2}$,
where $m_i$, $i=1,2$ are the masses of the (near degenerate) 
mass eigenstates. At the \DAF\ energy  
, for instance, 
$|\omega | \sim 10^{-4}|\overline{\zeta}|$, 
which lies within the sensitivity of the facility
for values of the average momentum 
transfer $\overline{\zeta} > 10^{-2}$ (which may be expected on account 
of naturalness), although this is actually a number that 
depends on the microscopic quantum theory of gravity, 
and hence is still elusive.
Nevertheless, in some concrete string-theory-inspired 
foam models examined in 
Ref. \cite{Mavromatos:2008bz}, 
this parameter is of the order $|\zeta | \sim \frac{m_1^2 + m_2^2}{k^2}$, 
and hence such models are in principle falsifiable 
at KLOE-2.

We close by mentioning that the $\omega$-effect can be
disentangled~\cite{Bernabeu:2003ym,Bernabeu:2005pm} 
experimentally from 
both, the C(even) background
- by means of different interference with the C(odd) resonant
contributions - and the decoherent evolution effects of space-time 
foam~\cite{Ellis:1983jz,Ellis:1995xd}, due to different structures 
in the relevant evolution equations of the reduced density matrices.

%
\subsection{$CPT$ violation and Lorentz-symmetry breaking}
\label{sec:DSQMLV}
This subsection addresses the prospects 
of searching for a different type of $CPT$ violation at KLOE-2:
as opposed to the situation described in Sec.~\ref{sec:DSQMCPT}, 
observable systems 
are taken to be governed by 
the standard laws of quantum mechanics
including unitary time evolution. 
It is  further assumed 
that possible $CPT$-violating effects 
beyond the Standard Model (SM) 
and General Relativity (GR)
can be described 
at low energies
within the framework of effective field theory (EFT).\footnote{The
SM and GR themselves are widely believed 
to be EFTs arising from underlying physics.
It would then seem contrived to presume 
that leading-order $CPT$-violating corrections 
lie outside EFT. 
Moreover, 
EFT has been successful 
in physics, 
and remains applicable 
even in the presence of
discrete backgrounds, 
such as in condensed-matter systems.}
This starting point 
together with a few mild mathematical assumptions
immediately implies that 
$CPT$ violation is necessarily accompanied by
Lorentz-symmetry breakdown
~\cite{Greenberg:2002uu,Greenberg:2003nv}, 
a result sometimes called the anti-$CPT$ theorem. 
In what follows,
$CPT$ and Lorentz-symmetry breakdown 
therefore can be treated on an equal footing.
Note that 
the converse of this statement, 
namely that Lorentz-symmetry breakdown 
always comes with $CPT$ violation, 
is not true in general. 
In the following we will discuss  
how the connection 
between $CPT$ and Lorentz-symmetry breaking
can be exploited 
for $CPT$-violation searches 
in neutral meson systems.

The above ideas 
form the foundation for the construction 
of the SME (c.f. Sect.\ref{sec:DSQMCPT}) 
~\cite{Kostelecky:2003fs,Colladay:1998fq,Colladay:1996iz}: 
one starts with the usual Lagrangians 
${\cal L}_{\rm SM}$ 
and ${\cal L}_{\rm GR}$ 
for the SM and GR, 
respectively, 
and then adds $CPT$- and Lorentz-violating corrections 
$\delta{\cal L}_{\rm CPTV/LV}$:
\begin{equation}
\label{sme}
{\cal L}_{\rm SME}={\cal L}_{\rm SM}+{\cal L}_{\rm GR}+\delta{\cal L}_{\rm CPTV/LV}\;.
\end{equation}
Here, ${\cal L}_{\rm SME}$ denotes the SME Lagrangian.
The corrections $\delta{\cal L}_{\rm CPTV/LV}$ 
are constructed by 
contracting conventional SM or GR fields 
of arbitrary mass dimension 
with $CPT$- and Lorentz--symmetry--breaking nondynamical vectors or tensors
to form quantities 
that transform as scalars
under coordinate changes.
A SME term 
may be assumed in the form: 
\begin{equation}
\label{sample}
\delta{\cal L}_{\rm CPTV/LV}\supset -a_\mu\overline{\psi}\gamma^\mu\psi\;,
\end{equation}
where $\psi$ is a SM fermion 
and $a_\mu$ 
controls the extent of $CPT$- and Lorentz--symmetry- violation. 
We remark that 
$a_\mu$ depends on the fermion species. 
It will be explained below that 
neutral-meson oscillations 
are sensitive to this particular type of coefficient.

SME coefficients for $CPT$ and Lorentz--symmetry breakdown,
like $a_\mu$ in 
(\ref{sample}), 
are assumed to be caused by 
more fundamental physics 
possibly arising at the Planck-scale. 
On phenomenological grounds, 
these coefficients must be minuscule, 
and theoretical considerations suggest 
that they might be suppressed by 
some power of the Planck mass.
Mechanisms for generating SME coefficients 
can be accommodated in various approaches 
to physics beyond the SM and GR, 
such as 
string theory~\cite{Kostelecky:2000hz,Kostelecky:1991ak,Kostelecky:1988zi},
various space-time-foam 
models~\cite{AmelinoCamelia:1997gz,Alfaro:1999wd,Klinkhamer:2003ec}, 
non-commutative field theory~\cite{Carroll:2001ws}, 
and cosmologically 
varying sca{\-}lars~\cite{Bertolami:2003qs,ArkaniHamed:2004ar,Kostelecky:2002ca}. 
In this sense, 
the SME is well motivated.

A number of studies 
have provided the SME with a firm theoretical 
foundation~\cite{Jackiw:1999yp,Kostelecky:2001jc,Kostelecky:2000mm,Lehnert:2004ri,Lehnert:2003ue,Altschul:2005mu,Bluhm:2008yt,Bluhm:2004ep}. 
To date, the SME has been employed 
to identify ultrahigh-precision $CPT$ and Lorentz--symmetry tests 
and has provided the basis for the analysis 
of numerous experimental $CPT$- and Lorentz-violation searches
in astrophysics~\cite{Kostelecky:2001mb,Xia:2007qs}, 
AMO (atomic, molecular, and optical) 
physics~\cite{Dehmelt:1999jh,Humphrey:2001wm,Muller:2003zzc}, 
 at colliders~
\cite{Kostelecky:1997mh,Nguyen:2001tg,Link:2002fg,Kostelecky:2008zz,Bossi:2008aa,Testa:2008xz,Aubert:2007bp,Bennett:2007yc,Hohensee:2009zk}, 
and with gravity~\cite{Battat:2007uh}. 
For practical reasons, 
most of these observational results 
constrain SME coefficients for stable or quasistable particles 
(i.e., photons, electrons, protons, and neutrons)~\cite{Kostelecky:2008ts}. 
Experimental tests of $CPT$ and Lorentz symmetry 
for unstable particles
are more difficult, 
and the majority of the corresponding SME coefficients 
remain unbounded. 
The extraction of experimental limits for such unstable particles 
therefore assumes particular urgency. 
Neutral-meson interferometry is currently 
the only feasible method to determine Planck-scale limits
on $CPT$ breakdown in the quark sector.

The next step is to employ the SME
to make predictions for neutral meson interferometry. 
To this end, 
recall that
an effective quantum-mechanical description 
of such a meson system 
is provided by a simple 2$\times$2 Hamiltonian $\Lambda$.
Indirect $CPT$ violation in this system occurs if and only if 
the difference 
$\Delta\Lambda \equiv \Lambda_{11} - \Lambda_{22}$ 
of the diagonal elements of $\Lambda$ is nonzero. 
It follows 
that $\Lambda$ contains
two real parameters for $CPT$ breakdown. 
The $w\xi$ formalism 
provides a convenient parametrization of $\Lambda$
that is independent of phase conventions,
valid for arbitrary-size $CPT$ and $T$ breaking, 
model independent, 
and expressed in terms of mass and decay rates insofar 
as possible~\cite{Kostelecky:2001ff}. 
In this formalism,
the $CPT$ violation determined by $\Delta\Lambda$
is controlled by the parameter $\xi_K$. 
On the other hand, 
the usual parametrization of $\Lambda$ 
depends on the phase convention 
and can be applied only if both $CPT$ and $T$ violation are small. 
In this conventional parametrization, 
$CPT$ violation is governed by the coefficient $\delta_K$.
The relation between $\xi_K$ and $\delta_K$ is given by
$\xi_K\approx2\delta_K$.

The remaining task is now 
to determine $\Delta\Lambda$
in terms of SME coefficients.
This is most conveniently achieved
with perturbative methods: 
the leading-order corrections to $\Lambda$ 
are expectation values of 
$CPT$- and Lorentz-violating interactions 
in the Hamiltonian 
evaluated with the unperturbed meson wave functions
$\ket{P^0}$, $\ket{\overline{P^0}}$. 
Note 
that the hermiticity of the perturbation Hamiltonian 
leads to real contributions 
~\cite{Kostelecky:1999bm,Kostelecky:1994rn}:
\begin{equation} 
\label{SMEresult}
\Delta\Lambda \approx \beta^\mu \Delta a_\mu\; ,  
\end{equation} 
where $\beta^\mu = \gamma (1, \vec{\beta})$ is the four-velocity 
of the meson state in the observer frame. 
In this equation,
we have defined $\Delta a_\mu = r_{q_1}a^{q_1}_\mu - r_{q_2}a^{q_2}_\mu$, 
where $a^{q_1}_\mu$ and $a^{q_2}_\mu$ 
are coefficients for $CPT$-- and Lorentz--invariance violation 
for the two valence quarks in the $P^0$ meson. 
These coefficients have mass dimension one, 
and they arise from Lagrangian terms 
of the form~(\ref{sample}),
more specifically
$- a^q_\mu \overline{q} \gamma^\mu q$, 
where $q$ denotes the quark flavor. 
The quantities $r_{q_1}$ and $r_{q_2}$ 
characterize normalization and quark-binding  
effects.

The result~(\ref{SMEresult}) 
illustrates  
that among the consequences of $CPT$ breakdown 
are the 4-velocity and the 4-momentum dependence 
of observables. 
This dependence on the direction 
is associated with Lorentz--invariance violation, 
in accordance with the above considerations 
involving the anti-$CPT$ theorem. 
It thus becomes apparent 
that the standard assumption of 
a constant parameter $\delta_K$ for $CPT$ violation 
is incompatible with unitary quantum field theory. 
More specifically, 
the presence of the 4-velocity in Eq.(\ref{SMEresult}) implies that 
$CPT$ observables will typically vary with the magnitude 
and orientation of the meson momentum. 
Additional effect is the motion of the laboratory:
the $CPT$- and Lorentz-violating SME coefficient $\Delta a_\mu$ 
is taken as space-time constant,
and it is conventionally specified
in the Sun-centered celestial equatorial frame.
But Earth-based laboratories move with respect to this frame 
so sidereal (and in principle seasonal) variations of observables
are possible. 
Thus, time, direction, and momentum binning 
is required for $CPT$ tests in neutral-mesons systems. 
Moreover, 
different laboratories have different orientations
in the celestial equatorial frame
and hence can be sensitive to different components of $\Delta a_\mu$.
\par
Another important fact is that 
the $CPT$ violation in each of the neutral-meson systems 
(e.g., $K$, $D$, $B_d$, $B_s$)
may be governed by a different $\Delta a_\mu$. 
Because of the distinct masses and decay rates, 
the physics of each system is distinct.  
Since each $\Delta a_\mu$ 
contains four components,
a complete experimental study of $CPT$ breaking 
requires four independent measurements in each 
neutral-meson system.
\par
In a high-energy fixed-target experiment, 
the momenta of the mesons 
are approximately aligned with the beam direction 
so that only the time stamp of the event 
is needed for directional information.
These measurements typically involve uncorrelated mesons, 
a fact that further simplifies their conceptual analysis. 
From the decomposition $\beta_\mu\Delta a^{\mu}  
=(\beta^0\Delta a^{0}-\Delta \vec{a}_{\parallel}\cdot\vec{\bm \beta}_{\parallel})
-(\Delta \vec{a}_{\perp}\cdot\vec{\bm \beta}_{\perp})$, 
where $\parallel$ and $\perp$ are taken 
with respect to the Earth's rotation axis, 
it is apparent  
that all four components of $\Delta a_{\mu}$ 
can be determined: 
the $\perp$ components via their sidereal variations 
and the sidereally constant piece in the first parentheses 
via its dependence on the momentum. 
However, 
the variation of $|\vec{\bm \beta}|$ with the energy is tiny, 
so it is difficult to disentangle 
the individual components $\Delta a^{0}$ and $\Delta \vec{a}_{\parallel}$. 
These ideas have been employed 
in experiments with the $K$ and $D$ systems. 
For the $K$ system, 
a linear combination of $\Delta a_0$ and $\Delta a_Z$ 
as well as a linear combination of $\Delta a_X$ and $\Delta a_Y$ 
have been bounded
to about $10^{-20}\,$GeV and $10^{-21}\,$GeV, 
respectively~\cite{Kostelecky:1997mh,Nguyen:2001tg}.
For the $D$ system, 
the FOCUS experiment has constrained 
both a linear combination of $\Delta a_0$ and $\Delta a_Z$ 
as well as $\Delta a_Y$ to roughly $10^{-16}$ GeV~\cite{Link:2002fg}.
\par
$CPT$ tests are also possible 
with correlated mesons 
at a symmetric collider, 
a set-up relevant for the KLOE-2 experiment.
Since kaons at the $\phi$-factory  
are monoenergetic, the analysis of energy 
dependence can not be pursued; nevertheless   
such a set-up has other crucial benefits: 
the wide angular distribution and the correlation of the mesons 
allows the extraction of 
limits 
on all four components of $\Delta a_{\mu}$.
\par
Searches for $CPT$ violation 
at symmetric-collider experiments 
with correlated kaons 
require analyses of the double-decay rate
of the quarkonium state 
for various final states $f_1$, $f_2$. 
With sufficient experimental resolution, 
the dependence of certain decays on the two meson momenta 
$\vec p_1$, $\vec p_2$ and on the sidereal time $t$ 
can be extracted 
by appropriate data binning and analysis. 
Note that 
different asymmetries can be sensitive to 
distinct components of $\Delta \Lambda$. 
\par
Consider 
the case of double-semileptonic decays 
of correlated kaon pairs.
Assuming the $\Delta S = \Delta Q$ rule, 
it can be shown 
that the double-decay rate $R_{l^+l^-}$ 
depends on the ratio:
\begin{equation}
\label{r+-}
\left| \frac{\eta_{l^+}}{\eta_{l^-}}\right| 
\approx 
1 - \frac {4\Re (i \sin\phi_{SW} ~ e^{i\phi_{SW}})} 
{\Delta m} \Delta\Gamma(\vec p) \Delta a_0 
\,.
\end{equation}
Here,
$\phi_{SW} \equiv \tan^{-1}(2\Delta m/(\Gamma_S - \Gamma_L ))$
is the superweak angle and $\Delta \Gamma(\vec p)$ stands for the 
difference of the \kl\ and \ks\ decay widths.
Note the absence 
of all angular and time dependence in Eq.~(\ref{r+-}). 
This fact arises 
because 
for a symmetric collider 
the relation 
$\vec{\bm \beta_1}\cdot\Delta\vec a = - \vec{\bm \beta_2}\cdot\Delta\vec a$ 
holds, 
which leads to a cancellation between the effects from each kaon. 
KLOE has exploited this idea 
to place limits on $\Delta a_0$ 
~\cite{Kostelecky:2008zz,Bossi:2008aa,Testa:2008xz}:
\begin{equation}
\Delta a_0 = (0.4\pm 1.8)\times 10^{-17}\hbox{ GeV}~.
\end{equation}
There are also other double-decay channels
suitable for $CPT$ tests,
in which there is no longer a cancellation of the spatial contributions 
of $\Delta a_{\mu}$.
Mixed double decays, 
in which only one of the two kaons has a $\xi_K$-sensitive mode,
are one example. 
Another possibility is the decay 
$K_S K_L\to\pi^+\pi^-\pi^+\pi^-$. 
This latter option has been chosen by
KLOE to place stringent 
limits on
all three spatial components of $\Delta a_{\mu}$
~\cite{Kostelecky:2008zz,Bossi:2008aa,Testa:2008xz}:
\begin{eqnarray}
\Delta a_X = (-6.3\pm 6.0)\times 10^{-18}\hbox{ GeV} \nonumber \\
\Delta a_Y = ( 2.8\pm 5.9)\times 10^{-18}\hbox{ GeV} \nonumber \\
\Delta a_Z = ( 2.4\pm 9.7)\times 10^{-18}\hbox{ GeV}~.
\end{eqnarray}
In fact the $\Delta a_{X,Y,Z}$ parameters 
can be evaluated performing a sidereal-time dependent analysis
of the asymmetry:
\begin{eqnarray}
\label{eq:deltaz2}
A(\Delta t)= \frac{
N^+-N^-}
{N^++N^-} ~,
\nonumber
\end{eqnarray}
with
$N^+=I\left(\pi^+\pi^-(+),\pi^+\pi^-(-);\Delta t>0\right)$ and
$N^-=I\left(\pi^+\pi^-(+),\pi^+\pi^-(-);\Delta t<0\right) ~,$
where the two identical final states are distinguished 
 by their emission in the forward ($\cos\theta>0$) or backward 
($\cos\theta<0$)
hemispheres
(denoted by the symbols $+$ and $-$, 
respectively), and $\Delta t$ is the time difference between $(+)$ and $(-)$ $\pi^+\pi^-$ decays.
\par
In an asymmetric collider, 
quarkonium is 
produced 
with a sizable net momentum. 
In this case, 
all four components of $\Delta a_\mu$ 
for the neutral meson system 
contribute to observable effects. 
As a consequence, 
appropriate data binning 
would also allow up to four independent $CPT$ measurements. 
The existing asymmetric $B_d$ factories BaBar and Belle 
are able to perform such measurements.
Results from the BaBar experiment 
bound various component combinations of $\Delta a_{\mu}$ for the $B_d$ meson 
to about $10^{-14}$ to $10^{-15}\,$GeV~\cite{Aubert:2007bp}.
The same study 
finds a $2.8\sigma$ signal for sidereal variations~\cite{Aubert:2007bp}.
While this significance level is still consistent with no effect, 
it clearly motivates further experimental $CPT$- and Lorentz-violation studies 
with neutral meson systems. 
\par
To summarize this subsection, 
$CPT$ violation, 
which comes with Lorentz--symmetry breakdown 
in unitary field theories,
is well motivated 
by various approaches to physics beyond the SM and GR.
Experimental $CPT$ and Lorentz--symmetry tests therefore are excellent tools 
in the search for new physics, 
possibly arising at the Planck scale. 
Such effects need not necessarily be correlated 
across various particles species, 
so measurements in different physical systems 
are typically inequivalent.
In the quark sector, 
neutral meson interferometry is currently 
the only 
precision method 
for $CPT$-violation searches.
While various experiments 
have placed tight constraints 
on $CPT$ breakdown 
in the $K$, $D$, and $B_d$ systems, 
only KLOE has been sensitive 
to all four $CPT$-violating SME coefficients, 
and KLOE-2 has the   
unique opportunity 
to perform a 
complete 
$CPT$ test in the kaon sector. 
%
\subsection{Other $CPT$-symmetry tests at the $\phi$ factory}
Valuable information on $CPT$ invariance is provided by the study 
  of \ks\ semileptonic decays at the $\phi$-factory where large samples of 
tagged, pure, and monochromatic \ks\ mesons can be 
isolated. 
In fact, when one kaon is detected at a time
$t_1 \gg \tau_S$, the decay 
amplitude of the entangled-kaon state as in Eq.(\ref{eq:state})   
factorizes and the system behaves as if the initial state 
was an incoherent mixture of states
$|K_S \rangle_{{\vec k}}\otimes |K_L \rangle_{-{\vec k}}$ and 
$|K_L \rangle_{{\vec k}}\otimes |K_S \rangle_{-{\vec k}}$.
Hence the detection of one kaon at large times tags one 
$|K_S \rangle$ state in the opposite direction. This is a unique
feature of the $\phi$-factory
exploited to select pure $K_S$ beams.
At KLOE, the $K_S$ is tagged by identifying the interaction 
of the $K_L$ in the calorimeter. 
In fact about $50\%$ of the 
$K_L$'s 
reaches the calorimeter
before decaying; their interactions (\kl\--crash)  
are identified
by a high--energy, neutral and delayed deposit in
the calorimeter, i.e., not associated to any 
charged track in the event and delayed of $\sim30$ ns 
($\beta_K\sim0.22$) with respect to  photons 
coming from the interaction region (IP).
In particular, 
$K_S\rightarrow \pi e \nu$ decays are selected 
requiring a $K_L$-crash and two tracks forming a vertex close 
to the IP and associated with two energy deposits in the calorimeter. 
Pions and electrons are identified using the time-of-flight technique. 
The number of signal events is normalized to the number 
of $K_S\rightarrow \pi^+\pi^-$ in the same data set. 

\par
The semileptonic \ks\ decays are interesting for 
various reasons: the CKM matrix
element \Vus\ 
can be accurately extracted from the 
measurement of the semileptonic decay widths (Sect. \ref{sec:SMtest}). 
From the measurement of the
lepton charge, the comparison between 
semileptonic asymmetry in \ks\ decays, $A_S$, and the asymmetry $A_L$ in
\kl\ decays provides significant tests of both, 
the $\Delta S = \Delta Q$ rule,
and $CPT$ invariance. 

KLOE has obtained a very pure sample of $\simeq$ 13,000 semileptonic \ks\
decay events with an integrated luminosity of 0.41 fb$^{-1}$ and
measured for the first time the partial decay rates for transitions 
to final states of each lepton charge, \gammo{\ks\!\toP\!\Pep\Ppim\Pnu} and 
\gammo{\ks\!\toP\!\Pem\Ppip\Pnubar}. The charge asymmetry 
\begin{eqnarray}
A_{S}&=&
\frac{
  \Gamma\left(\ks\toP\Ppim\Pep\nu\right) -
  \Gamma\left(\ks\toP\Ppip\Pem\bar{\nu}\right)
}{
  \Gamma\left(\ks\toP\Ppim\Pep\nu\right) +
  \Gamma\left(\ks\toP\Ppip\Pem\bar{\nu}\right)
}\nonumber \\
&=& (1.5 \pm 9.6_{\mbox{stat}}
\pm 2.9_{\mbox{syst}})\times 10^{-3} ~.
\label{eq:asy}
\end{eqnarray}
has been obtained \cite{Ambrosino:2006si}. 

Assuming $CPT$ invariance, 
$A_{S}\!=\!A_{L}\!=\!2\,\mathrm{Re}\,\epsilon\!\simeq\!3\!\times\!10^{-3}$, 
where $\epsilon$ is the parameter related 
to $CP$ violation in $K\!\leftrightarrow\!\overline{K}$ 
$\Delta S=2$ transitions. 
The difference between the charge asymmetries,
\begin{equation}
A_{S}-A_{L}=4\,\left(\mathrm{Re}\,\delta+\mathrm{Re}\,x_{-}\right),
\label{eq:rexm}
\end{equation}
signals $CPT$ violation either in the mass matrix ($\delta$ term), 
or in the decay amplitudes with $\Delta S\!\neq\!\Delta Q$ ($x_{-}$ term).
The sum of the asymmetries, 
\begin{equation}
A_{S}+A_{L}=4\left(\mathrm{Re}\,\epsilon-\mathrm{Re}\,y\right),
\label{eq:rey}
\end{equation}
is related to $CP$ violation in the mass matrix ($\epsilon$ term) 
and to $CPT$ violation in the $\Delta S\!=\!\Delta Q$ 
decay amplitude ($y$ term).
Finally, the validity of the $\Delta S\!=\!\Delta Q$ rule in $CPT$-conserving transitions can be tested through the quantity:
\begin{equation}
\mathrm{Re}\,x_{+} = \frac{1}{2}\frac{\Gamma(\DKSeIII)-\Gamma(\DKLeIII)}{\Gamma(\DKSeIII)+\Gamma(\DKLeIII)}.
\label{rex}
\end{equation}
In the SM, $\mathrm{Re}\,x_{+}$ is of the order of $G_{F} m_{\pi}^2\!\sim\!10^{-7}$,
being due to second-order weak transitions. 

Using the values of $A_L$,
$\mathrm{Re}\,\delta$, and $\mathrm{Re}\,\epsilon$ 
from other experiments \cite{Amsler:2008zzb}, KLOE obtained 
the following results \cite{Ambrosino:2006si}:
\begin{eqnarray}
\Re\, y  &=&( 0.4 \pm 2.5)\times 10^{-3}  
\nonumber \\
\Re\, x_- &=&(-0.8 \pm 2.5)\times 10^{-3}
\nonumber \\
\Re\, x_+ &=&(-1.2 \pm 3.6)\times 10^{-3}
~.
\end{eqnarray}
which are the most precise results on $CPT$ and 
$\Delta S\!=\!\Delta Q$ rule violation parameters in kaon 
semileptonic decays.
%


At KLOE-2, the integrated luminosity of 25 fb$^{-1}$ and   
better performance of the tracking system for decays close 
to the IP, beneficial for both the acceptance and the 
background rejection power,  
allow a total fractional uncertainty on the BR of 0.3\% and 
thus sensitivity for obtaining evidence of a non--zero
semileptonic asymmetry in \ks\  decays.  

The analogous measurements on $K_S\to\pi \mu \nu$ decays,
never obtained before, are more difficult
since poor $\pi$ - $\mu$ separation power at low energy 
implies larger contamination on both, the semileptonic decay sample from 
$K_S \to \pi^+ \pi^-$, and the $K_S\to\pi^{-(+)} \mu^{+(-)} \nu$ sample from 
 $K_S\to\pi^{+(-)} \mu^{-(+)} \nu$.  


A preliminary analysis of KLOE data has proven 
the capability to isolate the $K_S\to\pi \mu \nu$ events 
giving the first experimental evidence of the signal,  
and has shown the potentiality of reaching a BR precision
better than 2\% with a sample of 
2.5 fb$^{-1}$.
This translates to a precision of $\simeq$ 0.4\% 
on the BR($K_S\to\pi \mu \nu$) with KLOE-2. 

\subsection{Quantitative Bohr's complementarity}
\label{sec:DSQMeraser}
Bohr's complementarity principle or the closely related concept of
duality in interferometric or double--slit like devices are at the
heart of QM. The complementarity principle was
phrased by Niels Bohr in an attempt to express the most fundamental
difference between classical and quantum physics. According to this
principle, and in sharp contrast to classical physics, in quantum
physics we cannot capture all aspects of reality simultaneously, the
information 
obtained in one single setup 
being always limited.
By choosing the setup, e.g. the double--slit parameters, and thus the
quantum state under investigation, the predictability, i.e., the 
a--priori  
knowledge of the path 
 is simply calculated
(particle--like information), whereas the 
interference pattern (wave--like information) is observed by the
experiment. In the case of pure state, the sum of the squares of
these two quantities adds up to one, i.e., 
available information, particle--like and wave--like, is conserved.
This principle has been investigated both in theory and
experiment mainly for photons, electrons and neutrons propagating
through a double slit or through an interferometer. 
The neutral kaon system is optimal for testing Bohr's
complementarity: it is 
an interfering system with 
no need of experimental devices for effecting interference and 
at an energy scale     
not usually tested from other systems.
Moreover, the $CP$ violation
makes these kaonic systems special. Via Bohr's complementarity
relation new light can be shone upon $CP$ violation 
by showing that it moves 
information about reality to other aspects without violating the
complementarity principle, i.e., from predictability ${\cal P}$ to
coherence ${\cal V}$ and vice--versa, 
as discussed in Ref.~\cite{Hiesmayr:2007bt}.

The qualitative well--known statement that 
the observation
of an interference pattern and the acquisition of which--way
information are mutually exclusive 
has been
rephrased to a quantitative statement
\cite{Greenberger:1988zz,Englert:1996zz}:
\begin{eqnarray}\label{comp}
\vspace{-2mm}
{\cal P}^2(y)+{\cal V}_0^2(y)\leq 1\;,
\vspace{-2mm}
\end{eqnarray}
where the equality is valid for pure quantum states and the
inequality for mixed ones. ${\cal V}_0(y)$ is the fringe visibility
which quantifies the sharpness or contrast of the interference
pattern (the wave--like property) and can depend on an external
parameter $y$, whereas ${\cal P}(y)$ denotes the path
predictability, i.e., the 
a--priori knowledge one can have
on the path taken by the interfering system (the particle--like
property). It is defined by
\begin{equation}
\vspace{-2mm}
{\cal P}(y)\;=\;|p_I(y)-p_{II}(y)|\;,
\end{equation}
where $p_I(y)$ and $p_{II}(y)$ are the probabilities for taking each
path ($p_I(y)+p_{II}(y)=1)$. 
The predictability and visibility are in general dependent from 
external parameters. For example, in the usual double--slit
experiment the intensity is given by
\begin{equation}
I(y)\;=\; F(y)\;\big(1+{\cal V}_0(y) \cos(\phi(y))\big)\;,
\end{equation}
where $F(y)$ is specific for each setup and $\phi(y)$ is the
phase--difference between the two paths. The variable $y$
characterizes in this case the detector position, thus visibility
and predictability are $y$--dependent.
In Ref.~\cite{Bramon:2003aj} physical situations were investigated for which
the expressions of ${\cal V}_0(y), {\cal P}(y)$ and $\phi(y)$ can be
calculated analytically. These systems include interference patterns of
various types of double--slit experiments ($y$ is linked to
position), the oscillations due to particle mixing ($y$ is
linked to time) as in case of kaon systems, and also Mott scattering
experiments on identical particles or nuclei ($y$ is linked to a
scattering angle). All these two--state systems 
can be treated in a unified way via the generalized
complementarity relation. 
Here we investigate shortly the neutral kaon case 
while the reader can refer to Ref. \cite{Bramon:2003aj} 
for more details and discussion on other systems. 

The time evolution of an initial $K^0$ state 
can be described by (in the following $CP$ 
violation effects can safely be neglected)
\begin{eqnarray}\label{K-evolution-slit}
\label{singlekaon} |K^0 (t)\rangle &\cong&
\frac{e^{-\frac{\Delta\Gamma}{4} t}}
{\sqrt{2\cosh(\frac{\Delta\Gamma}{2} t)}}\;   
\left\lbrace e^{i \Delta m\cdot
t+\frac{\Delta\Gamma}{2} t} \, |K_S\rangle +
|K_L\rangle\right\rbrace\ \nonumber \\  
\end{eqnarray}
where $\Delta\Gamma=\Gamma_L-\Gamma_S<0\,$. The state of 
Eq.(\ref{K-evolution-slit}) can be interpreted as follows. The two mass
eigenstates $|K_S\rangle, |K_L\rangle$ represent the two slits. At
time $t=0$ both slits have the same width, as time evolves one slit
decreases as compared to the other, however, in addition the whole
double slit shrinks due to the decay property of the kaons. This
analogy gets more obvious if we consider for an initial $K^0$ the
probabilities for finding 
a $K^0$ or a $\bar K^0$ state at time $t$, i.e., the strangeness
oscillation: 
\begin{eqnarray}
P(K^0, t)&=& \left| \langle K^0 | K^0(t) \rangle \right|^2  =
\frac{1}{2} \biggl\lbrace
1+\frac{\cos(\Delta m \cdot t)}{\cosh(\frac{\Delta\Gamma}{2} t)}\biggr\rbrace\nonumber\\
P(\bar K^0, t)&=& \left| \langle \bar K^0 | K^0(t) \rangle \right|^2
= \frac{1}{2} \biggl\lbrace 1-\frac{\cos(\Delta m\cdot
t)}{\cosh(\frac{\Delta\Gamma}{2} t)}\biggr\rbrace\;.
\end{eqnarray}
We observe that the oscillating phase is given by $\phi(t)=\Delta m\cdot
t$ and the time dependent visibility by
\begin{eqnarray}\label{visibility}
{\cal V}_0(t)=\frac{1}{\cosh(\frac{\Delta\Gamma}{2} t)}\;,
\end{eqnarray}
maximal at $t=0$. The ``which width'' information,
analogously to the ``which way'' information in usual double--slit
experiments, can be directly calculated from Eq.(\ref{singlekaon})
and gives the following predictability, 
\begin{eqnarray}\label{predictability}
{\cal P}(t)&=&\left| P(K_S, t)-P(K_L,
t)\right|=\left|\;\frac{e^{\frac{\Delta \Gamma}{2} t} \;-\;
e^{-\frac{\Delta \Gamma}{2}t}}{2\cosh(\frac{\Delta\Gamma}{2} t)}\;
\right|\nonumber\\
&=&\left|\;\tanh\big(\frac{\Delta\Gamma}{2} t\big)\right|\;.
\end{eqnarray}
The larger the time $t$ is, the more probable is the propagation of
the $K_L$ component, because the $K_S$ component has died out, the
predictability converges to its upper bound $1$.
Both expressions, the predictability (\ref{predictability}) and the
visibility (\ref{visibility}), satisfy the complementary relation
(\ref{comp}) for all times $t$: 
\begin{eqnarray}\label{kaon-complement-relation}
{\cal P}^2(t)+{\cal V}_0^2(t) \;=\;
\tanh^2\big(\frac{\Delta\Gamma}{2}t\big) +
\frac{1}{\cosh^2(\frac{\Delta\Gamma}{2}t)} = 1\;
\end{eqnarray}
For time $t=0$ there is full interference, the visibility is ${\cal
V}_0(t=0)=1$, and we have no information about the lifetimes or
widths, ${\cal P}(t=0)=0$. This corresponds to the usual double--slit
scenario. For large times, i.e. $t\gg 1/\Gamma_S$, the kaon is most
probable in a long--lived state $K_L$ and no interference is
observed, we have information on the ``which width''. For times
between the two extremes we obtain partially information on ``which
width'' and on the interference contrast due to the natural
instability of the kaons. 
For pure systems however the 
information in Eq.(\ref{comp}) or 
Eq.(\ref{kaon-complement-relation}) is 
always maximal.

This strict formal analogy between the neutral kaon system 
and the double-slit device can be fully appreciated when exploited
to implement a quantum eraser experiment, which could be feasible at
KLOE-2, as briefly explained in the next subsection.
\subsubsection{The kaonic quantum eraser}\label{kaonicQE}
Photons intereference was demonstrated two 
hundred years ago by Thomas Young.
There is no physical reason why
even heavier particles should not interfere, and 
also experiments with very massive ``particles'' like the
fullerenes have proven this fundamental feature
of QM. 
Further studies have shown that the knowledge of the path
to the double slit is the reason why interference is lost. The
gedanken experiment of Scully and Dr\"uhl in 1982 
\cite{Scully:1982zz}
surprised the physics community demonstrating that 
if the knowledge of the particle path 
is erased, the interference is brought back again.

Since that work many different types of quantum erasures have been
analyzed and experiments were performed with atom interferometers
and entangled photons where the quantum erasure in the so-called
delayed--choice mode captures better the essence and the most
subtle aspects of the eraser phenomenon. In this case the meter, the
quantum system which carries the mark on the path, is a system
spatially separated from the interfering system, referred to as object system. 
The decision 
to erase or not the mark of the meter system 
---and therefore 
to observe or not the 
interference--- can be taken long after the measurement on the
object system has been completed. This was nicely phrased by
Aharonov and Zubairy in their \textit{Science} review ``\textit{Time
and the Quantum: Erasing the Past and Impacting the Future}''
\cite{Aharonov:2005dc} where also the kaon eraser 
proposed in Refs.~\cite{Bramon:2003bj,Bramon:2004zp} is presented.

\noindent The quantum erasure experiments 
were mostly based on 
some kind of double--slit device 
or on entangled states.
which are both available at \DAF\ .

All existing experiments with usual quantum systems as photons, based
on entanglement and double--slit devices, can be divided 
into two classes. In the first, the erasure of the ``which way''
information is fully--actively performed by an experimenter, i.e., 
choosing the appropriate experimental setup. In the second class 
erasure is obtained by a probabilistic process, e.g. a beam splitter. 
The erasure of the ``which way''
information is in this case partially active. 
Both options of erasure 
can be tested with the neutral kaon system. 
In addition, neutral kaons can be measured by a 
different, completely passive procedure.
Herewith we obtain four different erasure options, two more
than with usual quantum systems.
An example of active measurement is placing a matter block into 
the beam for the determination of kaon strangeness from final--state  
particle identification.
It is active because the experimenter chooses 
%
to measure the strangeness of the kaons 
during their time evolution and also the time when the measurement 
is performed.
But there is as well the option of simply waiting for the kaon decay.
If by chance a semileptonic decay is detected 
then 
the kaon strangeness is determined 
as well, 
without the experimenter's control on
the mode and the decay time, 
with a passive--measurement procedure.

In summary, these two plus two possibilities of quantum erasure
prove in a new way the very concept of a quantum eraser. 
 The proposed quantum--eraser 
procedures 
give the same probabilities even
regardless of the temporal ordering of the measurements. Thus kaonic
erasers can also be operated in the so called  
delayed--choice mode as discussed in
Refs.~\cite{Bramon:2004pc,Bramon:2003te}, 
shedding new 
light on the very nature of the quantum eraser and contributing to
clarify the eraser working principle hot--debated in literature.
\subsection{KLOE-2 prospects on kaon interferometry}
\par
\begin{figure}[htb]
\begin{center}
\vspace{-8mm}
\resizebox{0.8\columnwidth}{!}{
\includegraphics{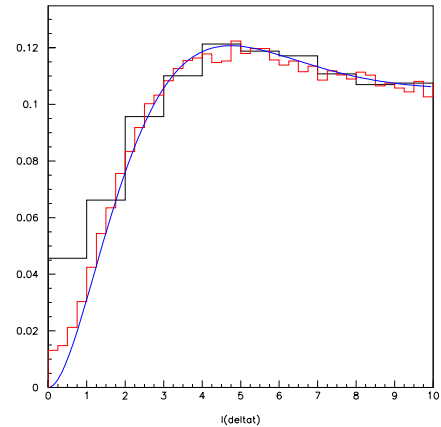}
}
\vspace{-1.5cm}
 \caption{Monte Carlo simulation of the 
 $I(\pi^+\pi^-,\pi^+\pi^-;\Delta t)$ 
as a function of 
$|\Delta t|$ (in $\tau_S$ units) with the 
            present KLOE resolution $\sigma_{\Delta t}\approx  \tau_S$ 
                (histogram with large bins), with an improved resolution 
                $\sigma_{\Delta t}\approx0.3\,\tau_S$ (histogram with small bins), and
            in the ideal case (solid line).
    \label{fig:resol} }
\end{center}
\end{figure}
\begin{figure}[htb]
\begin{center}
\resizebox{1.0\columnwidth}{!}{
\includegraphics{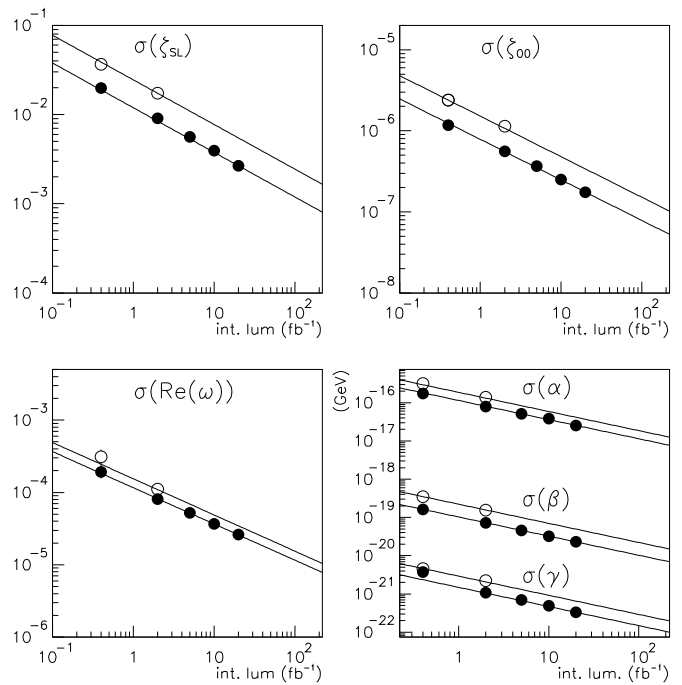}
}
 \caption{
The sensitivity to the  $\zeta_{SL}$,
            $\zeta_{0\bar{0}}$, $\Re$  $\omega$ and 
$\alpha$, $\beta$, $\gamma$ parameters
            with the present KLOE resolution 
                $\sigma_{\Delta t}\approx \tau_S$ (open circles), 
                and with the improved resolution
            $\sigma_{\Delta t}\approx0.3\,\tau_S$ 
expected with the IT at KLOE-2 (full circles).
    \label{fig:resol1} }
\end{center}
\vspace{-0.5cm}
\end{figure}

The decay mode 
$\phi \rightarrow K_S K_L  \rightarrow \pi^+\pi^- \pi^+\pi^-$ 
is very rich in physics. 
In general all decoherence effects including $CPT$--violation 
related to decoherence phenomena 
should manifest as a deviation from the QM  
prediction $I(\pi^+\pi^-, \pi^+\pi^-;$  $\Delta t=0)=0$.
Hence the 
reconstruction of events 
in the region $\Delta t \approx 0$, i.e., with vertices 
close to the IP, is crucial for precise determination 
of the parameters related to 
the $CPT$ violation and to the decoherence. The vertex resolution affects the 
$I(\pi^+\pi^-, \pi^+\pi^-; \- \Delta t)$ distribution precisely 
in that region as shown 
in Fig. \ref{fig:resol}, and impacts the measurement of the 
decoherence parameter
in two ways: 
i) reducing the sensitivity of the fit 
with respect to  
the accuracy expected from statistical fluctuations only; 
ii) introducing a  source of systematic uncertainty. 
The major upgrade planned at KLOE-2 is the addition of the 
vertex detector between the spherical beam pipe and
the drift chamber as presented in Sect. \ref{intro}, 
designed to reach an experimental resolution on charged vertices   
of $\sigma_{\Delta t} \approx0.3\,\tau_S$.
%
%
%
In Fig. \ref{fig:resol1}, the statistical uncertainty on 
several decoherence and $CPT$-violating parameters 
is shown as a function of the integrated 
luminosity for the case 
$\sigma_{\Delta t}\approx \tau_S$ (KLOE resolution), 
and for $\sigma_{\Delta t} \approx0.3\,\tau_S$ (KLOE-2 resolution 
with the inner tracker).
The improvement 
on vertex resolution leads to an increase of 
the experimental sensitivity by a factor of two. 
The physics program concerning interferometry 
is summarized in Tab. \ref{tab:modes1}, 
where the 
KLOE-2 sensitivity to the main parameters that can be 
extracted from 
$I(f_1,f_2;\Delta t)$
with   
an integrated luminosity of $ 25$ fb$^{-1}$ 
are listed  
and compared 
with the best 
published measurements.
\begin{table*}[!htb]
\begin{center}
  \caption{KLOE-2  
sensitivity on several parameters.
}
  \label{tab:modes1}
\vspace{3mm}
  \begin{tabular}{|c|c|c|c|c|}
    \hline
\textbf{f$_1$} & \textbf{f$_2$} & \textbf{Parameter} & \textbf{Present best measurement} 
& \textbf{KLOE-2 (25 fb$^{-1}$)} \\
    \hline
$K_S\rightarrow\pi e \nu$      &  
  & 
$A_S$ &
$(1.5\pm 11)\times 10^{-3}$ &
$\pm\,1 \times 10^{-3}$ \\
    \hline
$\pi^+\pi^-$ &  
$\pi l \nu$  & 
$A_L$ &
$(332.2\pm 5.8 \pm 4.7 )\times 10^{-5}$ &
$\pm\,4 \times 10^{-5}$ \\
    \hline
$\pi^+\pi^-$ &  
$\pi^0\pi^0$   &
$\Re{{\epsilon^{\prime}\over{\epsilon}}}$ &
$(1.65 \pm 0.26) \times 10^{-3}$ (PDG)&
$\pm\,0.3 \times 10^{-3} $ \\
    \hline
$\pi^+\pi^-$ &  
$\pi^0\pi^0$   &
$\Im{{\epsilon^{\prime}\over{\epsilon}}}$ &
$(-1.2 \pm 2.3) \times 10^{-3}$ (PDG)&
$\pm\,4 \times 10^{-3} $ \\
    \hline
%
$\pi^+ l^- \bar{\nu}$ & 
$\pi^- l^+ \nu$  &
$(\Re\delta+\Re x_-)$ & 
$\Re\delta=(0.25 \pm 0.23) \times 10^{-3}$ (PDG)&
$\pm\,0.3 \times 10^{-3} $ \\
& &  & $\Re x_-=(-4.2 \pm 1.7) \times 10^{-3}$ (PDG)& \\
    \hline
$\pi^+ l^- \bar{\nu}$ & $\pi^- l^+ \nu$  &
$(\Im\delta + \Im x_+)$ &
$\Im\delta=(-0.6 \pm 1.9) \times 10^{-5}$ (PDG)&
$\pm\,4 \times 10^{-3} $ \\
& & & $\Im x_+=(0.2 \pm 2.2) \times 10^{-3}$ (PDG)& \\
    \hline
$\pi^+\pi^-$ &  
$\pi^+\pi^-$  & 
$\Delta m$ &
$5.288\pm 0.043 \times 10^9 s^{-1}$ &
$\pm\,0.05 \times 10^9 s^{-1}$ \\
    \hline
$\pi^+\pi^-$ &
$\pi^+\pi^-$  & 
$\zeta_{SL}$ & 
 $(0.3\pm 1.9) \times 10^{-2}$ & 
$\pm\,0.2 \times 10^{-2} $ \\ 
    \hline
$\pi^+\pi^-$ &
$\pi^+\pi^-$  & 
$\zeta_{0\bar{0}}$ & 
$(0.1\pm 1.0)\times 10^{-6}$ & 
$\pm\,0.1 \times 10^{-6} $\\
    \hline
$\pi^+\pi^-$ &
$\pi^+\pi^-$ & 
$\alpha$ & 
$(-0.5\pm2.8)\times 10^{-17}$ GeV & 
$\pm 2\,\times 10^{-17}$ GeV\\
    \hline
$\pi^+\pi^-$ &
$\pi^+\pi^-$  & 
$\beta$ & $(2.5\pm2.3)\times 10^{-19}$ GeV & 
$\pm\,0.2 \times 10^{-19}$ GeV\\ 
    \hline
$\pi^+\pi^-$ &
$\pi^+\pi^-$  & 
$\gamma$ & $(1.1\pm2.5)\times 10^{-21}$ GeV & 
$\pm\,0.3 \times 10^{-21}$ GeV\\ 
& & & (compl. pos. hyp.) &  \\
& & & $(0.7\pm1.2)\times 10^{-21}$ GeV & $\pm\,0.2 \times 10^{-21}$ GeV \\
    \hline
$\pi^+\pi^-$ &
$\pi^+\pi^-$  & 
$\Re \omega$ & $(-1.6 ^{+3.0} _{-2.1} \pm 0.4)\times 10^{-4}$ & 
 $\pm\,3 \times 10^{-5} $ \\ 
    \hline
$\pi^+\pi^-$ &
$\pi^+\pi^-$  & 
$\Im\omega$ & $(-1.7  ^{+3.3} _{-3.0} \pm 1.2) \times 10^{-4}$  &
 $\pm\,4 \times 10^{-5} $ \\
    \hline
$K_{S,L}\rightarrow\pi e \nu$      &  
  & 
$\Delta a_0$ &
 (prelim.: $(0.4\pm 1.8)\times 10^{-17}$ GeV) &
$\pm\,2 \times 10^{-18}$ GeV \\
    \hline
$\pi^+\pi^-$ &
$\pi^+\pi^-$ &
$\Delta a_Z$ &
 (prelim.: $(2.4\pm 9.7)\times 10^{-18}$ GeV) &
$\pm\,1 \times 10^{-18}$ GeV \\
    \hline
$\pi^+\pi^-$ &
$\pi^+\pi^-$ &
$\Delta a_X$, $\Delta a_Y$ &
(prelim.: $<9.2\times 10^{-22}$ GeV) & 
$\pm\,6 \times 10^{-19}$ GeV \\
    \hline
\end{tabular}
\end{center}
\end{table*}
Although not exhaustive, the list is nevertheless indicative of the 
extent to which the experimental results at KLOE-2 can contribute 
to this field. 
%
%

%
%
%
\section{Low Energy QCD}
\label{sec:leqcd}
%
The interest for hadronic physics at low energy is not only related to 
the development of the EFT describing to some extent 
the non-perturbative phenomena, but also  
to precision physics involving light flavors u,d,s, where  
the sensitivity for testing the SM is often limited by the 
knowledge    
 of hadronic interactions (strong and electroweak).     
\par
The impact on the field of the measurements of semileptonic and 
leptonic kaon channels has already been discussed 
in Sect. \ref{sec:SMtest}. 
\par  
ChPT is the theoretical 
framework to study 
low-energy implications of the symmetry properties of the QCD theory
\cite{Weinberg:1978kz,Gasser:1983yg}. The pseudoscalar 
octet ($\pi$, K, $\eta$) is identified as the multiplet of 
massless Goldstone bosons 
associated to the chiral--symmetry breaking and the scalars are used 
as degrees of freedom 
of the effective field theory at energy scale 
$\lambda <$ M$_\rho$. 

%
%
%
%

This section is mostly dedicated to the measurements of non--leptonic  
and radiative decays of the \ks\ and the $\eta$ meson;
other significant results in the field can be achieved at KLOE-2 
with the study   
of events in the continuum, reviewed in 
Sects. \ref{sec:hadcs}--\ref{sec:gg}.

%
\newcommand{\la}{\langle}
\newcommand{\ra}{\rangle}
\newcommand{\Frac}[2]{\frac{\displaystyle #1}{\displaystyle #2}}

\subsection{Rare kaon decays}

The measurement of \ks\ decays at the $\phi$-factory has the unique 
feature to rely on pure \ks\ beams, tagged by the reconstruction of 
\kl\ decays and \kl\ interactions in the calorimeter. Background sources
are limited to the dominant \ks\ decay channels, $K_S \to \pi \pi$, 
 severely constrained by the \kl\ 4-momentum reconstruction.

With a target integrated luminosity of 25 fb$ ^{-1} $,
accessible channels include non-leptonic and radiative decays 
with BR 
down to 
 ${\mathcal O} (10 ^{-9} )$, 
 $ K_S \rightarrow 3 \pi $, $K_S\rightarrow \gamma \gamma$, 
$K_S\rightarrow \pi ^0 \gamma \gamma$,  
$K_S \rightarrow \pi ^0  \ell ^{+}\ell ^{-}$, 
$K_S \rightarrow \pi ^+ \pi ^-  \gamma $, 
$K_S \rightarrow \pi ^+ \pi ^-   e ^+ e ^- $. 


We can divide kaon decays into three categories:   
i) long--distance dominated (LD),  
ii) with comparable  short-- and long--distance contributions  and 
iii) short--distance dominated (SD) decays. 
Categories i) and ii) 
are those of interest for KLOE-2:  
the LD modes are 
addressed by the \ks\--decay analysis. 
They are properly described by ChPT
\cite{D'Ambrosio:1996nm,deRafael:1995zv,Pich:1995bw,Ecker:1994gg,D'Ambrosio:2003ef}
, whose $\Delta S=1$  lagrangian at ${\mathcal O} (p^4)$ reads:  
\beq
{ { \cal L}} _{\Delta S=1}={G_8 F^4 }
 \underbrace{
\la \lambda _6 
D_\mu U^\dag D^\mu U \ra}_{K \rightarrow \pi \pi/3\pi,\gamma\gamma } +
\underbrace{{G_8 F^2 } \sum N_i W_i}_{K^+ \rightarrow \pi^+
\gamma \gamma, \ K \rightarrow \pi \pi \gamma}+
 ... \label{eq:LW}
\eeq
$G_8$ is fixed   
by the  $K\rightarrow \pi \pi $ amplitudes; 
the first (${\cal L} _{\Delta S=1} ^2$) and the second  term (${\cal L} _{\Delta S=1} ^4$)  represent 
 the 
${\mathcal O} (p^2)$ and   ${\mathcal O} (p^4)$
weak lagrangian, respectively \cite{Ecker:1992de,D'Ambrosio:1997tb}.  
There are 37 counterterm coefficients, $N_i$'s, and  operators, $W_i$'s.
Unfortunately the low energy constants $N_i$'s (LECs) are both 
theoretically and phenomenologically  very poorly known.
New measurements of the non-leptonic and radiative decays 
are relevant to test the weak ChPT Lagrangian and help settling the LECs, 
which encode the underlying quark dynamics. \\
\indent
The Vector Dominance Model, by analogy to the treatment of the 
strong amplitudes, has been used to 
determine the relevant combination of LECs in processes with 
vector--meson--exchange contributions.
As an example, the 
results for  
radiative non-leptonic kaon decays   
are shown in Tab. \ref{tab:lecs}~\cite{Ecker:1992de,D'Ambrosio:1997tb}. 

\begin{table}
  \begin{center}
    \caption{Combinations of the counterterm coefficients in radiative non-leptonic kaon decays.}
    \label{tab:lecs}
$$
\begin{array}{l|l}
\mbox{Decay} & { {\cal L}} _{\Delta S=1} ^4  \\ \hline
K^+ \rightarrow \pi^+ l^+ l^- & N_{14}^r - N_{15}^r  \\
K_S\rightarrow \pi^0 l^+ l^- & 2 N_{14}^r   + N_{15}^r  \\
K^{\pm }\to \pi ^{\pm}\gamma\gamma   &N_{14}-N_{15}-2N_{18} \\
 K_{S}\to \pi ^{+}\pi ^{-}\gamma&N_{14}-N_{15}-N_{16}-N_{17} \\ 
K^{\pm }\to \pi ^{\pm }\pi ^{0}\gamma&N_{14}-N_{15}-N_{16}-N_{17} \\
 K_L \rightarrow \pi^+ \pi^- \  e^+ e^- & N_{14}^r + 2 N_{15}^r -3(N_{16}^r-N_{17}) \\
 K^+ \rightarrow \pi^+ \pi^0 e^+ e^- & N_{14}^r + 2 N_{15}^r -3(N_{16}^r-N_{17}) \\
  K_S \rightarrow \pi^+ \pi^- e^+ e^- & N_{14}^r  -  N_{15}^r -3(N_{16}^r+N_{17}) \\
\end{array}$$
\end{center}
\end{table}

%
%


\subsubsection{$ K \rightarrow 3 \pi $}
\label{thks3pi}
 
The amplitudes of $K\to 3 \pi$ decays 
can be approximated by  
a second--order polynomial in the pion kinetic energy T$_i$ using the  
X, Y variables \footnote{ Neglecting the pion mass difference, 
X=$\frac{2 m_k}{m_{\pi}^2}$(T$_{2}-$T$_{1}$); 
Y=$\frac{2 m_k}{m_{\pi}^2}$ (T$_{3}+m_{\pi}-\frac{1}{3} m_{k}$)}.
An alternative parametrization has been recently proposed 
\cite{Cabibbo:2005ez,Amsler:2008zzb} to evaluate 
how the Dalitz plot is affected by $\pi \pi$ 
re-scattering and in particular to 
extract the $\pi \pi$ scattering lenghts from the analysis of the cusp  
in $K^+ \to \pi^+ \pi^0 \pi^0$. 
Chiral expansion relates  
$K\to \pi \pi$ to $K\to 3 \pi$ transitions so that  significant 
constraints on the  values of the ${\mathcal O} (p^4)$ counterterms of 
Eq.(\ref{eq:LW}) have been obtained 
from the Dalitz plot of the experimental data, largely dominated by  $\Delta I=1/2$ transitions.


%
Lack of data on the subleading  $\Delta I=3/2$ amplitudes calls for the 
measurement of the $K_S \to \pi^+ \pi^- \pi^0$ decay at KLOE-2.
In fact, the  $K_S \to \pi^+ \pi^- \pi^0$ amplitude: 
\begin{equation} \begin{array}{lll}
A( K_S \rightarrow \pi ^+ \pi ^- \pi ^0 )&=&\frac{2}{3}b_2X-\frac{4}{3}d_2XY, 
\label{expan11} \end{array} \end{equation}
is directly related to the $\Delta I=3/2$, $L=1$
transitions, being the I=0 isospin final state 
suppressed by the high angular momentum barrier \cite{D'Ambrosio:1994km}. 


A preliminary but rather detailed search for this decay has been 
performed at KLOE with
740 pb$^{-1}$ of integrated luminosity. Main backgrounds come from
charged kaon events in which one $\pi^0$ mimics the $K_L$ interaction in
the calorimeter used to tag the $K_S$ decay, and from 
\ks\ $\to \pi^0 \pi^0$ 
events with one $\pi^0 \to e^+ e^- \gamma$.
Kinematic constraints are effective for background suppression. 
Final contaminations at the level of 20-30\% have been obtained 
selecting the signal sample 
with a global efficiency of 1-2\%.
KLOE-2 can eventually 
 select from 50 to few hundreds events, depending 
on the total integrated luminosity and also on the 
fraction of data collected after the  
installation of the 
detector upgrades, all of them beneficial 
 for improving the acceptance 
and the reconstruction capability of these events.   




The $CP$-violating transition $K_S \to \pi^0 \pi^0 \pi^0$ receives
contributions from I=1 final state only.
KLOE has obtained the upper limit BR($K_S\to 3\pi^0$)$\le 1.2\times 10^{-7}$ 
at 90\% C.L.,  
using a sample of 450 pb$^{-1}$ \cite{Ambrosino:2005iw}. 

The main background source is given by incorrectly reconstructed 
$K_S\to \pi^0 \pi^0$ decays, when 
the photon-energy deposit in the calorimeter is reconstructed 
as split clusters, or 
additional photons from machine background are 
assigned to the \ks\ decay. 
The kinematic constraints applied to fully reconstructed final states 
have allowed KLOE to improve by a factor of ten 
the upper limit on this channel and to increase the sensitivity 
on the $CP$ 
and $CPT$ tests using the unitary condition (Bell-Steinberger relation)  
as explained in Ref. \cite{Ambrosino:2006ek}.   
KLOE, with a \ks\ tagging efficiency of about 23\% and a 
selection efficiency on the signal of $\sim25\%$ 
has found 2 events to be compared with 3.13  
expected by Monte Carlo simulation of the 
background sources.
Since then, further improvements on the 
clustering procedure to recover erroneously-split photon clusters  
are proven to reduce 
contamination by a factor of six while 
leaving the signal efficiency unaffected. 
There are currently studies underway to evaluate 
the increase in acceptance 
from   
the installation of the crystal calorimeters (cf. Sec.\ref{intro}),
and in the \ks\ tagging efficiency resulting from the addition  
of the \kl\ $\to \pi^+ \pi^- \pi^0$ sample to the 
events with \kl\ interactions in the calorimeter used in the 
KLOE analysis.
In any case, as a conservative estimate, KLOE-2 can obtain 
an upper limit lower than 
$ 10^{-8}$ with one year of data taking. 
     
 
  

\subsubsection{$K_S\to  \gamma^{(*)}   \gamma^{(*)}  $   / $K_S \to  \pi^0 \gamma \gamma $   }
\label{thksgg}

\begin{figure}[htb]
\begin{center}
\psfig{file=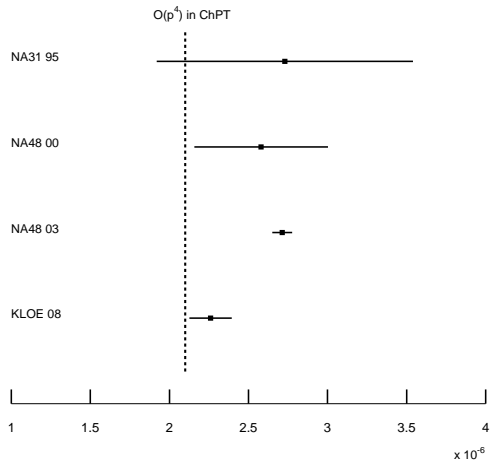,width=7.8cm}
\caption{ BR($K_S\to\gamma\gamma$) at ${\mathcal O} (p^4)$ in ChPT 
(dashed line) compared with the experimental results. 
It should be considered that 
from a naive dimensional analysis higher order 
theoretical contributions are expected 
to be suppressed by a factor 
m$_{k}^{2}$/(4$\pi$F$_\pi$)$^2\sim$0.2. 
}
\label{fig:ksgg}
\end{center}
\end{figure}

$K_S\rightarrow\gamma\gamma$ does not receive any  SD  
contribution while LD terms starts at 
 ${\mathcal O}(p^4)$  without counterterm structure.
This implies that:   i) we 
have only  one loop contribution 
 and ii) this contribution is scale-independent 
\cite{D'Ambrosio:1986ze,Buccella:1991ni}.
The 
$BR(K_S\rightarrow\gamma\gamma)$ = 2.1 $\times 10^{-6}$ 
is predicted at ${\mathcal O}(p^4)$ in terms of two known LECs of lowest order. 
From
a naive dimensional analysis higher order contributions are expected 
to be suppressed by a 
factor  m$_{k}^{2}$/(4$\pi$F$_\pi$)$^2\sim$0.2.
This process is the ideal test of the ChPT
(and in general of the EFT) at the quantum level.
At present, for 
the branching fraction, we have:
$$
\begin{array}{l}
{\rm ChPT \quad at \quad } {\mathcal O}(p^4)  \quad  2.1\times 10^{-6}\\
\\
{\rm NA48} \quad (2.713\pm 0.063)\times 10^{-6}\\ 
{\rm KLOE} \quad (2.26 \pm 0.13)\times 10^{-6}\\
\end{array}
$$
with the recent measurements differing by $\sim$3 $\sigma$'s 
\cite{Lai:2002sr,Ambrosino:2007mm}, as also shown in Fig. \ref{fig:ksgg}.
These results are based on different
experimental methods. At the hadronic machines, the branching ratio has
been obtained from the simultaneous measurement of $K_L$ and $K_S$ decays to
the same final state, $K_{S,L}\to \gamma \gamma$. The subtraction of 
the $K_L$ background has been performed on the basis 
of precision measurements of 
the ratio $BR(K_{L}\to \gamma \gamma)/BR(K_{L}\to 3 \pi^0)$.
At the $\phi$-factory, the $K_S$ decays are tagged by $K_L$ interactions 
in the calorimeter and the $K_{S}\to \gamma \gamma$ signal has to be   
separated from the 
$K_S\to \pi^{0}\pi^0$ decays with lost or unresolved photon clusters.
 
The KLOE result is fully compatible with  ${\mathcal O}(p^4)$ calculations in ChPT
\cite{D'Ambrosio:1986ze} thus severely constraining ${\mathcal O}(p^6)$ 
contributions to this process, while the NA48 measurement points to  
sizable higher--order corrections, 
${\frac{A({\mathcal O}(p^6))}{A({\mathcal O}(p^4))}}\sim$ 15\%. 
KLOE-2 will improve both sample size and data quality, the latter  
thanks to the detector upgrade, namely the crystal calorimeters CCALT  
important for the identification of \ks\ $\to \pi^0 \pi^0$ with photons  
at low angle with respect to the beam direction.

The KLOE-2 measurement can clarify the disagreement between KLOE and NA48 
\cite{Lai:2002sr,Ambrosino:2007mm}  
and help settling the   ${\mathcal O}(p^6)$  contributions 
to $A(K_L \rightarrow\pi^0\gamma\gamma)$, related by chiral symmetry only 
to the  
$K_S\rightarrow\gamma\gamma$ terms \cite{Buchalla:2003sj}. 
The  $ K_S \to \pi ^0 \gamma \gamma $ decay is dominated by pole 
contributions, from $\pi ^0$ and $\eta$: 
$ K_S \to \pi ^0 \lq\lq \pi^0 "  (\lq\lq \eta " )  \to  \pi ^0  \gamma
\gamma $. 
The diphoton invariant mass spectrum 
should be better assessed for a significant test of the pole dominance 
and for the study of the 
weak chiral vertex, $ K_S \to \pi ^0  \lq\lq P "$.  
  
The NA48 analysis of the $ K_S \to \pi ^0 \gamma \gamma $   
has provided the branching fraction and the diphoton 
invariant--mass ($m_{\gamma\gamma} $) spectrum \cite{Lai:2003vc} 
at high $z$ ($z=m_{\gamma\gamma} ^2/m_K ^2 $).
The result is based on 31 candidates, 
with an expected background of $13.7 \pm 3.2$ events:  
\beq  
BR(K_S \to \pi ^0 \gamma \gamma )_{z>0.2}= (4.9 \pm 1.6 \pm  0.9  ) \times
10^{-8},\eeq
to be compared with the theoretical prediction of  
$3.8 \times  10^{-8}$ \cite{Ecker:1987fm}.


At KLOE-2 the acceptance for $K_S\to\pi^0\gamma\gamma$ is 
about 60\%, as for $K_S\to\pi^0\pi^0$, and we expect 
10 events/ fb$^{-1}$ to be sorted out from the dominant 
$K_S\to \pi^0 \pi^0$ sample. 

\begin{figure}[h!]
  \begin{center}    
    \includegraphics[totalheight=2.5cm]{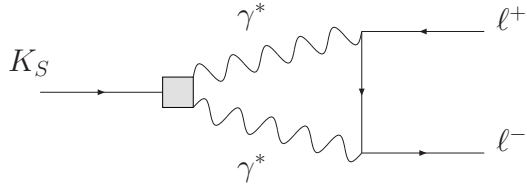}
    \caption{Long distance contribution to $\ks \rightarrow \ell^+ \ell^-$ process mediated by two-photon exchange.}
    \label{fig:ksee}
  \end{center}
\end{figure}

The \ks\ decays into lepton pairs ($e^+e^-$, $\mu^+\mu^-$) are due to 
$\Delta S=1$ flavor-changing 
neutral-current (FCNC) transitions.
Decay amplitudes receive contributions both from LD  
effects, dominated by the $\gamma \gamma$ 
intermediate state shown in Fig. \ref{fig:ksee}, 
and from SD interactions, due to box
and penguin diagrams via $W$, $Z$ exchange.
The SD contribution can be rather precisely evaluated in the SM,   
while the LD terms can be determined in ChPT. 
At lowest order,  ${\mathcal O} (p^4)$, one obtains~\cite{Ecker:1991ru}:
\begin{equation}
\begin{array}{rcl}
\Gamma(K_S \rightarrow \mu^+ \mu^-)/\Gamma(K_S \rightarrow \gamma \gamma) &
\simeq& 2 \times 10^{-6},\\
\Gamma(K_S \rightarrow  e^+ e^-)/\Gamma(K_S \rightarrow \gamma \gamma) &\simeq& 8 \times 10^{-9}, 
\end{array}
\end{equation}
which translate into  BR$(K_S \to \mu^+ \mu^- ) \simeq 5\times 10^{-12}$ 
and  BR$(K_S \to e^+ e^- ) \simeq 2\times 10^{-14}$. 
Significantly higher values would point to new physics (NP).
\par
Using the complete data set KLOE 
has obtained an upper limit on the electron channel, 
$BR(K_S\to e^+e^-)\leq 9\times 10^{-9}$\cite{Ambrosino:2008zi}, 
one order of magnitude more stringent than previous experimental results \cite{Angelopoulos:1997gu}. 
KLOE-2 can improve the sensitivity for both the leptonic channels to the $10^{-9}$ level.




\subsubsection{$K_{S} \rightarrow \pi ^{0}\ell ^{+}\ell ^{-}$ }

\label{thkspill}


\noindent

The measurements of BR$(K_{L}\to\pi^0l^+l^-)$ can 
clarify the mechanism of
quark-flavor mixing and $CP$-violation, both
within and beyond the SM
\cite{Buchalla:2003sj,Buras:2003dj}. In order to extract with high
precision the direct $CP$-violating component of the 
$K_L$ amplitude, it is necessary to determine from 
data the branching fraction of the $CP$-conserving transition
$K_S\to\pi^0l^+l^-$.

Any improvement on the present results 
on $K_{S,\pm}\to \pi ^{0,\pm}\ell ^{+}\ell ^{-}$ is desirable to assess  
the $CP$-violating term in  
$K_{L}\rightarrow \pi ^{0}e^{+}e^{-}$ thus obtaining the SD contribution 
with the 
accuracy needed for a significant test of the SM. 
The $CP$-conserving decays $K^\pm (K_{S}) \rightarrow \pi ^{\pm} 
(\pi ^{0})\ell ^{+}\ell ^{-}$ are dominated by the 
LD process $K\to\pi    \gamma ^*  \to \pi   \ell^+ \ell^-$ 
\cite{Ecker:1987qi,D'Ambrosio:1998yj}.
The decay amplitudes can be expressed in terms of one form 
factor $W_i(z)$ ($i=\pm,S$, 
$z=q^2/ M_K^2$) which  can be decomposed as the sum of 
a polynomial 
plus a
non-analytic term, $W_{i}^{\pi \pi}(z)$, generated  by  
a $\pi \pi $ loop. The non-analytic term is determined by   
the $K\rightarrow 3 \pi$ amplitude \cite{D'Ambrosio:1998yj}.
At $\mathcal{O(}p^{6})$ it reads: 
\begin{eqnarray}
W_{i}(z)\,=\,G_{F}M_{K}^{2}\,(a_{i}\,+\,b_{i}z)\,+\,W_{i}^{\pi \pi
}(z)\;,
\label{eq:Wp6}
\end{eqnarray}
where the  constants $a_{i}$ and $b_{i}$ parametrize local 
contributions starting at $\mathcal{O(}p^{4})$  
and $\mathcal{O(}p^{6})$, respectively. 
The most accurate measurements have been obtained with charged kaon beams by  
BNL-E865 \cite{Appel:1999yq}  and NA48/2 \cite{:2009pv}:
\begin{equation}
\begin{array}{ccr}
{\mathrm a_{+}}=-0.587 \pm 0.010 & 
{\mathrm \quad b_{+}}=-0.655 \pm 0.044 & \cite{Appel:1999yq} \\
{\mathrm a_{+}}=-0.578 \pm 0.016 & 
{\mathrm \quad b_{+}}=-0.779 \pm 0.066 & \cite{:2009pv}\\
\end{array}
\label{eq:ab+}
\end{equation} 
The experimental value of the ratio $b_{+}/a_{+}$ is bigger than 
the estimate from a naive dimensional analysis, 
 $b_{+}/a_{+}\sim $  
\par \noindent
${\mathcal O}[M_k^2/(4\pi F_\pi)^2] \sim 0.2$, 
thus pointing to 
large VMD contributions to the $K \to \pi l l$ decay .  

Chiral symmetry only cannot provide  
the \ks\ couplings $a_{S}$ and $b_S$ 
in terms of $a_{+}$ and $b_{+}$ \cite{Ecker:1987qi,D'Ambrosio:1998yj} and 
definite predictions 
on the \ks\ $\to \pi l l$ mode have to rely on model-dependent assumptions.  
Vector Dominance hypothesis leads to:
\begin{equation}
\begin{array}{l}
BR(K_S \rightarrow \pi^0 e^+ e^-) \approx  5.2 \times 10^{-9} \cdot a_{S}^2 \\
BR(K_S \rightarrow \pi^0 \mu ^+  \mu ^-) \approx 1.2  \times 10^{-9} \cdot a_{S}^2 \\
\end{array}
\label{eq:BRKS}
\end{equation} 
NA48 has obtained first evidence of the \ks\ $\to \pi l l $ decay for both 
leptonic channels on the basis of 
7 and 6 candidates  
\cite{Batley:2003mu,Batley:2004wg} in the $e e$ and $\mu \mu$ mode,
respectively. 
The experimental result on the ratio of the branching fractions is 
in agreement with Eq.(\ref{eq:BRKS}) from which they found:
 \begin{equation}
  | a_S|_{e e } = 1.06 ^{+0.26} _{-0.21} \pm 0.07 \qquad 
 |a_S|_{\mu \mu } = 1.54 ^{+0.40} _{-0.32} \pm 0.06.
\end{equation}
%
%
\par \noindent
A preliminary KLOE analysis 
of the $K_{S}\rightarrow \pi ^{0}e^{+}e^{-}$ 
based on 480 pb$^{-1}$, an equivalent 
amount of MC-simulated background and 2$\times$ 10$^5$ MC-signal events, 
has proven that 
the background sources, mostly from  
$K_{S}\rightarrow$ $\pi ^{0} \pi^0 \to $  $\gamma \gamma \gamma e^{+}e^{-}$,  
can be rejected keeping    
the overall efficiency to the  5\% level. 
With the analysis of the entire KLOE-2 data sample  
we can thus expect a measurement comparable with the NA48 result.
\subsubsection{$K\to \pi \pi \gamma $   / $K\to \pi \pi  e e$   }
\label{thkspipig}

Although radiative non--leptonic decays 
$K_L, K_S \to \pi^+\pi^-\gamma$ and 
$K^{\pm} \to \pi^{\pm}\pi^0\gamma$ are dominated by LD  
contributions, subleading SD terms in this case can be 
disentangled studying $CP$-violating observables, useful to test 
SM predictions \cite{Cappiello:2007rs}.
\par \noindent
The $K\rightarrow \pi \pi \gamma$ amplitudes 
contain two terms, from inner bremss\-tra\-hlung, IB, and direct
emission, DE. Due to the large contribution from low-energy photons 
collinear with pions, 
the IB amplitude is in general much larger than DE. 
The IB component is suppressed  by $CP$--invariance for $K_L$ and 
by  $\Delta  I=1/2$ rule 
for $K^{\pm}$, thus allowing the measurement of 
the DE term in these decays. 
Electric transitions only are allowed for the IB part, $E_{IB}$, 
while the DE component can be separated into electric  
($E1$)  and magnetic ($M1$) dipole terms \cite{D'Ambrosio:1992bf,D'Ambrosio:1994du}. 
Summing over photon
helicities cancels out the interference between 
electric and magnetic terms. 
At lowest order, ${\mathcal O} (p^2)$,
there is only the bremss\-tra\-hlung contribution.  
At ${\mathcal O} (p^4)$ both $E1$ and $M1$ transitions contribute. 
$E1$ and $M1$ are $CP$--allowed for charged kaons while $CP$-simmetry 
suppresses $E1$ transitions in the $K_L$ decays.
 
\begin{table}
  \begin{center}
    \caption{Experimental results on BR($k \to \pi \pi \gamma$) 
and contributing transitions.}
    \label{tab:two}
\vspace{-2mm}
$$
\begin{array}{ccc} 
K_S\rightarrow\pi^+\pi^-\gamma & \hskip5mm <9\cdot 10^{-5} &
\hskip5mm {E1} \\ 
K^+\rightarrow\pi^+\pi^0\gamma & \hskip5mm
(0.44\pm0.07)10^{-5}
& \hskip5mm M1,{E1} \\ 
K_L\rightarrow\pi^+\pi^-\gamma & \hskip5mm
(2.92\pm0.07)10^{-5}
&\hskip5mm {M1,} {\rm VMD} \\
\end{array}$$
\end{center}
\vspace{-3mm}
\end{table}
Present  experimental status 
is summarized in Tab. \ref{tab:two}.  
NA48/2 has recently obtained first evidence of the  
$K^{+}\rightarrow \pi ^{+}\pi ^{0}\gamma $ dipole electric 
transition,  $E1$, as shown in Tab. \ref{tab:tre}. 
The decay rate depends only on $T_{\pi}^{*}$, the kinetic pion
energy in the kaon rest frame. Integrating over $T_{\pi}^{*}$,
the rate reads:
\begin{equation}
\frac{d \Gamma}{d W} = \frac{d \Gamma (IB)}{ d W} \times
(1+{\it k_{INT}}\cdot W^2 + {\it k_{DE}}\cdot W^4),
\end{equation}
where 
$W^2 = \frac{(P_k^{*} P_\gamma^{*} )(P_\pi^{*} P_\gamma^{*})}{(m_K m_\pi
  )^2}$,  
$P_i^{*}$ is the particle 4--momentum and $INT$ denotes the interference
between IB-- and DE--photon emission.
%
%
Both  
$E1$ 
from the interference and $M1$ from the DE term, 
are extracted from a fit to the W spectrum. 
The weak counterterms contributing to $E1$ 
\cite{Ecker:1992de,D'Ambrosio:1997tb} 
can be evaluated from the experimental results \cite{Raggi:200739}. 
\begin{table}
  \begin{center}
    \caption{Contributions from direct emission and interference ($INT$)   
between IB and DE to 
radiative $K^{+}\rightarrow \pi ^{+}\pi ^{0}\gamma $ decays observed 
by NA48/2 \cite{Raggi:200739}.}    
     \label{tab:tre}
\vspace{3mm}
\hskip2.5cm $T_{\pi}^{\ast} \in \left[0, 80\right]$ MeV \\ 
\begin{tabular}{ccc}
\vspace{3mm}
$\frac {\Gamma(K^{+}\rightarrow \pi ^{+}\pi ^{0}\gamma)}{\Gamma(Tot)}_{DE}$&=&$(3.35{{\pm} }0.35\pm 0.25){{\times} }10^{-2}$\\
\vspace{3mm}
$\frac {\Gamma(K^{+}\rightarrow \pi ^{+}\pi ^{0}\gamma)}{\Gamma(Tot)}_{INT}$&=&$(-2.67{{\pm} }0.81\pm 0.73){{\times} }10^{-2}$\\
\end{tabular}
\end{center}
\vspace{-6mm}
\end{table}
under the assumption of 
$E1$ and $M1$ amplitudes 
independent from $W$. The presence of a form factor 
could affect the 
distribution and therefore simulate an interference term similar
to what observed by NA48/2 \cite{Cappiello:2007rs}.
The interference term can be further investigated by 
KLOE-2 both with the analysis of $K^\pm$ decays   
and looking at the $K_S\rightarrow\pi^+\pi^-\gamma$ 
channel which has 
the same counterterm combination  of the weak chiral lagrangian 
contributing to the
DE component.
Experimentally, 
$BR(K_S \to \pi^+\pi^-\gamma)(IB)/BR(K_S \to \pi^+\pi^-)=6.36 \times 10^{-3}$ 
with E$_{\gamma}^{*} >$ 20 MeV 
while only bounds  from E731 exist on the DE component 
\cite{Amsler:2008zzb}: 

$BR(K_S \to \pi^+\pi^-\gamma)$(DE, $E_{\gamma}^{*}>50$ MeV) $< 6 \times 10^{-5}$

$BR(K_S \to \pi^+\pi^-\gamma)$(INT, $E_{\gamma}^{*}>50$ MeV) $< 9 \times 10^{-5}$.
\noindent
From theory, $BR(K_S \to \pi^+\pi^-\gamma)$(INT, $E_{\gamma}^{*}>20$ MeV) = 10$^{-6} \div 10^{-5}$, 
reduced by a factor of 2.5 with a 50 MeV cutoff.
\par   
The background at KLOE mainly consists 
of accidental and spurious clusters from machine background and photon 
splitting, respectively. 
Both background sources can be safely removed by energy cuts 
and kinematic contraints 
coming from completely reconstructed final states. 
KLOE-2 can select a \ks\ tagged sample of $\sim 3 \times 10^8$
events/fb$^{-1}$  
which corresponds to a total counting of radiative decays of  
$\sim 8 \times 10^5$/fb$^{-1}$ 
with at least 120/fb$^{-1}$ events contributing to 
the interference. 
Since the dominant IB term, the interference and 
the contribution from counterterm combinations have a different 
dependence on E$_{\gamma}^*$  
so that a precision measurement of the photon energy distribution 
can disentangle 
the different components.
\par \noindent
Among other kaon radiative decays, accessible 
channels 
at KLOE-2 are the 
$K \to \pi \pi e^+ e^-$, particularly 
the $K_S \to \pi^+ \pi^- e^+ e^-$ mode. 
\par \noindent 
NA48 has obtained \cite{Lai:2003ad}
BR($K_s\to \pi^+ \pi^- e^+ e^- )= (4.69 \pm 0.30)\times 10^{-5}$ 
on the basis of 
676 events normalized to the $K_S \to \pi^0\pi^0\pi^0_{D}$, i.e., the kaon
decay into three neutral pions followed by the Dalitz decay of one 
of the pions.
With a preliminary analysis based on $\sim$1 fb$^{-1}$, KLOE has obtained  
 974 cleanly identified events 
providing the BR measurement with a statistical error of 
5\%.  
Differently from NA48, negligible errors are related
to the normalization which is done by $K_S$ tagging
and $K_S \to \pi^+\pi^-$ counting. The evaluation 
of systematics indicates
they could be brought to the per cent level, 
so that KLOE-2 can reduce the error 
to 1\%. 
The $K_S$ amplitude is dominated by the $CP$-even IB component 
needed to estimate the $CP$--violating term in the $K_L$
decay to the same final state. A test of the $CP$ symmetry in this channel 
can be 
performed through the measurement of the angular asymmetry 
between $\pi\pi$ and $e^+e^-$ planes. 

%
\subsection{Physics of $\eta$ and $\eta'$ mesons}
\label{sec:etaradec}

ChPT provides accurate description of the strong 
and electroweak interactions of the pseudoscalar mesons at low energies
\cite{Ecker:1994gg,Pich:1995bw,Bernard:2006gx,Bijnens:2006zp}.     

Same
final states 
as in the case of \kl\  
can be investigated with the decay of the isospin-singlet $\eta$ meson.
Due to weak $K_L-\eta$ mixing, understanding $\eta$ decays 
is a prerequisite for the calculation of SM contributions to
rare kaon decays such as $K_L\to\pi^0 e^+ e^-$
\cite{Ecker:1987qi,Flynn:1988gy}. CP violation in flavor--conserving 
processes can be tested in the 
$\eta$ decays to final states 
that are, as in the \kl\ case: 
$\eta\to\pi\pi$, $\eta\to\pi^0 e^+ e^-$ and 
$\eta\to\pi^+\pi^- e^+ e^-$ decay. In the latter, $CP$ violation 
could manifest in the angular asymmetry between 
the $\pi^+\pi^-$ and $e^+ e^-$ decay planes.

Conversion decays provide information
about transition form--factors needed, 
e.g., for the
light-by-light (LbL) contribution to the anomalous magnetic moment of the
muon. The $\eta(\eta')\to 3\pi$ decay is a valuable source of
information on light quark masses.

Most complications in treating the $\eta$ in ChPT are due to the mixing of
the $\eta$ and $\eta^\prime$ fields \cite{Bijnens:2005sj}. 
The $\eta^\prime$ is the most 
exotic meson of the pseudoscalar nonet, 
being identified with the $U(1)_A$
anomaly singlet ($\eta_0$) which is massive even in the chiral limit.  
It was observed by Witten \cite{Witten:1979vv} that in QCD with 
infinite number of colors (N$_c$) \cite{'tHooft:1973jz} the $\eta_0$ state 
is massless and the global $SU(3)_L\times SU(3)_R$ symmetry is replaced 
by $U(3)_L\times U(3)_R$.
Systematical treatment of the $1/N_c$ expansion, 
providing foundation for the $\eta$--$\eta^\prime$ mixing,
 was introduced in ChPT
by Kaiser and Leutwyler \cite{Kaiser:2000gs}. 

The $\eta^\prime$ 
decays to  
both vector and scalar mesons. 
 The dominance of the vector-meson coupling is demonstrated by     
radiative decays, 
$\eta'\to\rho\gamma$ and $\eta'\to\omega\gamma$.
Light scalar mesons $\sigma, f_0, a_0$ should  
play a significant role in the $\eta\pi\pi$ and $\pi\pi\pi$ decays  
\cite{Fariborz:1999gr,AbdelRehim:2002an,Beisert:2002ad} 
although the experimental evidence of the off-shell 
contributions from large--width scalars is, per se, 
more elusive. 
The treatment of the resonances is beyond the scope of standard ChPT,
 nevertheless 
continous progress in this field has been achieved over the years. 
The VMD model, for example, has been incorporated into the theory
\cite{Fujiwara:1984mp,Ecker:1988te} and unitarity constraints 
included by dispersion relations \cite{Anisovich:1996tx,Kambor:1995yc}.

In summary, 
the extension of ChPT stimulated by 
precision calculations of kaon and $\eta$ decays 
has also provided 
elegant and consistent methods 
for the treatment of $\eta^\prime$ physics.  
%
The dominant $\eta$ and $\eta'$ decays  
were recently analyzed 
in the framework of the ChPT extension as in Ref. \cite{Beisert:2003zd}.
%
The amplitudes of hadronic and anomalous decays 
have been calculated in Refs.\cite{Beisert:2002ad,Borasoy:2005du} and  
Refs.\cite{Borasoy:2004qj,Borasoy:2003yb}, respectively.
%
%
\subsubsection{Hadronic decays into three pions}

The decay widths of isospin--violating decays $\eta(\eta')\to3\pi$ are
comparable to those of the second--order electromagnetic decays
$\eta(\eta')\to\gamma\gamma$. The three pion decays are essentially
due to the QCD Lagrangian term proportional to $d$ and $u$ quark mass 
difference, the electromagnetic contribution being suppressed 
\cite{Sutherland:1966zz,Bell:1996mi}.
The decay width is thus sensitive to the mass difference of the quarks: 
$\Gamma_{\eta(\eta') \to 3\pi} \propto \Gamma_0\times (m_d-m_u)^2$,   
and the decay is suitable for precise determination of the ratios of 
the light quark masses \cite{Leutwyler:1996qg}. The calculations of the 
decay width in ChPT are based on the expansion in the quark masses and the 
calculations of $\Gamma_0$ in the isospin limit. 
The measurement of both,   
the value of the decay width, and the Dalitz--plot 
distribution are needed in order to test ChPT predictions and 
extract the fractional difference between light quark masses.  
The decay width is normalized using 
the two--photon width which can be measured at KLOE-2 
in $\gamma \gamma$ interactions, $e^+e^- \to e^+e^-\eta$. 
The theoretical predictions for $\Gamma_0$    
are tested comparing 
the results from the Dalitz plot distributions 
to the ratio of the two partial widths: 
%
\begin{equation}
  r_{\eta,\eta'} \equiv \frac{\Gamma(\eta,\eta'\to\pi^0\pi^0\pi^0)}
  {\Gamma(\eta,\eta'\to\pi^+\pi^-\pi^0)}.
  \label{eqn:r3pi}
\end{equation}
The decay amplitudes for $\eta, \eta' \to \pi^+ \pi^- \pi^0$ and 
$\eta, \eta' \to \pi^0 \pi^0 \pi^0$ are related by Bose symmetry,   
assuming $\Delta I=1$ transitions. This implies that, e.g., the relation
$r_{\eta,\eta'} \le 3/2$ 
holds in the limit $m_{\pi^+}=m_{\pi^0}$.


For a three body decay 
one can 
use symmetrized, dimensionless variables to parameterize the phase space:
\begin{equation}
  x \equiv \frac{1}{\sqrt{3}} \frac{T_1-T_2}{\langle T\rangle};\ \   
  y \equiv \frac{T_3}{\langle T\rangle}-1,
\end{equation}
where $T_1,T_2,T_3$ are the pion kinetic energies in the $\eta$ meson
rest frame and $\langle T\rangle$ is the average energy:
$\langle T\rangle=(T_1+T_2+T_3)/3=(m_\eta-m_1-m_2-m_3)/3$.    
Decays into three $\pi^0$ are conveniently described using polar 
coordinates $(\sqrt{z},\phi)$ in the $(x,y)$ plane:
\begin{equation}
  x = \sqrt{z}\sin\phi; \quad y=\sqrt{z}\cos\phi.
\end{equation}
The Dalitz plot has a sextant symmetry 
so that 
the range 
$0^\circ\le\phi<60^\circ$ contains all information on the decay. 
The variable $z$ is given by:
\begin{equation}
  z = x^2+y^2= \frac{2}{3} 
  \sum_{i=1}^3\left(\frac{T_i-\langle T\rangle}{\langle T\rangle}\right)^2, 
\quad 0\le z\le 1 .
  \label{eqn:z}
\end{equation}
%

\section*{$\eta \to 3\pi$}

The neutral decay $\eta\to 3\pi^0$  has been recently measured by 
Crystal Ball at BNL \cite{Tippens:2001fm}, KLOE \cite{Ambrosino:2007wi},
CELSIUS/WASA \cite{:2007iy}, WASA-at-COSY \cite{Adolph:2008vn},  
Crystal Ball at MAMI-B \cite{Unverzagt:2008ny} and Crystal Ball at MAMI-C 
\cite{Prakhov:2008ff}. The density of the Dalitz plot is nearly constant
and the results of the experiments are expressed by the lowest order 
expansion at the center:
\begin{equation}
  \mid \bar{{\mathcal A}}(z,\phi)\mid^2= c_0(1 + 2\alpha z +...).
  \label{eqn:alpha}
\end{equation}
The experimental values for the $\alpha$ parameter are consistent and
the average value is $\alpha=-0.0312\pm0.0017$, while 
ChPT calculations at NNLO, although with large uncertainty, predict 
a positive value for
$\alpha$, $\alpha = 0.013 \pm 0.032$ \cite{Bijnens:2007pr}.
A better agreement with experimental data was found by 
model--dependent calculations 
which include  
unitarity resummations, 
giving 
$\alpha = -0.014 $ \cite{Kambor:1995yc}, and 
$\alpha = -0.031 \pm 0.003$ \cite{Borasoy:2005du}.

The data for $\eta\to\pi^+\pi^-\pi^0$ decay are dominated by the recent
KLOE results with $1.3\times 10^6$ events in the Dalitz plot
\cite{Ambrosino:2008ht}. All other experiments, performed more than 
thirty years ago, had orders of magnitude lower statistics. The 
mechanism of the $\eta\to\pi^+\pi^-\pi^0$ decays is described by an 
expansion of the amplitude $A(x,y)$ around the center, $x=y=0$,  and 
the density of the Dalitz plot is expressed as:
\begin{equation}
  |{\mathcal A}(x,y)|^2\propto 1+ay+by^2+dx^2+fy^3+...
  \label{eqn:dalexpans}
\end{equation}
The LO ChPT amplitude is linear in the $y$ variable with $a=-1.039$ and
$b=a^2/4$. From the Dalitz plot density extracted by KLOE, 
$a$ = -1.090 $ \pm 0.005({\rm stat})^{+0.008}_{-0.019}({\rm syst})$  
 agrees within few per cent with the LO estimate, however higher order 
ChPT calculations, e.g.,  $a = -1.371$ at NLO, and $a = -1.271\pm 0.075$ 
at NNLO \cite{Bijnens:2007pr}, tend to make the
agreement worse, for $a$ and for all the parameters describing 
the $y$--dependence. 
On the other hand, the $d$ value,  
$d = 0.057\pm 0.006({\rm stat})^{+0.007}_{-0.016}({\rm syst})$, 
from KLOE is consistent 
with $d = 0.055\pm 0.057$ from the NNLO calculations. 
Experimental uncertainties 
are lower than the errors on theoretical    
calculations, 
nevertheless even higher experimental precision is needed  
to constrain also the imaginary part of the decay amplitude.
The KLOE large drift chamber allowing 
excellent momentum resolution 
on $\pi^\pm$ tracks 
has provided the most precise measurement of the Dalitz--plot
distribution of the $\eta\to\pi^+\pi^-\pi^0$.  
The analysis of KLOE data is based on 450 pb$^{-1}$ 
and the sample will be increased by an order of magnitude 
with the first year of KLOE-2 data-taking. 
The systematical uncertainty of the KLOE results 
is dominated by the event selection procedure.
To overcome this limit,   
new kinds of data-selection criteria 
will be devised for the larger data samples available. 

For the  ratio $r_{\eta}$ as in Eq.(\ref{eqn:r3pi}), the ChPT calculation  
at NNLO \cite{Bijnens:2007pr})
$r_{\eta} = 1.47$ 
is in good agreement with the measurements, 
$r_{\eta}=1.48\pm 0.05$ (world average) and 
$r_{\eta} = 1.432 \pm 0.026$ (from the PDG fit) \cite{Amsler:2008zzb}. 
However, the experimental 
values of the Dalitz--plot parameters, assuming  
$I = 1$ for the final state,
lead to  $r_{\eta}=1.54$, close to the LO result in ChPT,   
$r_{\eta} = 1.52$. Therefore a new
precision measurement 
of the $r_{\eta}$ ratio is needed to get a consistent overall picture of 
the $\eta\to 3\pi$ decays. This requires simultaneous analysis of both 
$\eta\to\pi^+\pi^-\pi^0$ and $\eta\to\pi^0\pi^0\pi^0$ decays 
leading to cancellation of some of the systematical effects in the ratio.

\section*{$\eta' \to 3 \pi$}

In the case of $\eta'$, it was argued that the decay $\eta' \to \pi\pi\pi$ 
can be related via $\pi-\eta$ mixing to the isospin--conserving 
$\eta' \to \eta\pi\pi$ decay and the ratio 
$BR(\eta' \to \pi\pi\pi) / BR(\eta' \to \eta\pi\pi)$ gives the $\eta-\pi^0$  
mixing angle or the quark--mass difference \cite{Gross:1979ur}. 
This scenario has been recently 
criticized 
\cite{Borasoy:2006uv} and 
it was shown that the relation between the two hadronic decays is  
more complicated \cite{Borasoy:2005du} and cannot be expressed by one 
mixing parameter only.
The same work \cite{Borasoy:2005du} emphasizes that  
a significant contribution to the $\eta'\to\pi^+\pi^-\pi^0$ 
from the $\rho^\pm\pi^\mp$
intermediate state 
and thus lower values of the  $r_{\eta'}$ ratio 
are expected.   
A fit to all data
(until 2006) on rates and Dalitz plot parameters of $\eta$ and $\eta'$
hadronic decays leads to 
$BR(\eta'\to \pi^+\pi^-\pi^0) \sim$0.01 \cite{Borasoy:2006uv}. 
Large contributions from 
$\rho^\pm\pi^\mp$ should also manifest 
in the Dalitz plot distribution. 
The $\eta' \to \pi^0\pi^0\pi^0$ decay has been 
observed by GAMS--2000  \cite{Binon:1984fe} on the basis of  
62 candidates and an expected background of 34 events. 
The measured partial width of about 300 eV is very close to  
$\Gamma(\eta\to\pi^0\pi^0\pi^0)$ but due to 
the different total widths for the contributions 
of isospin--conserving and radiative decays to the $\eta^\prime$, 
it leads to  $BR(\eta^\prime \to \pi^0\pi^0\pi^0)$ = 0.15\%. 
The Dalitz plot slope is consistent with zero 
within large errors: $\alpha = -0.1 \pm 0.3$ \cite{Alde:1987jt}. 
GAMS--$4\pi$ has recently collected $235\pm 45$ $\eta' \to \pi^0\pi^0\pi^0$ 
events, obtaining the $BR(\eta' \to \pi^0\pi^0\pi^0) = (1.8 \pm 0.4) 
\times 10^{-3}$ and $\alpha=-0.59\pm 0.18$ \cite{Blik:2008zz}.
\noindent
In 2008, CLEO \cite{:2008tb} 
has identified 20.2$^{+6.1}_{-4.8}$ 
$\eta' \to \pi^+\pi^-\pi^0$ decays corresponding to  
$BR(\eta' \to \pi^+\pi^-\pi^0)$ = $( 37^{+11}_{-9} \pm 4 ) \times 10^{-4}$. 
This result is much lower than predictions from Ref. \cite{Borasoy:2006uv}, 
pointing to a smaller contribution 
than expected from the $r_{\eta^\prime}$ measurement. 
This is also confirmed
by the relatively uniform Dalitz plot distribution found by the same 
experiment \cite{:2008tb}.
To solve this open question new data on 
$\eta'\to \pi^+\pi^-\pi^0$ providing stringent constraints on 
all hadronic $\eta$ and $\eta'$ 
decays \cite{Borasoy:2005du,AbdelRehim:2002an} are needed. 
In the first KLOE-2 run we expect
$\approx$8,000 events produced. The experimental challenge 
in this case is the background suppression, 
mostly from $\phi\to\pi^+\pi^-\pi^0$.

\subsubsection{$\eta / \eta' \to \pi^+ \pi^- \gamma$}

\begin{figure}
  \begin{tabular}{cc}
    \hspace{-0.6cm}
    \resizebox{0.26\textwidth}{!}{\includegraphics{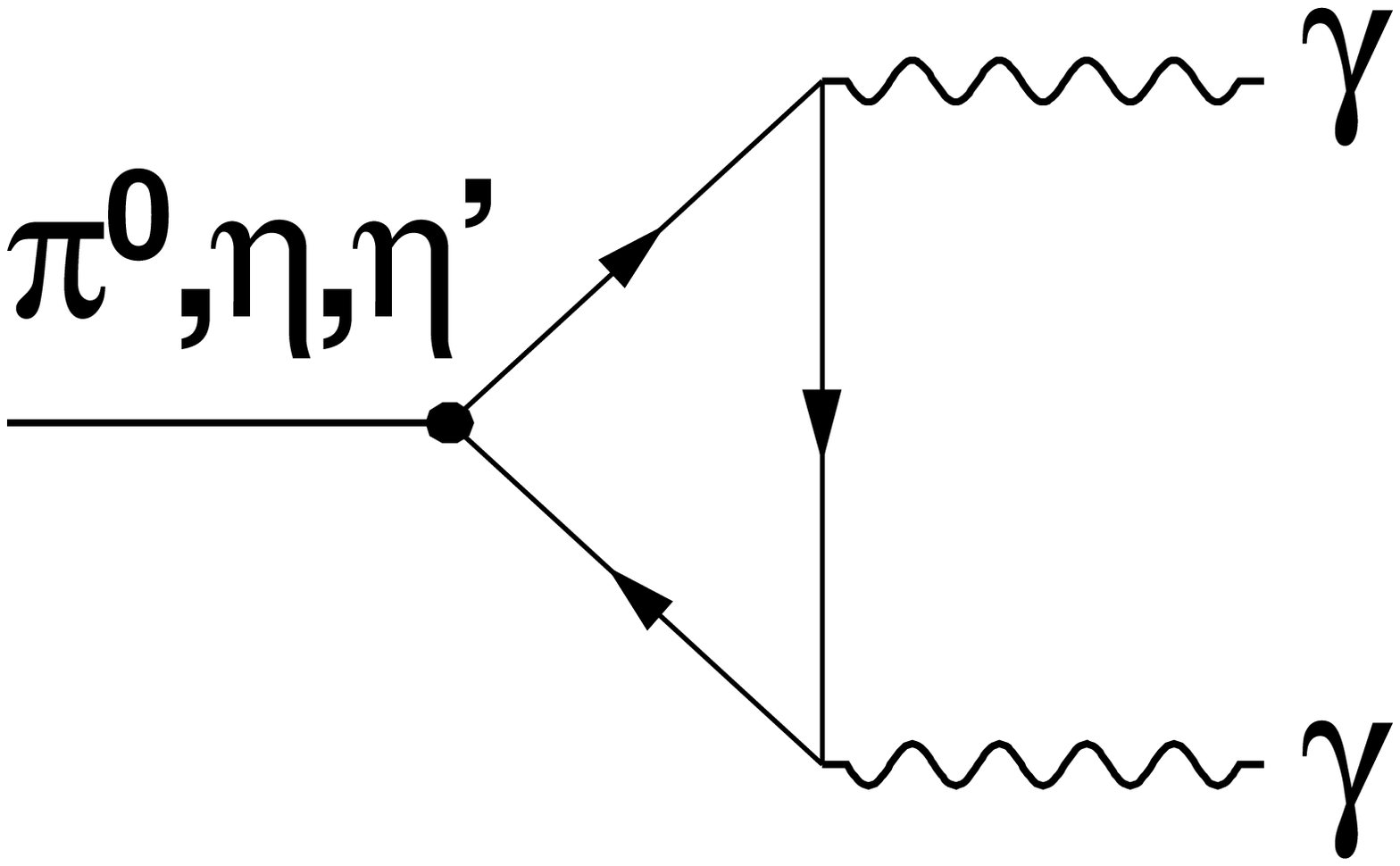}} &
    \hspace{-0.8cm}
    \resizebox{0.26\textwidth}{!}{\includegraphics{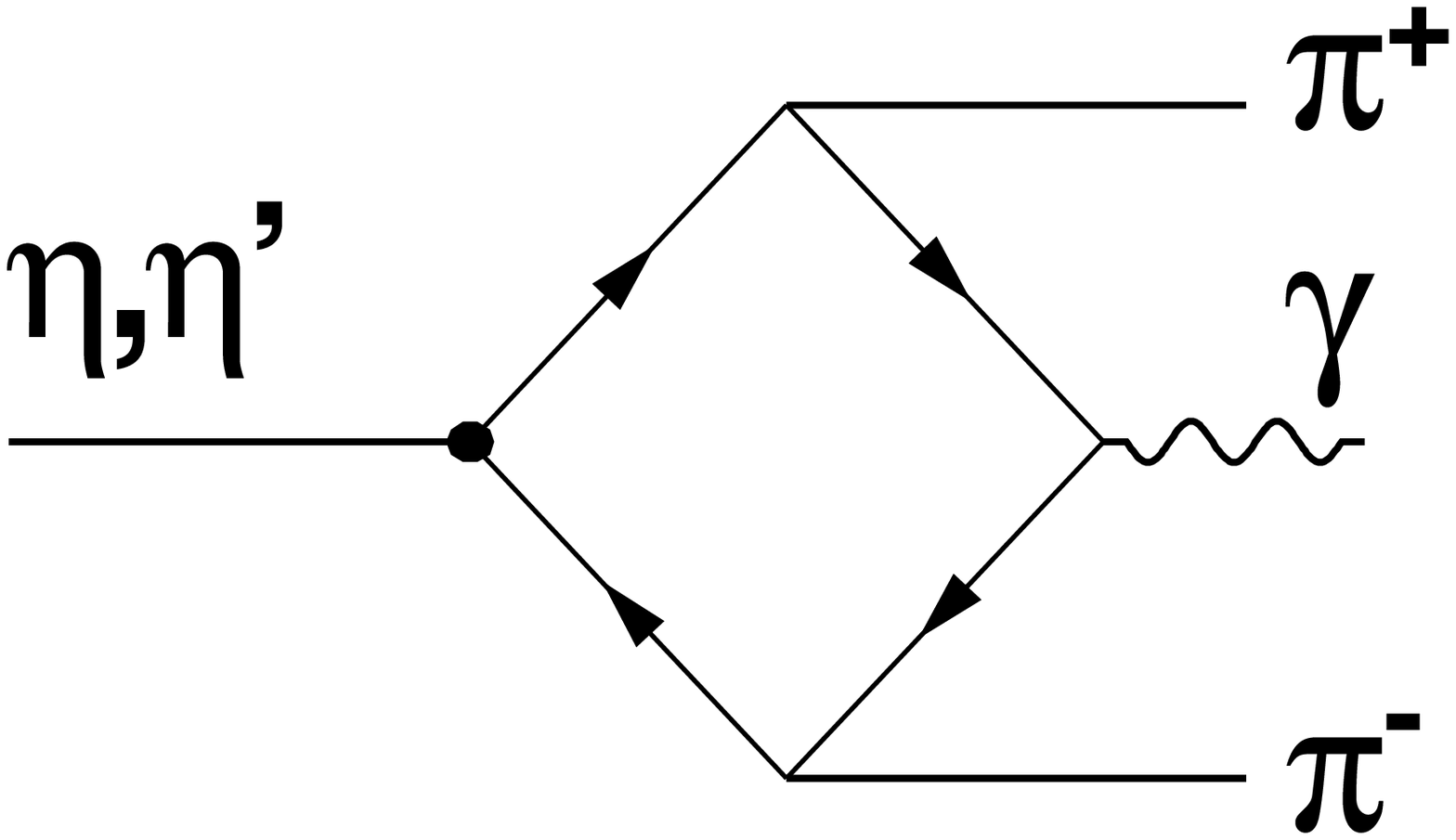}} \\
  \end{tabular}
  \caption{Diagrams for the triangle and box anomalies.}
  \label{picDiagFey}
\end{figure}

The decays $\eta / \eta' \to \gamma \gamma$ proceed through the triangle 
a\-no\-ma\-ly, and they are accounted for by the Wess-Zumino-Witten (WZW) 
term into the ChPT Lagrangian. The box anomaly is a higher term of WZW  
describing the direct coupling of three pseudoscalar mesons with the 
photon (Fig. \ref{picDiagFey}). 
%
The $\eta, \eta' \to \pi^+ \pi^- \gamma$ decays provide a good tool to 
investigate the box anomaly, which describes the not--resonant part 
of the coupling.
The invariant mass of the pions is a good observable to disentangle this
contribution from other possible resonant ones, e.g. from the $\rho$-meson. 
The kinematic range of the $\eta$ and $\eta'$ decays is above the 
chiral limit where the WZW coupling holds. 
%
Many efforts have been done to correctly describe contributions of the 
box anomaly to the two decays.
Some examples, which have implemented unitary effects via final-state (FS)  
interactions, are: HLS, where the hidden local symmetries have been used 
to evaluate the WZW term \cite{Benayoun:2003we}; the chiral unitary 
approach using Bethe-Salpeter equation \cite{Borasoy:2004qj}; the N/D 
structure matching the one loop corrections obtained with an Omnes function 
\cite{Holstein:2001bt}. 

The $\eta \to \pi^+ \pi^- \gamma$ decay has been measured in 1970s 
\cite{Gormley:1970qz,Layter:1973ti}. 
The analysis of the two data sets, 7,250 and 18,150 
events respectively, shows some contradiction. Theoretical papers 
trying to combine the two measurements have found discrepancies in data 
treatment and problems with obtaining consistent results 
\cite{Benayoun:2003we,Borasoy:2007dw}.
Therefore new measurements 
are needed 
to clarify the scenario.

KLOE has collected 
$5 \times 10^6$ $\eta \to \pi^+ \pi^- \gamma$ decays. 
This sample allows a detailed 
investigation of the di-pion invariant mass distribution. 
%
%
%
The main background is due to the $\phi \to  \pi^+ \pi^-\pi^0$, with a
natural signal to background ratio (S/B) of 1/200. The energy resolution 
for photons is improved exploiting the kinematic contraints to the 
$\phi\to\eta\gamma$
and $\eta\to\pi^+\pi^-\gamma$ decays, 
leading to a final S/B of
$\sim 20/1$ with a selection efficiency on the signal of about 60\%.
The preliminary analysis of $130$ pb$^{-1}$ provides $10^5$ signal events. 
Studies are under way to finalize the selection procedure in order to 
obtain an efficiency as independent as possible from the 
$\pi^+\pi^-$ invariant mass and thus 
an accurate investigation of the box anomaly effect.
%
%

%
%
Same considerations can be applied to the $\eta'$ meson, taking into 
account that in this case also the $\rho$-resonance 
has to be considered.   
Data related to the $\eta' \to \pi^+ \pi^- \gamma$ decays are much 
more recent: in 1997 the analysis of 7,392 events provided by Crystal 
Barrel gave evidence for the box anomaly in the invariant mass of pions 
\cite{Abele:1997yi}, while in 1998 the L3 collaboration found that 
their data (2,123 events) were well described by the resonant 
contribution \cite{Acciarri:1997yx}.
The relevance of the box anomaly asks for more data, in particular 
from the next scheduled run of KLOE-2 
when simultaneous analyses of the $\eta$ and $\eta'$ radiative decays 
can be performed.
The data already collected at KLOE contain 
$1.5 \times 10^5$ $\eta' \to \pi^+ \pi^- \gamma$ decays.
The feasibility of this measurement has been evaluated  
on the KLOE sample.   
With an estimated analysis efficiency of about $20\%$, it is possible
to obtain with KLOE-2/step-0 a result on the basis of 
$\sim 10^5$ selected decays. 
%


\subsubsection{Pseudoscalar form factors}
\label{sec:ff}
\begin{figure}
  \begin{center}
    \resizebox{45mm}{30mm}{\includegraphics{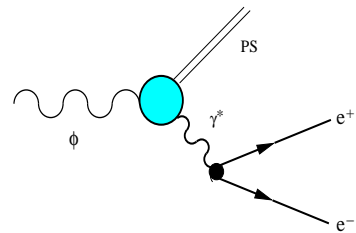}}
    \end{center}
  \caption{Diagram of the pseudoscalar production at the $\phi$ factory.} 
  \label{fig:eeprod}
\end{figure}
Pseudoscalar production at the $\phi$ factory (Fig. \ref{fig:eeprod}) 
associated to internal conversion of the photon into a lepton pair allows 
the measurement of the form factor ${\cal F}_P(q_1^2=M_\phi^2, q_2^2 > 0)$ 
of the pseudoscalar $P$ 
in the kinematical region 
of interest for the VMD model. The vector meson dominance assumption generally 
provides a good description of the photon coupling to hadrons but in the 
case of the transition form factor of the $\omega$ 
meson \cite{Dzhelyadin:1980tj,:2009wb} where standard VMD fails 
in predicting the strong rise of the coupling in 
the $(0.4< M_{l^+ l^-} < 0.6)$ GeV region. 
Recently, a model for implementing systematic corrections to the 
standard VMD calculations has been proposed \cite{Terschluesen:2010ik} which 
correctly describes the $\omega \to \pi^0 \mu^+ \mu^-$ experimental results, 
and predicts deviation from standard VMD for the 
$\phi \to \eta e^+ e^- $ decay spectrum. The only existing data 
on the latter process come from the SND experiment at Novosibirsk 
which has measured the M$_{ee}$ invariant mass distribution on the 
basis of 74 events \cite{Achasov:2000ne}. 
The accuracy on the shape does not allow any  
conclusion on the models. The predictions can be tested with the 
analysis of KLOE and KLOE-2/step0 data, e.g.,    
selecting the $\eta$ sample by means of the reconstruction of 
$\eta \to \pi^+ \pi^- \pi^0$ decays. 
This kind of procedure has been 
tested on a 90 pb$^{-1}$ sample obtaining, as a figure of merit 
for the KLOE-2 reach,
 a clean sample with a 
selection efficiency of   
15\% which 
translates to 12,000 events/fb$^{-1}$.

\begin{figure}
  \begin{center}
    \hspace{-0.5cm}
    \resizebox{40mm}{30mm}{\includegraphics{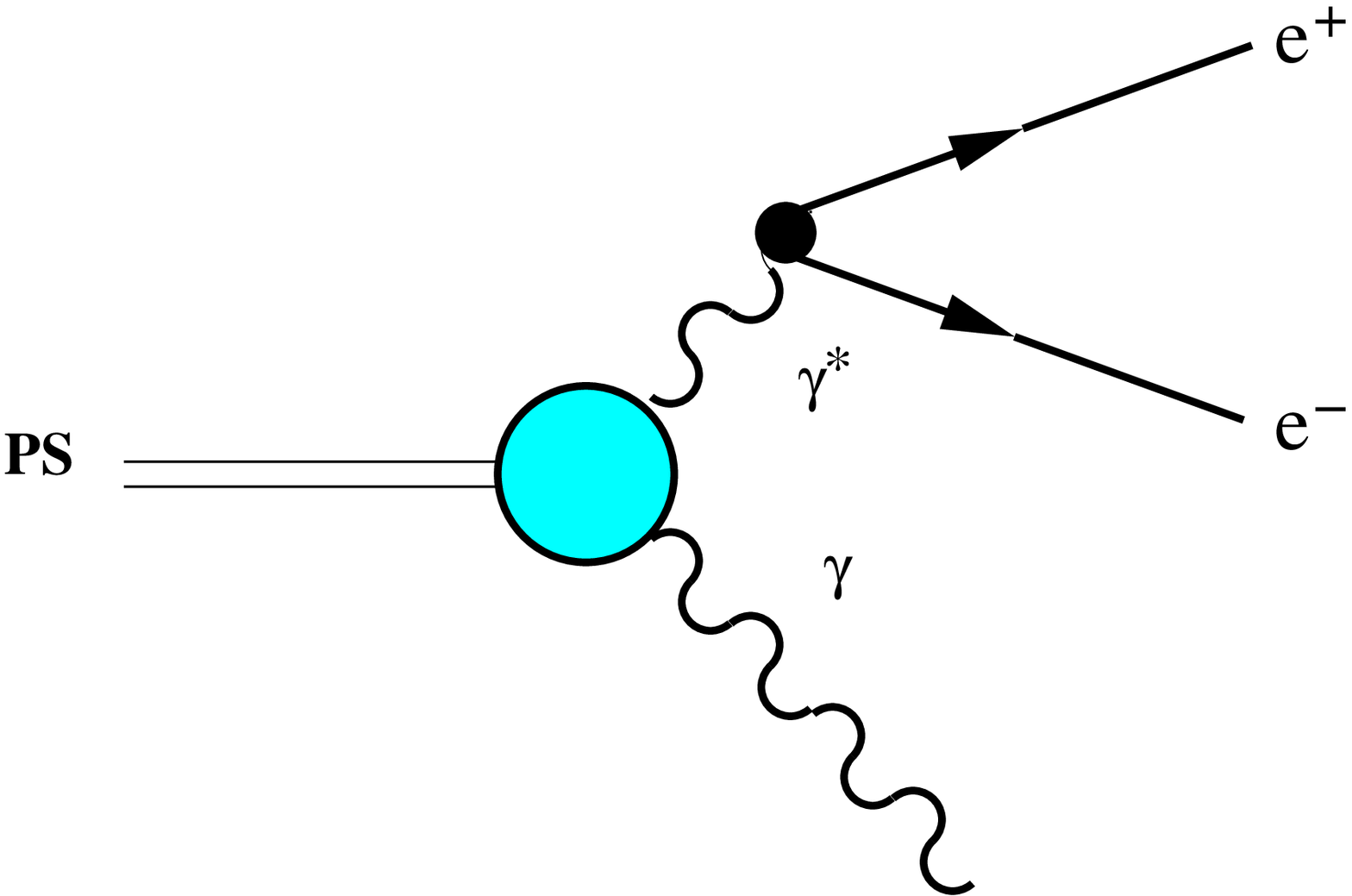}}
    \resizebox{40mm}{30mm}{\includegraphics{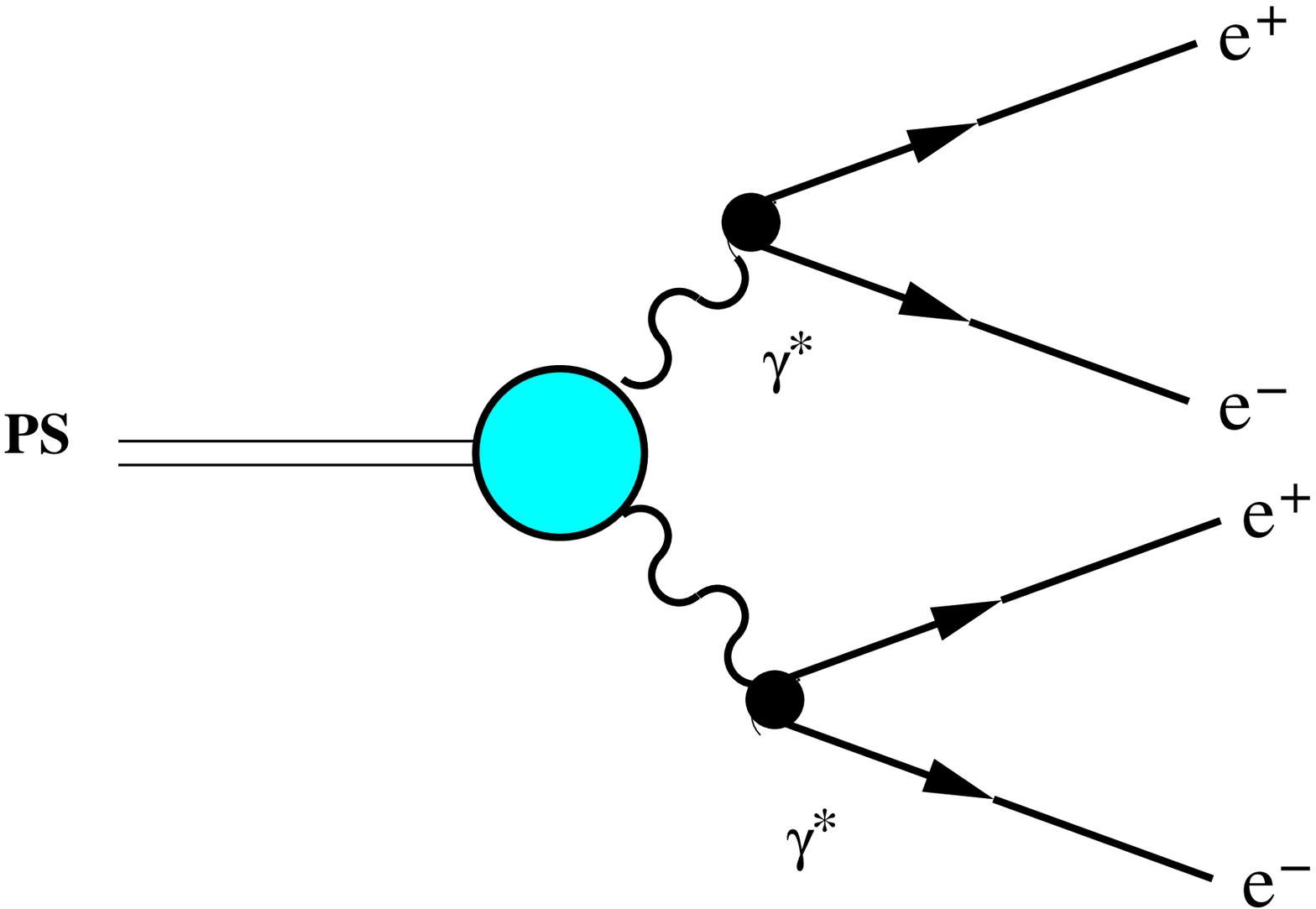}}
    \end{center}
  \caption{Diagram for pseudoscalar Dalitz (left) and double Dalitz 
    (right) decays.} 
  \label{fig:decay}
\end{figure}

Pseudoscalar form factors in the time-like region, where $q^2 > 0$, can 
also be 
accessed through Dalitz or double Dalitz decays (Fig. \ref{fig:decay}),
which allows the study of ${\cal F}_P(q_1^2,0)$ and ${\cal F}_P(q_1^2,q_2^2)$
respectively.
Several parametrisations have been proposed, gui\-ded by theoretical 
constraints and phenomenological models.
The form factors at $q^2>0$ are not well measured and 
the analysis of both $\pi^0$ and $\eta$ meson decays deserves 
more efforts. 
For the $\pi^0$, 
both ${\cal F}_{\pi^0}(q^2,0)$ and ${\cal F}_{\pi^0}(q_1^2,q_2^2)$ 
are extracted with large errors \cite{MeijerDrees:1992qb,Abouzaid:2008cd}, 
while in the $\eta$ case only the $\eta \to \mu^+ \mu^- \gamma$ form 
factor has been measured \cite{Dzhelyadin:1980kh,:2009wb}.

A copious number of $\pi^0$ will be produced at KLOE-2 through the 
decays $\phi \to \pi^+ \pi^- \pi^0$ and $K_L \to \pi^+ \pi^- \pi^0$.
For both processes    
the $\pi^0$ momentum can be obtained from precision measurements 
of the charged--particle tracks.
In the first case the event topology is simpler and the $\pi^0$ 
momentum can be determined with  $< 1$ MeV resolution using the
measured $\pi^+$ and $\pi^-$ tracks and energy-momentum conservation. 
The $\pi^0$ decay vertices at the IP are 
are precisely and efficiently reconstructed.
%
%
The sample can be enlarged using 
the $K_L \to \pi^+ \pi^-\pi^0$ decay, even though 
the resolution on $K_L$ momentum and 
vertex position are worse than in the $\phi \to \pi^+ \pi^- \pi^0$ mode. 
With 10 fb$^{-1}$ taking into account the analysis efficiency, a sample of 
$10^8$ $\pi^0 \to e^+ e^- \gamma$ and 180,000 
$\pi^0 \to e^+ e^- e^+ e^-$ events is expected, larger than what has been 
used in previous experiments. 

\begin{figure}
  \begin{center}
    \resizebox{0.8\columnwidth}{!}{\includegraphics{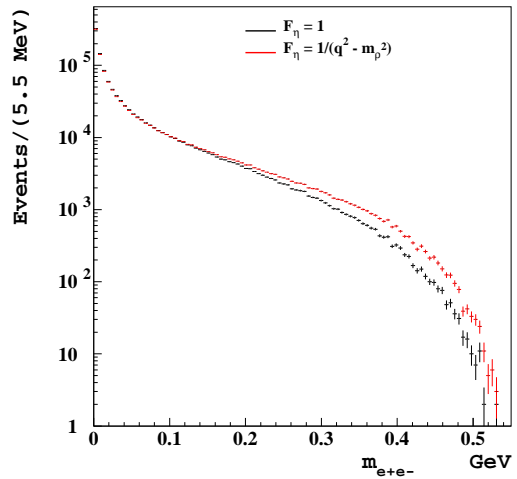}}
  \end{center}
  \caption{The $m_{e^+ e^-}$ distribution  
    of  $\eta\to e^+ e^- \gamma$ events, for point-like and pole approximation 
 of the ${\cal F}_\eta(q^2,0)$ form factor.}
  \label{fig:etaDalitzForm}
\end{figure}

In the $\eta$ case we expect $3 \times 10^5$ $\eta \to e^+ e^- \gamma$ 
and $2.8 \times 10^4$ $\eta \to e^+ e^- e^+ e^-$ events.
As a figure of merit of the sensitivity to the $\eta$ form factors with the
first KLOE-2 run, 
in Fig. \ref{fig:etaDalitzForm} 
the $m_{e^+e^-}$ distribution of the  $\eta \to e^+ e^- \gamma$ events 
is shown for point-like and pole approximation 
of the 
${\cal F}_\eta(q^2,0)$ form factor,
assuming
an overall analysis efficiency of 10\%.
%
%
KLOE-2 can also provide the first measurement of the 
$m_{e^+ e^-}$ distribution from the analysis of the 
four--electrons mode which is expected to be almost independent 
from the form factor parametrisation \cite{Bijnens:1999jp}.
%


\subsubsection{$\eta$ decays into four charged particles}

With ${\mathcal O}(10)$ fb$^{-1}$ KLOE-2 can access the 
first three decays summarized in Tab.~\ref{tab:eta4c}.
\begin{table}
  \begin{center}
    \caption{Measured and expected BRs for $\eta$ into four charged 
      particles.}
    \label{tab:eta4c}
    \begin{tabular}{lcc}
      \hline
      Decay                  & BR$_{\rm exp}$           & BR$_{\rm theo}$ \\
      \hline
$\eta \to \pi^+ \pi^- e^+ e^-$ & 
      $(26.8\pm 1.1) 10^{-5}$\cite{Ambrosino:2008cp} & 
      $25.7 \div 36 \cdot 10^{-5}$ \\
$\eta \to e^+ e^- e^+ e^-$     & 
      $<6.9 \cdot 10^{-5}$\cite{Akhmetshin:2000bw} & 
      $2.5 \div 2.6 \cdot 10^{-5}$ \\
$\eta \to \mu^+ \mu^- e^+ e^-$ &
      -- & 
      $1.6 \div 2.2 \cdot 10^{-5}$ \\
$\eta \to \pi^+ \pi^- \mu^+ \mu^-$ & 
      -- & 
      $8 \cdot 10^{-7}$ \\
      \hline
    \end{tabular}
  \end{center}
\end{table}
These decays allow the study of the $\eta$ meson internal structure, 
measuring the virtual-photon 4-momentum via the invariant mass of the 
lepton pair \cite{Landsberg:1986fd}. This is important   
for the connections with light-by-light physics~\cite{Bijnens:1999jp}.
%
In particular, the comparison of the kinematical distributions of the 
$\eta \to e^+ e^- e^+ e^-$ 
to $\eta \to \mu^+ \mu^- e^+ e^-$ can provide significant  
information on the underlying processes.
These decays can also probe some physics beyond the SM:
CP violation in $\eta \to \pi^+ \pi^- e^+ e^-$ \cite{Gao:2002gq} 
and light CP-odd Higgs of the NMSSM 
(next--to--minimal supersymmetric SM) in 
$\eta \to \pi^+ \pi^- \mu^+ \mu^-$ \cite{He:2008zw}.
The latter could provide up to 1-2 events per fb$^{-1}$, not 
sufficient 
to be investigated at KLOE-2. 
The $\eta \to \pi^+ \pi^- e^+ e^-$ decay has been already studied with  
KLOE \cite{Ambrosino:2008cp}, which measured an asymmetry $A_\phi$ in the 
angle between the $\pi\pi$ and $ee$ decay planes 
consistent with zero to the 3\% precision level, while theoretical 
predictions allow this quantity to be up to 2\%.
%
The largest contribution to the uncertainty of this measurement  
comes from the statistical error.
In KLOE the minimal transverse momentum of  reconstructed tracks,
$P_{T_{\rm min}}$, is 23 MeV. 
It limits the selection efficiency to $\sim$8\%.
%
%
The installation of the inner tracker, 
in the second phase of the KLOE-2 experiment 
would reduce $P_{T_{\rm min}}$ to 
16 MeV  
improving at the same time the tracking resolution, 
beneficial for the rejection of 
photon-conversion events mimicking the signal~\cite{Ambrosino:2008cp}.
With a sample of 20 fb$^{-1}$ and the increase in the efficiency 
from both acceptance and easier background rejection, KLOE-2  
could obtain $A_\phi$ with 0.8--1\% precision.
\subsubsection{$\eta \to \pi^0 \gamma \gamma$}

\begin{figure}
  \begin{center}
    \resizebox{0.2\textwidth}{!}{\includegraphics{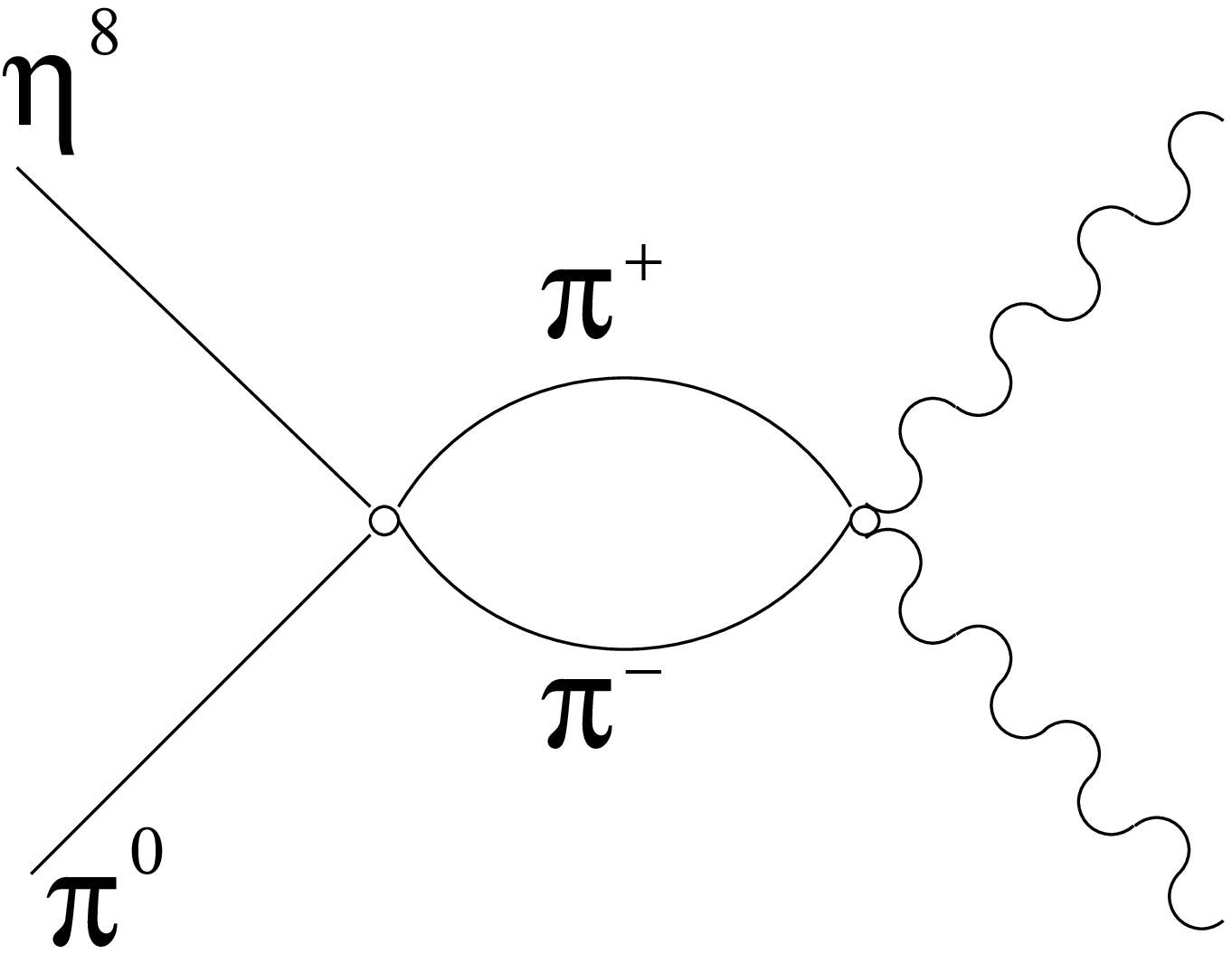}}
    \hspace{0.5cm}
    \resizebox{0.2\textwidth}{!}{\includegraphics{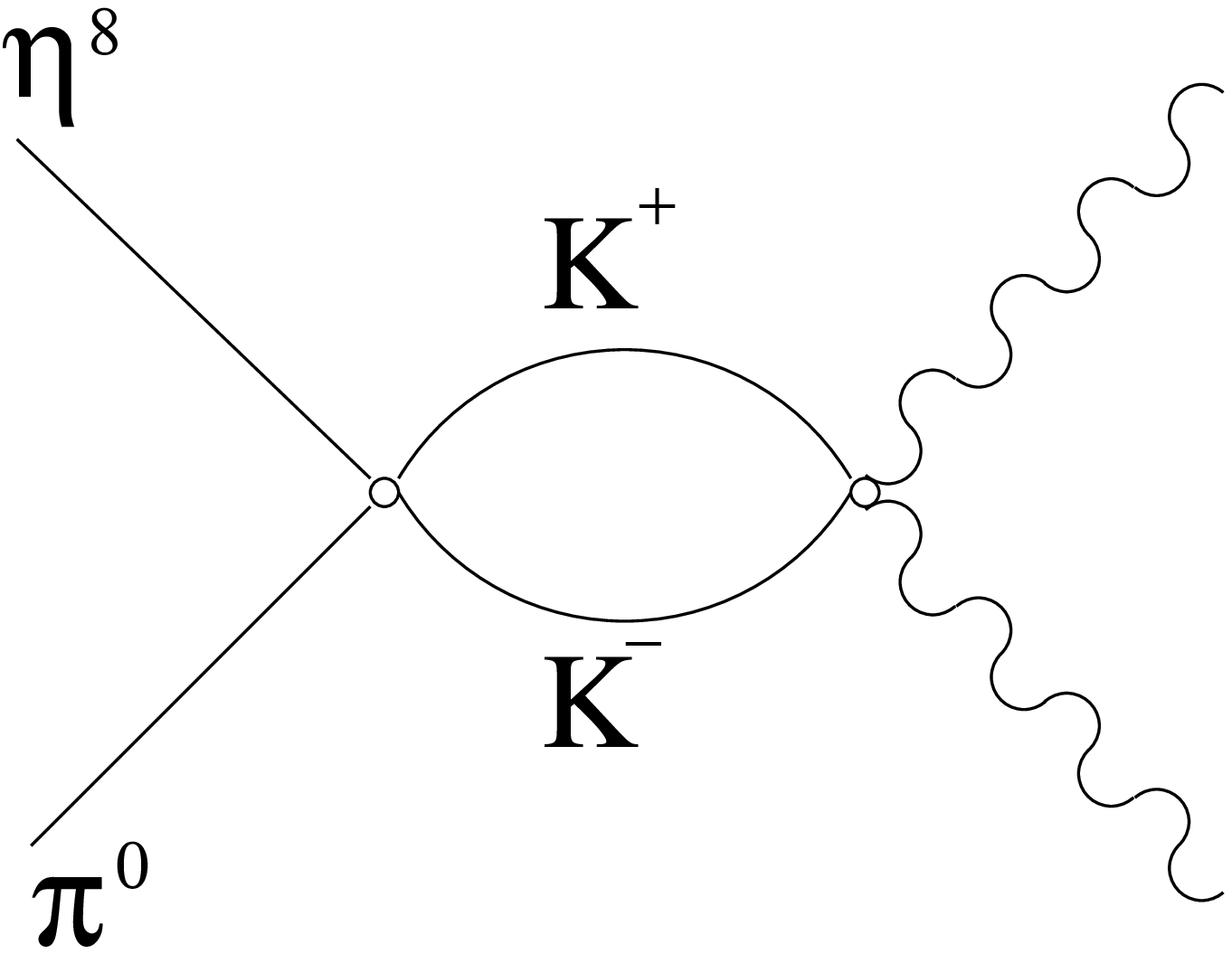}}
  \end{center}
  \caption{Meson loop diagrams in the $\eta \to \pi^0 \gamma \gamma$ 
    decay.}
  \label{fig:loopdiagrams}
\vspace{-0.3cm}
\end{figure}

\begin{figure}
  \begin{center}
    \resizebox{0.8\columnwidth}{!}{\includegraphics{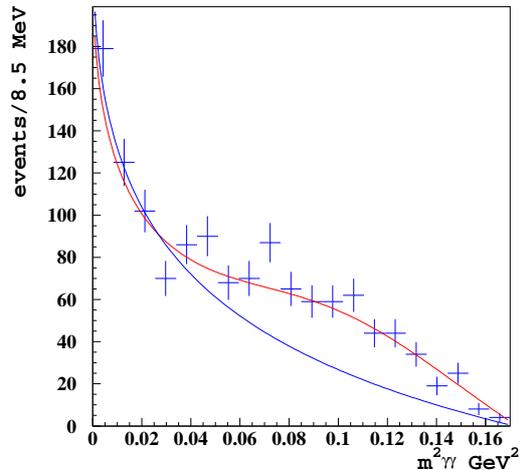}}
  \end{center}
  \caption{The $m^2_{\gamma \gamma}$ distribution in 
    $\eta \to \pi^0 \gamma \gamma$ decays expected at KLOE-2 for the 
    VMD+$a_0(980)$ model, with 
    constructive and 
    destructive 
    interference term. 
Crosses are the simulated experimental 
    data assuming 5\% constant efficiency as a function of 
$m^2_{\gamma \gamma}$ and 
    constructive interference.} 
  \label{fig:experiments2}
\end{figure}

The $\eta \to \pi^0 \gamma \gamma$ decay is very interesting both from 
the experimental and theoretical point of view. 
Big theoretical efforts to evaluate the  $BR(\eta \to \pi^0 \gamma \gamma)$ 
were originally motivated by the large branching fraction obtained by 
experiments dating back 40 years.
ChPT terms \cite{Ametller:1991dp} start at order ${\mathcal O}(p^4)$ 
through the pion and 
kaon loop diagrams shown in Fig. \ref{fig:loopdiagrams}.
%
The $\eta^8 \pi^0 \pi^+ \pi^-$ vertex and 
kaon-loop diagrams  
are suppressed by G-parity conservation and 
the large kaon mass, respectively. Therefore, 
relatively sizable contributions at higher order are expected.
At ${\mathcal O}(p^6)$ the 
counterterms of the chiral lagrangian can not be determined by 
experimental input only so that 
the prediction of the decay rate goes beyond pure ChPT constrained 
by experimental data. In Ref. \cite{Ametller:1991dp} large $N_C$ expansion 
and resonance saturation were used to determine the 
${\mathcal O}(p^6)$ LECs in the Lagrangian, obtaining 
$\Gamma(\eta \to \pi^0 \gamma \gamma) = 0.18\ $eV. 
The contribution 
from $a_0(980)$ was also estimated, 
$\Gamma(\eta \to \pi^0 \gamma \gamma)_{a_0} = 0.02$ eV but the 
relative phase with respect to the VMD term is unknown. 
The uncertainty from the sign of the interference term 
%
%
is quoted as the dominant contribution to the theoretical error at ${\mathcal O}(p^6)$, 
$\Gamma(\eta \to \pi^0 \gamma \gamma) = 0.18\ \pm 0.02\ $ eV or 
$BR(\eta \to \pi^0 \gamma \gamma) = (1.40 \pm 0.14) \times 10^{-4}$ . 
Higher order terms in ChPT have been also studied, e.g. in Ref.\cite{Ametller:1991dp} 
where all-order VMD counterterms have been added, obtaining 
$\Gamma(\eta \to \pi^0 \gamma \gamma) = ( 0.42 \pm 0.2 ) \ $ eV with the uncertainty again  
mostly from the sign of the interference term.   
%
On the experimental side, several measurements have been 
published since 1970s with smaller and smaller values of the branching fraction, 
decreased by three order of magnitude from 1967 to 1981 when  
%
%
the GAMS--2000 experiment 
measured $BR(\eta \to \pi^0 \gamma \gamma) = (7.1 \pm 1.4) \times 10^{-4}$.
In 2005 the 
AGS/Crystal Ball collaboration has obtained an even lower value, 
$BR(\eta \to \pi^0 \gamma \gamma) = (3.5 \pm 0.9) \times 10^{-4}$, updated 
to $BR(\eta \to \pi^0 \gamma \gamma) = (2.2 \pm 0.5) \times 10^{-4}$ 
in year 2008 with a revised analysis aiming at the optimization of the 
selection criteria and fitting procedure of the  
experimental spectra. 
The prelimary KLOE result  
\cite{DiMicco:2005rv} is 
$BR(\eta \to \pi^0 \gamma \gamma) = (8 \pm 3) \times 10^{-5}$, 
1.8\,--$\sigma$ lower than the ChPT prediction at ${\mathcal O}(p^6)$
and 2.2\--$\sigma$ lower than the AGS/CB measurement. 
Using the KLOE preliminary result on the branching fraction and the  
analysis efficiency obtained  
of $\sim 5\%$, 1,300 $\eta \to \pi^0 \gamma \gamma$ 
events are expected from the first year of data-taking 
at KLOE-2, thus allowing 
an accuracy of 3\% . 
Moreover, KLOE-2 can provide the  
the $m_{\gamma \gamma}$ distribution   
%
%
with sufficient precision to solve the ambiguity connected to the sign 
of the interference 
between VMD and scalar terms as 
shown in Fig. \ref{fig:experiments2}.

\subsubsection{$\eta -\eta'$ mixing and gluonium content}
\label{sec:gluonium}
The $\eta'$ meson, being almost a pure SU(3)$_{\small \rm Flavor}$ 
singlet, is considered a good candidate to host a gluon condensate.
\noindent
KLOE \cite{Ambrosino:2009sc} has extracted the $\eta'$ gluonium content 
and the $\eta$-$\eta'$ mixing angle 
according to the model of Ref.
\cite{Rosner:1982ey}. 
The $\eta$ and $\eta'$ wave functions can be decomposed in three terms: 
the $u,d$ quark wave function 
$\left | q \bar{q} \right > = \frac{1}{\sqrt{2}} 
\left(\left | u\bar{u} \right >+\left |d\bar{d} \right >\right)$,
the $\left | s \bar{s}\right >$ component and the gluonium 
$\left| GG \right>$.  The wave functions are written as:
\begin{eqnarray}
  \left |\eta'\right >  & = &  
  \mathrm{cos}\psi_G\,\mathrm{sin}\psi_P \left |q \bar{q} \right > +  
  \mathrm{cos}\psi_G\,\mathrm{cos}\psi_P \left |s \bar{s} \right > + 
  \mathrm{sin}\psi_G \left |GG \right> \nonumber \\
  \left |\eta \hphantom{'} \right>  & = & 
  \mathrm{cos}\psi_P \left |q \bar{q} \right >  - 
  \mathrm{sin}\psi_P \left |s \bar{s}  \right > \label{eq:etaquark}
\end{eqnarray}
where $\psi_P$ is the $\eta$-$\eta'$ mixing angle and 
$Z^2_G = \mathrm{sin}^2 \psi_G$ is the gluonium fraction in the 
$\eta'$ meson. The $Z^2_G$ parameter can be interpreted as the mixing 
with a pseudoscalar glueball. In Ref.\cite{Cheng:2008ss} it has been 
identified with the $\eta$(1405).
The gluonium fraction has been extracted fitting the widths of the magnetic 
dipole transition $V \to P \gamma$, where $V$ are the vector mesons
$\rho, \omega,\phi$ and $P$ the pseudoscalar mesons $\pi^0, \eta, \eta'$,  
together with the $\pi^0 \to \gamma \gamma$ and $\eta' \to \gamma \gamma$ 
partial widths. 
In particular, the KLOE measurement of 
$R_{\phi} = BR(\phi \to \eta' \gamma)/BR(\phi \to \eta \gamma)$ has 
been used, 
related to the gluonium content and mixing angle by \cite{Ambrosino:2006gk}:  
%
\begin{equation}
  \label{eq:mixing}
  R_{\phi } = \mathrm{cot}^{2}\psi_{P} \mathrm{cos}^2 \psi_{G} \left(
    1-\frac{m_s}{{\bar m}}\frac{Z_{q}}{Z_{s}}\frac{\mathrm{tan} \psi_V}{\mathrm{sin}2\psi_{P}}\right )^2
  \left( \frac{p_{\eta^{\prime}}}{p_{\eta}}  \right )^3
\end{equation}
where $p_{\eta'}$ and $p_{\eta}$ are the momenta of the $\eta'$ and 
$\eta$ meson respectively, $m_{s}/\bar{m} = 2m_{s}/(m_u+m_d)$ is the mass ratio of the  
constituent quarks 
and $\psi_{V}$ is the $\phi$-$\omega$  
mixing angle. The $Z_{q}$ and $Z_{s}$ parameters take into account hadronic 
uncertainties \cite{Escribano:2007cd}.
The evidence at 3\,--$\sigma$ level of 
a gluonium component in the $\eta'$ 
has been obtained, $Z^2_G = (12 \pm 4)$\%. A detailed study of the fit 
has shown that the measurement of the ratio 
$\Gamma(\eta' \to \gamma \gamma)/\Gamma(\pi^0 \to \gamma \gamma)$ is 
the most sensitive to the gluonium fraction. 

The theoretical frameworks to explain $\eta' \to \gamma \gamma$ 
and 
$V \to P \gamma$ transitions are slightly different, the first being related 
to quark-antiquark annihilation to two photons, the second 
to 
spin--flip transitions between the vector and pseudoscalar meson. 
Therefore it is important to obtain both, a sensitivity to the gluonium 
fraction 
independent from the $\eta' \to \gamma \gamma$ decay,  
and a new measurement of 
the $\eta' \to \gamma \gamma$ branching fraction. 
The $\eta' \to \gamma \gamma$ branching fraction is $\sim$2\%. Using 
BR($\phi \to \eta' \gamma$) $ \sim$6 $\times 10^{-5}$ and 
$\sigma(e^+ e^- \to \phi) = 3.3\, \mu$b, we expect 4 events/pb$^{-1}$.  
Unluckily this signal 
is overwhelmed by the $e^+ e^- \to \gamma \gamma$ background. With the 
available sample after one year running at KLOE-2, we will able 
to perform a selection based 40,000 events, sufficient to improve the 
fractional accuracy on BR($\eta^\prime \to \gamma \gamma$) to the 
per cent level.   

The $BR(\eta' \to \gamma \gamma)$ 
 limits the accuracy on the 
$\eta'$ decay width 
while the $\eta' \to \pi^+ \pi^- \eta$ and the $\eta' \to \pi^0 \pi^0 \eta$ 
branching ratios dominate the systematic error on the $R_{\phi}$ measurement 
because the $\eta'$ is identified through 
these decays. 
%
\begin{table}
  \begin{center}
    \caption{Fit to the gluonium content in $\eta'$ assuming 1\% error 
      on the $\eta'$ branching fractions. The 
      $\eta' \to \gamma \gamma/\pi^0 \to \gamma \gamma$ constraint is
      used (not used) in the left (right) column.}
    \begin{tabular}{c|c|c}
      & with $\eta' \to \gamma \gamma/\pi^0 \to \gamma \gamma$ & 
      without $\eta' \to \gamma \gamma/\pi^0 \to \gamma \gamma$ \\
      \hline
      
      $Z^2_{G}$ & 0.11 $\pm$ 0.03 & 0.11 $\pm$ 0.04 \\
      $\psi_P$ & (40.5 $\pm$ 0.6)$^\circ$ & 40.5 $\pm$ 0.6 \\
      $Z_{NS}$ & 0.93 $\pm$ 0.02 & 0.93 $\pm$ 0.03\\
      $Z_S$ & 0.83 $\pm$ 0.05 &   0.83 $\pm$ 0.05 \\
      $\psi_V$ & (3.32 $\pm$ 0.08)$^\circ$ & (3.32 $\pm$ 0.09) \\
      $m_s/\bar{m}$ & 1.24 $\pm$ 0.07 & 1.24 $\pm$ 0.07 \\
      \hline
    \end{tabular}
    \label{tab:KLOE2}
  \end{center}
\end{table}
Table \ref{tab:KLOE2} shows how the evidence for a gluonium content in the 
$\eta'$ could be confirmed also independently from the 
$\Gamma(\eta^\prime \to \gamma \gamma)$/$\Gamma(\pi^0 \to \gamma \gamma)$ 
ratio  if the fractional accuracy on the 
$\eta^\prime$ partial widths is brought to 1\% level. We have assumed 
that 
the actual branching fractions and the correlation matrix of the measurements 
are those obtained from the present fit.   
%

\begin{figure}
  \begin{center}
    \resizebox{0.8\columnwidth}{!}{\includegraphics{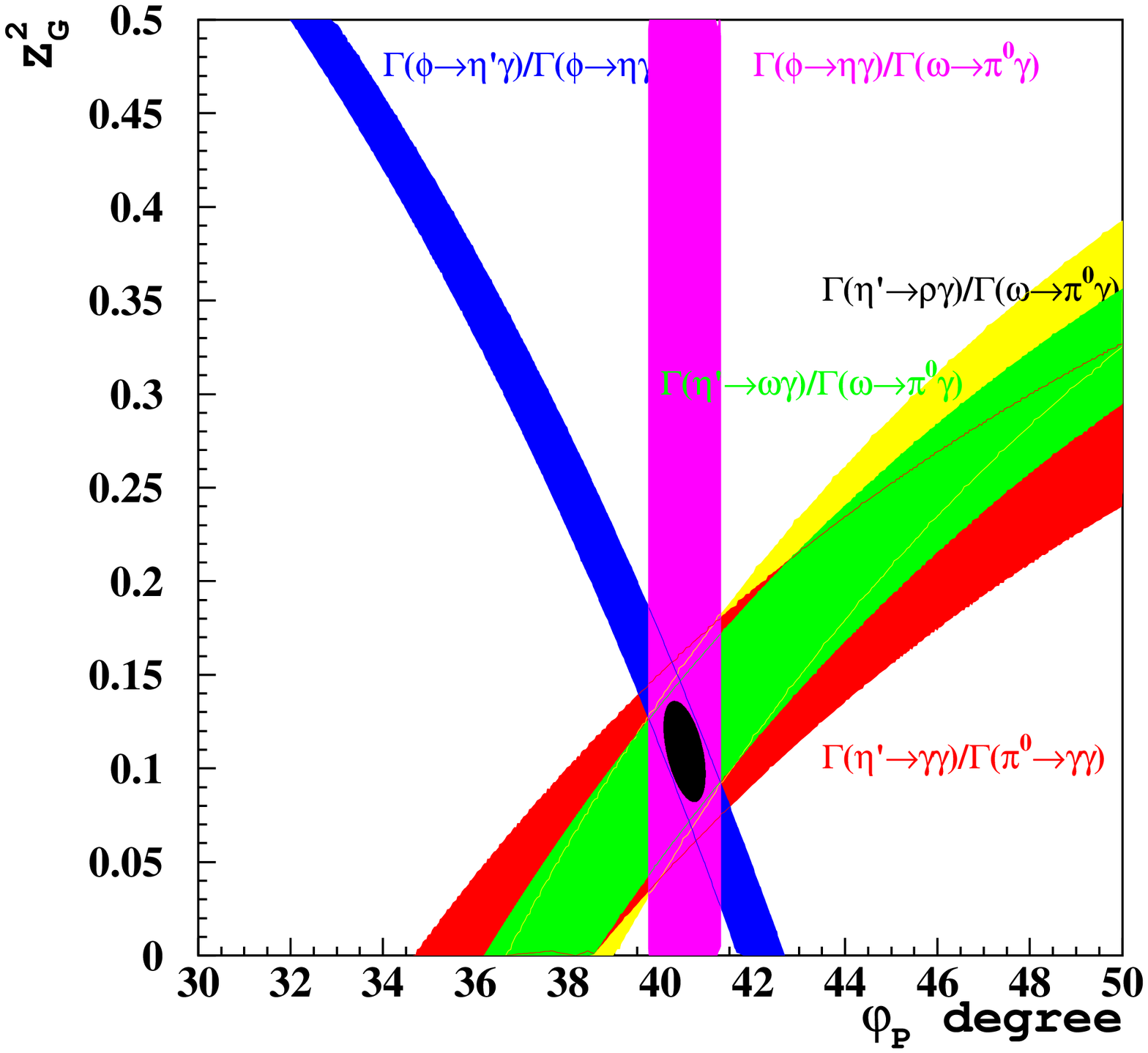}}
    \resizebox{0.8\columnwidth}{!}{\includegraphics{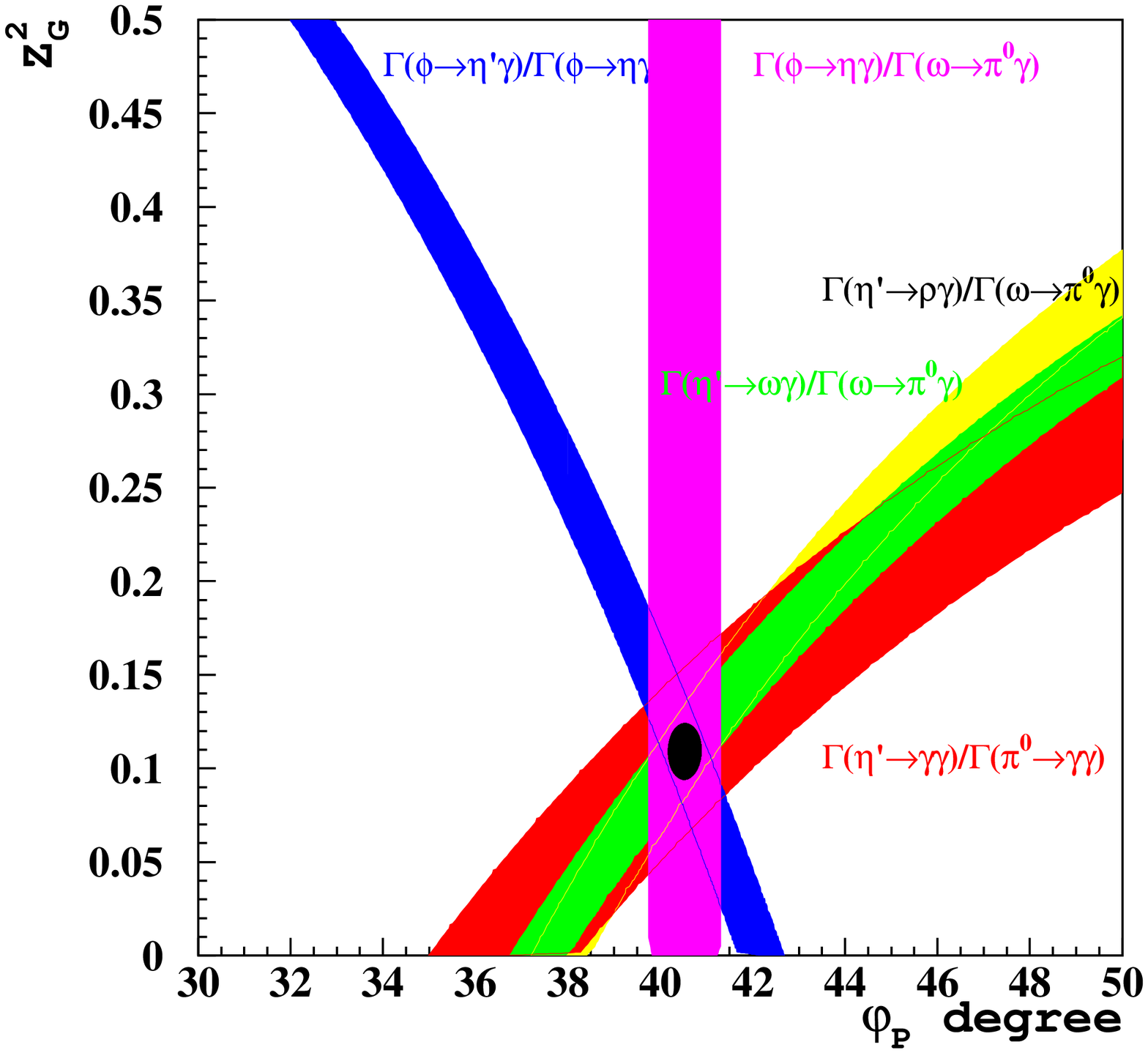}}
  \end{center}
  \caption{The 68\% C.L. region in 
the $\psi_P,Z^2_G$ plane.
Top: only the $\eta^\prime$ branching ratios are 
    improved to 1\% precision. 
Bottom: the $\eta'$ total width is also lowered to 1.4\%.} 
  \label{fig:contKLOE2}
\end{figure}

With the KLOE-2 data-taking above the $\phi$ peak, e.g., 
at  $\sqrt{s} \sim 1.2$ GeV,  
it is possible to measure the $\eta'$ 
decay width $\Gamma(\eta' \to \gamma \gamma)$ through 
$\sigma(e^+ e^- \to e^+ e^- (\gamma^* \gamma^*) \to e^+ e^- \eta')$, 
as discussed in Sect. \ref{sec:gg}.
The measurement to 1\% level of both the cross section and the 
$BR(\eta' \to \gamma \gamma)$ 
would bring the fractional error on the 
$\eta'$ total width, 
$\Gamma_{\eta'} = \Gamma(\eta' \to \gamma \gamma) /
BR(\eta' \to \gamma \gamma)$, 
%
%
to $ \sim$1.4\%.
In Fig. \ref{fig:contKLOE2} the 68\% C.L. region in 
the $\psi_P,Z^2_G$ plane obtained with the improvements discussed in this subsection is shown.
The comparison of the top to bottom panels makes evident how the fit 
accuracy increases  
with the  precision measurement of the $\eta^\prime$ total width.

\subsection{Low--Mass Scalars}
\label{sec:scalar}

The radiative processes $\phi\rightarrow\gamma S,$ where $S=f_{0}(980),$
$a_{0}(980),$ $\sigma\equiv f_{0}(600),$ are followed by a decay of $S$ into
two pseudoscalar mesons $(PP)$. The 
analysis of the $M_{PP}$ invariant mass distribution in  
$\phi\rightarrow\gamma PP$ transitions  
is sensitive to the nature of light scalar states. 
If they are tetraquarks, quarkonia or molecules 
is one of the open questions of 
low-energy QCD. 
Interestingly, the tetraquark assignment \cite{Jaffe:1976ig,Maiani:2004uc} 
naturally explains the mass pattern and decay widths, although different
instanton-driven \cite{Hooft:2008we} or quark-antiquark 
processes \cite{Giacosa:2006rg} have been considered.

Two kinds of theoretical models 
for the $\phi\rightarrow\gamma S\rightarrow\gamma PP$ 
have been analyzed: (i) the `no structure' approach based on a point-like 
$\phi\gamma S$ coupling 
\cite{Isidori:2006we,Black:2006mn,Giacosa:2008st}; 
(ii) 
the kaon-loop coupling of the  $\phi$ to $\gamma S$ 
\cite{Achasov:2005hm}. Future analyses will 
determine which 
way is dominant and 
the decay amplitudes of the scalar states. Moreover, the precision study 
of the $S\rightarrow\gamma\gamma$ reactions at KLOE-2 
as presented in Sect. \ref{sec:gg} would  
add 
information relevant for the underlying quark dynamics.

The solution of the scalar puzzle, which has stimulated many other theoretical 
studies, e.g. 
\cite{Bramon:2002iw,Escribano:2003xa,Oller:2002na,Ivashyn:2009te},   
can also shed light on the 
mechanism of chiral symmetry restoration at non--zero temperature and 
density, where scalar mesons, which carry the quantum numbers of 
the vacuum, play a central role \cite{Heinz:2008cv}.

\subsubsection{Structure of $f_0(980)$ and $a_0(980)$}

\begin{figure}
  \begin{center}
    \resizebox{0.8\columnwidth}{!}{\includegraphics{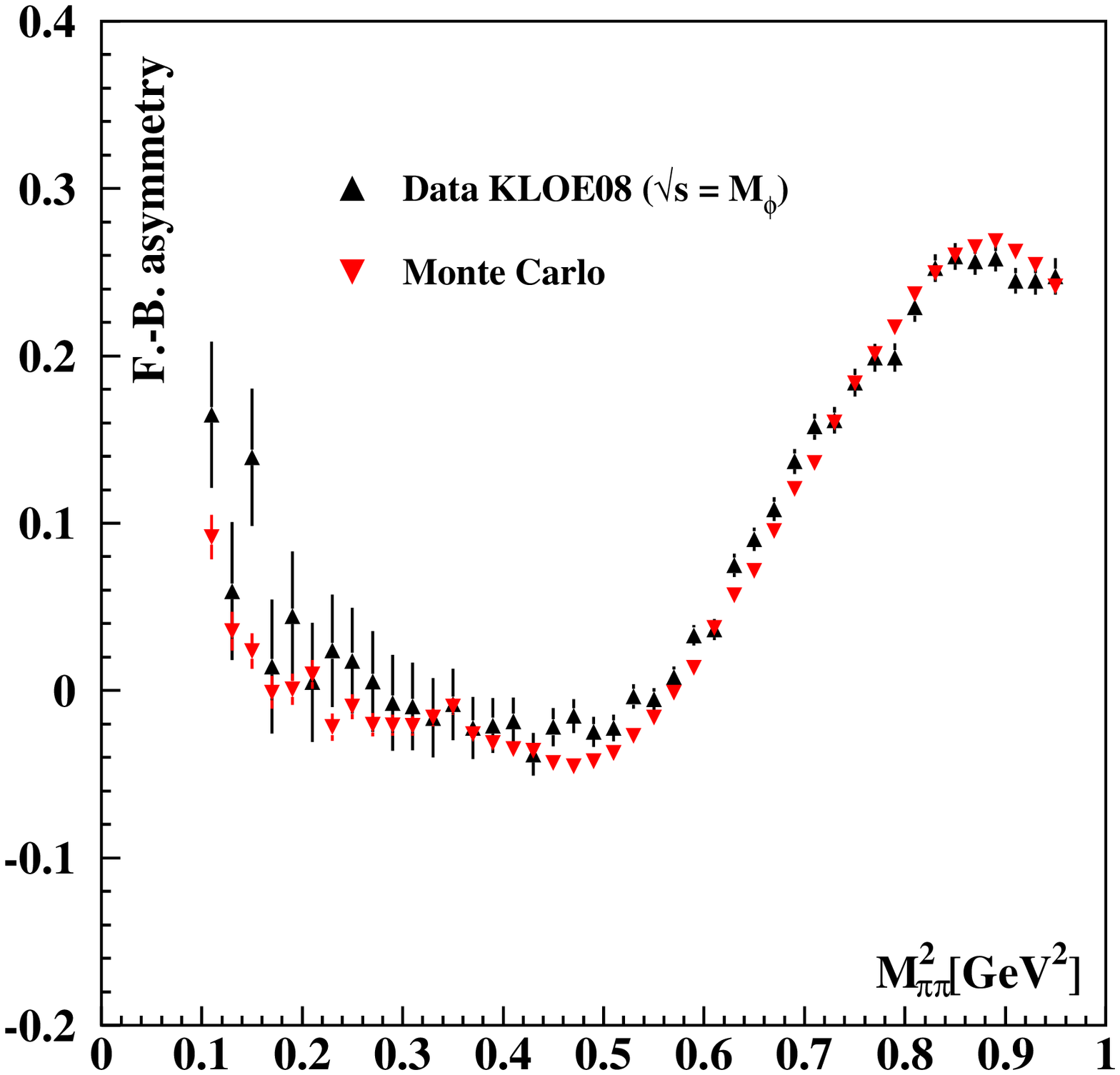}}
    \resizebox{0.8\columnwidth}{!}{\includegraphics{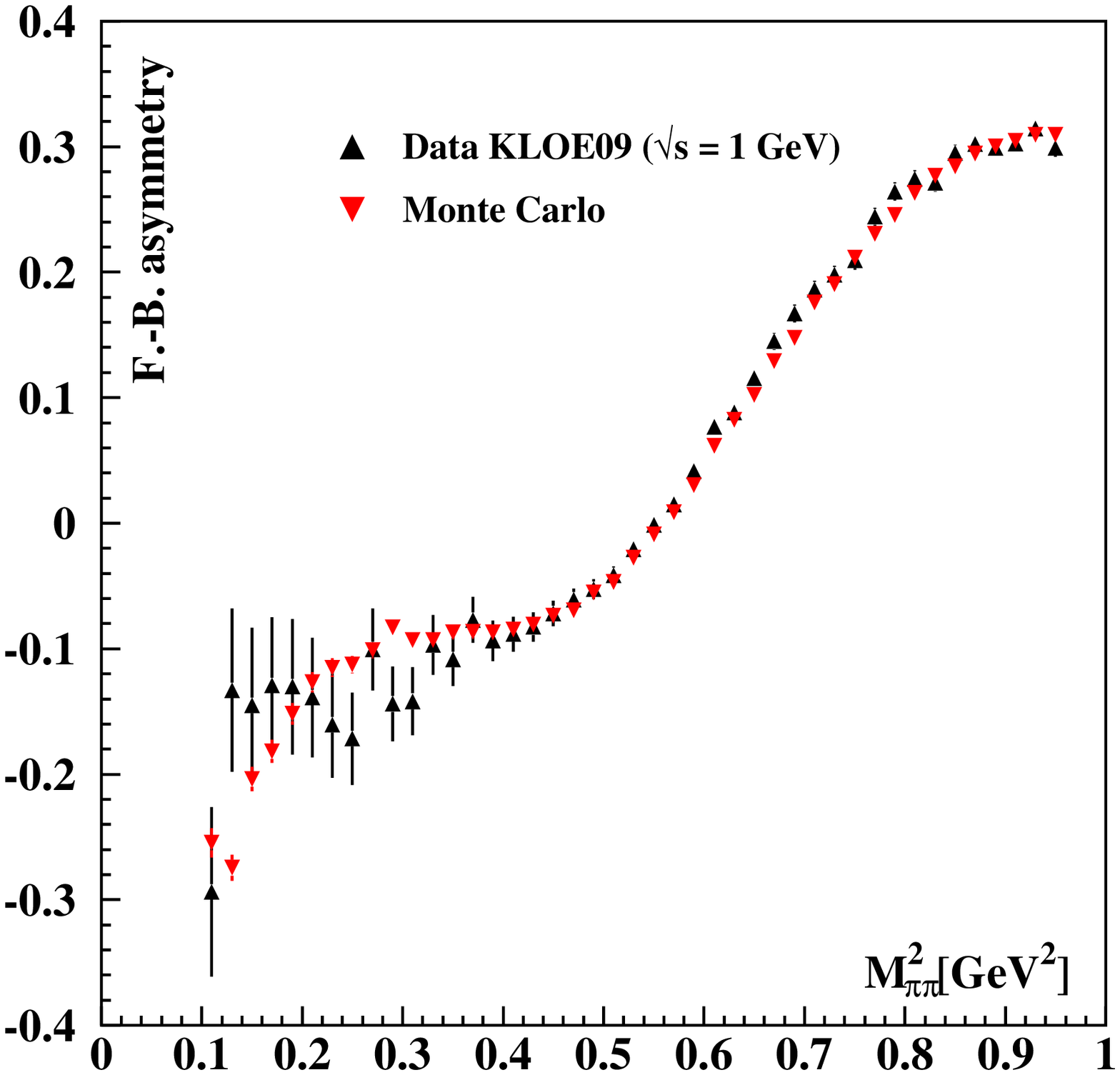}}
  \end{center}
  \caption{Forward-backward asymmetry, $A_{FB}$, 
 as a function of $M_{\pi\pi}^2$ in 
 $e^+ e^-\to \pi^+ \pi^- \gamma$ events 
    for data at $\sqrt{s} = m_{\phi}$ (top) and at 
    $\sqrt{s} = 1$ GeV (bottom).} 
  \label{fig:chargeasymmetry}
\end{figure}

The KLOE experiment has measured the branching ratio and the spectral 
function of the decays $\phi \to S \gamma \to \pi^0 \pi^0 \gamma,\quad   
\pi^0 \eta \gamma,\quad \pi^+ \pi^- \gamma$ 
\cite{Ambrosino:2006hb,Ambrosino:2009py,Ambrosino:2005wk} obtaining  
within the Kaon Loop phenomenological model  
the coupling of scalar mesons to the $\pi\pi$ and $\eta \pi^0$ 
pairs.
While in the latter only the $a_0(980)$ can contribute, in the 
$\pi\pi$ channel both, the $f_0(980)$, and the $\sigma$ whose existence has 
been debated since a long time, are expected.
%
KLOE has fitted the $\pi^0 \pi^0$ invariant mass fixing the parameters of 
the $\sigma$ meson from other measurements and phenomenological fits. 
Data from $\pi^+ \pi^-$ phase shift were also taken into account. 
%
%
%
For the charged channel, due to the large background from the 
$e^+ e^- \to \pi^+ \pi^- \gamma$ dominated by the $\rho$ pole, no 
sensitivity to the $\sigma$ meson is shown by the fit. The present 
analysis of KLOE data is based on $\sim$500 pb$^{-1}$.
KLOE-2 will have a 20 times larger sample, so that the $\sigma$ could 
also show up in the $\pi^+ \pi^-$ channel.
Moreover, the $\pi^+ \pi^-$ forward-backward asymmetry ($A_{FB}$) is expected to be 
affected by the presence of the $\sigma$. This asymmetry is 
generated by the different behaviour under C-conjugation of the 
$\pi^+ \pi^-$ amplitude for initial state radiation (ISR) with respect 
to the other contributions (final state radiation (FSR) 
and $e^+ e^- \to \phi \to$ ($f_0(980) + \sigma$) $\gamma$. 
In Fig. \ref{fig:chargeasymmetry} $A_{FB}$ is shown for two sets of KLOE 
data, collected at $\sqrt{s} = m_{\phi}$ and $\sqrt{s} = 1$ GeV. 
In the second case no scalar should be present due to the low value of 
the $e^+ e^- \to \phi$ cross section. 
A combined fit to the invariant mass spectrum and asymmetry 
is of interest at KLOE-2
for the study of the properties of 
the $f_0(980)$ and $\sigma$ mesons.

The set of scalar couplings to the pseudoscalars can be used to determine 
the structure of the scalar meson in the naive $SU(3)$ hypothesis. In 
Tab. \ref{tab:su3couplings}, the SU(3) predicted couplings with different 
quark structure of the pseudoscalars is shown compared with the KLOE 
results.
\begin{table*}
  \begin{center}
    \caption{Comparison of the measured couplings to the SU(3) 
      predictions for different quark-structure hypotheses.}
\vspace{2mm} 
    \begin{tabular}{|c|c|c|c|c|}
      \hline
      & KLOE & \multicolumn{3}{|c|}{SU(3)} \\
      \hline
      &      & 4q & q$\bar{q}$ $f_0 = s \bar{s}$ & $q\bar{q}$ $f_0 = (u\bar{u}+d\bar{d})/\sqrt{2}$ \\
      \cline{3-5}
      $(g_{a_0 K^+K^-}/g_{a_0 \eta\pi})^2$ & 0.6 - 0.7 & 1.2-1.7 & 0.4 & 0.4 \\
      $(g_{f_0 K^+K^-}/g_{f_0 \pi^+ \pi^-})^2$ & 4.6 - 4.8 & $>> 1$ &  $>> 1$ & 1/4 \\
      $(g_{f_0 K^+K^-}/g_{a_0 K^+ K^-})^2$ & 4 - 5 & 1 & 2 & 1 \\
    & & & & \\
    \hline
    \end{tabular} 
    \label{tab:su3couplings}
  \end{center}
\end{table*}
%
The $4$--quark model predicts values larger than observed for 
$g_{a_0K^+K^-}/g_{a_0 \eta\pi}$ while the $2$--quark model can be accommodated 
only if $f_0 = s \bar{s}$, which is disfavoured by the mass degeneracy
of the two mesons. 
The ratio of $f_0$ to $a_0$ couplings to $K^+ K^-$ is higher with 
respect to all predictions.
\subsubsection{Structure of the $\sigma$ meson}
\begin{figure}
  \begin{center}
 \resizebox{0.8\columnwidth}{!}{\includegraphics{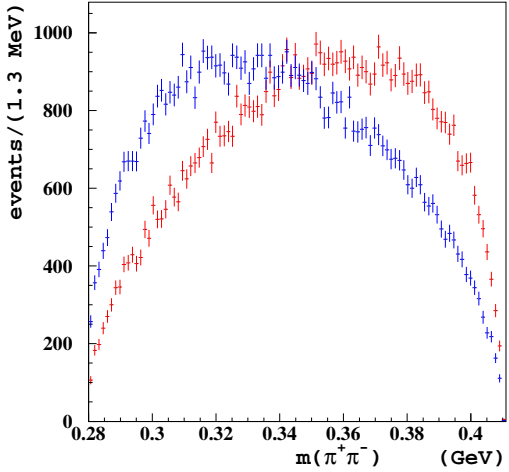}}
  \end{center}
\vspace{-2mm}
  \caption{The $m_{\pi^+ \pi^-}$ distribution 
in the $\eta' \to \eta \pi^+ \pi^-$ 
    decay with the $\sigma$ meson (right--centered distribution) and without 
(left--centered distribution) contribution.} 
  \label{fig:etapsigmeson}
\end{figure}

In the $\eta'\to \pi^+ \pi^- \eta$ and $\eta' \to \pi^0 \pi^0 \eta$, 
the $\pi\pi$ system is produced mostly with scalar quantum 
numbers. Indeed, the available kinetic energy of the $\pi^+ \pi^-$ 
pair is [0, 137] MeV, suppressing high angular momentum contribution.
Furthermore, the exchange of vector mesons is forbidden by G-parity 
conservation. For these reasons, only scalar mesons can participate to 
the scattering amplitude. The decay can be mediated by the $\sigma$, 
$a_0(980)$ and $f_0(980)$ exchange and by a direct contact term due to 
the chiral anomaly \cite{Fariborz:1999gr}. 
The scalar contribution can be determined by fitting the Dalitz plot of 
the $\eta' \to \pi \pi \eta$ system. The golden channel for KLOE-2 is the 
decay chain $\eta' \to \pi^+ \pi^- \eta$, with $\eta \to \gamma \gamma$.
The signal can be easily identified requiring the $\eta$ and $\eta'$ 
invariant masses. Such final state was already studied at KLOE to measure
the branching fraction of the $\phi \to \eta' \gamma$ decay 
\cite{Aloisio:2002vm}. The analysis efficiency was 22.8\%, with 10\% 
residual background contamination. 
With ${\mathcal O}(10)$ fb$^{-1}$, we expect 
80,000 fully reconstructed events. Following Ref.\cite{Fariborz:1999gr} 
we have written the matrix element of the decay and built a MC generator. 
In Fig. \ref{fig:etapsigmeson} the $m_{\pi^+ \pi^-}$ invariant mass 
distribution is shown with and without the $\sigma$ contribution with the 
expected KLOE-2 statistics. 
Good sensitivity to the parameters 
of the $\sigma$ meson is therefore expected from the 
study of this channel.

\subsubsection{$\phi \to K^0 \bar{K}^0 \gamma$}

The $\phi$ meson can decay 
to the $a_0(980)$ and $f_0(980)$ scalars with the emission of one photon.
%
%
Both of them couples to $K^+ K^-$ and $K^0 \bar{K}^0$, so that the 
decay $\phi \to K^0\bar{K}^0\gamma$ is expected to proceed through 
an $[f_0(980) + a_0(980)] \gamma$ intermediate state. SU(2) relates 
the coupling constant to the pseudoscalars in a clean way: 
$g_{f_0 \pi^+ \pi^-}   = 2 g_{f_0 \pi^0 \pi^0}$, 
$g_{f_0 K^0 \bar{K}^0} =   g_{f_0 K^+ K^-}$, 
$g_{a_0 K^0 \bar{K}^0} = - g_{a_0 K^+ K^-}$.
Due to the opposite sign in $g_{f_0 K^0\bar{K}^0} \cdot g_{f_0 K^+K^-}$ 
and $g_{a_0 K^0\bar{K}^0} \cdot g_{a_0 K^+K^-}$, destructive interference 
between $f_0(980)$ and $a_0(980)$ is expected. The scalars decay in an 
even combination of 
$\left |K^0 \right > \left |\bar{K}^0\right >$:  
$\frac{|K^0>\bar{K}^0>+|\bar{K}^0>K^0>}{\sqrt{2}}$ = $ 
\frac{|K_S>|K_S>+|K_L>K_L>}{\sqrt{2}}$.
The channel with two $K_S$ in the final state can be easily identified 
through the $K_S \to \pi^+ \pi^-$ decay, looking for 4 tracks pointing 
to the IP. The main background comes from $\phi \to K_S K_L$ 
events with a $CP$--violating decay $K_L \to \pi^+ \pi^-$. 
In KLOE, the $K_S$ decay path is 6 mm, while for the $K_L$ is $\sim 3.4$ m. 
Therefore the $K_L$ vertices are uniformly distribuited in a small region 
around the IP. Cuts on the vertex position are the most 
effective way 
to remove $K_S K_L$ background. 
KLOE \cite{Ambrosino:2009rg} has already analyzed a sample of 2.2 fb$^{-1}$ 
of data, observing 5 events with $3.2 \pm 0.7$ expected background. The
result is $BR(\phi \to K^0\bar{K}^0 \gamma) < 1.9 \times 10^{-8}$, 
at 90\% C.L.. Scaling 
these numbers with the KLOE-2 statistics we expect to reach a sensitivity 
of $BR(\phi \to K^0\bar{K}^0 \gamma ) < 1 \times 10^{-8}$.
The inner tracker will provide in the second phase of the experiment 
three times better vertex  
resolution which is beneficial for the rejection capability and together 
with 20 fb$^{-1}$ of integrated luminosity could lead to first 
observation of the decay.     
%


%

%
\section{Physics in the Continuum: $\sigma_{had}$}
\label{sec:hadcs}
%

In this section we discuss the physics reach of the  
DA$\mathrm{\Phi}$NE running outside the $\phi$--meson peak. 
We consider a maximal energy of $\sqrt{s}=2.5$ GeV with a luminosity 
of $\sim$10$^{32}$ cm$^{-2}$ s$^{-1}$ already exceeded 
by DA$\mathrm{\Phi}$NE at the $\phi$ peak.
With such a machine one can collect an integrated luminosity of 
${\mathcal O}(10)$ fb$^{-1}$ between 1 and 2.5 GeV in a few years 
of data taking.
This high statistics,  much larger than what collected at
 any collider in this energy range, 
would allow major improvements in physics, 
with relevant implications for the precision tests of the SM, such as the  
$g-2$ of the muon and the effective fine-structure constant 
at the $M_Z$ scale,  
 $\alpha_{em}(M^2_Z)$. 
The only direct competitor 
is VEPP-2000 at Novosibirsk, 
which will cover 
the center-of-mass energy range between 1 and 2 GeV with two experiments. 
VEPP-2000 
is expected to start by year 2010 with a luminosity 
from $10^{31}$cm$^{-2}$s$^{-1}$ at 1 GeV to  
$10^{32}$cm$^{-2}$s$^{-1}$ at 2 GeV, as presented in more detail in 
Sect. \ref{sec:vepp2000}.
Other indirect competitors are the higher--energy 
$e^+e^-$ colliders ($\tau$-charm and B-factories) which in principle
can cover the same 
energy range by means of radiative return.
However, due to the photon emission, 
the ``equivalent'' luminosity produced by these machines
 in the region between 1 and 2.5 GeV is
much less than what proposed in the 
KLOE-2 programme.

In the following subsections we present the main physics motivations for the 
 off-peak running.
 We start with the improvements on 
the cross sections $\sigma(e^+e^-$ $\to$ {\it hadrons}) 
in a wide center-of-mass energy range, from the
$\pi\pi$ threshold up to 2.5 GeV 
discussing the implications for precision
tests of the SM (Sect. \ref{subsubsec:SMTESTS}) and vector--meson
spectroscopy (Sect. \ref{Vecto}).
The physics reach with the study of $\gamma \gamma$ processes is presented  
in Sect. \ref{sec:gg}. 
%
\subsection{SM precision tests and $\sigma_{had}$ at low energy}
\label{subsubsec:SMTESTS}

The comparison of the SM 
predictions with precision data served in the last few
decades as an invaluable tool to test this theory at the quantum
level. It has also provided stringent constraints on many NP 
scenarios.  The remarkable agreement between the
precision measurements of electroweak observables and their 
SM predictions is a striking experimental confirmation of the 
theory, even if there are a few observables where the agreement is not
so satisfactory.  On the other hand, the Higgs boson has not yet been
observed and there are clear phenomenological facts (e.g., dark
  matter, matter-antimatter asymmetry in the universe) as well as
strong theoretical arguments hinting at the presence of physics
beyond the SM. Future colliders, such as the 
  LHC or the ILC 
will hopefully answer many such questions offering at the same time,
great physics potential, and a new challenge to provide even more
precise theoretical predictions.

\begin{figure}[ht]
\begin{center}
\resizebox{0.8\columnwidth}{!}{
\includegraphics[width=8cm]{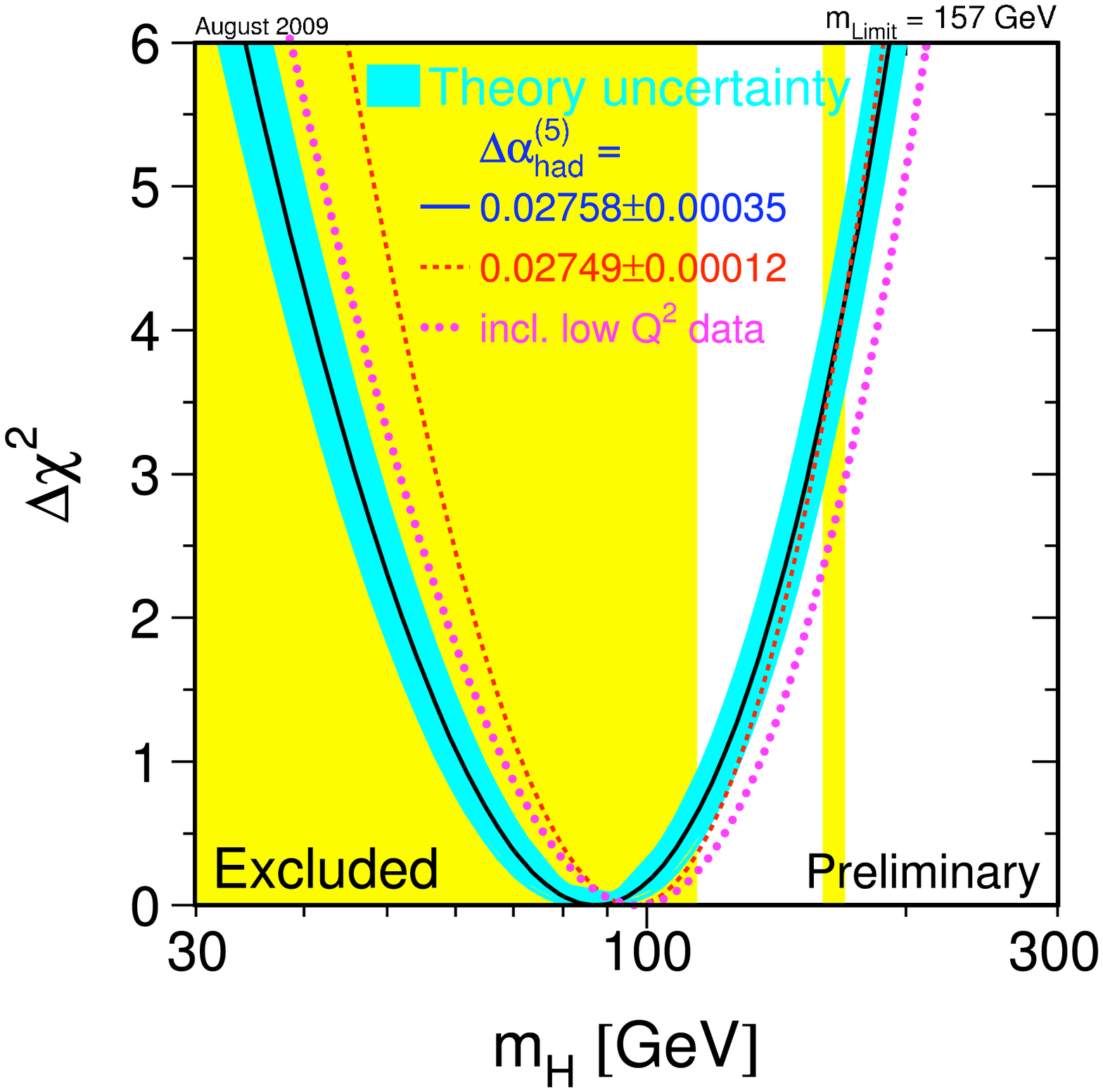}}
\caption{The line is the result of the Electroweak Working Group fit
  using all data~\cite{LEPEWWG}; the band represents an estimate of
  the theoretical error due to missing higher order corrections.  The
  vertical bands show the 95\% C.L. exclusion limits on
  $M_{\scriptscriptstyle{H}}$ from the direct searches.}
\label{fig:blueband}
\end{center}
\end{figure}

Precise SM predictions require precise input
parameters. Among the three basic input parameters of the electroweak
({\small EW}) sector of the SM -- the fine-structure constant
$\alpha$, the Fermi coupling constant $G_F$ and the mass of the $Z$
boson -- $\alpha$ is by far the most precisely known, determined
mainly from the anomalous magnetic moment of the electron with an
amazing relative precision of 0.37 parts per billion
(ppb) \cite{Hanneke:2008tm}. However, physics at non-zero squared
momentum transfer $q^2$ is actually described by an effective
electromagnetic coupling $\alpha(q^2)$ rather than by the low-energy
constant $\alpha$ itself. The shift of the fine-structure constant
from the Thomson limit to high energy involves non-perturbative
hadronic effects which spoil this fabulous precision. Indeed, the
present accuracies of these basic parameters
are~\cite{Hanneke:2008tm,Amsler:2008zzb,Jegerlehner:2003rx,Jegerlehner:2006ju,Jegerlehner:2008rs}
\begin{eqnarray}
  \frac{\delta \alpha}{\alpha} \sim 4 \times 10^{-10}, &&
  \frac{\delta \alpha(M_{\scriptscriptstyle{Z}}^2)}
{\alpha(M_{\scriptscriptstyle{Z}}^2)} \sim {\mathcal O}(10^{-4}), 
\nonumber \\  
  \frac{\delta M_{\scriptscriptstyle{Z}}}
{M_{\scriptscriptstyle{Z}}} \sim 2 \times 10^{-5},  &&
  \frac{\delta G_F }{G_F} \sim 4 \times 10^{-6}. \nonumber
\end{eqnarray}
The relative uncertainty of $\alpha(M_{\scriptscriptstyle{Z}}^2)$ is
roughly one order of magnitude worse than that of
$M_{\scriptscriptstyle{Z}}$, making it one of the limiting factors in
the calculation of precise {\small SM} predictions.
The effective fine-structure constant at the scale
$M_{\scriptscriptstyle{Z}}$ 
$\alpha(M_{\scriptscriptstyle{Z}}^2)$
 plays a
crucial role in basic {\small EW} radiative corrections of the {\small
  SM}. An important example is the {\small EW} mixing parameter
$\sin^2 \!\theta$, related to $\alpha$, $G_F$ and
$M_{\scriptscriptstyle{Z}}$ via the Sirlin
relation~\cite{Marciano:1980pb,Sirlin:1980nh,Sirlin:1989uf,Degrassi:1990tu,Ferroglia:2001cr}
\begin{equation}
  \sin^2 \!\theta_{\scriptscriptstyle{S}} \cos^2 \!\theta_{\scriptscriptstyle{S}} = 
  \frac{\pi \alpha}
{\sqrt 2 G_F M_{\scriptscriptstyle{Z}}^2 (1-\Delta r_{\scriptscriptstyle{S}})},
\label{eq:sirlin}
\end{equation}
where the subscript $S$ identifies the renormalization scheme. $\Delta
r_{\scriptscriptstyle{S}}$ incorporates the universal correction
$\Delta \alpha(M_{\scriptscriptstyle{Z}}^2)$, large contributions that
depend quadratically on the top quark mass~\cite{Veltman:1977kh} (which led
to its indirect determination before the discovery of this quark at
the Tevatron~\cite{Abe:1995hr,Abachi:1995iq}), plus all remaining quantum
effects. 
Neglecting the latter sub-leading corrections,
{\small SM} predictions of precision observables which have been
measured at {\small LEP} would fail at the 10--$\sigma$ level.
In the {\small SM}, $\Delta r_{\scriptscriptstyle{S}}$ depends on
various physical parameters such as $\alpha$, $G_F$,
$M_{\scriptscriptstyle{Z}}$, $M_{{\scriptscriptstyle{W}}}$,
$M_{\scriptscriptstyle{H}}$, $m_f$, etc., where $m_f$ stands for a
generic fermion mass. As $M_{\scriptscriptstyle{H}}$, the mass of the
Higgs boson, is the only relevant unknown parameter in the {\small
  SM}, important indirect bounds on this missing ingredient can be set
by comparing the calculated quantity in Eq.(\ref{eq:sirlin}) with the
experimental value of $\sin^2 \!\theta_{\scriptscriptstyle{S}}$. These
constraints can be easily derived using the simple formulae of
Refs. \cite{Degrassi:1997iy,Degrassi:1999jd,Ferroglia:2002rg,Awramik:2003rn,Awramik:2004ge,Awramik:2006uz}, which relate the effective {\small EW} mixing
angle $\sin^2 \!\theta_{\rm eff}^{\rm lept}$ (measured at {\small LEP}
and {\small SLC} from the on-resonance asymmetries) with
$\Delta\alpha(M_{\scriptscriptstyle{Z}}^2)$ and other experimental
inputs such as the mass of the top quark. It is important to note that an
error of $\delta \Delta\alpha(M_{\scriptscriptstyle{Z}}^2) = 35 \times
10^{-5}$~\cite{Burkhardt:2005se} in the effective electromagnetic
coupling constant dominates the uncertainty of the theoretical
prediction of $\sin^2 \!\theta_{\rm eff}^{\rm lept}$, inducing an
error $\delta(\sin^2 \!\theta_{\rm eff}^{\rm lept}) \sim 12 \times
10^{-5}$ (which is not much smaller than the experimental one
$\delta(\sin^2 \!\theta_{\rm eff}^{\rm lept})^{\scriptscriptstyle \rm
  EXP} = 16 \times 10^{-5}$ determined by {\small LEP} and {\small
  SLD}~\cite{leplsd:2005ema,LEPEWWG}) and affecting the Higgs boson mass
bound. Moreover, as measurements of the effective {\small EW} mixing
angle at a future linear collider may improve its precision by one
order of magnitude~\cite{Weiglein:2004hn}, a much smaller value of
$\delta\Delta\alpha(M_{\scriptscriptstyle{Z}}^2)$ will be required 
as discussed in the next subsection.
It is therefore crucial to assess all viable
options to further reduce this uncertainty.  The latest global fit of
the {\small LEP} Electroweak Working Group, which employs the complete
set of {\small EW} observables, leads at present to the value
$M_{\scriptscriptstyle{H}} = 87^{+35}_{-26}$GeV and the 
95\% C.L. 
upper limit of 
157~GeV  (Fig. \ref{fig:blueband}) \cite{LEPEWWG}. This limit increases to 
186~GeV when including the {\small LEPII} direct search lower limit of
114~GeV.
%
\subsection{The fine-structure constant at the $M_Z$ scale}
\label{subsubsec:ALPHAEFF}
Let us examine the running of the effective
fine-structure constant to the scale $M_{\scriptscriptstyle{Z}}$, that
can be defined by
$ \Delta 
\alpha(M_{\scriptscriptstyle{Z}}^2) = 4\pi\alpha \mbox{Re}[\Pi^{(f)}_{\gamma
  \gamma}(0)- \Pi^{(f)}_{\gamma \gamma}(M_{\scriptscriptstyle{Z}}^2)],
$
where $\Pi^{(f)}_{\gamma\gamma}(q^2)$ is the fermionic part of the
photon vacuum polarization function (with the top quark
decoupled). Its evaluation includes hadronic contributions where
long-distance {\small QCD} 
cannot be calculated analytically.
These contributions cause the aforementioned dramatic loss of
accuracy, by several orders of magnitude which occurs moving from the
value of $\alpha$ at vanishing momentum transfer to that at
$q^2=M_{\scriptscriptstyle{Z}}^2$. The shift $\Delta
\alpha(M_{\scriptscriptstyle{Z}}^2)$ can be divided in two parts:
$\Delta\alpha(M_{\scriptscriptstyle{Z}}^2) = \Delta\alpha_{\rm
  lep}(M_{\scriptscriptstyle{Z}}^2) + \Delta \alpha_{\rm
  had}^{(5)}(M_{\scriptscriptstyle{Z}}^2)$. The leptonic contribution
is calculable in perturbation theory and known up to three--loops:
$\Delta\alpha_{\rm lep}(M_{\scriptscriptstyle{Z}}^2) = 3149.7686\times
10^{-5}$~\cite{Steinhauser:1998rq}. 
The hadronic contribution $\Delta \alpha_{\rm
  had}^{(5)}(M_{\scriptscriptstyle{Z}}^2)$ of the five quarks
($u$, $d$, $s$, $c$, and $b$) can be computed from hadronic $e^+ e^-$
annihilation data via the dispersion relation~\cite{Cabibbo:1961sz}: 
\begin{equation} 
  \Delta \alpha_{\rm had}^{(5)}(M_{\scriptscriptstyle{Z}}^2) = 
-\left(\frac{\alpha M_{\scriptscriptstyle{Z}}^2}{3\pi}
  \right) \mbox{Re}\int_{4m_\pi^2}^{\infty} ds 
\frac{R(s)}{s(s- M_{\scriptscriptstyle{Z}}^2
  -i\epsilon)},
\label{eq:delta_alpha_had}
\end{equation}
where $R(s) = \sigma^{(0)}(s)/(4\pi\alpha^2\!/3s)$ and
$\sigma^{(0)}\!(s)$ is the total cross section for $e^+ e^-$
annihilation into any hadronic state, with vacuum polarization  
and initial--state {\small QED} 
corrections subtracted off.  In the 1990s, detailed evaluations of
this dispersive integral have been carried out by several
authors~\cite{Eidelman:1995ny,Burkhardt:1995tt,Martin:1994we,Swartz:1995hc,Davier:1997vd,Davier:1998si,Kuhn:1998ze,Groote:1998pk,Erler:1998sy,Martin:2000by,Alemany:1997tn}. More recently, some of these analyses
were updated to include new $e^+ e^-$ data -- mostly from 
CMD-2~\cite{Akhmetshin:2003zn} and BESII~\cite{Bai:2001ct}
-- obtaining:
$
\Delta\alpha_{\rm had}^{(5)} = 2761 \, (36) \times 10^{-5}
$
\cite{Burkhardt:2001xp},
$
\Delta\alpha_{\rm had}^{(5)} = 2757 \, (36) \times 10^{-5}
$
\cite{Jegerlehner:2001wq},
$
\Delta\alpha_{\rm had}^{(5)} = 2755 \, (23) \times 10^{-5}
$
\cite{Hagiwara:2003da}, and
$
\Delta\alpha_{\rm had}^{(5)} = 2749 \, (12) \times 10^{-5}
$
\cite{deTroconiz:2004tr}. 
The reduced uncertainty of the latter result has been obtained
making stronger use of theoretical inputs.
The reduction, by a factor of two, of the uncertainty quoted
in Ref. \cite{Eidelman:1995ny}   
($70 \times 10^{-5}$) with
respect to that in~\cite{Jegerlehner:2001wq} 
($36 \times 10^{-5}$) is mainly due to
the data from BESII. The latest updates,
$
\Delta\alpha_{\rm had}^{(5)} = 2758 \, (35) \times 10^{-5}
$
\cite{Burkhardt:2005se}, 
$
\Delta\alpha_{\rm had}^{(5)} = 2768 \, (22) \times 10^{-5}
$
\cite{Hagiwara:2006jt}, and
$
\Delta\alpha_{\rm had}^{(5)} = 2759.4 \, (21.9) \times 10^{-5}
$
\cite{Jegerlehner:2008rs}, include also the KLOE 
measurements \cite{Aloisio:2004bu}.  
In particular, the latest 
CMD-2~\cite{Akhmetshin:2006wh,Akhmetshin:2006bx,Aulchenko:2006na} and 
SND~\cite{Achasov:2006vp} data where included in the derivations of
Refs.~\cite{Hagiwara:2006jt,Jegerlehner:2008rs}, and the
BaBar~\cite{Aubert:2007ur,Aubert:2007ef,Aubert:2007uf} data were also
used in that of Ref.~\cite{Jegerlehner:2008rs}.
Table~\ref{tab:future} from Ref. \cite{Jegerlehner:2001wq} shows that an
uncertainty $\delta \Delta\alpha_{\rm had}^{(5)} \sim 5 \times
10^{-5}$ needed for precision physics at a future linear collider 
requires the measurement of the hadronic cross section with a
precision of ${\mathcal O}(1\%)$ from $\pi \pi$ threshold up to 
the $\Upsilon$ peak.
\begin{table}[h]
\begin{center}
\caption{\label{tab:future} The uncertainties $\delta
\Delta\alpha_{\rm had}^{(5)}$ (first column) and the errors induced by these
uncertainties on the theoretical {\small SM} prediction for $\sin^2
\!\theta_{\rm eff}^{\rm lept}$ (second column). The third column indicates
the corresponding requirements on the $R$ measurement.}
\vspace{3mm}
 \renewcommand{\arraystretch}{1.4}
 \setlength{\tabcolsep}{1.6mm}
{\footnotesize
\begin{tabular}{|c|c|c|c|}
\hline
$\delta \Delta\alpha_{\rm had}^{(5)} \!\times \! 10^{5} $ 
& $\delta(\sin^2 \!\theta_{\rm eff}^{\rm lept}) \! \times \! 10^{5}$  
& Request on $R$\\
\hline \hline
22   &  7.9 & Present \\
\hline
7   &   2.5 & $\!\delta R/R \sim  1\%$ up to ${J/\psi}$\\
\hline   
5   &   1.8 & $\delta R/R \sim  1\%$ up to $\Upsilon$\\
\hline   
\end{tabular}
}
\end{center}
\end{table}

As advocated in Ref. \cite{Jegerlehner:1999hg}, the dispersion integral
as in Eq.(\ref{eq:delta_alpha_had}) can be calculated in a 
different way: it is
sufficient to calculate $\Delta
\alpha^{(5)}_{\mathrm{had}}(s)$ not directly at $s=M_Z^2$, but at some
much lower scale $s_0=-M_0^2$ in the Euclidean region, which is chosen
such that the difference $\Delta\alpha^{(5)}_{\mathrm{had}}(M_Z^2)-
\Delta \alpha^{(5)}_{\mathrm{had}}(-M_0^2)$ can be reliably calculated
using perturbative QCD (pQCD). In Eq.(\ref{eq:delta_alpha_had}) pQCD is
used to compute the high energy tail, including some perturbative
windows at intermediate energies. An extended use of pQCD is
possible by monitoring the validity of pQCD via the Adler function,
essentially the derivative of $\Delta
\alpha^{(5)}_{\mathrm{had}}(s)$ evaluated in the spacelike region:
$\frac{D(Q^2)}{Q^2}=-\frac{3\pi}{\alpha}\,\frac{d \Delta
\alpha_\mathrm{had}}{d q^2}|_{q^2=-Q^2}$.
Using a full-fledged state-of-the-art pQCD prediction for the Adler function
one finds that $\Delta\alpha^{(5)}_{\mathrm{had}}(-M_Z^2)-
\Delta \alpha^{(5)}_{\mathrm{had}}(-M_0^2)$ can be neatly calculated from
the predicted Adler function~\cite{Eidelman:1998vc} 
for $M_0 \sim$2.5 $\mbox{GeV}$ as a
conservative choice. Also the small 
missing $\Delta\alpha^{(5)}_{\mathrm{had}}(M_Z^2)-
\Delta \alpha^{(5)}_{\mathrm{had}}(-M_Z^2)$ terms can safely be
calculated in pQCD. The crucial point is that pQCD is used in a fully
controlled manner, away from thresholds and resonances. There are three
points to note: i) this strategy allows a more precise determination
of $\Delta\alpha^{(5)}_{\mathrm{had}}(M_Z^2)$ than the direct method
based on Eq.(\ref{eq:delta_alpha_had}); ii) it requires a 
precise QCD calculation and relies on a very precise determination of
the QCD parameters $\alpha_s$, $m_c$ and $m_b$ 
\cite{Kuhn:2007vp};
iii) the method relies mainly on the precise cross section
measurements at low energy which at the same time are 
needed to reduce the uncertainty on the prediction of the muon
$g-2$.  Thus projects such as KLOE-2 are 
crucial for a better
determination of the effective fine structure constant and the muon
$g-2$~\cite{Jegerlehner:2008rs}.
\subsection{The muon $g$$-$$2$}
\label{subsubsec:GMINUS2}
During the last few years, in a sequence of measurements 
of increasing precision, the E821 collaboration at the BNL has
determined $a_{\mu} = (g_{\mu}-2)/2$ with a fabulous relative
precision of 0.5 parts per million (ppm)~\cite{Brown:2000sj,Brown:2001mga,Bennett:2002jb,Bennett:2004pv,Bennett:2006fi}, allowing
us to test all sectors of the {\small SM} and to scrutinize viable
alternatives to this
theory~\cite{Czarnecki:2001pv,Stockinger:2006zn,Stockinger:2008zz}. The present
world average experimental value is
$
    a_{\mu}^{\mbox{$\scriptscriptstyle{\rm EXP}$}}  = 
               116 \, 592 \, 089 \, (63) \times 10^{-11} 
               ~(0.54~\mbox{ppm})
$~\cite{Bennett:2004pv,Bennett:2006fi,Roberts:2010cj}.
This impressive result is still limited by statistical errors, and a
proposal to measure the muon $g$$-$$2$ to a precision of 0.14 ppm has
recently been submitted to FNAL~\cite{Carey:2009zz}. But how precise
is the theoretical prediction?

The {\small SM} prediction of the muon $g$$-$$2$ is conveniently divided into
{\small QED}, electroweak ({\small EW}) and hadronic (leading- and
higher-order) contributions:
$
    a_{\mu}^{\scriptscriptstyle \rm SM} = 
         a_{\mu}^{\scriptscriptstyle \rm QED} +
         a_{\mu}^{\scriptscriptstyle \rm EW}  +
         a_{\mu}^{\mbox{$\scriptscriptstyle{\rm HLO}$}} +
         a_{\mu}^{\mbox{$\scriptscriptstyle{\rm HHO}$}}.
$  
The {\small QED} prediction, computed up to four (and estimated at
five) loops, currently stands at
\newline
$a_{\mu}^{\scriptscriptstyle \rm QED} = 116584718.10(16)
\times 10^{-11}$\cite{Kinoshita:2005zr,Aoyama:2007dv,Aoyama:2007mn,Laporta:1992pa,Laporta:1996mq,Passera:2006gc,Kataev:2006yh,Passera:2004bj}, 
while the {\small EW} effects, suppressed by a factor
$(m_{\mu}^2/M_{\scriptscriptstyle{W}}^2)$, provide
$a_{\mu}^{\scriptscriptstyle \rm EW} = 154(2) \times 10^{-11}$
\cite{Czarnecki:2002nt,Czarnecki:1995sz,Czarnecki:1995wq}.

\begin{figure}[ht]
\begin{center}
\resizebox{0.9\columnwidth}{!}{
\includegraphics{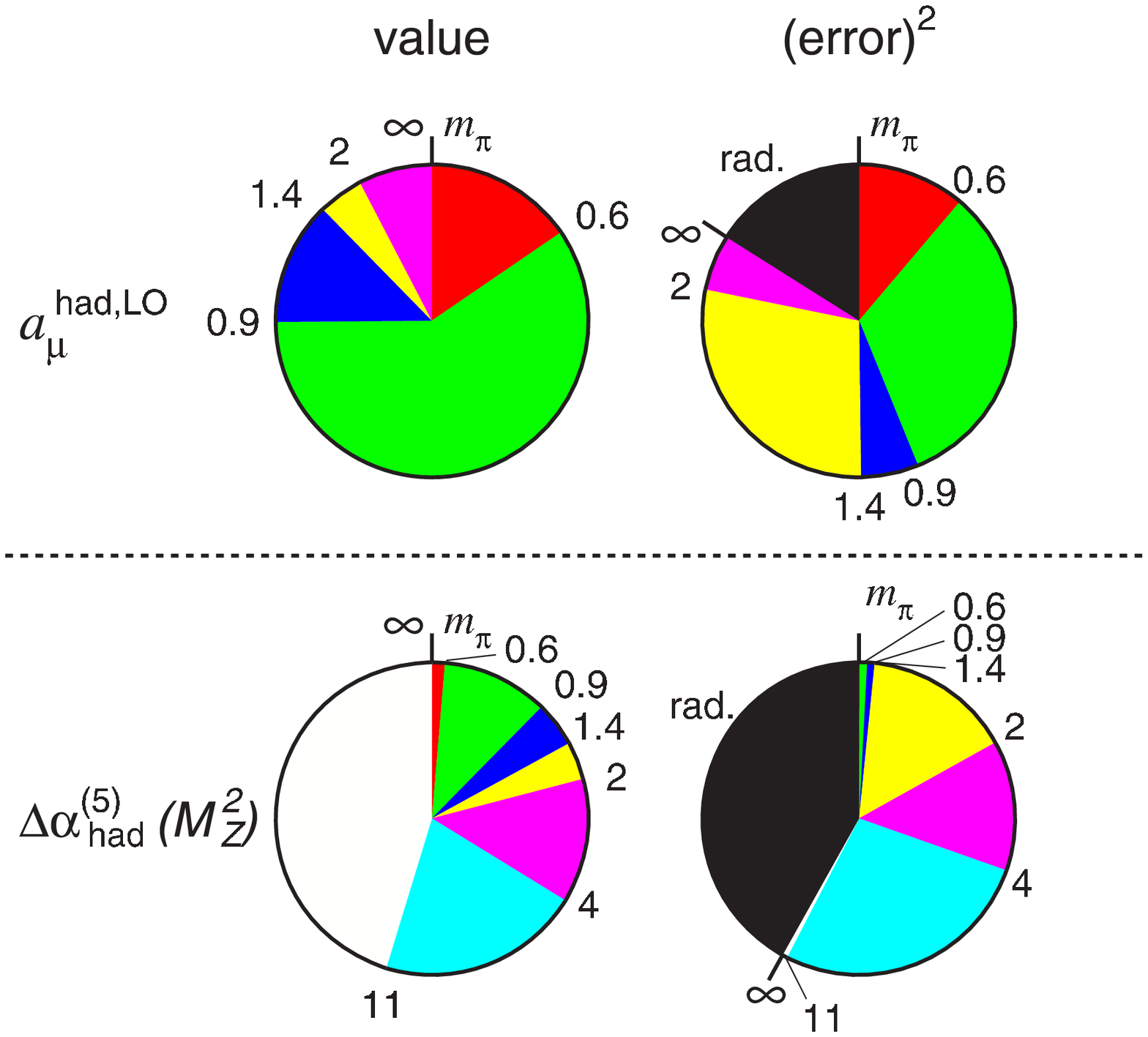}}
\caption{The pie diagrams show the fractions of the total contributions and
  $({\rm errors})^2$ from various energy intervals in the dispersion
  integrals in Eqs.(\ref{eq:delta_alpha_had}),(\ref{eq:amu_had}). The
  diagrams for the LO hadronic contribution to $g-2$, shown in the first
  row, correspond to sub-contributions with energy boundaries at 0.6, 0.9,
  1.4, 2 GeV and $\infty$, whereas for the hadronic contribution to the
  effective fine-structure constant, shown in the second row, the boundaries
  are at 0.6, 0.9, 1.4, 2, 4, 11.09 GeV and $\infty$.  In the $({\rm
  error})^2$ diagrams, the contributions arising from the treatment of the
  radiative corrections to the data are also
  included~\cite{Teubner:2008zz}.}
\label{fig:pies}
\end{center}
\end{figure}

As in the case of the effective fine-structure constant at the scale
$M_{\scriptscriptstyle{Z}}$, the {\small SM} determination of the
anomalous magnetic moment of the muon is presently limited by the
evaluation of the hadronic vacuum polarization and, in turn, by our
knowledge of the low-energy total cross section for $e^+ e^-$
annihilation into hadrons. Indeed, the hadronic leading-order
contribution $a_{\mu}^{\mbox{$\scriptscriptstyle{\rm HLO}$}}$, due to
the hadronic vacuum polarization correction to the one-loop diagram,
involves long-distance {\small QCD} effects which cannot be computed
perturbatively. However, using analyticity and unitarity, it was shown
long ago that this term can be computed from hadronic $e^+ e^-$
annihilation data via the dispersion integral~\cite{Gourdin:1969dm}
\begin{eqnarray}
      a_{\mu}^{\mbox{$\scriptscriptstyle{\rm HLO}$}} &=& 
      (1/4\pi^3)
      \int^{\infty}_{4m_{\pi}^2} ds \, K(s) \sigma^{(0)}\!(s) \nonumber \\
                                                                  &=&
      (\alpha^2/3\pi^2)
      \int^{\infty}_{4m_{\pi}^2} ds \, K(s) R(s)/s,
\label{eq:amu_had}
\end{eqnarray}
where $K(s)$ is a  kernel function  which decreases monotonically for increasing~$s$. This
integral is similar to the one entering the evaluation of the hadronic
contribution $\Delta \alpha_{\rm had}^{(5)}(M_{\scriptscriptstyle{Z}}^2)$ in
Eq.(\ref{eq:delta_alpha_had}). Here, however, the weight function in the
integrand gives a stronger weight to low-energy data. Figure \ref{fig:pies}
from Ref. \cite{Teubner:2008zz} shows the fractions of the total
contributions and $({\rm errors})^2$ from various energy intervals in the
dispersion integrals for $a_{\mu}^{\mbox{$\scriptscriptstyle{\rm HLO}$}}$
and $\Delta \alpha_{\rm had}^{(5)}(M_{\scriptscriptstyle{Z}}^2)$.
%
An important
role among all $e^+ e^-$ annihilation measurements is played by the
precise data collected by the CMD-2~\cite{Akhmetshin:2003zn,Akhmetshin:2006wh,Akhmetshin:2006bx,Aulchenko:2006na}
 and SND~\cite{Achasov:2006vp}
experiments at the VEPP--2M collider in Novosibirsk for the
$e^+e^-\rightarrow \pi^+\pi^-$ cross section in the energy ranges
$\sqrt{s} \in [0.37,1.39]$ GeV and $\sqrt{s} \in [0.39,0.98]$ GeV,
respectively.
In 2004 the KLOE experiment 
has obtained a precise measurement of $\sigma(e^+e^-\rightarrow
\pi^+\pi^-)$ via the initial-state radiation (ISR) method at
the $\phi$ resonance~\cite{Aloisio:2004bu}.  This cross
section was extracted for $\sqrt{s}$ between 0.59 and 0.97 GeV with a
systematic error of 1.3\% and a negligible statistical one. More
recently  KLOE published a new measurement with larger statistics and a systematic
 error of 0.9\%~\cite{Ambrosino:2008en}. KLOE, CDM-2 and SND give consistent
 contributions to $a_{\mu}$, and 
%
%
%
the evaluations of the dispersive integral in Eq.(\ref{eq:amu_had})
are in good agreement:
$
      a_{\mu}^{\mbox{$\scriptscriptstyle{\rm HLO}$}} = 
      6894 \, (42)_{\rm exp} (18)_{\rm rad} \times 10^{-11}
$\cite{Hagiwara:2006jt},
$
      a_{\mu}^{\mbox{$\scriptscriptstyle{\rm HLO}$}} =  
      6909 \, (39)_{\rm exp} (20)_{\rm th} \times 10^{-11}
$\cite{Davier:2007ua,Eidelman:2007zz},
$
      a_{\mu}^{\mbox{$\scriptscriptstyle{\rm HLO}$}} =
      6903.0 \, (52.6) \times 10^{-11} 
$\cite{Jegerlehner:2008zz,Jegerlehner:2009ry}.

The $e^+e^-\rightarrow \pi^+\pi^-(\gamma)$ process via
ISR has also been studied by BaBar, which has recently 
reported a new measurement of the $\pi^+\pi^-(\gamma)$ cross section from threshold up to 3 GeV~\cite{Aubert:2009fg}.
The BaBar data have been included in a new evaluation of 
the lowest order hadronic contribution to $a_{\mu}$: 
$     
 a_{\mu}^{\mbox{$\scriptscriptstyle{\rm HLO}$}} =
      6955.0 \, (41.0) \times 10^{-11} 
$\cite{Davier:2009zi}.

The term $a_{\mu}^{\mbox{$\scriptscriptstyle{\rm HLO}$}}$ can
alternatively be computed incorporating hadronic $\tau$-decay data,
related to those of hadroproduction in $e^+e^-$ collisions via isospin
symmetry~\cite{Alemany:1997tn,Davier:2002dy,Davier:2003pw}.  
Unfortunately, even if isospin
violation corrections are taken into 
account~\cite{Marciano:1988vm,Sirlin:1981ie,Cirigliano:2001er,Cirigliano:2002pv},
the $\tau$--based value is higher than the $e^+e^-$--based one, 
leading to a small 
($\sim$2 $\sigma$) $\Delta a_{\mu}$ difference \cite{Davier:2009ag}. 
As the $e^+e^-$ data are more
directly related to the $a_{\mu}^{\mbox{$\scriptscriptstyle{\rm
      HLO}$}}$ calculation than the $\tau$ ones, the latest analyses
do not include the latter. Also, 
recently studied
additional isospin-breaking corrections somewhat reduce the difference
between these two sets of data (lowering the $\tau$-based
determination)~\cite{FloresBaez:2006gf,FloresBaez:2007es,Davier:2009ag}, and a new analysis of the pion form factor
claims that the $\tau$ and $e^+e^-$ data are consistent after isospin
violation effects and vector meson mixing are
considered~\cite{Benayoun:2007cu,Benayoun:2009im,Benayoun:2009fz}.
\par
The higher-order hadronic term is further divided into two parts:
$
     a_{\mu}^{\mbox{$\scriptscriptstyle{\rm HHO}$}}=
     a_{\mu}^{\mbox{$\scriptscriptstyle{\rm HHO}$}}(\mbox{vp})+
     a_{\mu}^{\mbox{$\scriptscriptstyle{\rm HHO}$}}(\mbox{lbl}).
$
The first one, 
$-98\, (1) \times 10^{-11}$\cite{Hagiwara:2006jt},
is the ${\mathcal O}(\alpha^3)$ contribution of diagrams containing hadronic
vacuum polarization insertions~\cite{Krause:1996rf}. The second term,
also of ${\mathcal O}(\alpha^3)$, is the hadronic light-by-light contribution; as
it cannot be determined from data, its evaluation relies on specific
models. Three major components of
$a_{\mu}^{\mbox{$\scriptscriptstyle{\rm HHO}$}}(\mbox{lbl})$ can be
identified: charged-pion loops, quark loops, and pseudoscalar ($\pi^0$,
$\eta$, and $\eta'$) pole diagrams. The latter ones dominate the final
result and require information on the electromagnetic form factors of the
pseudoscalars (c.f. Sect. \ref{sec:etaradec} and Sect. \ref{sec:lble}).
Recent determinations of $a_{\mu}^{\mbox{$\scriptscriptstyle{\rm
      HHO}$}}(\mbox{lbl})$ vary between
$80(40) \times 10^{-11}$\cite{Knecht:2001qf,Knecht:2001qg}
and
$136(25) \times 10^{-11}$\cite{Melnikov:2003xd}.
The latest ones,
$105(26) \times 10^{-11}$\cite{Prades:2009tw} and
$116(39) \times 10^{-11}$\cite{Nyffeler:2009tw,Jegerlehner:2009ry},
lie between them. If we add the second of these two results to the
leading-order hadronic contribution, for example the value of
Refs. \cite{Jegerlehner:2008zz,Jegerlehner:2009ry}, and the rest of the
{\small SM} contributions, we obtain
$     a_{\mu}^{\mbox{$\scriptscriptstyle{\rm SM}$}}= 116591793 (66)  
\times 10^{-11}$.
The difference with the experimental value is then
$\Delta a_{\mu} = a_{\mu}^{\mbox{$\scriptscriptstyle{\rm EXP}$}} -
a_{\mu}^{\mbox{$\scriptscriptstyle{\rm SM}$}}$ = 296 (91) $\times 10^{-11}$,
i.e., 3.3 standard deviations (all errors were added in quadrature).
Slightly higher discrepancies are obtained employing the values of the
leading-order hadronic contribution reported in
Refs.~\cite{Davier:2007ua,Eidelman:2007zz} or \cite{Hagiwara:2006jt}.
Recent reviews of the muon $g$$-$$2$ can be found in
Refs.~\cite{Passera:2004bj,Jegerlehner:2008zz,Jegerlehner:2009ry,Jegerlehner:2007xe,Miller:2007kk,Passera:2005mx,Passera:2007fk,Davier:2004gb,Knecht:2003kc,Prades:2009qp}.
\par
Hypothetical errors in the {\small SM} prediction that could explain
the present discrepancy with the experimental value were discussed in
Ref.~\cite{Passera:2008jk,Passera:2010ev}. 
The authors concluded that none of them looks
likely. In particular, a hypothetical increase of the hadroproduction
cross section in low-energy $e^+e^-$ collisions could bridge the muon
$g$$-$$2$ discrepancy, but it was shown to be unlikely in view of current
experimental error estimates. If, nonetheless, this turns out to be
the explanation of the discrepancy, it was shown that the 95\% C.L.
upper bound on the Higgs boson mass is then reduced to about 130 GeV
which, in conjunction with the experimental 114.4 GeV 95\% C.L. lower
bound, leaves a narrow window for the mass of this fundamental
particle.
\par
The analysis of this section shows that while the {\small QED} and
{\small EW} contributions to the anomalous magnetic moment of the muon
appear to be ready to rival the fore\-casted precisions of future
experiments, much effort will be needed to reduce the hadronic
uncertainty.  This effort is challenging but possible, and certainly
well motivated by the excellent opportunity the muon $g$$-$$2$ is
providing us to unveil (or constrain) NP effects.  Once
again, a medium-term program of hadronic cross section measurements is
clearly essential. 
\subsection {$\sigma_{had}$ measurements at low energy}
\label{sec:rstatus}
\begin{figure}[h]
\begin{center}
\resizebox{0.8\columnwidth}{!}{
\includegraphics{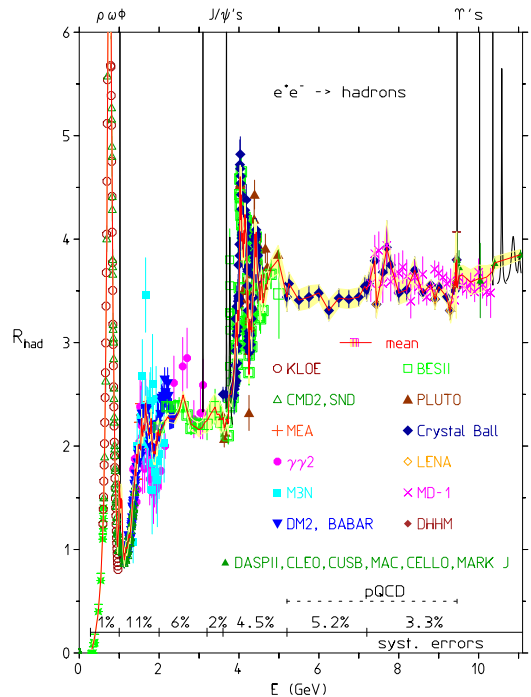}}
\caption{An updated compilation of $R$ measurements.
 In the bottom line the overall uncertainties of the different
regions are reported.} 
\label{rfj}
\end{center}
\end{figure}

In the last years big efforts on $e^+e^-$ data
in the energy range below a few GeV led to a substantial reduction 
in the hadronic uncertainty on $\Delta\alpha^{had}$ and $a_{\mu}^{had}$.
Figure \ref{rfj} shows an up-to-date compilation of this data.
The main improvements have been achieved
 in the region below 5 GeV: between 2 and 5 GeV 
(where the data are now closer to the prediction of 
pQCD), the BESII collaboration reduced the error 
to $\sim$7\%~\cite{Bai:2001ct}
(before it was $\sim$15\%); between 1 and 4.5 GeV 
BaBar measured various final states with more than two hadrons with a
 systematic accuracy between 5\% and 15\%, 
as shown in Tab. \ref{babar_r};
below 1 GeV, the CMD-2 and SND collaborations at Novosibirsk, 
KLOE at Frascati and BaBar at Stanford  measured
the pion form factor in the energy range around the $\rho$ peak with a
systematic error of $0.8\%$, $1.3\%$, $0.9\%$, and $0.5\%$, respectively.
\begin{table}
\caption{Systematic accuracy on several processes studied by 
BaBar 
in the energy range $\sqrt{s} <$ 4.5 GeV using ISR. 
Integrated luminosity is 232 fb$^{-1}$ for all processes 
but 3$\pi$ and $2\pi^+2\pi^-$ 
where only 89 fb$^{-1}$ were used. 
}
\vspace{2mm}
\label{babar_r}
\begin{center}
\begin{tabular}{|c|c|c|}
\hline
Process & Systematic accuracy & Reference \\
\hline
\hline
$\pi^+\pi^-\pi^0$ &(6-8)\% &  \cite{Aubert:2004kj}\\
$2\pi^+2\pi^-$ & 5\% & \cite{Aubert:2005eg}\\  
$2\pi 2\pi^0$  & (8-14)\%  & \cite{Denig:2008zz}\\
$2\pi^+2\pi^-\pi^0$& (8-11)\%  &  \cite{Aubert:2007ef}\\
$2\pi^+2\pi^-\eta$ &   7\%     &  \cite{Aubert:2007ef} \\
$3\pi^+3\pi^- + 2\pi^+2\pi^-2\pi^0$ & (6-11)\% & \cite{Aubert:2006jq}  \\
$KK\pi$          &  (5-6)\% &  \cite{Aubert:2007ym}\\
$K^+K^-\pi\pi$    & (8-11)\%  & \cite{Aubert:2007ur} \\
\hline
\end{tabular}
\end{center}
\end{table}
The CMD-2 and SND collaborations at Novosibirsk and BESII in Beijing  
were performing the hadronic cross section measurements in a 
traditional way, i.e., by varying the $e^+e^-$ beam energies.
KLOE, BaBar, and more recently Belle 
used the method of radiative return 
\cite{Arbuzov:1998te,Binner:1999bt,Benayoun:1999hm,Rodrigo:2001jr,Kuhn:2002xg,Rodrigo:2001kf,Czyz:2002np,Czyz:2003ue}, 
as reviewed in Refs. \cite{Kluge:2008fb,Actis:2009gg}.
Figure \ref{rfj} shows that, despite the recent progress, 
the region between 1 and 2 GeV is still poorly known, with a fractional 
accuracy of $\sim$10\%. Since about 50\% of the error squared, 
$\delta^2$ $a_{\mu}^{had}$,   
comes from this region (and about  70\% 
of $\delta^2 \Delta \alpha^{(5)}_{\mathrm{had}}(-M_0^2)$), it is 
evident how desirable an improvement on this region is.
%
\subsection{Vector-meson spectroscopy}
\label{Vecto}
Cross sections of exclusive final states are also
important for the spectroscopy of vector mesons whose properties provide
fundamental information on interactions of light quarks. 
The $\omega(1420)$, $\rho(1450)$, $\omega(1650)$, $\phi(1680)$, and
$\rho(1700)$ have been found between 1 and 2~GeV~\cite{Amsler:2008zzb} 
but even their basic parameters ($M,~\Gamma,~\Gamma_{ee}$)
are poorly known.
The mass of these  states  is in satisfactory
agreement with 
QCD which actually predicts
three sets of vectors $\rho,~\omega,~\phi$ 
from 1 to 2~GeV~\cite{Godfrey:1985xj}.
Recent studies of $e^+e^- \to \pi^+\pi^-\pi^0$ by
SND~\cite{Achasov:2003ir} and BaBar~\cite{Aubert:2004kj} as well as
$e^+e^- \to 2\pi^+2\pi^-\pi^0$ by CMD-2~\cite{Akhmetshin:2000wv} and
BaBar~\cite{Aubert:2007ef}, have significantly affected the 
$\omega(1420)$ and $\omega(1650)$ parameters. In the
$\pi^+\pi^-\pi^0$ final state the cross sections measured by
SND and BaBar are in good agreement at $\sqrt{s} <1.4$~GeV, whereas
above this energy the BaBar cross section is more than two times
larger than what previously measured by DM2~\cite{Antonelli:1992jx}.
The estimated values of the leptonic width
of the $\omega(1420)$ and $\omega(1650)$ 
states are significantly higher than 
what found in Ref. \cite{Godfrey:1985xj}.
The $\rho(1450)$ and $\rho(1700)$ are expected to predominantly decay into
four pions. Various measurements of both $2\pi^+2\pi^-$ and
$\pi^+\pi^-2\pi^0$ agree with each 
other~\cite{Bisello:1990du,Akhmetshin:2004dy,Achasov:2005rg,Aubert:2005eg}
showing one broad structure only, while three interfering $\rho$'s
have been observed in the two--pion decay of the 
$\tau$-lepton~\cite{Fujikawa:2008ma}.
A peculiar interference
pattern or the existence of additional exotic states close to the
regular $q\bar{q}$ states and mixed with 
them~\cite{Barnes:1996ff,Balazs:1998sb} can explain these results.
It is also possible that the $\rho(1450)$ and $\rho(1700)$ have
different decay modes.
BaBar has recently reported the observation of two structures at
$\sim$~1.5~GeV~\cite{Aubert:2007ym}. The broad one in the $K K^*$ state
(width of $418 \pm 25 \pm 4$~MeV) could be interpreted as the
overlapping of $\rho(1450)$ and $\rho(1700)$ mesons, but the one in the 
$\phi\pi^0$ final state with a smaller
width of $144 \pm 75 \pm 43$~MeV, should have a different origin. 
Its properties are very
close to those of the $C(1480)$ observed more than 20 years ago
in $\pi^- p$ collisions~\cite{Bityukov:1986yd}.
The $\phi\pi^0$ cross section
also exhibits some narrow structures at 1.9~GeV,
which could be the $\rho(1900)$ earlier observed in the $3\pi^+3\pi^-$ state
photoproduction~\cite{Frabetti:2003pw}, in the total $R$ at 
$e^+e^-$ collider~\cite{Antonelli:1996xn} as well as  in both  
$3\pi^+3\pi^-$ and
 $2\pi^+2\pi^-2\pi^0$ cross sections measured by DM2~\cite{Castro:1994pi} and
BaBar~\cite{Aubert:2006jq} (although
with a larger width). 
The statistical significance of these
findings is not sufficient 
to exclude that 
 an OZI--violating decay mode of the $\rho(1700)$ 
was observed~\cite{Aubert:2006jq}.
BaBar also reports clear evidence for the $\rho(2150)$ meson in the
$\eta^{\prime}(958)\pi^+\pi^-$ and $f_1(1285) \pi^+\pi^-$
cross sections~\cite{Aubert:2007ef}.
Finally, 
BESII has recently reported evidence for a very broad
($\sim$800 MeV) $K^+K^-$ state at about 1.5~GeV produced
in $J/\psi$ decays~\cite{Ablikim:2006hp}, with
isovector quantum numbers which have been obtained from a 
partial-wave analysis.
These measurements can hardly be reconciled
with what was observed to both $K^+K^-$ and $K^0_SK^0_L$
final states.
\par
Evidence for the $\phi(1680)$ was only based on an 
old observation of the structure in the $K^0_S K^{\pm} \pi^{\mp}$
final state by DM1~\cite{Mane:1982si}. BaBar has recently 
confirmed the resonance in both 
$K K^*$ and $\phi\eta$~\cite{Aubert:2007ym} modes  
while Belle has found it in the 
$\phi\pi^+\pi^-$ final state~\cite{Shen:2009zze}. 
There is, however, a conflicting result
of FOCUS~\cite{Link:2002mp}, which observes a structure with similar
parameters in the $K^+K^-$ and does not observe it in the $K K^*$
channel.
\par
BaBar has observed a new isoscalar resonance in the
$K^+K^-\pi^+\pi^-$ and $K^+K^-\pi^0\pi^0$ final states (predominantly
$\phi f_0(980)$) at 2.12~GeV~\cite{Aubert:2005eg,Aubert:2007ur}. 
This result is confirmed by BESII 
with the analysis of 
the same final state in $J/\psi \to \eta\phi f_0(980)$ 
decay~\cite{Ablikim:2007yt}
and by Belle~\cite{Shen:2009zze}.
\par
There are still many puzzles and unknown things 
calling for more theoretical and experimental efforts. 
Some progress, limited by the number of events, is expected from the 
ISR studies at BaBar and Belle. 
A real breakthrough can happen with the direct scanning  
at lower-energy colliders, 
VEPP--2000 and DA$\mathrm{\Phi}$NE. 
\par 
Before discussing the improvements on the 
hadronic cross section measurements expected from KLOE-2 in the 
region [2$m_\pi$--2.5 GeV], we review the status of 
the main competitor, VEPP-2000.
\subsection{VEPP-2000 prospects}
\label{sec:vepp2000}
The VEPP-2000 
$e^+e^-$ collider at the Budker Institute of Nuclear Physics (BINP)  
in Novosibirsk will cover the energy range up to 2 GeV 
which accounts for the $\sim$92\% of the  hadronic contribution to the 
$g-2$ of the muon.  
Two different detectors will take data at VEPP-2000: the upgraded Spherical 
Neutral Detector (SND) and newly constructed Cryogenic Magnetic Detector 
(CMD-3).
\noindent
SND is a non-magnetic detector, whose basic part is a 
three--layer electromagnetic calorimeter consisting of NaI(Tl) crystals, 
13.4 radiation--length long. 
The calorimeter covers nearly 90\% of the full solid angle.
Charged-particle tracking is provided by 
a cylindrical drift chamber. Particle identification 
is performed with aerogel Cherenkov counters of high
density ($n = 1.13$).

CMD-3 is a magnetic detector based on a cylindrical drift chamber
(DC) to measure coordinates, angles and
momenta of charged particles.
The DC is surrounded by a
Z-chamber (ZC) to
measure the coordinate along the beam direction.
DC and ZC are placed inside the thin
($\sim$ 0.18 $X_{0}$) superconducting solenoid with a magnetic field of 1.3 T.
Coordinates and energies of photons are measured by the barrel system based
on the liquid Xe calorimeter ($7X_{0}$) and CsI ($8X_{0}$) crystals. 
BGO crystals ($13X_{0}$) are used in the end cap calorimeter. 


As a rule, the pion form factor is calculated via the 
ratio $N_{\pi\pi}/N_{ee}$ 
which is directly measured using collinear events. 
The trigger and reconstruction
inefficiencies cancel out partly in this ratio. 
Geometrical efficiencies $\epsilon_{ee}$ and $\epsilon_{\pi\pi}$ are 
calculated via Monte-Carlo simulation together with corrections for the 
detector imperfections. To achieve a systematic error less than 0.1\%  
the beam energy must be measured with $10^{-4}$ fractional uncertainty 
or better. 
The variation of the derivative with the energy, 
$d|F_{\pi}(\sqrt{s})|^{2}/d\sqrt{s}/|F_{\pi}(\sqrt{s})|^{2}
\cdot \Delta \sqrt{s}$ ($\Delta E = 10^{-3} E $) 
does not exceed $\pm$ 1\% but in the energy range near $\omega$ 
and $\phi$ mesons where it is about 6\%.    

Radiative corrections are included in the Monte Carlo generator (MCGPJ)
 by means of the Structure Function approach.
The estimated theoretical error on the process 
$e^+e^- \to \pi^+\pi^-,$ $K^+K^-$ (with point-like pions and kaons)
is 0.2\%. 
A similar (or even better) accuracy is expected for channels with neutral 
particles in the final state because only ISR contributes 
to the cross sections.

In a few years CMD--3 and SND experiments at VEPP--2000 will provide 
new precision results 
on the measurements of  the exclusive hadronic cross sections below 2 GeV.
Progress is particularly expected in the $e^+e^- \to \pi^+\pi^-$ channel, 
where a systematic uncertainty of about 0.4\% or even better will be
achieved. 

\subsection{Improving $\sigma_{\pi \pi}$ below 1 GeV with KLOE-2}
\label{sec:lowhadcs}
The region below 1 GeV is dominated by the two-pion channel which 
accounts for 70\% of the contribution to $a_{\mu}^{had}$, and for
40\% to the total error $\delta^2 a_{\mu}$ as shown in Fig. \ref{fig:pies}. 
How can this error be reduced?
Let us consider  the region around the $\rho$ and $\pi^+\pi^-$ 
threshold:
\noindent
i) the $\pi^+\pi^-$ region between 0.5 and
1 GeV has been studied by different experiments.
CMD-2
and SND
have performed an energy
scan at the $e^+e^-$ collider VEPP-2M ($\sqrt{s}\in$ [0.4--1.4]
GeV) with $\sim$10$^6$ and $\sim$4.5$\times 10^6$ events respectively, and
systematic fractional errors from 0.6\% to 4\% in the cross sections,
depending on $\sqrt{s}$. 
The pion form factor has also been 
measured by KLOE using ISR, and more recently by 
BaBar. 
KLOE collected  more than 3.1 million events, corresponding
 to an integrated luminosity of 240 pb$^{-1}$, leading to a relative error
of 0.9\% in the energy region [0.6--0.97] GeV dominated by systematics.
BaBar has presented a $\pi^+\pi^-(\gamma)$ cross section 
measurement based on half a million selected events. 
The pion form factor is obtained by the ratio 
$\pi^+\pi^-(\gamma)$ to $\mu^+\mu^-(\gamma)$ which allows 
a systematic error of 0.5\% in the $\rho$ region 
increasing to 1\% outside; 
%
ii) the threshold region [2$m_\pi$--0.5 GeV] provides 13\% of the total
$\pi^+\pi^-$ contribution to the muon anomaly: 
$a_\mu^{\rm HLO}$ [2$m_\pi$--0.5$\quad$GeV] = $(58.0\pm2.1) 
\times$ 10$^{-10}$. To overcome the lack of
precision data at threshold energies, 
the pion form factor 
is extracted from a
parametrization based on ChPT, constrained from
spacelike data~\cite{Amendolia:1986wj}.
The most effective way to measure the cross section near the 
threshold in the timelike region is
provided by ISR events, where the emission of an energetic photon allows
the study of the two pions at rest.  However at DA$\mathrm{\Phi}$NE, 
the process
$\phi\to\pi^+\pi^-\pi^0$ with one missing photon is hundreds of
times more frequent than the signal, and therefore a precision measurement
requires an accurate evaluation of the background. Furthermore, 
irreducible background due to 
$\phi\to\pi^+\pi^-\gamma$ is also present while running at the $\phi$--
resonance peak.  The background issue can be largely overcome by 
taking data at
$\sqrt{s}<M_\phi$: KLOE has analysed about  
200 pb$^{-1}$ of integrated luminosity at 1 GeV and a new measurement 
of the pion form factor 
has been recently presented \cite{Muller:2009pj}.

\begin{figure}[htu]
\begin{center}
\resizebox{0.8\columnwidth}{!}{
\includegraphics{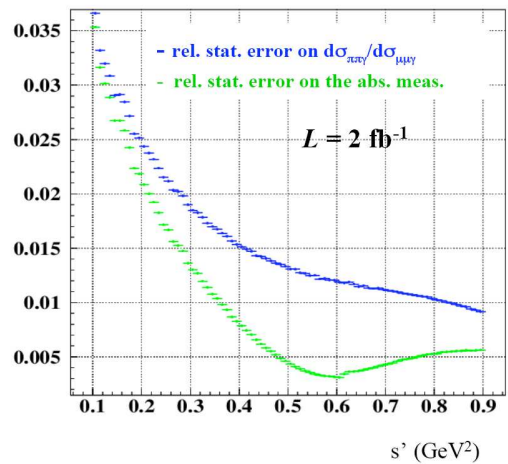}}
\caption{\label{error2} The fractional statistical
error on the cross section 
${\rm d}\sigma_{\pi\pi\gamma}/{\rm d}M^2_{\pi\pi}$ and on the ratio
${\rm d}\sigma_{\pi\pi\gamma}/{\rm d}\sigma_{\mu\mu\gamma}$ as a function
of the $\pi\pi$ invariant mass 
squared, $s^\prime$ = $M_{\pi\pi}^2$ (bin width = 0.01 GeV$^2$) 
for an integrated luminosity of 2 fb$^{-1}$ at 1 GeV. }
\end{center}
\end{figure}
\noindent
Figure \ref{error2} shows the statistical precision that can be
reached in the region below 1 GeV with an integrated luminosity of 
2 fb$^{-1}$ at 1 GeV for each bin of $\Delta s$ = 0.01 GeV$^2$. 
This luminosity 
leads to a statistical error on $a_\mu^{\rm HLO}$ of a few
per mill. The experimental systematic error could be kept 
at the same level (now is at~1\%) by taking data at 1 GeV, 
where the background conditions
(mainly from $\phi\to\pi^+\pi^-\gamma$ and $\phi\to\pi^+\pi^-\pi^0$)  
which especially affect the threshold region, are much more favourable.
The theoretical error, of the order of 0.6\%, dominated by the 
uncertainty on the radiator function (=0.5\%), 
could be reduced to a few per mill in the future. 
The dependence on the theory 
is much reduced by the measurement of the  
$\pi\pi\gamma/\mu\mu\gamma$ ratio where 
the main uncertainty on the radiator function and 
vacuum polarization cancel out in the ratio so that 
with an integrated luminosity of 2 fb$^{-1}$ at 1 GeV a 
fractional 
error of ~0.4\% could be reached. 
\subsection{Improving $\sigma_{had}$ above 1.02 GeV with KLOE-2}
\label{sec:highhadcs}
The region [1--2.5] GeV, with an uncertainty on $\sigma_{had}$ of 
roughly 11\%, 
is the most poorly known, and contributes about 
55\% of the uncertainty on $a_{\mu}^{\mbox{$\scriptscriptstyle{\rm HLO}$}}$ 
and 40\% of the error on 
$\Delta\alpha^{(5)}_{had}(M_Z^2)$~\cite{Jegerlehner:2008rs}. 
In this region BaBar has published results on 
$e^+ e^-$ into three and four hadrons, obtained with an integrated luminosity 
of 89 fb$^{-1}$~\cite{Aubert:2004kj}.
\begin{figure}[h]
\begin{center}
\resizebox{0.85\columnwidth}{!}{
\includegraphics{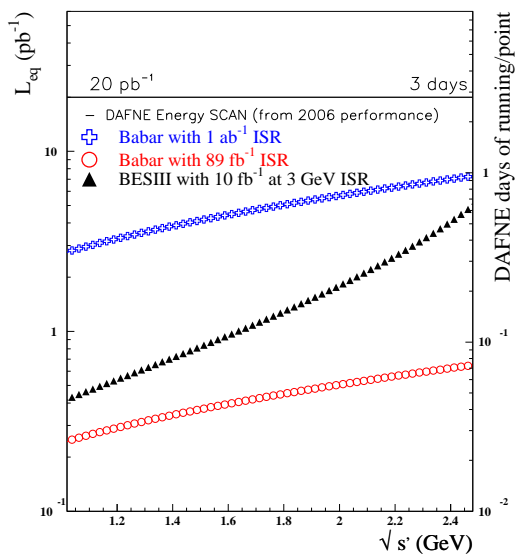}}
\vspace{-0.5cm}
\caption{\label{fig:w2} Comparison of the 
event yield in terms of ``equivalent'' luminosity for 
 BaBar with 89 fb$^{-1}$ (circle); BaBar with 1 ab$^{-1}$ (cross); 
BESIII using 10 fb$^{-1}$  at 3 GeV (triangle). 
A polar angle of the photon larger than $20^\circ$ 
and a bin width of 25 MeV have been assumed.}
\end{center}
\vspace{-1cm}
\end{figure}
KLOE-2 can improve both the exclusive and inclusive measurements.
At $10^{32}$cm$^{-2}$s$^{-1}$ luminosity,   
a scan in the region from 1 to 2.5 GeV aiming at   
an integrated luminosity of 20 pb$^{-1}$ per point corresponds 
to few days of data taking for each energy bin. 
By assuming an energy step of 25 MeV, the whole
region would be thus scanned in one year of data taking.
As shown in Fig. \ref{fig:w2}, the statistical yield 
 would be one order of magnitude higher 
than what achieved with 1 ab$^{-1}$ at BaBar, and better than 
what expected at BESIII with 10 fb$^{-1}$ at 3 GeV.
\begin{figure}[h]
\begin{center}
\vspace{-0.5cm}
\resizebox{0.85\columnwidth}{!}{
\includegraphics{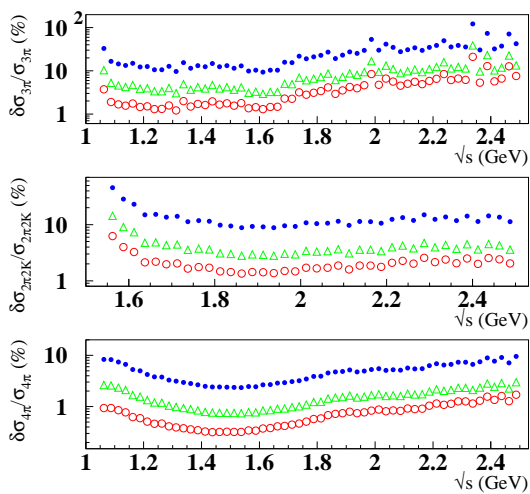}}
\vspace{-1.cm}
\caption{\label{fig:impactscan} Comparison of the statistical accuracy in
the cross-section among 
an energy scan with 20 pb$^{-1}$ per 
point ($\circ$) at KLOE-2; the published BaBar results ($\bullet$), 
BaBar with 890 pb$^{-1}$ of integrated luminosity  
(triangle) for the $\pi^+\pi^-\pi^0$
(top), $\pi^+\pi^-K^+K^-$ (middle) and $2\pi^+ 2\pi^-$ (bottom) channels. 
An energy step of 25 MeV is assumed.}
\end{center}
\end{figure}
Figure \ref{fig:impactscan}  shows the statistical error on the 
$\pi^+\pi^-\pi^0$, $2\pi^+ 2\pi^-$ and $\pi^+\pi^-K^+K^-$ channels 
which can be
 achieved by  an energy scan of 20 pb$^{-1}$ per point,
compared  
with Babar published (89 fb$^{-1}$) and tenfold 
(890 fb$^{-1}$) statistics.
The energy scan allows a statistical accuracy at   
1\% level for most of the energy points.
In addition, KLOE-2 can benefit of the high machine luminosity and 
use ISR as well. 
%
\par 
The energy scan for a precision measurement of the cross sections 
requires the knowledge of the energy of the beams circulating in the collider 
with a fractional accuracy of ${\mathcal O}({10^{-4}})$, i.e., $\sim$100 keV 
at 1 GeV.
Two well estabilished techniques 
exist: the resonant depolarization (RD) technique and  
Compton backscattering (CBS) of laser photons against the electron beam, 
as reviewed 
in Ref. \cite{Blinov:2009zza}.     
The RD technique provides an accuracy of 1-3 keV 
and is based on polarized electrons,  
a requirement not easy to fullfil at the low-energy colliders 
and 
making in addition 
the energy determination a very time--consuming task. 
The CBS method is less accurate, but 
does not need of any special condition on the beams. For this reason, 
it seems more suitable for DA$\mathrm{\Phi}$NE.
The measurement procedure 
consists of the acquisition 
of  the order of 10$^6$ events providing an accurate determination of the   
photon Compton spectrum.
In one acquisition hour at VEPP--4M 
\cite{Blinov:2009zza} 
a statistical accuracy on the beam energy of 70--100 keV 
has been obtained with the CBS method.
The simultaneous measurement of the energy of the scattered electron
can be used to improve the precision on the beam energy. This is what 
has been done at the GRAAL beam at the ESRF machine in Grenoble. 
The GRAAL beam  is a Compton--backscattered 
beam with maximal photon energy of 1.5 GeV 
($\omega_{l}$ = 3.51 eV, $E_{e}$ = 6 GeV), mainly devoted to photonuclear  
reactions  \cite{Bartalini:2005wx}. At GRAAL the determination of 
the $\gamma$-ray energy is achieved by tagging 
the scattered electrons which  
are separated from the primary beam by 
machine optics (bending magnets and quadrupoles) 
downstream to the laser--beam interaction region.  The tagging detector 
consists of a silicon $\mu$strip placed in a suitable position 
along the machine lattice,  and measuring the 
displacement of these electrons from the 
orbit of nominal--energy particles.
From the displacement, the energy of the electron is inferred  
and thus the photon one. 
The advantage of this approach lies in the great accuracy 
of the position measurements of the silicon strip detectors. 
A precision in the energy beam 
to the 10 keV level was actually achieved with 
the $\mu$strip detectors~\cite{Gurzadyan:2004rx}.   
 
\subsection{Summary on $\sigma_{had}$ measurements}
In summary, precision tests of the SM 
require a more accurate knowledge of the hadronic
cross section in the whole energy range, from the  2$m_{\pi}$ threshold
 to 2.5 GeV. The region between 1 and 2.5 GeV is at
present the most poorly known and is crucial especially 
for the hadronic corrections  
to the effective fine structure constant at the $M_Z$ scale. 
In order to improve
the theoretical accuracy of $a^{{\rm HLO}}_{\mu}$ 
the precision measurement of the hadronic cross section close 
to the $\pi \pi$ threshold is also required. 
In both regions 
KLOE-2 can give important contributions bringing the accuracy on 
$a^{{\rm HLO}}_{\mu}$ to about $3\times$10$^{-10}$. The result 
can be achieved by measuring both i) the ratio $\pi^+ \pi^- (\gamma)$ 
to $\mu^+ \mu^- (\gamma)$ with 0.4\% accuracy at 1 GeV with ISR and 
ii) the hadronic cross section in the [1--2.5] GeV region with 
[1--2]\% fractional error. This would represent a factor of two improvement 
on the error of $a^{{\rm HLO}}_{\mu}$, needed to clarify 
the 3--$\sigma$ discrepancy in conjuction with  
the proposed muon $g-2$ experiments 
at FNAL \cite{Carey:2009zz} and J-PARC \cite{jparcLoIamu}.
%
%
%
%
%

%
\section{Physics in the Continuum: $\gamma \gamma$ Processes}
\label{sec:gg}
%


%
\def\bgea{\begin{eqnarray}}
\def\enea{\end{eqnarray}}
\newcommand{\FF}{{\cal F}_{P^{*}\gamma^*\gamma^*}}
\newcommand{\FFc}{{\cal F}_{P^{*}\gamma^*\gamma}}
\newcommand{\kloe}{K{\kern-.07em LOE} }
\newcommand{\gaga}{\ensuremath{\gamma\gamma}}
The upgrade of the KLOE detector with the installation of four
stations~\cite{Babusci:2009sg} to tag electrons and positrons from the
reaction 
\begin{equation}
e^+e^-\to e^+e^-\gamma^*\gamma^*\to e^+e^- X,
\label{eq:gg-stat1}
\end{equation}
gives the opportunity to investigate $\gamma\gamma$ physics at \DAF\ .
Since it was realised in 1970s that significant production rates could be
achieved~\cite{Brodsky:1970vk,Brodsky:1971ud}, the two photon processes have
been investigated at most of the $e^+e^-$ colliders with the limitations
imposed by the rate of these subleading processes at the luminosity
reached by past--generation colliders.\par
In the $\gamma \gamma$ processes $C=+1$ hadronic states can be produced.
In the energy region accessible with the KLOE-2 taggers the two
$\gamma^*$ can be considered quasi-real so that only $J^{PC}=0^{\pm +}$,
$2^{\pm +}$ quantum numbers are allowed~\cite{Yang:1950rg}.  
If no cut is applied to the final--state leptons, the Weizs\"acker--Williams
or Equivalent Photon Approximation (EPA)~\cite{Brodsky:1971ud} can be used
to understand the main qualitative features of the process 
(\ref{eq:gg-stat1}). 
Then the event yield, $N_{eeX}$, can be evaluated according to:
\begin{equation}
N_{eeX} = L_{ee}\int\frac{{\rm d F}}
{{\rm d W_{\gamma\gamma}}}\,
\sigma_{\gamma\gamma\to X}({\rm W_{\gamma\gamma}})\,
{\rm d W_{\gamma\gamma}}~,
\label{eq:gg-stat2}
\end{equation}
where $W_{\gamma\gamma}$ is the invariant mass of the two quasi--real
photons, $L_{ee}$ is the integrated luminosity, and d$F$/d$W_{\gaga}$ is
the $\gamma\gamma$ flux function:
\begin{figure}[htbp]
\begin{center}
\resizebox{0.85\columnwidth}{!}{
\includegraphics{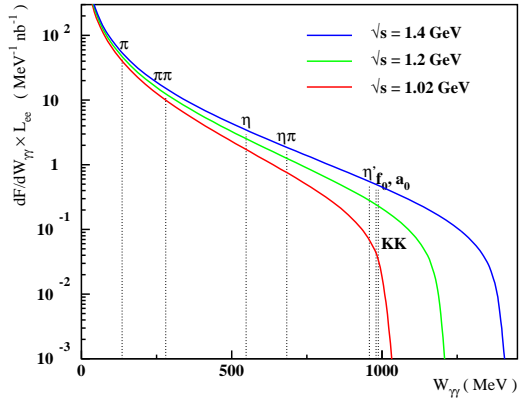}
}
\vspace{-1.5cm}
\caption{Differential $\gamma \gamma$ flux function as a 
function of the center-of-mass energy.}
\label{fig:gg-stat1}
\end{center}
\end{figure}
\begin{equation}
\frac{{\rm d}F}{{\rm d}W_{\gaga}}~ ~ = ~~
\frac{1}{W_{\gaga}}\ \left(\frac{2\alpha}{\pi}\right)^2\
\left(\ln\frac{E_b}{m_e}\right)^2 f(z)
\label{eq:gg-stat3}
\end{equation}
where $E_b$ is the beam energy and 
\begin{equation}
f(z) = (z^2+2)^2\ \ln\frac{1}{z}
-(1-z^2)\ (3+z^2)\;, \quad
z = \frac{\rm W_{\gamma\gamma}}{2E_b}.
\label{eq:fz}
\end{equation}
Figure \ref{fig:gg-stat1} shows the flux function multiplied
by an integrated luminosity $L_{ee}=1\mbox{ fb}^{-1}$,
as a function of the $\gamma\gamma$ invariant mass for three different
center-of-mass energies. 
This plot demonstrates the feasibility of the detection of the final
states $\pi^+\pi^-$, $\pi^0\pi^0$, $\pi^0\eta$ whose cross--sections 
are of the order of or larger than 1 nb 
\cite{Boyer:1990vu,Marsiske:1990hx,Oest:1990ki,Mori:2006jj,Uehara:2008pf}, 
and the identification of
the resonances produced in these channels. 
Running \DAF\ at center-of-mass energies up to 1.4 GeV
can eventually complete the low energy spectroscopy studies with the
detection of the heavier $K^+K^-$ and $\bar{K}^0 K^0$ final states. 
Single pseudoscalar ($X=\pi^0$, $\eta$ or $\eta^\prime$) production is also
accessible and would improve the determination of the two--photon decay 
widths of these mesons,
relevant for the measurement of the pseudoscalar mixing angle
$\varphi_P$, and the measurement of the valence gluon content in the
$\eta^\prime$ wavefunction.
Moreover, it would be possible to measure the transition form factors 
$\mathcal{F}_{X\gamma^*\gamma^*}(q^2_1,q^2_2)$ as a function of the 
momentum of the virtual photons, $q^2_1$ and $q^2_2$.
The interest in such form factors is rising again 
in connection with the theoretical evaluation of the hadronic 
light-by-light
contribution to the muon magnetic anom\-a\-ly.
\subsection{Scalar resonances in two photon collisions}
The two photon production of hadronic resonances is often advertised as one of
the clearest ways of revealing their
composition~
\cite{Klempt:2007cp,Pennington:2007zy,Barnes:1985cy,Barnes:1992sg,Achasov:1987ks,Achasov:2006cq,Hanhart:2007wa,Branz:2008ha,Volkov:2009pc,Mennessier:2008kk,Mennessier:2008mc,Achasov:2007qr,Achasov:2009ee,Giacosa:2008xp,Giacosa:2007bs,vanBeveren:2008st}.
For instance, the nature of the isoscalar scalars seen in $\pi\pi$
scattering below 1.6 GeV, namely the $f_0(600)$ or $\sigma$, $f_0(980)$,
$f_0(1370)$ and $f_0(1510)$ mesons, remains an
enigma~\cite{Klempt:2007cp,Pennington:2007zy}.  
While models abound in which some are $\bar{q}q$, some $\bar{qq}qq$, sometimes one is a 
$\bar{K}K$-molecule, and one a glueball~\cite{Klempt:2007cp}, definitive statements
are few and far between. Their two photon couplings will help unraveling
the enigma.

\begin{figure}[htbp]
\begin{center}
\resizebox{0.8\columnwidth}{!}{
\includegraphics{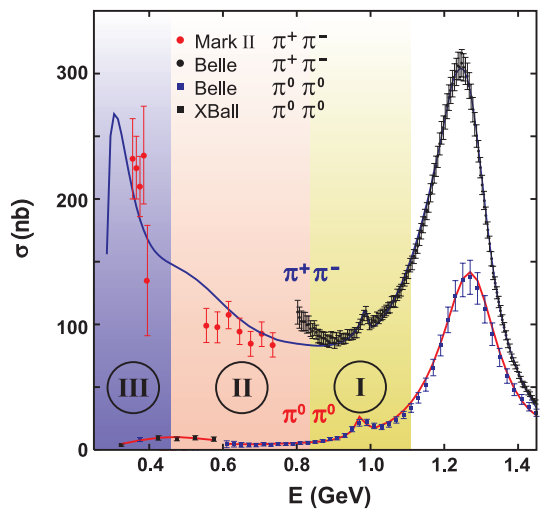}
}
\caption{Cross-section results for 
$\,\gamma\gamma\to\pi^+\pi^-\,$ from Mark~II~\cite{Boyer:1990vu} (below 800
MeV) and Belle~\cite{Mori:2006jj} above, integrated over 
$|\cos \theta^*|\,\le\,0.6$, and for $\gamma\gamma\to\pi^0\pi^0$ from Crystal
Ball~\cite{Marsiske:1990hx} below 600 MeV and Belle~\cite{Uehara:2008pf}
above, integrated over 
$|\cos \theta^*|\,\le\,0.8$. $E$ is the $\gamma\gamma$ c.m. energy. The
curves are from the (as yet) unpublished Amplitude Analysis that includes
the data of Ref.~\cite{Uehara:2008pf}. The three shaded bands delineate the
energy regions where KLOE-2 can make a contribution discussed in the text.} 
\label{fig:xsects}
\end{center}
\vspace{-5mm}
\end{figure}

\begin{figure}[htbp]
\begin{center}
\resizebox{0.40\columnwidth}{!}{
\includegraphics{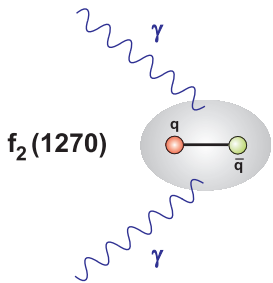}
}
\caption{Illustration of two photons coupling to a relatively long-lived
$\bar{q}q$ state. The photons predominantly couples to the electric charge of the constituents.}
\label{fig:f_2}
\end{center}
\end{figure}

The ability of photons to probe such structure naturally depends on the
photon wavelength. This is readily illustrated by looking at the 
integrated cross-sections from Mark II~\cite{Boyer:1990vu}, Crystal
Ball~\cite{Marsiske:1990hx} and Belle~\cite{Mori:2006jj,Uehara:2008pf}
shown in Fig.~\ref{fig:xsects} for $\gamma\gamma\to\pi^+\pi^-$,
$\pi^0\pi^0$. 
It is worthwhile to consider these processes from the crossed channel
viewpoint in which the photon scatters off a pion. 
At low energies the photon has long wavelength, and so sees the
whole hadron and couples to its electric charge. Thus the photon sees the
charged pions. 
The $\pi^+\pi^-$ cross-section is large and its value is a 
measure of the electric charge of the pion. In contrast, the neutral pion
cross-section is small. However, as the energy increases the photon
wavelength shortens and recognises that the pions, whether charged or
neutral, are made of the same charged constituents, namely quarks, and Ê
causes these to resonate, as illustrated in Fig.~\ref{fig:f_2}. Thus both channels reveal the well-known 
$\bar{q}q$ tensor meson, the $f_2(1270)$, seen as a peak in Fig.~\ref{fig:xsects},
with its production cross-section related to the average charge squared of its
constituents.
\par
\begin{figure}[htbp]
\vspace{1mm}
\begin{center}
\resizebox{0.40\columnwidth}{!}{
\includegraphics{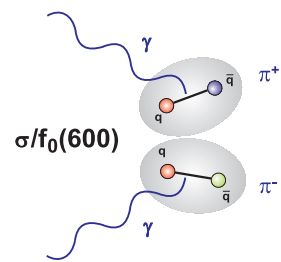}
}
\caption{Illustration of two photons coupling to a state that quickly decays to $\pi\pi$. In the case of the $\sigma/f_0(600)$, the photons coupling to the final state dominates over any coupling to the {\it intrinsic} make-up of the state,
regardless of its composition.
}
\label{fig:sigma}
\end{center}
\end{figure}

However, at lower energies, 500-1000 MeV, the photon wavelength is longer. 
States like the $\sigma$ are so short-lived that they very rapidly
disintegrate into two pions and when these are charged, the photons couple
to these, Fig.~\ref{fig:sigma}. 
The intrinsic make-up of the state, whether $\bar{q}q$, $\bar{qq}qq$ or
glueball, is Êobscured by the large coupling to the pions to which the
$\sigma$ decays. Data reveal a similar situation applies to the heavier,
and seemingly much narrower, $f_0(980)$. This state has equally large
hadronic couplings and is only narrow because it sits just below $\bar{K}K$
threshold, to which it strongly couples. 
Experimental results discussed below suggest the two photons largely see
its meson decay products too, regardless of whether the $f_0(980)$ is
intrinsically a $\bar{K}K$ molecule or
not~\cite{Barnes:1985cy,Barnes:1992sg,Achasov:1987ks,Achasov:2006cq,Hanhart:2007wa,Achasov:2007qr,Achasov:2009ee,Giacosa:2008xp,Giacosa:2007bs,vanBeveren:2008st}. ÊÊ

In the future, strong coupling QCD will eventually predict the two photon
couplings of these states according to models of their composition. While
theorists work on this, the experimental task 
is to determine the resonance couplings.
To extract these reliably requires a partial wave separation.
In principle one needs data with full angular coverage with polarised
photon beams. But the available data have no polarisation information and
in the two photon center-of-mass frame at most 80\% angular acceptance,
less in the case of $\pi^+\pi^-$ because of the difficulty of separating
these from the scattered $e^+$ and $e^-$. Thus even for the large
$f_2(1270)$ signal seen so prominently in the integrated cross-sections of
Fig.~\ref{fig:xsects}, determining its two photon width is not so easy. One
must separate the $\pi\pi$ amplitude into components with definite spin,
helicity and isospin.

The $\pi^+\pi^-$ cross-section near threshold is dominated by the one pion
exchange Born amplitude, producing the enhancement seen in
Fig.~\ref{fig:xsects}. 
Being controlled by $I=1$ exchange in the crossed channels means that at
low energies $\pi\pi$ production in $I=0$ and $I=2$ must be comparable in
all partial waves~\cite{Pennington:2008xd}. 
Therefore data on both final states, $\pi^+\pi^-$ and $\pi^0\pi^0$ should
be analized at the same time.

The era of high luminosity $e^+e^-$ colliders with their intense programme
of study of heavy flavour decays has, as a by-product, yielded two photon
data of unprecedented statistics. 
The Belle collaboration~\cite{Mori:2007bu,Mori:2006jj} 
has published results on $\gamma\gamma\to\pi^+\pi^-$ in 5 MeV bins above
800 MeV. These show a very clear peak for the $f_0(980)$,
Fig.~\ref{fig:xsects}. 
Belle~\cite{Mori:2006jj}, analysing just their integrated cross-section,
find its radiative width to be $205 ^{+95+147}_{-83-117}$ eV.  
The large errors reflect the many ways of drawing a background Êwhether in the $I=0$ $S$-wave where the resonance appears or in 
the other partial waves: remember that without full angular coverage the partial waves are not orthogonal, and so interferences occur. 
Despite these uncertainties, a number of theoretical predictions have now honed in on [$0.2-0.3$] keV for
the radiative width of the $f_0(980)$, whether it is a $\bar{K}K$ molecule
or a $\bar{qq}qq$ state~\cite{Hanhart:2007wa,Branz:2008ha,Volkov:2009pc,Giacosa:2008xp,Giacosa:2007bs,vanBeveren:2008st}. 
\par
The only way to make sense of the real uncertainties is to perform an Amplitude Analysis.
A key role is played by the general $S$-matrix properties of analyticity, unitarity and crossing symmetry. ÊWhen these are combined with the 
low energy theorem for Compton scattering, these anchor the partial wave amplitudes close to $\pi\pi$
threshold~\cite{Morgan:1991zx,Lyth:1971pm,Mennessier:1982fk,Mennessier:1986cp} as described in
Refs.~\cite{Pennington:1992vw,Pennington:1995fv,Boglione:1998rw,Pennington:2008xd} and so help to make up for 
the lack of full angular coverage in experiments. Crucially, unitarity
imposes a connection between the $\gamma\gamma\to\pi\pi$ partial wave
amplitudes and the behaviour of hadronic processes with $\pi\pi$ final
states.  
Below 1 GeV the unitarity sum is saturated by the $\pi\pi$ 
intermediate state, while above it the $\bar{K}K$ channel is critically 
important. Beyond [1.4--1.5] GeV multipion processes start 
to contribute as $\rho\rho$ threshold is passed. Little is known about the 
$\pi\pi\to\rho\rho$ channel in each partial wave. 
Consequently attention should be restricted to the region below 1.4 GeV,
where $\pi\pi$ and $\bar{K}K$ intermediate states dominate.
The hadronic scattering amplitudes for $\pi\pi\to\pi\pi$ and 
$\bar{K}K\to\pi\pi$ are known and so enable the unitarity constraint to be 
realised in practice and in turn allow an Amplitude Analysis to be undertaken.
\par 
Such an analysis has been performed~\cite{Pennington:2008xd} incorporating 
all the world data and its key angular
information~\cite{Boyer:1990vu,Mori:2006jj,Marsiske:1990hx,Behrend:1992hy,Bienlein:1992en}. 
 Since the $\pi\pi$ system can be formed in both $I=0$ and $I=2$ final
states, the $\pi^+\pi^-$ and $\pi^0\pi^0$ channels have to be treated 
simultaneously. Though there are now more than 2000 datapoints in the
charged channel below 1.5 GeV, there are only 126 in the neutral channel, and
they have to be weighted more equally to ensure that the isospin components
are reliably separable.
\par
These world data can then be fitted adequately by a range of
solutions~\cite{Pennington:2008xd}: a range, in which there remains a
significant 
ambiguity in the relative amount of helicity zero $S$ and $D$ waves,
particularly above 900 MeV.
The acceptable solutions have a $\gamma\gamma$ width for the $f_0(980)$
(determined from the residue at the pole on the nearby unphysical sheet,
being the only unambiguous measure) of between 96 and 540 eV, with 450 eV
favoured: a significantly larger value than predicted by many current
models~
\cite{Achasov:1987ks,Achasov:2006cq,Hanhart:2007wa,Branz:2008ha,Volkov:2009pc,Giacosa:2008xp,Giacosa:2007bs,vanBeveren:2008st}.
Of course, the experimental value includes the coupling of the $f_0(980)$
to its $\pi\pi$ and $\bar{K}K$ decay products and their final state 
interactions (the analogue of Fig.~\ref{fig:sigma}), not necessarily
included in all the theoretical calculations. 

The fits accurately follow the lower statistics data from Mark
II~\cite{Boyer:1990vu} and CELLO~\cite{Behrend:1992hy} (see the detailed
figures in 
Ref.~\cite{Pennington:2008xd}). However, they do not describe the Belle
$\pi^+\pi^-$ data between 850 and 950 MeV, as seen in Fig. \ref{fig:xsects}. This \lq\lq
mis-fit'' is even more apparent in the angular distributions. In the
charged pion channel, there is always a large $\mu^+\mu^-$
background. Though the Belle data have unprecedented statistics, the
separation of the $\pi^+\pi^-$ signal is highly sensitive to the $\mu$-pair
background. This may well be responsible for Êthe apparent distortion below
1~GeV in Fig.~\ref{fig:xsects}. 

Since that analysis, Belle have more recently published
results~\cite{Uehara:2008pf} (both integrated and differential cross-sections) on
$\pi^0\pi^0$ production in 20 MeV bins, Fig.~\ref{fig:xsects}. Again these reveal the
$f_0(980)$ as a small peak, rather than the shoulder seen in earlier much
lower statistics data from Crystal
Ball~\cite{Marsiske:1990hx,Bienlein:1992en}. A new Amplitude 
Analysis has been started, which significantly changes the solution space,
pushing the allowed amplitudes to those with a larger radiative width for
the $f_0(980)$. However, the final solutions are not yet available. 

Though the Belle experiment represents an enormous stride in two photon
statistics, there remains room for KLOE-2 to make a significant contribution
in each of the three energy regions displayed as bands in
Fig.~\ref{fig:xsects}. Ê
\begin{itemize}
\item[I.] Ê{\bf 850-1100 MeV}: accurate measurement of the $\pi^+\pi^-$ and
$\pi^0\pi^0$ cross-sections (integrated and differential) are crucially
still required, with clean $\mu\mu$ background separation. In addition,
any information just above 1~GeV on $\bar{K}K$ production would
provide an important constraint on the coupled channel Amplitude Analyses
described above. Moreover $\pi^0\eta$ studies will complement the results
from Belle~\cite{:2009cf}. 
\item[II.] {\bf 450-850 MeV}: this is the region where the
$\sigma$ pole lies. This is a region almost devoid of precision
$\gamma\gamma$ data and so allows a range of
interpretations~
\cite{Mennessier:2008kk,Mennessier:2008mc,Pennington:2006dg,Oller:2007sh,Bernabeu:2008wt}. ÊÊ
Given the importance of the $\sigma$ for our understanding of strong
coupling QCD and the nature of the vacuum, it is crucial to measure
$\pi\pi$ production in this region in both charge
modes~\cite{Pennington:2007eg,Pennington:2007yt}.  
\item[III.] {\bf 280-450 MeV}: though this region is controlled by the Born
  amplitude with corrections computable by the first few orders of ChPT, 
  it is the domain that anchors the partial wave
  analyses described here. 
The Mark II experiment~\cite{Boyer:1990vu} is the only one that has made a special measurement of the normalised 
cross-section for the $\pi^+\pi^-$ channel near threshold. As seen in Fig.~\ref{fig:xsects}, their data have very large error-bars.
\end{itemize}

\subsubsection{$e^+e^-\to e^+e^- \pi\pi$ at KLOE-2}
\label{sec:sigma}
KLOE-2 can improve the experimental 
precision in all three regions of Fig. \ref{fig:xsects}, 
contributing to the solution of the open questions 
on low--energy hadron physics 
with the study of 
$e^+ e^- \to e^+ e^- \pi \pi$.
This process is a clean electromagnetic probe to 
investigate 
the nature of the $\sigma$ meson through the analysis of the  
$\pi\pi$ invariant mass which is expected 
to be plainly affected by the presence of 
the scalar meson. 
 A precision measurement of the cross-section of $\gamma\gamma\to\pi^+\pi^-$
and $\gamma\gamma\to\pi^0\pi^0$ would also complete the information from
previous experiments allowing the determination of the 
$\gamma\gamma$ couplings
of the $\sigma$ and $f_0(980)$ from the partial wave analysis
discussed above. 

\begin{figure}[htb]
\begin{center}
\resizebox{.85\columnwidth}{!}{
\includegraphics{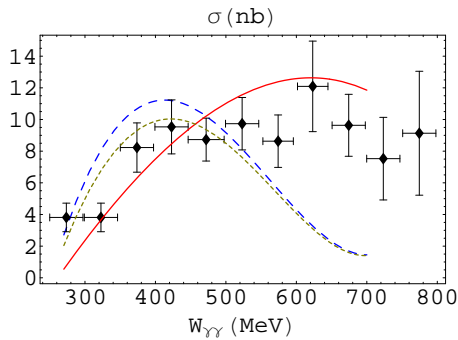}
}
\caption{$\sigma(\gamma\gamma\to\pi^0\pi^0)$: Crystal Ball data compared
with a two-loop ChPT prediction (solid curve), and with the cross-section of
$\gamma\gamma\to\sigma\to\pi^0\pi^0$ with (dotted) and without (dashed)
Adler zero, obtained according to Ref.~\cite{Nguyen:2006sr}.} 
\label{fig:crystball}
\end{center}
\end{figure}

From the experimental point of view the final state with neutral pions,
$e^+ e^-\to e^+ e^-\pi^0\pi^0$, is cleaner than $e^+ e^-\to e^+
e^-\pi^+\pi^-$, the latter being 
affected by the large background from
$e^+ e^-\to e^+e^-\mu^+\mu^-$ 
and from
the ISR process $e^+ e^-\to\pi^+\pi^-\gamma^* \to \pi^+\pi^-\ e^+ e^-$. 
In Fig. \ref{fig:crystball}, the low--energy part of the
$\gamma\gamma\to\pi^0\pi^0$ cross-section corresponding to the energy
regions II and III of Fig. \ref{fig:xsects} is shown.
The large uncertainties in the data do not allow any conclusion 
to be drawn about the
existence of a resonance-like structure in the region [400--500] MeV.

\begin{figure}[htbp]
\resizebox{0.5\textwidth}{!}{
\includegraphics{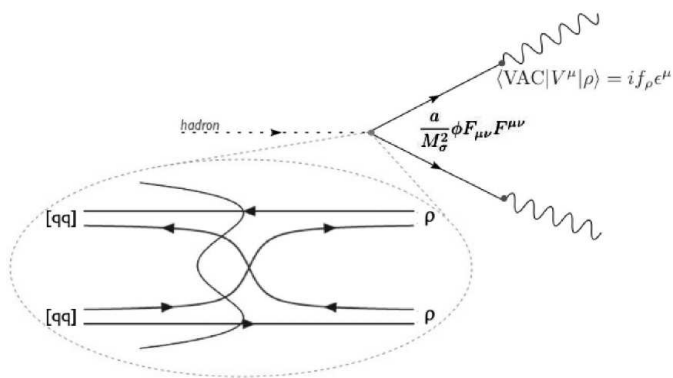}
}
\caption{Schematic view of the $\gamma\gamma\to\sigma$ transition.}
\label{fig:sigma1}
\end{figure}

All the studies to assess the KLOE-2 potentiality in the analysis 
of $e^+ e^-\to e^+ e^-\pi^0\pi^0$ were  
performed with 
the MC generator~\cite{Nguyen:2006sr},  
based on the full matrix element
calculation over the four-body phase space. 
The $\sigma\gamma\gamma$ coupling depicted in Fig. \ref{fig:sigma1} is
obtained assuming vector meson dominance: the $\sigma$ meson 
decays to $\rho\rho$ with transitions $\rho$-$\gamma$ whose strength is
described by VMD. 
The underlying dynamics in $\sigma\to\rho\rho$ is similar to
$\sigma\to\pi\pi$ assuming~\cite{Maiani:2004uc} 
the $\sigma$ as a bound state of two
diquarks: the process is described by the tunneling probability for a quark
to escape its diquark shell and bind with an anti-quark of the anti-diquark
to form a standard $q\bar{q}$ meson. 

\begin{figure}[htbp]
\begin{center}
\resizebox{0.3\textwidth}{!}{
\includegraphics{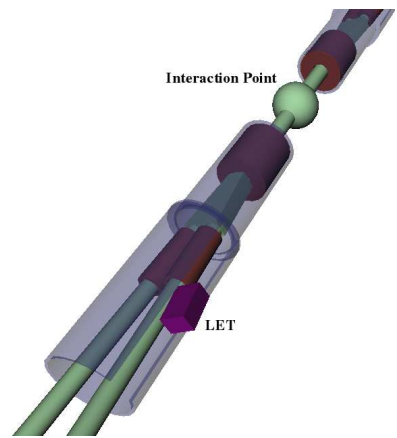}
}
\caption{View of one of the two LET detectors (the most external box) on the
\DAF\ beam--line.} 
\label{fig:let}
\end{center} 
\end{figure}

\begin{figure}[htbp]
\begin{center}
\resizebox{.8\columnwidth}{!}{\includegraphics{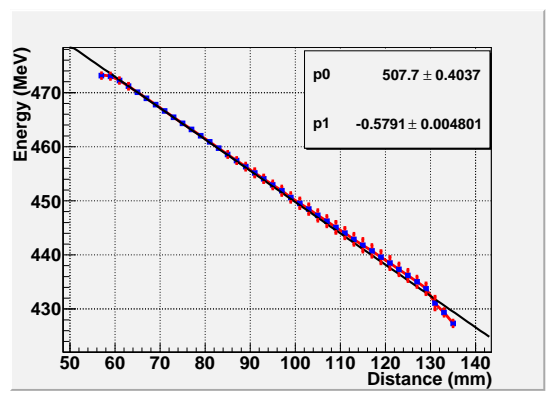}} 
\caption{Energy vs deviation from the nominal orbit for leptons
arriving on the HET.} 
\label{fig:het_lin}
\end{center} 
\end{figure}

Running at $\sqrt{s} = M_{\phi}$ the electron tagger is essential to reduce the
large background, mainly from $e^+e^-\to\phi\to K_S K_L$, with the $K_L$
escaping from the detector and the $K_S$ decaying to neutral pions, 
$K_S\to\pi^0\pi^0$.
The tagger is also important to close the kinematics of the process 
 $e^+ e^- \to e^+ e^- \pi \pi$ allowing 
a precise reconstruction of the $\gamma\gamma$
invariant mass (W$_{\gamma\gamma}$).
Leptons from $\gamma\gamma$ interactions, with $E <$510 MeV, follow a path  
through the machine optics different from the orbit of the circulating beams. 
The trajectories of these off--momentum electrons have been studied by means 
of a MC simulation, to evaluate their exit point from the \DAF\  beam pipe
and to find proper location for the tagger devices.
This study is based on BDSIM~\cite{Bdsim:MC}, a GEANT4 extension toolkit
capable of simulating particle transport in the accelerator
beamline. 
Particles coming from the IP with energies from 5 to 510 MeV have been
simulated with 0.5 MeV step.
The results 
clearly indicate the need to place two different
detectors in different regions on both sides of the IP: the Low Energy
Tagger (LET) to detect leptons with energy between 150 and 400 MeV and
the High Energy Tagger (HET) 
for those 
with energy greater than 420
MeV. 
The LET region is one meter from the IP,  
inside the KLOE magnetic field
(Fig.~\ref{fig:let}). 
In the LET region the correlation between the energy and the
position of the leptons is weak. 
For this reason the LET detector is a crystal calorimeter able to measure
the electron energy with a resolution better than 10\% over the range 
[150--400] MeV.
The HET detector is located just at the exit of the first bending dipole. 
In this position the off--momentum electrons escaping from the beam--pipe
show a clear correlation between energy and deviation from the nominal orbit
(Fig.~\ref{fig:het_lin}). 
Therefore the energy of the particles can be obtained 
from the 
displacement with respect to the machine orbit measured by a position detector.
The preliminary results of the MC simulation of the tagger stations 
show that the whole 
interesting range of the $\gamma\gamma$ invariant mass (zones II and III of 
Fig. \ref{fig:xsects}) is covered by either the LET/LET or the LET/HET 
coincidences, the latter signaling events from $\pi \pi $ 
threshold to 800 MeV with maximal acceptance over the 
$M_{\gamma \gamma}$ range from 
[600--760] MeV. A LET/LET coincidence indicates instead a $\gamma \gamma$ 
event at  
$M_{\gamma \gamma} < 350 $  MeV with acceptance decreasing as 
invariant mass increases.    
Using both, data coming from the taggers, and from the \kloe\ detector, 
the background can be reduced without relevant signal loss.

\begin{figure}[htbp]
\begin{center}
\resizebox{.8\columnwidth}{!}{
\includegraphics{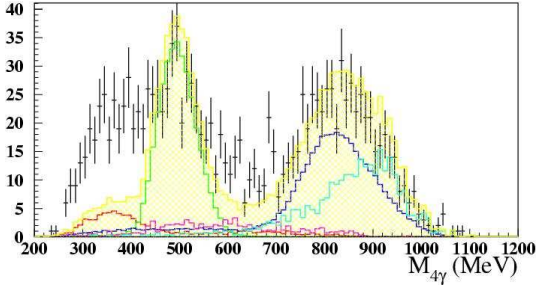}
}
\caption{Four--photon invariant mass: crosses are data, histograms are
several sources of background evaluated from MC.}
\label{fig:pi0pi0}
\end{center}
\end{figure}

For feasibility studies and to optimize the selection algorithms 
of $\gamma\gamma\to\pi^0\pi^0$, about 10 pb$^{-1}$ of KLOE data 
(i.e., without $e^\pm$ taggers) taken at $\sqrt{s}=1$ GeV have
been used.
Four photons  
compatible with the hypothesis of $\pi^0\pi^0$ decay with 
small ($\sim$80 MeV)  
total transverse momentum, were asked for.
The distribution of the four--photon invariant mass, $M_{4\gamma}$, 
shows an
excess of events below 400 MeV (Fig.~\ref{fig:pi0pi0}) 
not explained by $\phi$ decays
or other processes in the continuum.\\ 
The excess is due to genuine $e^+e^-\pi^0\pi^0$ events and after proper
normalization provides the cross--section. 
The studies point out that KLOE-2 with an integrated luminosity 
at the $\phi$ peak of $L=$ 5 fb$^{-1}$ can measure 
the $\gamma\gamma\to\pi^0\pi^0$ cross-section with the same energy 
binning of Fig. \ref{fig:crystball} obtained from 
Crystal Ball \cite{Marsiske:1990hx},
reducing the statistical uncertainty in each bin to 2\%. 

\subsection{Single pseudoscalar final states}
At leading order 
only two diagrams contribute to 
the  $e^+e^- \to e^+e^- P$ 
process at the $e^+e^-$ colliders: 
\bgea
e^+e^- &\to& e^+e^- \gamma^\ast \gamma^\ast \to e^+e^- P \quad
(t-\text{channel)},
\label{eq:tchannel} \\
e^+e^- &\to& \gamma^\ast \to \gamma^\ast P \to e^+e^- P \quad
(s-\text{channel)}.
\label{eq:schannel}
\enea
The t-channel with quasi-real photons dominates the 
production cross--section, however in general   
the $s$ 
and the interference contributions are not negligible.  
The two--photon decay widths of $\pi^0,~\eta$ and $ ~\eta^{\prime}$ can be
extracted  from the cross--section of the process (\ref{eq:tchannel}), 
of interest 
among other issues   
to determine the pseudoscalar mixing angle and
improve 
on the KLOE analysis
of the gluonium content in the $\eta^\prime$, as discussed 
in Sect. \ref{sec:gluonium}. 
\par \noindent
Moreover,   
the transition form factors
$\mathcal{F}_{P\gamma^*\gamma^*}(m^2_P,q_1^2,q_2^2)$ at spacelike
momentum transfers can be measured from the same process (\ref{eq:tchannel}). 
They are important to discriminate among different phenomenological  
models, 
relevant for 
the hadronic light-by-light scattering contribution to the $g-2$ of the 
muon~\cite{Jegerlehner:2009ry,Prades:2009pn}.

The $s$-channel contribution of Eq.(\ref{eq:schannel}) instead 
provides information on 
the timelike form factors, 
a complementary and 
promising opportunity 
to discriminate among alternative models of the underlying dynamics. 
\subsubsection{Two photon width of pseudoscalar mesons}
The two--photon decay width $\Gamma(P\to\gamma\gamma)$
can be measured from the cross-section which  
in the narrow--width approximation 
reads~\cite{Brodsky:1971ud}: 
\begin{equation}
\label{eq:gg-stat4}
\sigma_{e^+e^-\to e^+e^-P} \,=\,
\frac{16 \alpha^2 \Gamma(P\to\gamma\gamma)}{m_P^3}\
\left(\ln\frac{E_b}{m_e}\right)^2 f(z_P)\;, 
\end{equation}
where $m_P$ 
is the mass 
of the pseudoscalar meson, 
$z_P=\frac{m_P}{2E_b}$ and $f(z_P)$ is given by Eq.(\ref{eq:fz}).
Equation (\ref{eq:gg-stat4}) is obtained by neglecting
the $P\gamma\gamma$ transition form factors,
an approximation valid for quasi-real photons. 
Table \ref{tab:gg-stat1} shows the cross--section of pseudoscalar
meson production at different center-of-mass energies. 
\begin{table}[htbp]
\caption{Cross--sections of pseudoscalar
meson production at different $\sqrt{s}$ values computed
using Eq.(\ref{eq:gg-stat4}).}
\label{tab:gg-stat1}
\vspace{3mm}
\begin{center}
\renewcommand{\arraystretch}{1.5}
\begin{tabular}{|l||c|c|c||}
\hline
$\sqrt{s}$ (GeV) & $1.02$ & $1.2$ & $1.4$\\
\hline
$\sigma_{e^+e^-\to e^+e^-\pi^0}$ (pb) & 271 & 317 & 364\\
\hline
$\sigma_{e^+e^-\to e^+e^-\eta}$ (pb) & 45 & 66 & 90\\
\hline
$\sigma_{e^+e^-\to e^+e^-\eta^\prime}$ (pb )& 4.9 & 20 & 40\\
\hline
\end{tabular}
\end{center}
\end{table}
\renewcommand{\theenumi}{\roman{enumi}}
The $\Gamma(P\to\gamma\gamma)$ 
is used as input for:
\begin{enumerate}
\item the pseudoscalar mixing angle ($\varphi_P$) determination from 
\cite{Ambrosino:2006gk},  
{\small
\begin{eqnarray*}
\frac{\Gamma(\eta\to\gamma\gamma)}{\Gamma(\pi^0\to\gamma\gamma)} =
\left(\frac{m_\eta}{m_{\pi^0}}\right)^3\frac{1}{9}\left(
5\cos\varphi_P-\sqrt{2}
\frac{\bar{m}}{m_s}\sin\varphi_P\right)^2 & & \\
\frac{\Gamma(\eta^\prime\to\gamma\gamma)}{\Gamma(\pi^0\to\gamma\gamma)} =
\left(\frac{m_{\eta^\prime}}{m_{\pi^0}}\right)^3\frac{1}{9}\left(
5\sin\varphi_P+\sqrt{2}
\frac{\bar{m}}{m_s}\cos\varphi_P\right)^2 & &
\end{eqnarray*}
}
\item the test of the valence gluon content ($Z_G^2=\rm{sin}^2\phi_G$) in the
$\eta^\prime$ wavefunction~\cite{Kou:1999tt,Ambrosino:2006gk}, 
$$
\frac{\Gamma(\eta^\prime\to\gamma\gamma)}{\Gamma(\pi^0\to\gamma\gamma)} =
\left(\frac{m_{\eta^\prime}}{m_{\pi^0}}\right)^3\frac{1}{9} 
\cos^2\phi_G \cdot  
$$
$$
\left(
 5\sin\varphi_P + \sqrt{2}
\frac{f_n}{f_s}\cos\varphi_P\right)^2
$$
\end{enumerate}
\subsubsection*{${e^+e^-\to e^+e^-\pi^0}$}
For an efficient collection of this kind of events where the $\pi^0$ 
is mostly detected by one of the end--cap calorimeters, 
KLOE-2 has to overcome some limitations of the KLOE data--taking 
originated from the triggering criteria and software event filtering.
The latter can be improved modifying the rejection procedure for events 
with unbalanced energy deposit in one of the end--cap calorimeters.
The trigger requirement of having at least  
two energy deposits
not in the same end--cap calorimeter 
produces a distortion in the $\pi^0$ energy spectrum 
and limits the  global efficiency  
for ${e^+e^-\to e^+e^-\pi^0}$ to 20\%. 
Modifications in the KLOE trigger  which somewhat 
soften the requirements on the energy deposit in the calorimeter 
for events tagged by the LET stations are under evaluation.
In addition, the $e^\pm$ taggers at KLOE-2 allow the suppression of the background from
electroproduction of $\pi^0$'s ($\sigma(e N\to
e\pi^0X)\sim\mathcal{O}(1\,\mu\mathrm{b})$).  
\subsubsection*{${e^+ e^- \to e^+ e^- \eta}$}
Two different final states, 
$e^+ e^- \gamma\gamma$ and $e^+ e^- \pi^+\pi^-\pi^0$ can be considered.
The KLOE analysis of the latter process, based on an integrated luminosity
of $240\mbox{ pb}^{-1}$, is underway using data taken at $\sqrt{s}=1$ GeV,
off the $\phi$ meson peak.
Selection criteria consist of two prompt photons and two tracks with opposite
curvature.
Main background processes are:
\begin{itemize}
\item $e^+e^-\to\eta(\to\pi^+\pi^-\pi^0)\gamma$, with the monochromatic photon
lost in the beam pipe;
\item double radiative Bhabha events (cross-section of about 100
nb);
\end{itemize}
These amount to about 50\% of fractional background in the
final sample, that can be reduced with both, analysis refinements, 
and using $e^\pm$ taggers.

Due to the large background from $e^+e^-\to\gamma\gamma(\gamma)$,
information from $e^\pm$ taggers is more crucial for
the analysis of $e^+e^-\to e^+e^-\eta$ with
$\eta\to\gamma\gamma$.
\begin{table}[htbp]
\caption{Maximal energy of the $e^+e^-$ in the
final state for different center-of-mass energies.}
\vspace{2mm}
\label{tab:pseudo2} 
\begin{center}
\renewcommand{\arraystretch}{1.5}
\begin{tabular}{|l||c|c|c||}
\hline
$\sqrt{s}$ (GeV) & $1.02$ & $1.2$ & $1.4$\\
\hline
$\mathrm{E}^{max}_e(\pi^0)$ (MeV) & 500 & 592 & 693\\
\hline
$\mathrm{E}^{max}_e(\eta)$ (MeV) & 363 & 475 & 593\\
\hline
$\mathrm{E}^{max}_e(\eta^\prime)$ (MeV) & 60 & 218 & 373\\
\hline
\end{tabular}
\end{center}
\vspace{-0.8cm}
\end{table}
\subsubsection*{${e^+e^-\to e^+e^-\eta^\prime}$}
At $\sqrt{s}=1.02$ GeV, 
the $e^+e^-$ in the final state have less than 60 MeV momentum and are 
out of the acceptance of the tagging devices.
Table \ref{tab:pseudo2} shows the maximal energy for 
the $e^+e^-$ in the final state related to different
pseudoscalar--meson productions and $\sqrt{s}$ values.
Taking into account the $e^+e^-\to e^+e^-\eta^\prime$ cross--section, 
the background due to annihilation processes 
(scaling with $\sim$1/s), and the $e^+e^-$ energy spectrum, the ideal
conditions for data analysis would be obtained running at $\sqrt{s}=1.4$ GeV.  
\begin{table}[htbp]
\vspace{-4mm}
\caption{Event yield for the dominant $\eta^\prime$
decay channels obtained at $\sqrt{s}=$1.4 GeV and $L=1$ fb$^{-1}$.}
\label{tab:pseudo3}
\vspace{2mm} 
\begin{center}
\renewcommand{\arraystretch}{1.5}
\setlength{\tabcolsep}{1.5mm}
\begin{tabular}{|l|c|c|r|}
\hline
~ & Final state & ${\eta^\prime}$ fraction (\%) & events \\ 
\hline
$\pi^+\pi^-\gamma$ & $\pi^+\pi^-\gamma$ & $29.4\pm0.9$ & 12,000\\
\hline
$\pi^+\pi^-\eta$ & 
$\pi^+\pi^- \gamma \gamma$ & $17.5\pm0.5 $ & 7,000\\
\hline
$\pi^0\pi^0\eta$ & 
$\pi^0\pi^0 \pi^+\pi^-\pi^0 $& $4.7\pm0.3$ & 2,000 \\
\hline
$\omega\gamma$ & 
$\pi^+\pi^-\pi^0 \gamma $ & $2.7\pm0.3$ & 1,200 \\
\hline
$\gamma\gamma$ & $\gamma \gamma$ & $2.10\pm0.12$ &  800\\
\hline
\end{tabular}
\end{center}
\vspace{-2mm}
\end{table}
Radiative return to the $\phi$ meson is not a limiting factor,
since the cross-section ranges from 10 pb at $\sqrt{s}=1.2$ GeV
to 4 pb at $\sqrt{s}=1.4$ GeV.
Table \ref{tab:pseudo3} shows the event yields 
achieved
at $\sqrt{s}=1.4$ GeV with an integrated luminosity 
of 1 fb$^{-1}$ \footnote{At 
$\sqrt{s}=1.2$ GeV the yield is half than at $\sqrt{s}=1.4$ GeV}. 
Final states with no more than two charged tracks have been considered 
because they are experimentally clean and thus can be selected 
with high efficiency. 
Each decay chain provides $\sigma\propto\Gamma(\eta^\prime\to
X)\times\Gamma(\eta^\prime\to\gamma\gamma)$, 
thus we can obtain a precision measurement of $\Gamma(\eta^\prime\to\gamma\gamma)$ 
from the analysis of the dominant branching fractions of 
Tab. \ref{tab:pseudo3}, imposing $\sum BR_i = 1$. 
 
\subsubsection*{Monte Carlo for the KLOE-2 experiment}
There are several Monte Carlo generators for $\gamma \gamma$ physics
which can be considered in the forthcoming 
analyses of KLOE-2 data:
\begin{enumerate}
\item a code written by A.~Courau~\cite{Courau:1984ia}
used, e.g., in Ref. \cite{Alexander:1993rz,Bellucci:1994id} for the $e^+ e^-
\to e^+ e^- \pi^0\pi^0$, and in Ref. \cite{Ong:1999gp} for the $e^+ e^- \to
e^+ e^- \pi^0$ process; 
\label{mc:courau}
\item a code specifically developed by 
F. Nguyen {\it et al.}~\cite{Nguyen:2006sr} 
for the KLOE analysis of the $e^+ e^-\to e^+
  e^- \pi^0\pi^0$ (Sect. \ref{sec:sigma}); 
\label{mc:npp}
\item TREPS written by S.~Uehara~\cite{Uehara:1996di} 
and used by the Belle 
  collaboration \cite{Mori:2007bu,Mori:2006jj};
\label{mc:treps}
\item TWOGAM developed by D.~M.~Coffman,
used by CLEO \cite{Gronberg:1997fj} for the $e^+ e^- \to e^+ e^- P$, with $P
=\pi^0,\eta,\eta^\prime$;
\item GGResRC used by the BaBar collaboration \cite{Aubert:2009mc} and  
based on the Ref.    
 \cite{Brodsky:1971ud} for single--pseudoscalar production, and on
Ref.\cite{Budnev:1974de} for the two--pion final state; 
\item EKHARA developed by H. Czyz and 
  collaborators \cite{Czyz:2003gb,Czyz:2005ab,Czyz:2006dm};
\item GALUGA and GaGaRes written by G. A. Schuler \cite{Schuler:1997ex}, and 
F. A. Berends and R. van Gulik \cite{Berends:2001ta}, respectively,    
for the study of heavy resonances from $\gamma \gamma$ interaction   
at the LEPII. 
\end{enumerate}

The latter two generators describe high-energy physics and 
adapting them to KLOE-2 is not straightforward.  
Most of the others are based on the EPA.
EPA is useful to identify the main properties of the two photon processes,
however it leads
to some discrepancies with respect to the exact formulation  
as shown in Ref. \cite{Brodsky:1971ud}, especially for
single--pseudo\-scalar final states.
As KLOE-2 aims at the measurement 
of the form factors with both photons off--shell, MC simulations 
beyond EPA are desirable to guarantee the accuracy of the theoretical
description.  

Basic requirements for a MC generator for 
$e^+e^- \to e^+e^- P$ 
studies not restricted to the region where the photons 
are quasi--real, are:
\begin{enumerate}
\item not to use EPA,
\item include $s$- and $t$-channel as well as their interference,
\item have the flexibility to allow user--defined form factors, 
\item have the flexibility to allow KLOE-2 specific kinematical cuts in
  order to speed up the generation procedure. 
\end{enumerate}
%
Some work is in progress \cite{Czyz:inprep} to develop the EKHARA code on
this line. 
In the present version, EKHARA provides a generic description of
the $e^+e^-\to e^+e^- \pi^+\pi^-$ process. As it was developed for
the background studies for the KLOE measurement of the pion form
factor~\cite{:2008en}, several modifications are needed before 
it can be used for two photon
physics, but it already contains all relevant contributions to 
the $e^+e^-\to e^+e^-\pi^+\pi^-$ amplitude.
\subsubsection{Contribution to Light-by-Light scattering} 
\label{sec:lble}
\begin{figure}[htbp]
\begin{center}
\resizebox{0.25\textwidth}{!}{
\includegraphics{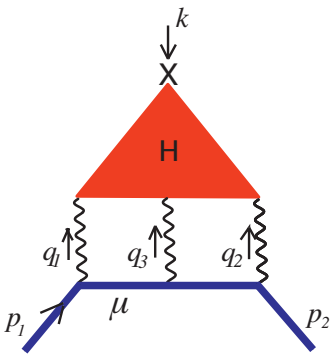}
}
\caption{Hadronic light-by-light scattering contribution.}
\label{fig:1}
\end{center}
\end{figure}

Figure \ref{fig:1} shows one of the six possible photon momenta configuration
of the hadronic light-by-light (HLbL) contribution to the muon anomalous
magnetic moment $a_{\mu}=(g_\mu-2)/2$.
This configuration is described by the vertex function 
\bea
\label{Mlbl}
\dis{\Gamma^\mu} (p_2,p_1)
&=&  e^6
\int {{\rm d}^4 q_1 \over (2\pi )^4}
\int {{\rm d}^4 q_2\over (2\pi )^4} Ê
{\Pi^{\mu\nu\rho\sigma} (k,q_1,q_3,q_2) 
\over q_1^2\, q_2^2 \, q_3^2} \nonumber \\ &\times& Ê
\gamma_\nu (\not{\! p}_2+\not{\! q}_2-m_\mu )^{-1} 
\gamma_\rho (\not{\! p}_1-\not{\! q}_1-m_\mu )^{-1} \gamma_\sigma \, 
\nonumber \\ 
\eea
where $k$ is the momentum of the
photon that couples to the external magnetic source,
$k=p_2-p_1=-q_1-q_2-q_3$ and $m_\mu$ is the muon mass. 
The dominant contribution to the hadronic four-point Êfunction 
\bea
\label{four}
\Pi^{\rho\nu\alpha\beta}(k,q_1,q_2,q_3)&=& \nonumber \\
i^3 \int {\rm d}^4 x \int {\rm d}^4 y
\int {\rm d}^4 z \, {\rm e}^{i (-q_1 \cdot x + q_2 \cdot y + q_3 \cdot z)}
&& \nonumber \\
\times \, \langle 0 | T \left[ ÊV^\mu(0) V^\nu(x) V^\rho(y) V^\sigma(z)
\right] |0\rangle && 
\eea
comes from the three light quark 
$(q = u,d,s)$ components of the electromagnetic current
$V^\mu(x)=\left[ \bar{q} \hat{Q} \gamma^\mu q \right](x)$
where $\hat{Q}$ denotes the quark electric charge matrix. 
In the limit $k \to 0$, current conservation implies:
\bea
\dis{\Gamma^\mu} (p_2,p_1) =
- \frac{a^{\rm HLbL}}{4 m_\mu} \, 
\left[\gamma^\mu, \gamma^\nu \right] \, q_\nu \, .
\eea

The main contributions to $a^{\rm HLbL}$ are summarized in
Ref.~\cite{Prades:2009tw} and reported in the following.   
Recent reviews of previous work 
~\cite{Hayakawa:1995ps,Hayakawa:1996ki,Hayakawa:1997rq,Bijnens:1995xf,Bijnens:1995cc,Bijnens:2001cq,Knecht:2001qf,Knecht:2001qg,Melnikov:2003xd} 
are in Refs.
~\cite{Prades:2008zz,Bijnens:2007pz,Bijnens:2007fp,DeRafael:2008iu,Hertzog:2007hz,Miller:2007kk,Jegerlehner:2009ry,Jegerlehner:2008zzb,Jegerlehner:2007xe}. 
The discussion in Ref. \cite{Prades:2009tw}, 
 according to an $1/N_c$
expansion~\cite{deRafael:1993za}, 
leads 
to 
the 
conclusions : 
\begin{itemize}
\item \emph{Contribution from $\pi^0, \eta$ and $\eta'$ exchange}:
implementing a new Operator Product Expansion (OPE) constraint
into a neutral pion exchange model, 
the authors of Ref.~\cite{Melnikov:2003xd} obtained 
$(11.4 \pm 1.0) \times 10^{-10}$.
Within the extended Nambu-Jona Lasinio (ENJL) mo\-del 
the momenta higher than a certain cutoff are accounted separately
via quark loops~\cite{Bijnens:1995xf,Bijnens:1995cc,Bijnens:2001cq} while 
they are included 
in the OPE--based model. 
Assuming that the bulk of high energy quark loops is associated with 
pseudoscalar exchange,  
in Ref.~\cite{Bijnens:1995xf,Bijnens:1995cc,Bijnens:2001cq} 
$(10.7 \pm 1.3) \times 10^{-10}$ is obtained. 
Taking into account 
these results,
the authors of Ref.\cite{Prades:2009tw} quote as 
central value the one in~\cite{Melnikov:2003xd} with the error as evaluated  
in Ref.\cite{Bijnens:1995xf,Bijnens:1995cc,Bijnens:2001cq}:
\be
a^{\rm HLbL}(\pi^0,\eta,\eta') = (11.4 \pm 1.3) \times 10^{-10} \, .
\ee
\item \emph{Contribution from pseudovector exchange}:
the analysis done in Ref.\cite{Prades:2009tw} 
suggests that the errors quoted within 
the large $N_c$ ENJL model are underestimated. 
Taking the average of the estimates and increasing the 
uncertainty to cover both models, Ref.\cite{Prades:2009tw} reports:  
\be
a^{\rm HLbL}({\rm pseudovectors}) = (1.5 \pm 1.0 ) \times 10^{-10} \, .
\ee
\item \emph{Contribution from scalar exchange}:
the ENJL model should give a reliable estimate 
of the large $N_c$ contributions, 
In Ref.\cite{Prades:2009tw} therefore the result from
~\cite{Bijnens:1995xf,Bijnens:1995cc} is kept but with a larger conservative
error to cover for unaccounted higher-energy resonances providing negative
contributions: 
\be
a^{\rm HLbL}({\rm scalars}) = -(0.7 \pm 0.7 )\times 10^{-10} \, .
\ee
\item \emph{Contribution from dressed pion and kaon loops}:
the NLO in $1/N_c$ contributions are 
the most complicated to calculate. 
In particular, the charged pion loop shows
a large instability due to model dependence. 
This and the contribution of higher-energy resonance loops 
are taken into account in Ref. \cite{Prades:2009tw} by 
choosing the central value as the full VMD result of 
Ref.\cite{Bijnens:1995xf,Bijnens:1995cc} with again a conservative error:
\be
a^{\rm HLbL}(\pi^+-{\rm dressed \,\, loop}) = -(1.9 \pm 1.9 ) \times
10^{-10} \, . 
\ee
\end{itemize}

Adding the contributions above  
as well as the small charm quark contribution and 
combining errors in quadrature, 
$a^{\rm HLbL}({\rm charm}) = 0.23 \times 10^{-10}$,
one gets the estimate~\cite{Prades:2009tw}:
\be
a^{\rm HLbL} = (10.5 \pm 2.6) \times 10^{-10} \, .
\ee

The proposed new $g_\mu-2$ experiment aims at 
an accuracy of $1.4 \times 10^{-10}$ calling 
for a considerable improvement on the present
calculations. The use of further theoretical and experimental constraints 
could result in reaching such accuracy soon. In particular, imposing 
as many as possible short-distance QCD constraints
~\cite{Hayakawa:1995ps,Hayakawa:1996ki,Hayakawa:1997rq,Bijnens:1995xf,Bijnens:1995cc,Bijnens:2001cq,Knecht:2001qf,Melnikov:2003xd} 
has already result in a better understanding of the dominant $\pi^0$
exchange. 
At present, none of the light-by-light hadronic parametrization satisfies 
fully all short-distance QCD constraints. 
In particular, this requires the inclusion of infinite number of narrow 
states for other than two-point functions and two-point functions with soft 
insertions~\cite{Bijnens:2003rc}. 
The dominance of certain momenta configurations
can help to minimize the effects of short distance QCD constraints
not satisfied, as in the model of Ref. \cite{Melnikov:2003xd}.

Recently, the $\pi^0$--exchange contribution to  $a^{\rm HLbL}$ 
has been evaluated from the off-shell form factor 
$\mathcal{F}_{\pi\gamma^*\gamma^*}$   
\cite{Dorokhov:2008pw,Nyffeler:2009tw},  
obtaining results 
similar to the ones quoted above.
How to take off-shellness effects consistently in the full 
four-point function of Eq.(\ref{four}) remains however an open question.

More experimental information on the decays $\pi^0 \to \gamma \gamma^*$,
$\pi^0 \to \gamma^* \gamma^*$ and $\pi^0 \to e^+ e^-$ 
(with radiative 
corrections included
~\cite{RamseyMusolf:2002cy,Kampf:2005tz,Kampf:2009tk,Kampf:2009kg}) 
can also help to confirm the results on neutral--pion exchange.

The $\gamma \gamma$ physics program at KLOE-2 will be 
well suited to study these processes 
contributing to solve the present model dependence of $a^{\rm
  HLbL}$. 

\begin{figure}[htbp]
\begin{center}
\resizebox{.5\textwidth}{!}{
\includegraphics{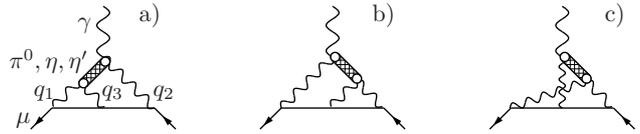}
}
\end{center}
\caption{Pseudoscalar exchange contribution to the light by light
  scattering.} 
\label{fig:pseudoscalar_exchange} 
\end{figure}

In fact,  the dominant contribution from pseudoscalar ($P$) exchange 
(Fig. \ref{fig:pseudoscalar_exchange}) can be written in terms of two
form factors, $\FFc$ and $\FF$ with one and two off-shell photons
respectively.   
\begin{eqnarray}
&& a^{\mathrm{HLbL}}(P) = Ê- e^6
\int {d^4 q_1 \over (2\pi)^4} {d^4 q_2 \over (2\pi)^4}
\, \frac{1}{q_1^2 q_2^2} \times \nonumber \\
&& \frac{1}{(q_1 + q_2)^2[(p+ q_1)^2 - m_\mu^2][(p - q_2)^2 -
m_\mu^2]} \times
\nonumber \\
&& Ê\left[
{\FF(q_2^2, q_1^2, q_3^2) \ \FFc(q_2^2, q_2^2, 0) \over q_2^2 -
m_{\pi}^2} \ T_1(q_1,q_2;p) + \nonumber \right. \\
&& Ê\left. {\FF(q_3^2, q_1^2, q_2^2) \
\FFc(q_3^2, q_3^2, 
0) \over q_3^2 - m_{\pi}^2} \ T_2(q_1,q_2;p) \right] 
\label{a_pion_2}
\end{eqnarray}
$T_1$ and $T_2$ are polynomial function of the photon and pseudoscalar
four momenta. 
\begin{figure*}[!tb]
\begin{tabular}{ccc}
\resizebox{.3\textwidth}{!}{\includegraphics{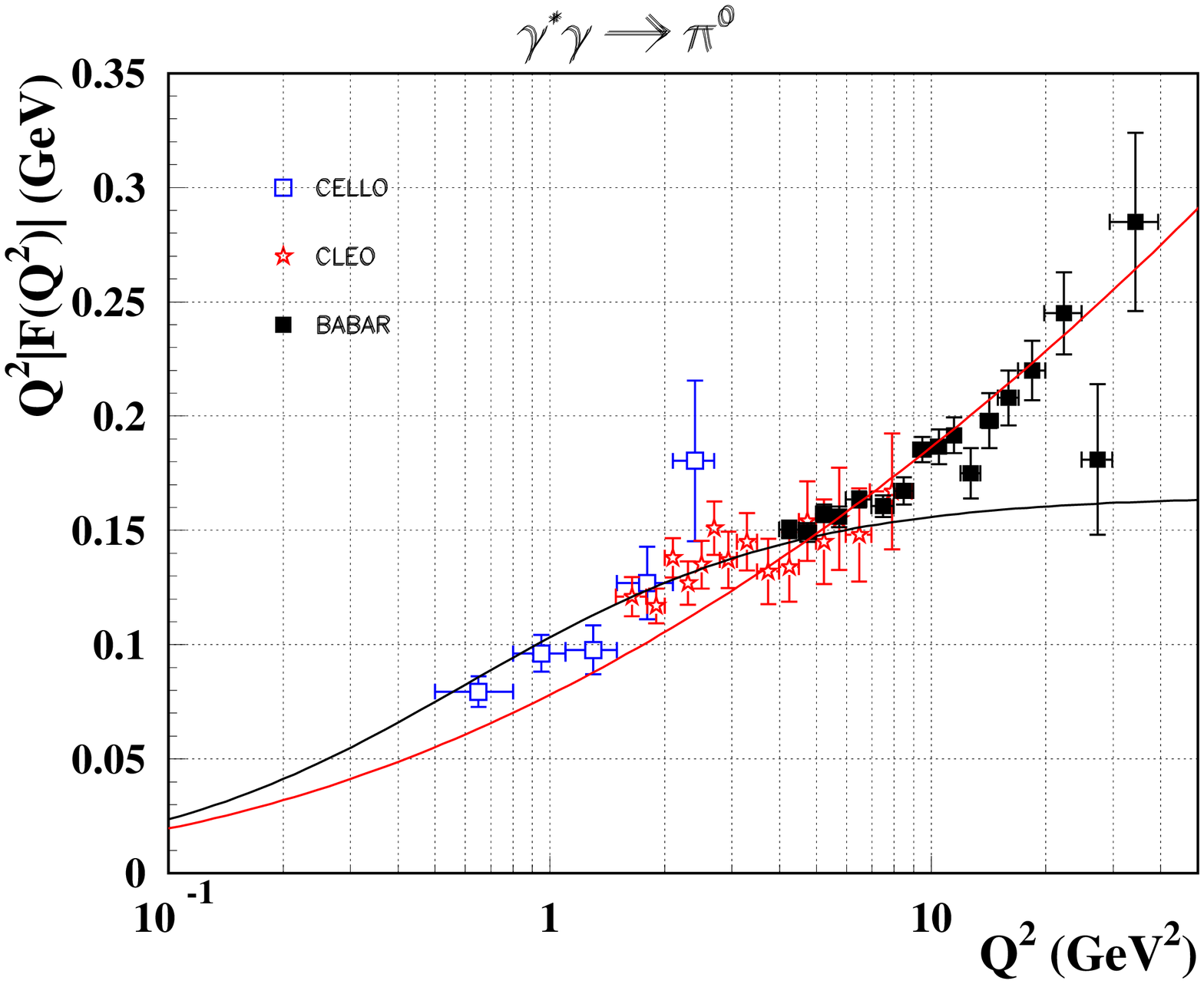}} & 
\resizebox{.3\textwidth}{!}{\includegraphics{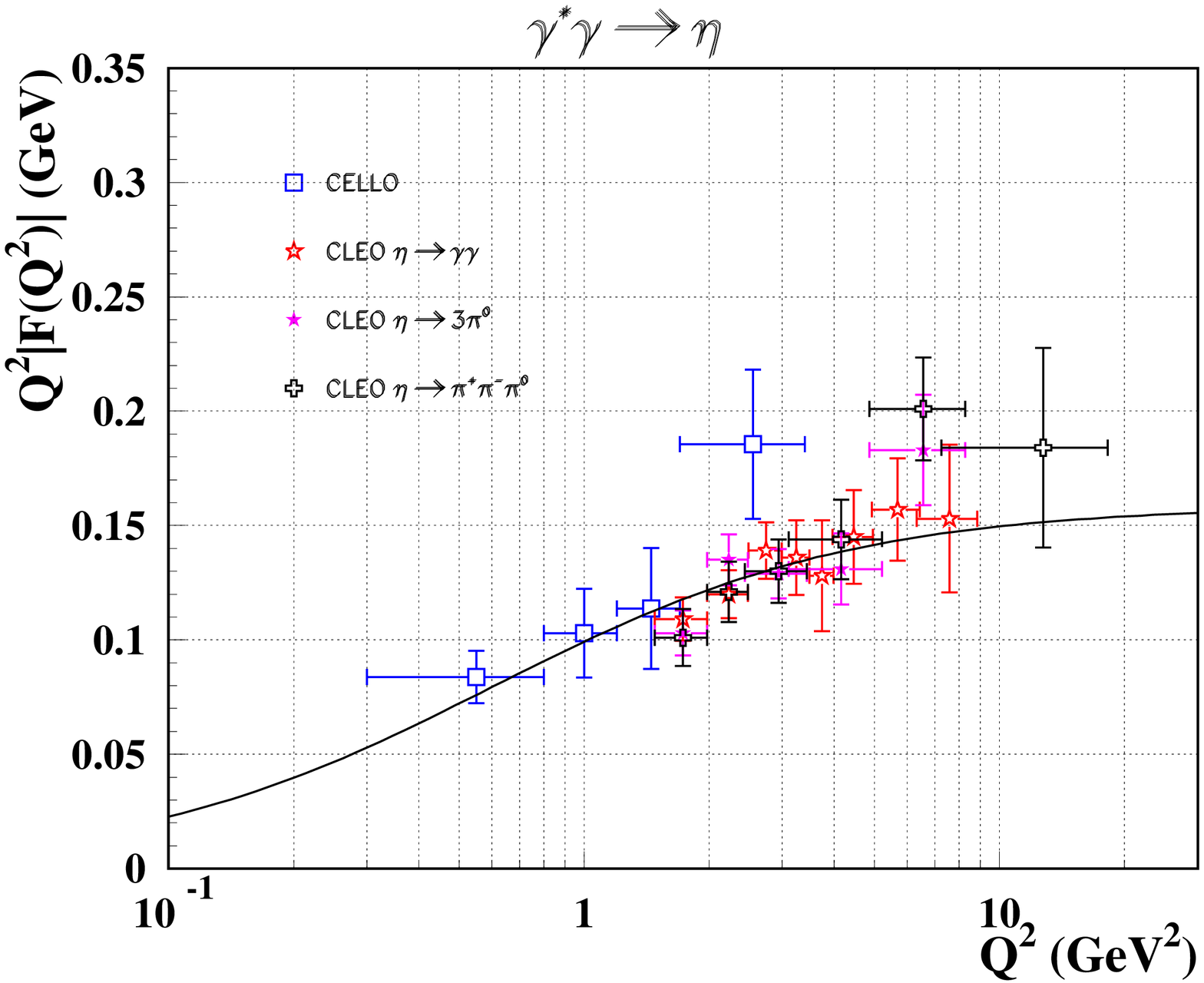}} & 
\resizebox{.3\textwidth}{!}{\includegraphics{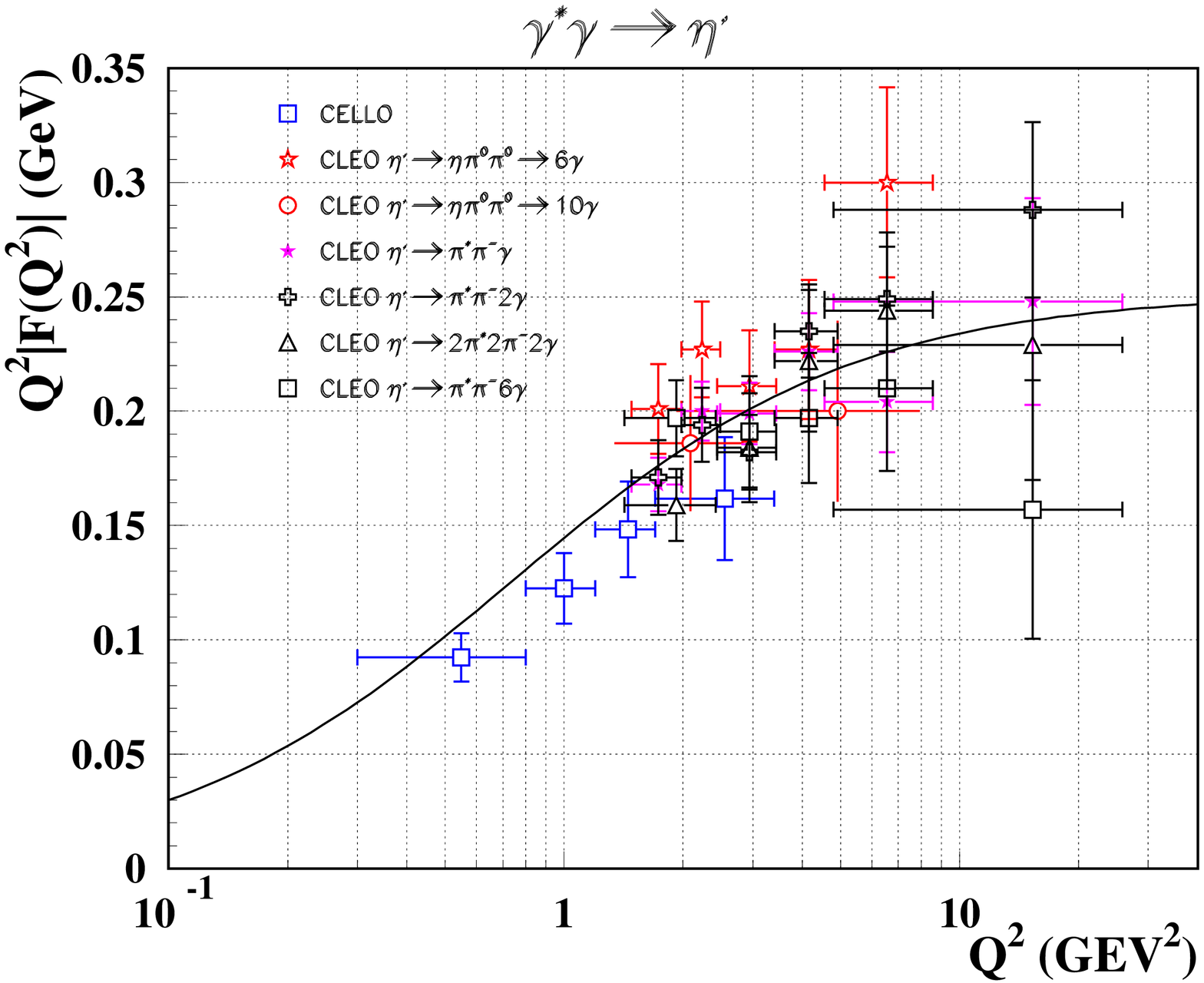}} \\
\end{tabular}
\caption{Left: the $\pi^0$ transition form factor as measured
by the CELLO, CLEO and BaBar experiments. The curve showing an asymptotic 
limit at 160 MeV is from CLEO parametrization \cite{Gronberg:1997fj} 
while the other is from the $\mathcal{F}_{\pi\gamma\gamma^*}(m^2_P,Q^2,0)$ 
expression given in Ref. \cite{Dorokhov:2009dg}. Center: the $\eta$ transition 
form factor as measured by CELLO and by CLEO in the specified 
$\eta$ decay channels.
Right: the $\eta^\prime$ form factor as measured by CELLO and by CLEO in the 
specified $\eta^\prime$ decay channels. The curves in the central and 
left panel  
represent the CLEO parametrization 
for the form factors \cite{Gronberg:1997fj}.}
\label{fig:pseudo1}
\end{figure*}

The functional form and the value of the form factors are theoretically
unknown. 
Several models have been used to evaluate them~\cite{Jegerlehner:2009ry},
experimental information can be used to constrain both the shape and the
value of the form factors.
 
The form factor at negative $q^2$ appears in the production cross
section of $\pi^0$, $\eta$ and $\eta^{\prime}$ mesons in the reaction $e^+
e^- \to e^+ e^- P$. 
By 
detecting one electron at large angle with respect to the beams, 
 $\mathcal{F}_{P\gamma\gamma^*}(m^2_P,Q^2,0)$ with one quasi--real and one
virtual spacelike photon ($Q^2=-q^2$) can be measured for the on--shell
pseudoscalar.  
 These form factors, as reviewed in Ref. \cite{Dorokhov:2009jd}, 
have been measured by the CELLO~\cite{Behrend:1990sr}, 
CLEO~\cite{Gronberg:1997fj} and 
recently BaBar ~\cite{Aubert:2009mc} collaborations in the range 
($1<Q^2<40$) GeV$^2$ using single--tagged samples. 
The experimental data are summarized in Fig. \ref{fig:pseudo1}.   
The region below 1 $GeV^2$ is still poorly measured and  
KLOE-2 can cover this region for both  
$\pi^0$ and $\eta$ mesons. 
Furthermore, by selecting events in which both $e^+$ and $e^-$
are detected by the drift chamber 
KLOE-2 can provide experimental information on form factors 
$\mathcal{F}_{P\gamma^* \gamma^*}(m^2_P,Q_1^2,Q^2_2)$, with two
virtual photons.


\section{Hidden WIMP Dark Matter}
%

\label{sec:DMquest}
In recent years, several astrophysical observations
have failed to find easy interpretations in terms of standard 
astrophysical and/or particle physics sources. 
A non exhaustive list of these observations includes 
the 511 keV gamma-ray signal from the galactic center observed by 
the INTEGRAL satellite~\cite{Jean:2003ci}, the excess in the 
cosmic ray positrons reported by PAMELA~\cite{Adriani:2008zr}, the total
electron and positron flux measured by ATIC~\cite{Chang:2008zzr}, Fermi~\cite{Abdo:2009zk}, 
and HESS \cite{Collaboration:2008aaa,Aharonian:2009ah}, 
and the annual modulation of the DAMA/LIBRA signal~\cite{Bernabei:2005hj,Bernabei:2008yi}.     

An intriguing feature of these observations is that 
they suggest 
\cite{Pospelov:2007mp,ArkaniHamed:2008qn,Alves:2009nf,Pospelov:2008jd,Hisano:2003ec,Cirelli:2008pk,MarchRussell:2008yu,Cholis:2008wq,Cholis:2008qq,ArkaniHamed:2008qp}) the existence of  
a WIMP dark matter particle belonging to a secluded gauge sector  
under which the SM particles are 
uncharged.
An abelian gauge field, the $U$ boson with mass near the GeV scale, 
couples the secluded sector to the SM through its kinetic mixing with the 
SM hypercharge gauge field\footnote{The 
possible existence of a light $U$ boson and the coupling to 
the SM particles were already postulated
several years ago~\cite{Fayet:1980ad,Fayet:1980rr,Holdom:1986eq}, 
in the framework of 
supersymmetric extensions of the SM. 
This boson can have both vector and axial-vector 
couplings to quarks and leptons; however, axial couplings are  
strongly constrained by data, leaving room for vector couplings
only.}
The kinetic mixing parameter, $\kappa$, can naturally be 
of the order 10$^{-4}$--10$^{-2}$. 

\begin{figure}[th]
\begin{center}
\includegraphics[width=.4\textwidth]{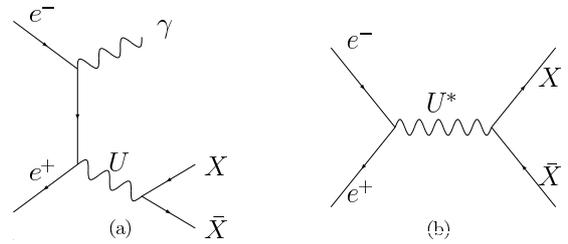}
\caption{\label{fig:diagrams} Production modes of light 
secluded-sector particles $X\bar{X}$ at \DAF.
Right: Production through an off-shell $U$.
Left:  Production of an on-shell $U$ and a photon --- the $U$ may subsequently decay into lighter
hidden-sector particles.}
\end{center}
\end{figure}

Annihilation of DM into the $U$ boson, which decays 
into charged leptons and is light enough to kinematically forbid 
a decay that produces anti-protons, can explain the electron 
and/or positron excesses, and the absence of a similar effect in the PAMELA 
anti-proton data.  
The condition $m_{U} \ll m_{\rm WIMP}$, where m$_{\rm WIMP}$ 
is the mass of the DM 
naturally leads to an enhanced WIMP 
annihilation cross section in the galaxy, as required by the data.  
If the DM is also charged under a non-abelian group, higgsed or confined near the 
GeV scale, 
then its spectrum naturally implements an inelastic DM 
scenario \cite{TuckerSmith:2001hy}, thereby explaining the annual modulation
signal reported by DAMA/LIBRA \cite{Bernabei:2005hj,Bernabei:2008yi} 
and reconciling it with the null results of other experiments 
\cite{ArkaniHamed:2008qn,Alves:2009nf,TuckerSmith:2001hy,Chang:2008gd}.
In this case, there could be many particles, such as hidden sector higgs and gauge 
bosons or hidden sector mesons, baryons, and glueballs, with 
masses near the GeV scale.\\ \indent
A dramatic consequence of the above hypotheses is that observable effects 
can readily be induced in $\mathcal{O}(\mbox{GeV}$)--energy \epem\ colliders, 
such as \DAF, and/or present and future B factories 
\cite{Batell:2009yf,Essig:2009nc,Reece:2009un,Bossi:2009uw,Borodatchenkova:2005ct,Yin:2009mc} 
and fixed target experiments 
\cite{Bjorken:2009mm,Batell:2009di,Essig:2010xa,Freytsis:2009bh} (c.f. also 
\cite{Strassler:2006im,Schuster:2009au,Strassler:2006qa,Han:2007ae,Strassler:2008bv,Baumgart:2009tn,Gunion:2005rw,McElrath:2005bp,Dermisek:2006py,Galloway:2008yh}).   
In particular, the $U$ boson can be directly produced at these 
low-energy facilities, 
or hidden sector particles at the GeV scale could be produced  
through an off-shell 
$U$ boson as in Fig. \ref{fig:diagrams}. \\ \indent  
\vskip 5mm
In a very minimal scenario, in addition to the $U$, it is natural to have a secluded Higgs 
boson, the $h'$, which spontaneously breaks the secluded gauge symmetry. 
Both the $U$ and the $h'$ can be produced at \DAF, 
and even in this minimal scenario a rich phenomenology is possible.  
The mass of the $U$ and $h'$ are both free parameters, and the 
possible decay channels can be very different depending on which 
particle is heavier.   
One thus has to consider different detection strategies for the 
two cases $m_{h'} < m_{U}$ and $m_{h'} > m_{U}$.  

In order for the $U$ and the $h'$ to be produced at \DAF, 
their masses need to be less than m$_{\phi}$.
This is somewhat a disadvantage with respect to the $B$-factories. 
Moreover, the integrated luminosity of the 
$B$-factories is larger than that 
of \DAF.  However, the production cross--sec\-tions 
scale as $1/s$, where $s$ is the squared center-of-mass energy, 
which partially compensates for the lower luminosity of  \DAF. 
In addition, there are scenarios with a confined secluded gauge group,  
in which searches at \DAF\ are easier than at the $B$-factories. 
A rather comprehensive discussion of these effects can be found 
in Refs. \cite{Batell:2009yf,Essig:2009nc}. 
In particular, the authors of Ref. \cite{Essig:2009nc} explicitely discuss 
the physics potential of \DAF.
It turns out that a wide spectrum of physics channels can be studied 
with KLOE, with rather different detection potentials to the $B$-factories, as 
further discussed 
in Ref. \cite{Bossi:2009uw}. 
\begin{figure}[th]
\begin{center}
\includegraphics[width=.80\columnwidth]{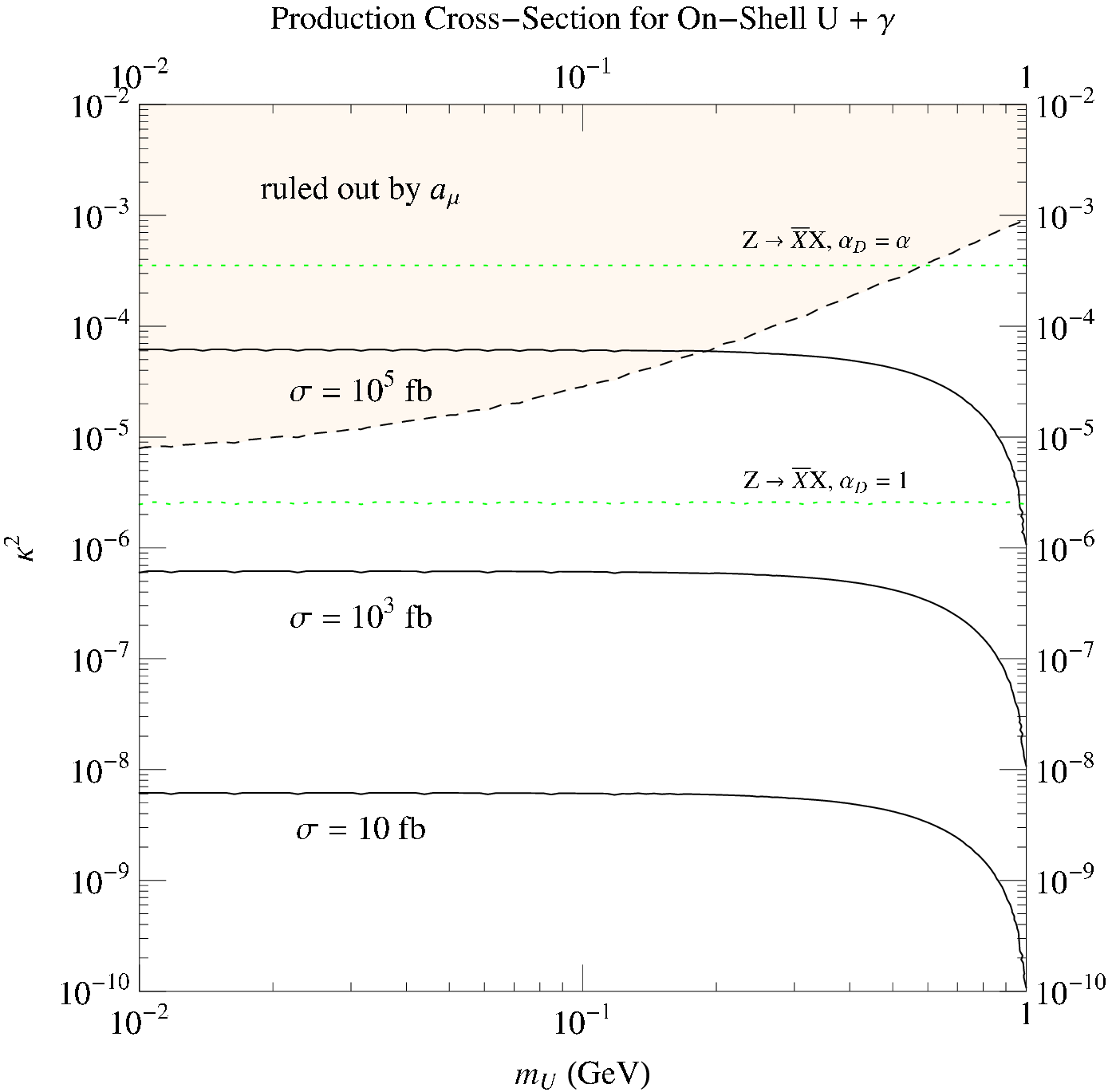}
\includegraphics[width=.80\columnwidth]{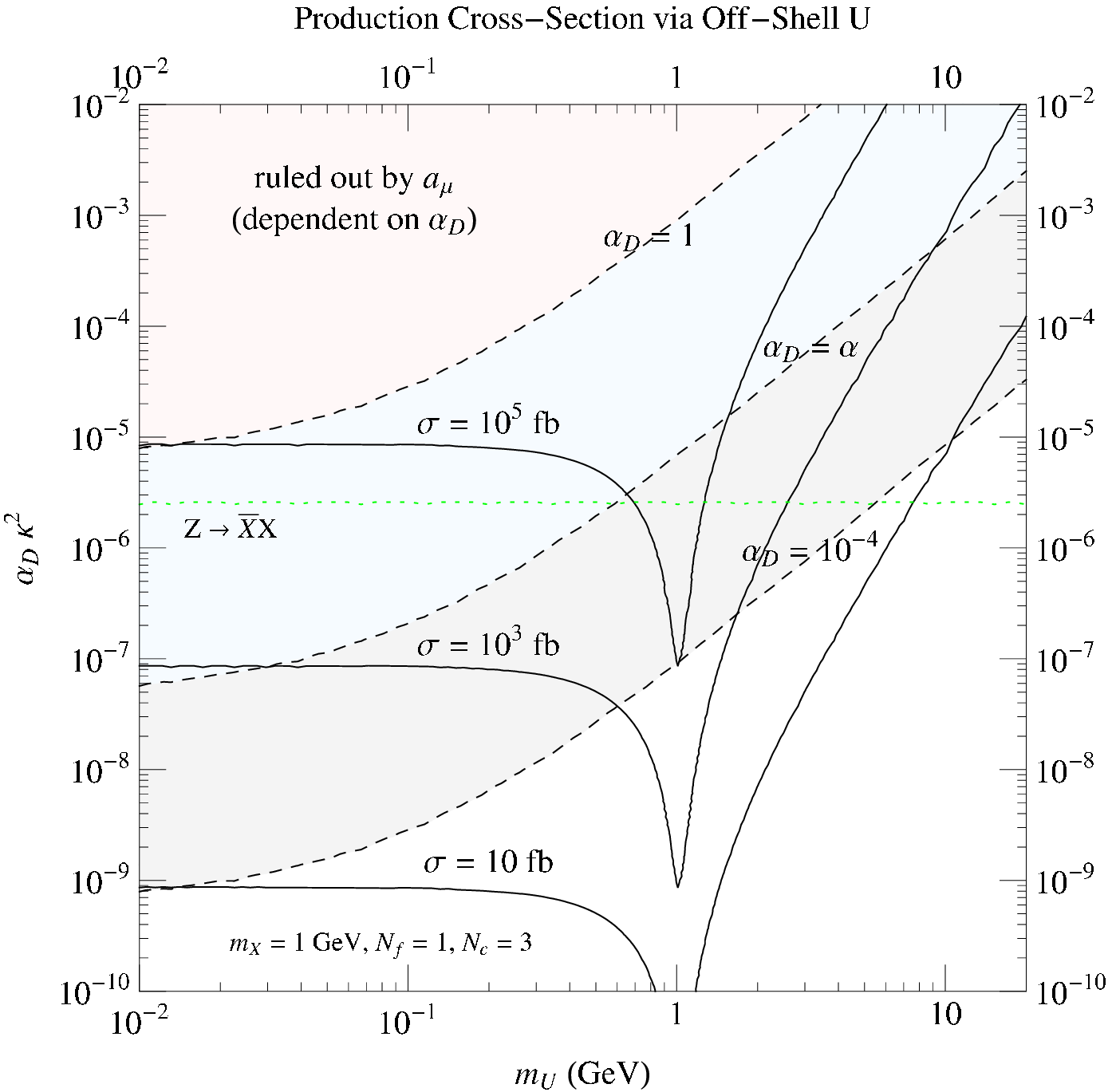}
\caption{\label{fig:DaphneCrossSection} 
Cross-section at \DAF\ for the production of an on-shell $U$ 
and a photon as a function of 
$\kappa^2$ and $m_{U}$ (top), and for the production of hidden-sector particles $X,\overline{X}$ through 
an off-shell $U$ as a function of $\alpha_D\kappa^2$(bottom). See text for more details.}
\end{center}
\end{figure}

One simple and interesting process to be studied is \epem\ $\rightarrow U\gamma$. 
The expected cross section can be as high as $\mathcal{O}({\rm pb})$ 
at \DAF\ energies, as shown in  
Fig. \ref{fig:DaphneCrossSection} (top). The on-shell boson can decay
into a lepton pair, giving rise to a $l^{+}l^{-}\gamma$ signal.
In the figure, 
black lines correspond to fixed cross-sections.
The constraints on the couplings of a 
new $U(1)$ gauge group kinetically mixing with hypercharge from measurements of
the muon anomalous magnetic dipole moment (shaded regions) 
\cite{Pospelov:2008zw} are also shown.
The 
dotted lines correspond to the lower bounds on the range of couplings that could be probed
by a search at LEP for rare $Z$-decays to various exotic final states, assuming that branching ratios as low
as $10^{-5}$ can be probed.  Original figures and more details can be found in \cite{Essig:2009nc}.
This decay channel was already addressed 
 in Ref. \cite{Borodatchenkova:2005ct} in which the possible existence 
of a low-mass ($\sim$10-100 MeV) $U$ boson was postulated in the framework
of models with scalar DM candidates~\cite{Boehm:2003hm}.   
The most relevant physics background comes from the  
QED radiative process with equal final state, which can be rejected by cutting 
on the invariant mass of the lepton pair~\cite{Borodatchenkova:2005ct}. 
A possible instrumental background that has to be taken
into consideration for the electron channel is due to
 \epem\ $\rightarrow \gamma\gamma$ events with 
subsequent conversion of one of the two photons on the beam pipe, 
with a probability of $\sim$2$\%$, that is particularly relevant for
low $U$-boson masses. The conclusion of~\cite{Bossi:2009uw} is that a reasonable
background rejection can be obtained for m$_{U}\ge$ 500 MeV. 
\par
The insertion of the inner tracker can be rather beneficial for 
the \epem\ $\rightarrow U\gamma\rightarrow l^+l^-\gamma$
case, since it will provide a better definition of the pair production vertex. 
A quantitative statement on this issue, however, needs the use of 
a detailed Monte Carlo simulation, which is at present unavailable.     
For the muon channel, the above mentioned background is not present. 
However, one has to take into account the physical process 
\epem\ $\rightarrow \pi\pi\gamma$, that is relevant since 
$\pi$-$\mu$ separation is non--trivial at \DAF\ energies. 
In general, the proposed detector upgrade for the second phase of 
KLOE-2 should be beneficial 
increasing acceptance  
both for charged tracks, thanks to the IT, and for photons, thanks 
to the forward calorimeters. \\ \indent
In Fig. \ref{fig:DaphneCrossSection} (bottom), the cross section for 
the production of an off-shell $U$ with 
subsequent decay of the $U$ into lighter
hidden-sector particles is shown. 
In this case, the cross section is a function of the 
product between $\kappa$ and the coupling
constant of the dark symmetry $\alpha_D$. The signature of these events 
is more complex than 
the previous one, depending on the available channels for the hidden-sector particles. The case
of multilepton channels will be briefly discussed further on. \\ \indent 
Assuming the existence of the $h'$, a particularly interesting 
process from the experimental point of view is the $h'$-strahlung, 
\epem\ $\rightarrow Uh'$, which can be observed 
at KLOE if $m_{U}$+$m_{h'}<m_{\phi}$.
As stated above, the signature of this process heavily depends on the
relation between $m_{U}$ and $m_{h'}$. 
Assuming that the $h'$ is lighter than the $U$ boson it 
turns out to be very long-lived \cite{Batell:2009yf} so that the signature 
of the process will be a lepton pair, generated by the $U$ boson decay, plus 
missing energy. \\ \indent 
There are several advantages for this type of signal. 
Firstly, there are no other physical processes with the same signature. 
The background due to QED $l^{+}l^{-}\gamma$  events with a 
 photon lost by the calorimeter, is highly suppressed  
due to the high detection efficiency of this device. Moreover, this 
kind of background would give rise to a missing momentum equal to 
the missing energy, while in the case of the signal these two quantities 
will be sizeably different, due to the non-zero $h'$ mass. In this case, 
differently from the one discussed above 
the key ingredient is the very high resolution of the DC as compared to
the calorimeter one. A third advantage in terms of both background 
rejection and detection efficiency is that the leptons tend to be produced at large 
angle (the angular distribution of the process is proportional 
to sin$^{3}\theta$, for m$_{h'}$, m$_{U} \ll \sqrt{s}$).
Finally, for a wide choice of m$_{U}$ and m$_{h'}$ 
the trigger efficiency should exceed 90$\%$.
\par \noindent 
The only physical process that can give rise to a dangerous background
at \DAF\ is the process $\phi \rightarrow K^{0}_{S}K^{0}_{L}$ followed
by a $K^{0}_{S} \rightarrow \pi^{+}\pi^{-}$ decay and the $K^{0}_{L}$
flying through the apparatus without interacting. This decay chain is 
relevant only for the $U \rightarrow \mu^{+}\mu^{-}$ channel and its 
contribution can be well calibrated by using the events in which the $K^{0}_{L}$
is observed in the apparatus. 
\par
In the case $m_{h'}>m_{U}$, the dark higgs frequently decays to a pair 
of real or virtual $U$'s. In this case one can observe events with 6 leptons
in the final state, due to the $h'$-strahlung process, or 4 leptons and 
a photon, due to the  \epem\ $\rightarrow h'\gamma$ reaction. 
\begin{figure}[th]
\includegraphics[width=0.8\columnwidth]{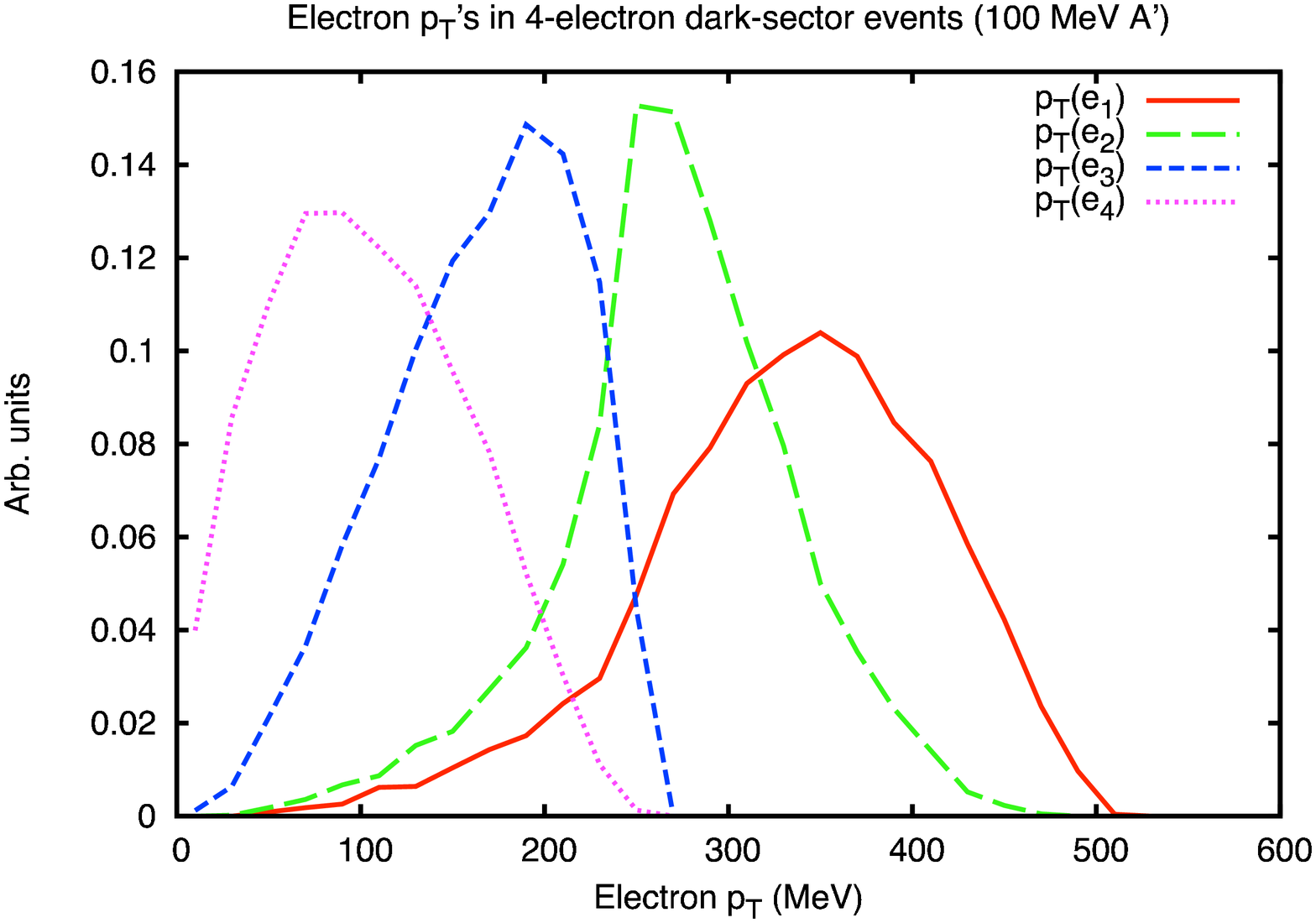}
\includegraphics[width=0.8\columnwidth]{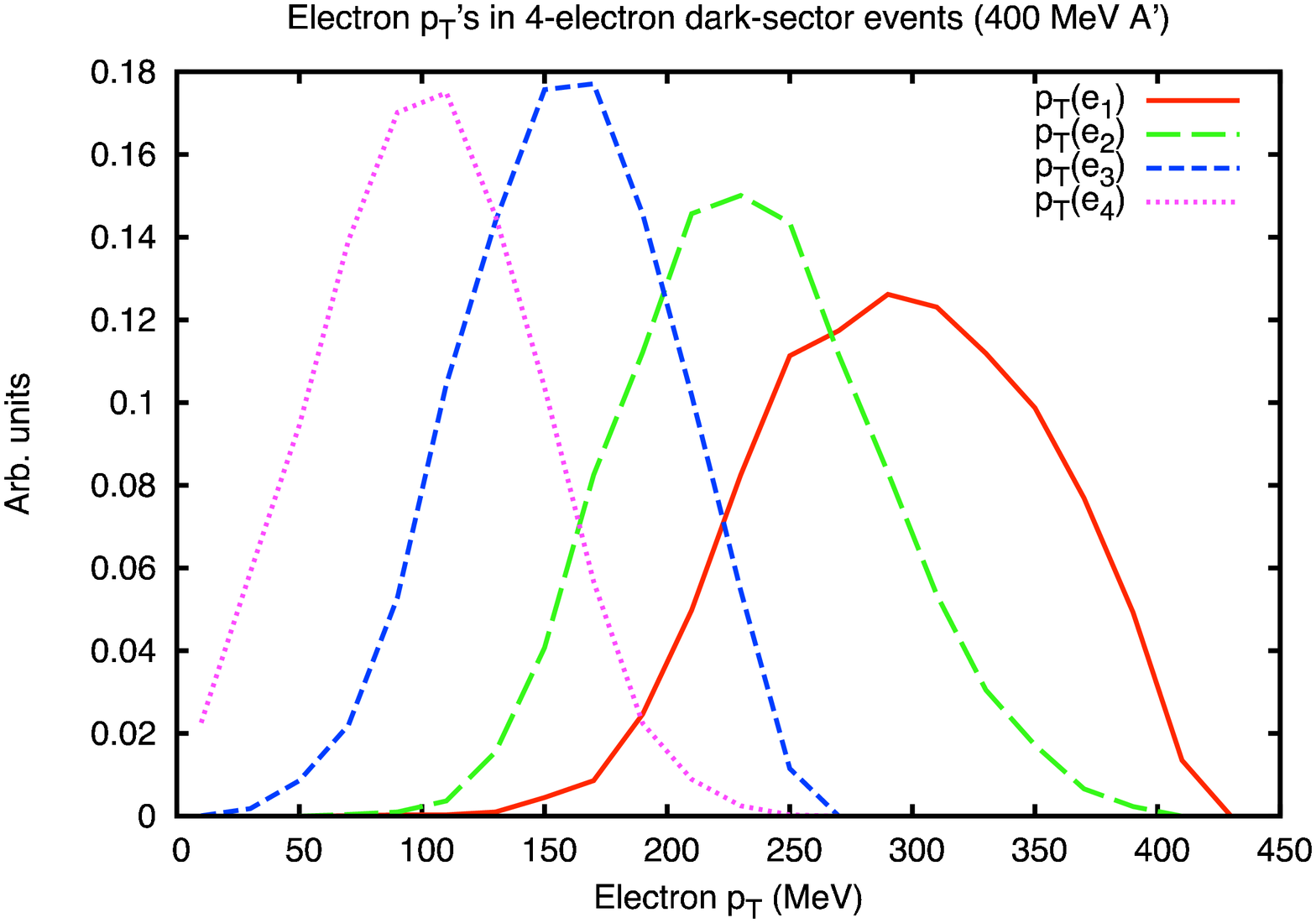}
\caption{\label{fig:leptonPts} 
Electron transverse momentum distributions for the process 
$e^+e^-\to U^* \to W_D W_D \to e^+e^-e^+e^-$ for two  
$U$-boson masses of 100 MeV (left) and 400 MeV (right). 
(Here $W_D$ refers to two massive hidden sector gauge bosons that 
mix with the $U$-boson.)}
\end{figure}
\begin{figure}[thbp]
\resizebox{0.8\columnwidth}{!}{
\includegraphics[angle=-90]{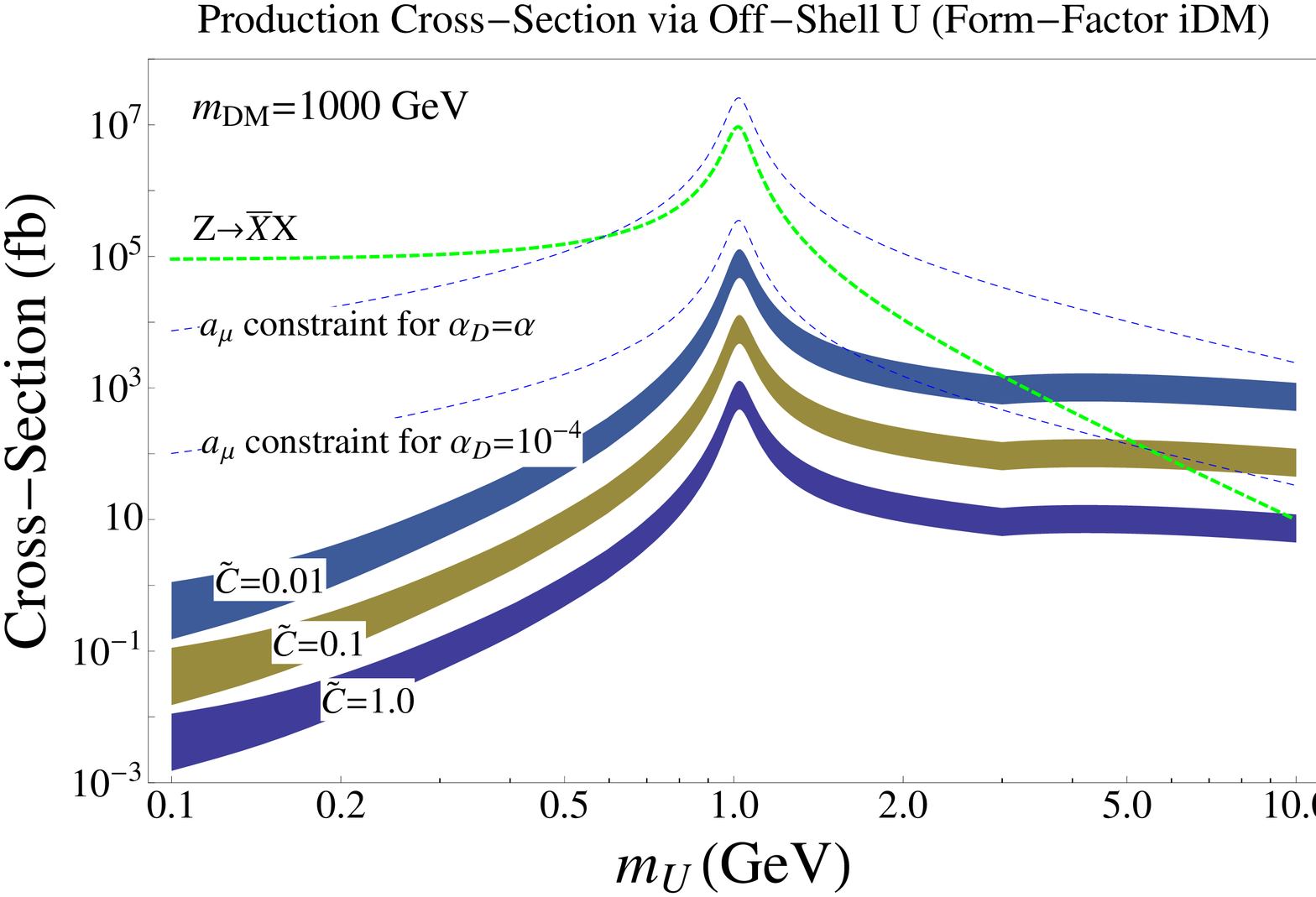}}
\resizebox{0.8\columnwidth}{!}{
\includegraphics[angle=-90]{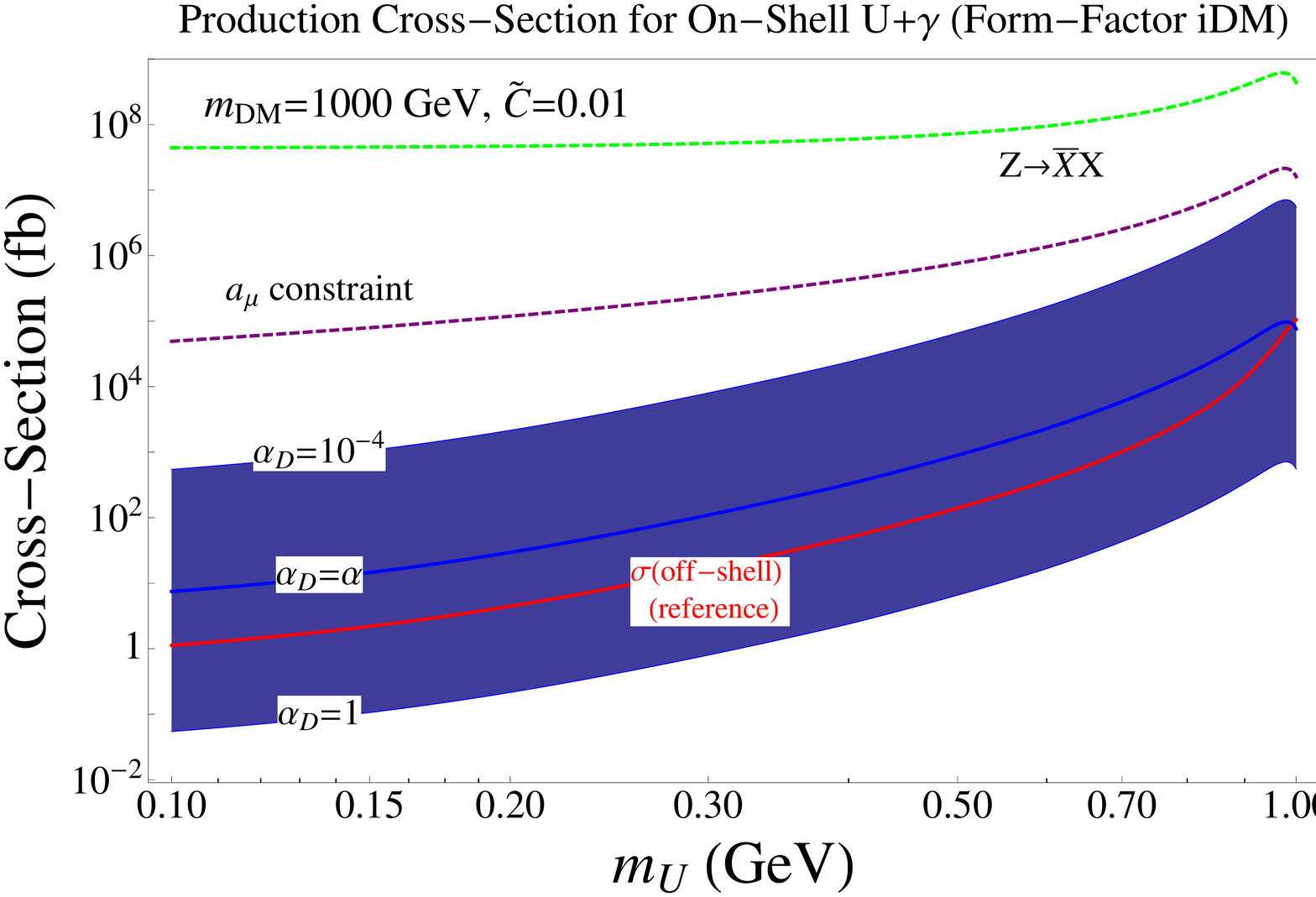}}
\caption{\label{fig:CompositeCrossSections} Inclusive cross-section at
  \DAF\ for production of dark--sector states, $X$, through
  an off-shell $U$ (top) and for production of an on-shell $U$ with
  a photon (bottom) as a function of $m_{U}$, after normalizing 
the couplings to the observed DAMA/LIBRA modulation rate 
assuming ``form-factor-suppressed
  dipole inelastic dark matter scattering'' --- 
see \cite{Essig:2009nc} for all the details.}
\end{figure}
Albeit very spectacular, the occurrence of these kind of events 
at KLOE is limited by the 
center--of--mass energy, 
especially
for the muon channel. 
In Fig. \ref{fig:leptonPts}, we show the electron transverse
momentum distributions for the process 
$e^+e^-\to U^* \to W_D W_D \to e^+e^-e^+e^-$ for two  
$U$-boson masses of 100 MeV (top) and 400 MeV (bottom). 
For such low multiplicity electron events, it is possible to completely 
reconstruct the event.  
However, the higher the multiplicity, the lower is the 
value of the minimal transverse momentum for the charged particles, resulting
in a possible sizable loss of acceptance. In this regard, 
the insertion of the IT 
would be clearly beneficial.  
In order to make more quantitative statements, however, a detailed
Monte Carlo simulation is needed for the higher multiplicity lepton signals. 
\\ \indent 
It is important to stress that multilepton events are 
expected also 
in more complex models with respect to the one we have so far 
taken into consideration. 
For instance in models with a confined sector, such as the one of 
Ref. \cite{Alves:2009nf}, one has hidden sector quarks which can be produced 
through an off-shell $U$ boson.  
The cross-section for this process can be very large, 
of $\mathcal{O}$(pb) and larger as shown in Fig. \ref{fig:DaphneCrossSection} (bottom).  
Once produced, these hidden-sector quarks can 
confine into hidden-sector vector mesons, that can easily decay back to 
leptons, and hidden-sector scalar mesons that cannot, so that 
a possible resulting signal is multileptons plus missing energy.
Recently, the BaBar collaboration has searched for 
4 lepton events in the three possible channels, 
4$e$, 4$\mu$ and 2$e$2$\mu$. A preliminary upper limit on
 $\alpha_D\kappa^2$ of order 10$^{-9}$, slightly dependent on the 
W$_{D}$ gauge boson mass, has been set for $M_{W_D}$ 
in the range [0.5--5] GeV~\cite{Aubert:2009pw}. 
\par
We briefly discuss the implications for \DAF\ of taking seriously 
the inelastic dark 
matter (iDM) explanation of DAMA/LIBRA and the null results of the other 
direct detection experiments \cite{TuckerSmith:2001hy,Chang:2008gd}.  
In proposals to generate the iDM splitting among the dark matter states from new
non-Abelian gauge interactions in a $\sim$ GeV-mass hidden sector, it is
also natural to assume that scattering is mediated by the $U$-boson
\cite{ArkaniHamed:2008qn}.
In such models, the signal rate
and spectrum reported by DAMA/LIBRA constrain the couplings of the $U$
to the SM electromagnetic current.
This allows us to estimate the inclusive cross-section for production of
hidden-sector states at \DAF\  for a given iDM model.
The expected cross-sections can be $\mathcal{O}$(pb), at least for  
$m_{U} \sim $ GeV. 
However, since the expected production cross-sections 
scale as $\sim m_{U}^4$, very low-mass $U$'s may evade detection in
existing data. 
The expected cross-sections are shown in Fig. \ref{fig:CompositeCrossSections}.    
The coefficient $\tilde{C}$ in these 
figures roughly measures the 
  effective strength of the DM coupling to nuclei, 
and thus affects the predicted production cross-section at \DAF.  
  Also shown are the constraints on the couplings of a 
new $U(1)$ gauge group kinetically mixing with hypercharge from measurements of
the muon anomalous magnetic dipole moment \cite{Pospelov:2008zw}.
The green dotted lines correspond to the lower bounds on the range 
of couplings that could be probed by a search at LEP for rare $Z$-decays 
to various exotic final states, assuming that branching ratios as low 
as $10^{-5}$ can be probed;
$m_{DM}$ is set to 1 TeV.  The upper plot, where $m_X=0.2$ GeV, 
the number of hidden-sector flavors, $N_f=1$, and the number   
 of colors in the hidden sectors, $N_c=3$, 
  shows various choices of $\tilde{C}=\{0.01,0.1,1.0\}$; in the 
  bottom one, $\alpha_D=\{1,\alpha,10^{-4}\}$ with 
  $\tilde{C}=0.01$.   
  Note that the off-shell production cross-section 
  scales linearly with the number of dark flavors $N_f$ and dark  
  colors $N_c$.  On both plots constraints from 
  $a_{\mu}$ as well as the rare $Z$ decay sensitivity region are also shown.
The details involved in generating this figure are discussed in \cite{Essig:2009nc}, to which 
we refer the interested reader.
\par
The $U$ boson can also be produced in radiative decays of neutral 
mesons, where kinematically
allowed. This possibility has been discussed in detail in~\cite{Reece:2009un}, 
where several different processes are taken into consideration. 
It turns out that the $\phi \rightarrow \eta U$ decays can potentially probe
 couplings down to $\kappa \sim$10$^{-3}$, i.e., can cover most of 
the parameter range interesting for the theory.
\par
The $U$ boson can be observed by its decay into a
lepton pair, while the $\eta$ can be tagged by its two-photon or 
$\pi^{+}\pi^{-}\pi^{0}$ decays. 
This channel is very attractive from the experimental point of view, since in the 
final state one has to reconstruct the decays of {\it two} particles, 
which greatly improves background rejection.
A possible instrumental background comes from the 
standard radiative process, $\phi \rightarrow \eta \gamma$, 
with $\gamma$ conversions on the beam pipe. 
As for the previously discussed similar cases, 
the same experimental considerations apply here.      

\begin{figure}[!h]
\begin{center}
\resizebox{0.75\columnwidth}{!}{
\includegraphics{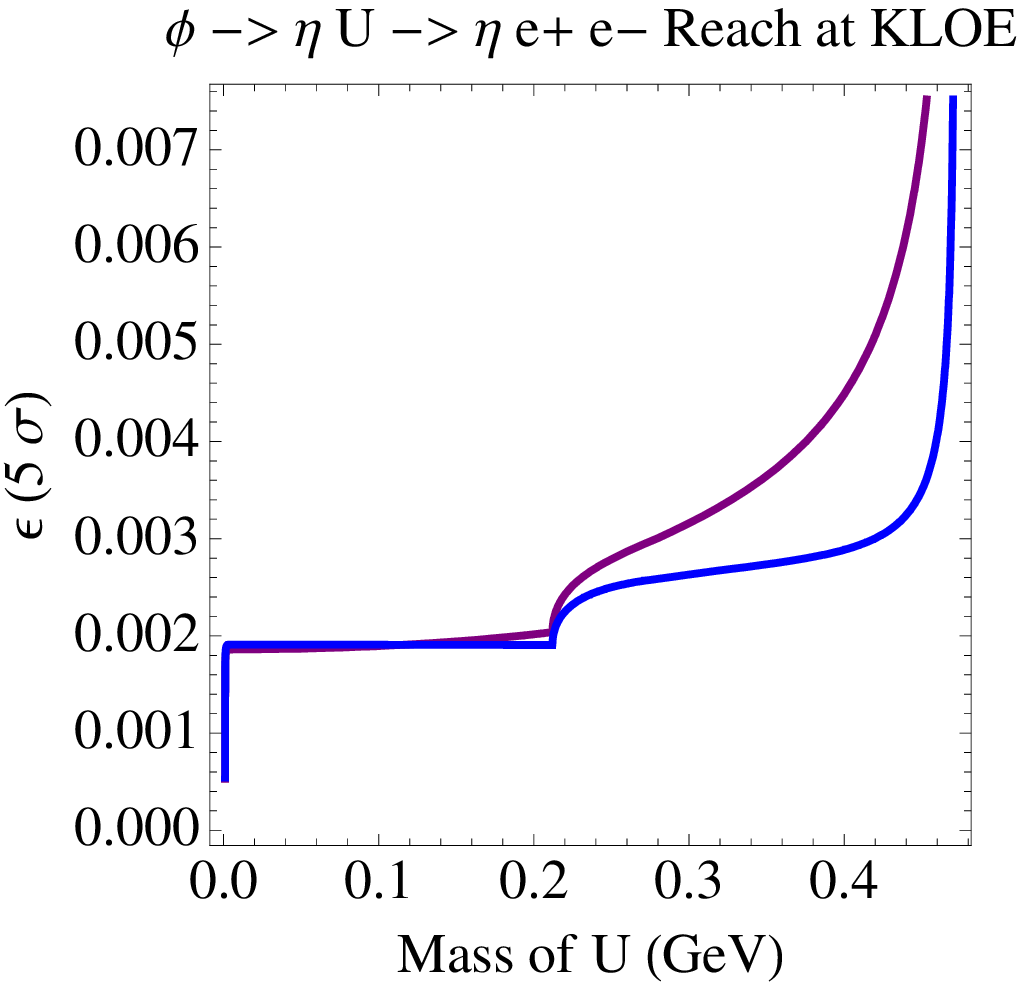}}
\vspace{4mm}
\resizebox{0.75\columnwidth}{!}{
\includegraphics{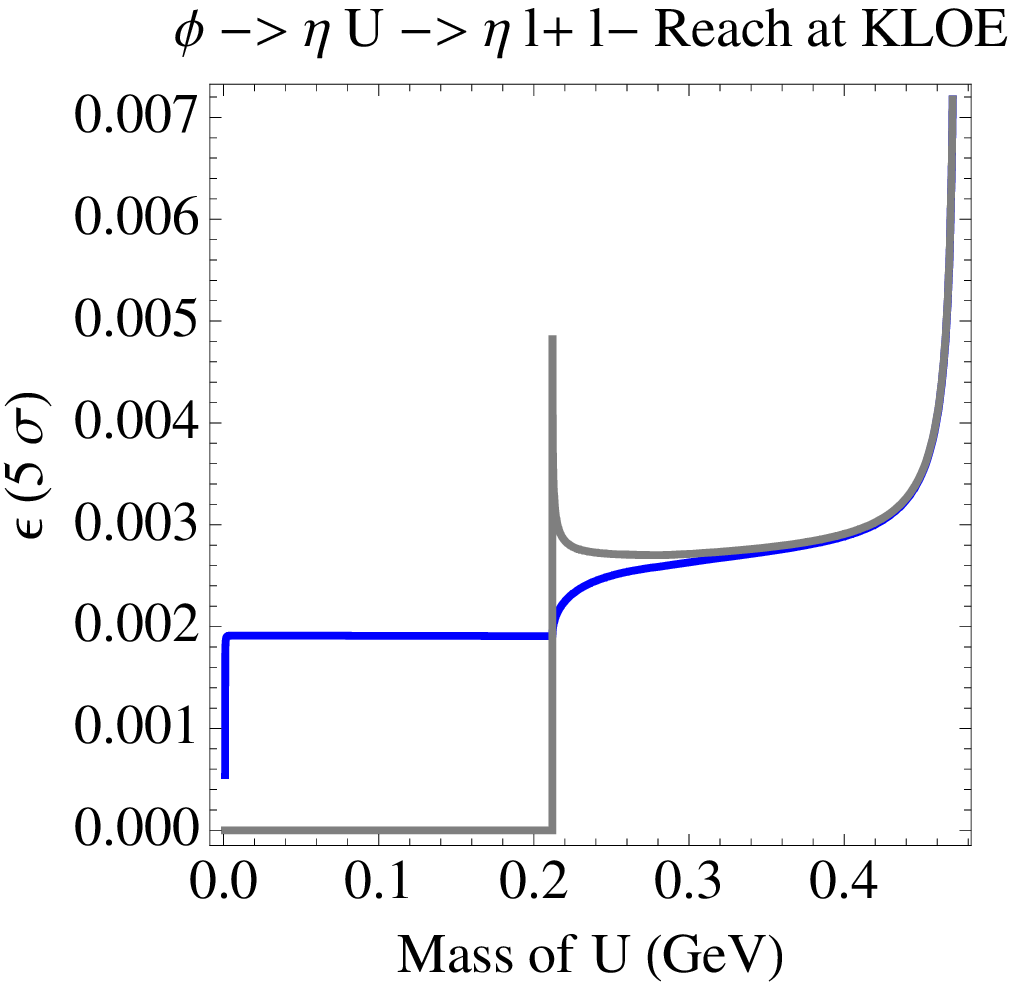}
}
\end{center} 
\caption{Reach for $U \to e^+ e^-$ at KLOE in the 
process $\phi \to \eta U$.  
Top: KLOE reach in $e^+e^-$. 
The upper curve is for constant form factor 
$F_{\phi\eta\gamma^*}(q^2) = 1$, 
whereas the lower curve is for the 
single-pole fit $F_{\phi\eta\gamma^*}(q^2) = 1/(1 - 3.8 {\rm GeV}^{-2}
q^2)$ 
from Ref. \cite{Achasov:2000ne}. 
Bottom: The reach in $e^+ e^-$ 
with single-pole form factor as above is compared with 
the corresponding reach in $\mu^+ \mu^-$ (gray curve).} \label{fig:KLOEreach}
\end{figure}

In Figure \ref{fig:KLOEreach} we present the reach of the 
U-boson using $\phi \to \eta U$. Specifically, 
we estimate the number of background events at given $m_U^2$ 
by taking the shape of the background distribution and 
normalizing it to the PDG branching fraction.  
Given this number of background events (in a window of size 
determined by momentum resolution of 0.4\% ), the value 
of $\epsilon$ for which $S/\sqrt{B} = 5$ is plotted.  
We emphasize that this is only an estimate of possible reach, 
assuming 8 billion total $\phi$'s. While it is simple to scale 
for different integrated luminosity, as the reach is proportional 
to $\mathcal{L}^{1/4}$, the actual reach requires detailed study 
fully including detector effects.   
Different choices of form factor significantly modify the reach 
near the upper end of the kinematically accessible range for $m_U$, 
but have little effect at the low end. The kink in the curve just above 
0.2 GeV is from the sudden drop in $U \to e^+ e^-$ branching fraction 
when the $\mu^+ \mu^-$ decay mode opens up. At around 0.28 GeV, 
the decay to $\pi^+ \pi^-$ opens up. We estimate its branching ratio 
using a vector meson dominance model of the pion form factor, 
$F_\pi(q^2) = \frac{1}{1 - q^2/m_\rho^2}$, which is approximately 
valid since we are probing $m_U \ll m_\rho$. 
The branching fraction to pions remains small ($\approx$14\%) 
even at $m_U \approx m_\phi - m_\eta$, so this mode has only 
a small effect on the reach. After the di-muon threshold, 
the combined reach of $e^+ e^-$ and $\mu^+ \mu^-$ is similar 
to the $e^+ e^-$ reach below the threshold.

\section{Summary}
\label{sec:fine}
%
The scientific program at the upgraded $\phi$--factory 
deserves the attention of particle physicists involved in several 
different experimental and theoretical projects.

Lattice QCD and kaon physics are entering a new phase in which 
experimental measurements and theoretical predictions are sufficiently 
precise to be sensitive to NP. 
One of the milestones of KLOE-2 is the improvement of the 
experimental inputs to \Vus\  
and thus the sensitivity of the unitarity 
test discussed in Sect. \ref{sec:SMtest}, 
for which unquenched lattice calculations are in progress 
to reduce the uncertainty on the form factors. 
Interesting enough, 
precision studies of kinematical distribution of the kaon semileptonic decays 
can independently provide the $f_K$/$f_\pi$/$f_+$(0) value for the 
validation of new lattice calculations. 

The accomplishment of the entire experimental program, with 
the completion of the detector upgrades in year 2011 followed by 
a data-taking campaign to integrate a luminosity of 20 fb$^{-1}$, 
would allow extensive, unique tests 
of CPT invariance and QM, 
investigating on fundamental questions 
of interest for the quantum theory of gravity.   

The program is not limited to kaon physics. Low-energy QCD 
issues are also addressed with the analysis 
of $\eta$ and $\eta^\prime$ decays and the study of 
two-photon production of hadronic resonances. 

Precision measurements of (multi)hadronic cross sections 
from $\pi \pi$ threshold to 2.5 GeV are required for the calculation 
of the hadronic corrections to the fine-structure constant at the $M_Z$ scale 
and for improving the theoretical accuracy of the anomalous magnetic 
moment of the muon.
Progress in this 
sector is necessary for precision physics at the ILC and to solve 
the 3--$\sigma$ discrepancy between the measurement and the SM prediction 
of $a_\mu$, for which new experiments at the FNAL and J-PARC have been 
proposed.   

The unique opportunity to investigate with proper sensitivity 
the elusive production of new particles 
is crucial for probing 
the hypothesis of a low-mass DM sector.

The list of measurements is not exhaustive but based on 
several focussing questions 
triggered by both the experimental and theoretical sides,
chosen as benchmarks for planning and operating  
KLOE-2 at \DAF.  
%
\section*{Acknowledgments}
%
%
%
We thank all 
the other 
participants to the workshop on the 
physics programme at the upgraded \DAF\ for their suggestions and useful 
discussions. The helpful and enjoyable collaboration of the KLOE 
colleagues is gratefully acknowledged. 
This work was supported in part 
by the European Union under contracts of the FP6 Marie
Curie Research and Training Networks: \emph{FLAVIAnet} 
(MRTN--CT--2006--035482), \emph{UniverseNet} (MRTN--CT--2006--035863),\\ 
  and \emph{HepTools} (MRTN--CT--2006--035505), and by 
the FP7 Research Infrastructure \emph{HadronPhysics2}
(INFRA--2008--227431).
The authors from Jagiellonian University (Cracow) and Soltan Institute 
for Nuclear Studies (Warsaw) were partially supported 
by MNiSW Grant No. 0469/\-B/H03\-/2009\-/37. 
B.C. Hiesmayr gratefully acknowledges 
the fellowship MOEL 428. 
J.Prades contribution was partially supported by MICINN, 
Spain and FEDER, European Commission
(EC)-- Grant No. FPA2006-05294, by the Spanish 
Consolider-Ingenio 2010 Programme
CPAN -- Grant No. CSD2007-00042, and by Junta de Andaluc\'{\i}a Grants
 No. P05-FQM 347 and P07-FQM 03048. 

\bibliographystyle{epj}
\bibliography{KLOE2physics}

\begin{thebibliography}{581}

\bibitem{Milardi:2009zz}
C.~Milardi et~al., ICFA Beam Dyn. Newslett. \textbf{48}, 23 (2009)

\bibitem{Zobov:2009zz}
M.~Zobov et~al., ICFA Beam Dyn. Newslett. \textbf{48}, 34 (2009)

\bibitem{KLOE2EoI}
R.~Beck et~al. (KLOE-2) (2006), {KLOE-2 Public Documents - K2PD-1},
  \texttt{http://www.lnf.infn.it/kloe2/}

\bibitem{Adinolfi:2002uk}
M.~Adinolfi et~al., Nucl. Instrum. Meth. \textbf{A488}, 51 (2002)

\bibitem{Adinolfi:2002zx}
M.~Adinolfi et~al., Nucl. Instrum. Meth. \textbf{A482}, 364 (2002)

\bibitem{Adinolfi:2002hs}
M.~Adinolfi et~al. (KLOE), Nucl. Instrum. Meth. \textbf{A492}, 134 (2002)

\bibitem{Adinolfi:2002me}
M.~Adinolfi et~al., Nucl. Instrum. Meth. \textbf{A483}, 649 (2002)

\bibitem{Aloisio:2004ig}
A.~Aloisio et~al., Nucl. Instrum. Meth. \textbf{A516}, 288 (2004)

\bibitem{Bossi:2008aa}
F.~Bossi, E.~De~Lucia, J.~Lee-Franzini, S.~Miscetti, M.~Palutan (KLOE), Riv.
  Nuovo Cim. \textbf{031}, 531 (2008), \texttt{0811.1929}

\bibitem{KLOE2LoI}
R.~Beck et~al. (KLOE-2) (2007), {Letter of Intent for the KLOE-2 Roll-in},
  \texttt{{LNF-07/19(IR)}}

\bibitem{Archilli:2010xb}
F.~Archilli et~al. (KLOE-2) (2010), \texttt{1002.2572}

\bibitem{Marciano:1987ja}
W.J. Marciano, A.~Sirlin, Phys. Rev. \textbf{D35}, 1672 (1987)

\bibitem{Hagiwara:1995fx}
K.~Hagiwara, S.~Matsumoto, Y.~Yamada, Phys. Rev. Lett. \textbf{75}, 3605
  (1995), \texttt{hep-ph/9507419}

\bibitem{Kurylov:2001zx}
A.~Kurylov, M.J. Ramsey-Musolf, Phys. Rev. Lett. \textbf{88}, 071804 (2002),
  \texttt{hep-ph/0109222}

\bibitem{Cirigliano:2009wk}
V.~Cirigliano, J.~Jenkins, M.~Gonzalez-Alonso, Nucl. Phys. \textbf{B830}, 95
  (2010), \texttt{0908.1754}

\bibitem{Cirigliano:2007xi}
V.~Cirigliano, I.~Rosell, Phys. Rev. Lett. \textbf{99}, 231801 (2007),
  \texttt{0707.3439}

\bibitem{Cirigliano:2007ga}
V.~Cirigliano, I.~Rosell, JHEP \textbf{10}, 005 (2007), \texttt{0707.4464}

\bibitem{Amsler:2008zzb}
C.~Amsler et~al. (Particle Data Group), Phys. Lett. \textbf{B667}, 1 (2008)

\bibitem{Ambrosino:2009rv}
F.~Ambrosino et~al. (KLOE), Eur. Phys. J. \textbf{C64}, 627 (2009),
  \texttt{0907.3594}

\bibitem{Antonelli:2008jg}
M.~Antonelli et~al. (FLAVIAnet working group on kaon decays) (2008),
  \texttt{0801.1817}

\bibitem{Masiero:2005wr}
A.~Masiero, P.~Paradisi, R.~Petronzio, Phys. Rev. \textbf{D74}, 011701 (2006),
  \texttt{hep-ph/0511289}

\bibitem{Cirigliano:2004pv}
V.~Cirigliano, H.~Neufeld, H.~Pichl, Eur. Phys. J. \textbf{C35}, 53 (2004),
  \texttt{hep-ph/0401173}

\bibitem{Cirigliano:2001mk}
V.~Cirigliano, M.~Knecht, H.~Neufeld, H.~Rupertsberger, P.~Talavera, Eur. Phys.
  J. \textbf{C23}, 121 (2002), \texttt{hep-ph/0110153}

\bibitem{Andre:2004tk}
T.C. Andre, Annals Phys. \textbf{322}, 2518 (2007), \texttt{hep-ph/0406006}

\bibitem{Cirigliano:2008wn}
V.~Cirigliano, M.~Giannotti, H.~Neufeld, JHEP \textbf{11}, 006 (2008),
  \texttt{0807.4507}

\bibitem{Ademollo:1964sr}
M.~Ademollo, R.~Gatto, Phys. Rev. Lett. \textbf{13}, 264 (1964)

\bibitem{Leutwyler:1984je}
H.~Leutwyler, M.~Roos, Z. Phys. \textbf{C25}, 91 (1984)

\bibitem{Bijnens:2003uy}
J.~Bijnens, P.~Talavera, Nucl. Phys. \textbf{B669}, 341 (2003),
  \texttt{hep-ph/0303103}

\bibitem{Bernard:2007tk}
V.~Bernard, E.~Passemar, Phys. Lett. \textbf{B661}, 95 (2008),
  \texttt{0711.3450}

\bibitem{Becirevic:2004ya}
D.~Becirevic et~al., Nucl. Phys. \textbf{B705}, 339 (2005),
  \texttt{hep-ph/0403217}

\bibitem{Jamin:2004re}
M.~Jamin, J.A. Oller, A.~Pich, JHEP \textbf{02}, 047 (2004),
  \texttt{hep-ph/0401080}

\bibitem{Cirigliano:2005xn}
V.~Cirigliano et~al., JHEP \textbf{04}, 006 (2005), \texttt{hep-ph/0503108}

\bibitem{Kastner:2008ch}
A.~Kastner, H.~Neufeld, Eur. Phys. J. \textbf{C57}, 541 (2008),
  \texttt{0805.2222}

\bibitem{Tsutsui:2005cj}
N.~Tsutsui et~al. (JLQCD), PoS \textbf{LAT2005}, 357 (2006),
  \texttt{hep-lat/0510068}

\bibitem{Dawson:2006qc}
C.~Dawson, T.~Izubuchi, T.~Kaneko, S.~Sasaki, A.~Soni, Phys. Rev. \textbf{D74},
  114502 (2006), \texttt{hep-ph/0607162}

\bibitem{Brommel:2007wn}
D.~Brommel et~al. (QCDSF), PoS \textbf{LAT2007}, 364 (2007), \texttt{0710.2100}

\bibitem{Lubicz:2009ht}
V.~Lubicz, F.~Mescia, S.~Simula, C.~Tarantino (European Twisted Mass), Phys.
  Rev. \textbf{D80}, 111502 (2009), \texttt{0906.4728}

\bibitem{Boyle:2007qe}
P.A. Boyle et~al., Phys. Rev. Lett. \textbf{100}, 141601 (2008),
  \texttt{0710.5136}

\bibitem{Boyle:2007wg}
P.A. Boyle, J.M. Flynn, A.~Juttner, C.T. Sachrajda, J.M. Zanotti, JHEP
  \textbf{05}, 016 (2007), \texttt{hep-lat/0703005}

\bibitem{Aubin:2004fs}
C.~Aubin et~al. (MILC), Phys. Rev. \textbf{D70}, 114501 (2004),
  \texttt{hep-lat/0407028}

\bibitem{Bazavov:2009bb}
A.~Bazavov et~al. (2009), \texttt{0903.3598}

\bibitem{Follana:2007uv}
E.~Follana, C.~Davies, G.P. Lepage, J.~Shigemitsu (HPQCD), Phys. Rev. Lett.
  \textbf{100}, 062002 (2008), \texttt{0706.1726}

\bibitem{Durr:2010hr}
S.~Durr et~al. (2010), \texttt{1001.4692}

\bibitem{Blossier:2009bx}
B.~Blossier et~al., JHEP \textbf{07}, 043 (2009), \texttt{0904.0954}

\bibitem{Aubin:2008ie}
C.~Aubin, J.~Laiho, R.S. Van~de Water (2008), \texttt{0810.4328}

\bibitem{Beane:2006kx}
S.R. Beane, P.F. Bedaque, K.~Orginos, M.J. Savage, Phys. Rev. \textbf{D75},
  094501 (2007), \texttt{hep-lat/0606023}

\bibitem{Allton:2008pn}
C.~Allton et~al. (RBC-UKQCD), Phys. Rev. \textbf{D78}, 114509 (2008),
  \texttt{0804.0473}

\bibitem{Aoki:2008sm}
S.~Aoki et~al. (PACS-CS), Phys. Rev. \textbf{D79}, 034503 (2009),
  \texttt{0807.1661}

\bibitem{FLAG:Colangelo}
G.~Colangelo et~al. (FLAVIAnet lattice averaging group - FLAG), PoS
  \textbf{KAON09}, 029 (2009)

\bibitem{Ambrosino:2005ec}
F.~Ambrosino et~al. (KLOE), Phys. Lett. \textbf{B632}, 43 (2006),
  \texttt{hep-ex/0508027}

\bibitem{Ambrosino:2007xm}
F.~Ambrosino et~al. (KLOE), JHEP \textbf{02}, 098 (2008), \texttt{0712.3841}

\bibitem{Ambrosino:2005fw}
F.~Ambrosino et~al. (KLOE), Phys. Lett. \textbf{B632}, 76 (2006),
  \texttt{hep-ex/0509045}

\bibitem{Ambrosino:2008pz}
F.~Ambrosino et~al. (KLOE), Phys. Lett. \textbf{B666}, 305 (2008),
  \texttt{0804.4577}

\bibitem{Ambrosino:2006si}
F.~Ambrosino et~al. (KLOE), Phys. Lett. \textbf{B636}, 173 (2006),
  \texttt{hep-ex/0601026}

\bibitem{Ambrosino:2006sh}
F.~Ambrosino et~al. (KLOE), Eur. Phys. J. \textbf{C48}, 767 (2006),
  \texttt{hep-ex/0601025}

\bibitem{Ambrosino:2005vx}
F.~Ambrosino et~al. (KLOE), Phys. Lett. \textbf{B626}, 15 (2005),
  \texttt{hep-ex/0507088}

\bibitem{Ambrosino:2007xz}
F.~Ambrosino et~al. (KLOE), JHEP \textbf{01}, 073 (2008), \texttt{0712.1112}

\bibitem{Ambrosino:2006gn}
F.~Ambrosino et~al. (KLOE), Phys. Lett. \textbf{B636}, 166 (2006),
  \texttt{hep-ex/0601038}

\bibitem{Ambrosino:2007yza}
F.~Ambrosino et~al. (KLOE), JHEP \textbf{12}, 105 (2007), \texttt{0710.4470}

\bibitem{Testa:2008xz}
M.~Testa et~al. (KLOE) (2008), \texttt{0805.1969}

\bibitem{Ambrosino:2008ct}
F.~Ambrosino et~al. (KLOE), JHEP \textbf{04}, 059 (2008), \texttt{0802.3009}

\bibitem{Ambrosino:2006up}
F.~Ambrosino et~al. (KLOE), Phys. Lett. \textbf{B638}, 140 (2006),
  \texttt{hep-ex/0603041}

\bibitem{Adinolfi:2003ca}
M.~Adinolfi et~al. (KLOE), Phys. Lett. \textbf{B566}, 61 (2003),
  \texttt{hep-ex/0305035}

\bibitem{Ambrosino:2007mm}
F.~Ambrosino et~al. (KLOE), JHEP \textbf{05}, 051 (2008), \texttt{0712.1744}

\bibitem{Ambrosino:2005iw}
F.~Ambrosino et~al. (KLOE), Phys. Lett. \textbf{B619}, 61 (2005),
  \texttt{hep-ex/0505012}

\bibitem{Ambrosino:2008zi}
F.~Ambrosino et~al. (KLOE), Phys. Lett. \textbf{B672}, 203 (2009),
  \texttt{0811.1007}

\bibitem{Aloisio:2003jn}
A.~Aloisio et~al. (KLOE), Phys. Lett. \textbf{B597}, 139 (2004),
  \texttt{hep-ex/0307054}

\bibitem{bib:tauL_prelim}
S.~Bocchetta et~al. (KLOE), PoS \textbf{KAON09}, 006 (2009)

\bibitem{bib:nim_qcalt}
M.~Martini et~al. (KLOE-2) (2009), \texttt{0906.1133}

\bibitem{Bernard:2006gy}
V.~Bernard, M.~Oertel, E.~Passemar, J.~Stern, Phys. Lett. \textbf{B638}, 480
  (2006), \texttt{hep-ph/0603202}

\bibitem{Bernard:2009zm}
V.~Bernard, M.~Oertel, E.~Passemar, J.~Stern, Phys. Rev. \textbf{D80}, 034034
  (2009), \texttt{0903.1654}

\bibitem{Bernard:2007cf}
V.~Bernard, M.~Oertel, E.~Passemar, J.~Stern, JHEP \textbf{01}, 015 (2008),
  \texttt{0707.4194}

\bibitem{Callan:1966hu}
C.G. Callan, S.B. Treiman, Phys. Rev. Lett. \textbf{16}, 153 (1966)

\bibitem{Dashen:1969bh}
R.F. Dashen, M.~Weinstein, Phys. Rev. Lett. \textbf{22}, 1337 (1969)

\bibitem{Gasser:1984ux}
J.~Gasser, H.~Leutwyler, Nucl. Phys. \textbf{B250}, 517 (1985)

\bibitem{Aubert:2007jh}
B.~Aubert et~al. (BaBar), Phys. Rev. \textbf{D76}, 051104 (2007),
  \texttt{0707.2922}

\bibitem{Epifanov:2007rf}
D.~Epifanov et~al. (Belle), Phys. Lett. \textbf{B654}, 65 (2007),
  \texttt{0706.2231}

\bibitem{Moussallam:2007qc}
B.~Moussallam, Eur. Phys. J. \textbf{C53}, 401 (2008), \texttt{0710.0548}

\bibitem{Jamin:2006tk}
M.~Jamin, A.~Pich, J.~Portoles, Phys. Lett. \textbf{B640}, 176 (2006),
  \texttt{hep-ph/0605096}

\bibitem{Jamin:2008qg}
M.~Jamin, A.~Pich, J.~Portoles, Phys. Lett. \textbf{B664}, 78 (2008),
  \texttt{0803.1786}

\bibitem{Boito:2008fq}
D.R. Boito, R.~Escribano, M.~Jamin, Eur. Phys. J. \textbf{C59}, 821 (2009),
  \texttt{0807.4883}

\bibitem{Aston:1987ir}
D.~Aston et~al., Nucl. Phys. \textbf{B296}, 493 (1988)

\bibitem{Abouzaid:2009ry}
E.~Abouzaid et~al. (KTeV) (2009), \texttt{0912.1291}

\bibitem{Boito:2009pv}
D.R. Boito, R.~Escribano, M.~Jamin, PoS \textbf{EFT09}, 064 (2009),
  \texttt{0904.0425}

\bibitem{Lai:2007dx}
A.~Lai et~al. (NA48), Phys. Lett. \textbf{B647}, 341 (2007),
  \texttt{hep-ex/0703002}

\bibitem{DescotesGenon:2005pw}
S.~Descotes-Genon, B.~Moussallam, Eur. Phys. J. \textbf{C42}, 403 (2005),
  \texttt{hep-ph/0505077}

\bibitem{RamseyMusolf:2007yb}
M.J. Ramsey-Musolf, S.~Su, S.~Tulin, Phys. Rev. \textbf{D76}, 095017 (2007),
  \texttt{0705.0028}

\bibitem{Davier:2005xq}
M.~Davier, A.~Hocker, Z.~Zhang, Rev. Mod. Phys. \textbf{78}, 1043 (2006),
  \texttt{hep-ph/0507078}

\bibitem{Marciano:2004uf}
W.J. Marciano, Phys. Rev. Lett. \textbf{93}, 231803 (2004),
  \texttt{hep-ph/0402299}

\bibitem{Towner:2007np}
I.S. Towner, J.C. Hardy, Phys. Rev. \textbf{C77}, 025501 (2008),
  \texttt{0710.3181}

\bibitem{Marciano:1999ih}
W.J. Marciano, Phys. Rev. \textbf{D60}, 093006 (1999), \texttt{hep-ph/9903451}

\bibitem{Marciano:2007zz}
W.J. Marciano, PoS \textbf{KAON}, 003 (2008)

\bibitem{Isidori:2006pk}
G.~Isidori, P.~Paradisi, Phys. Lett. \textbf{B639}, 499 (2006),
  \texttt{hep-ph/0605012}

\bibitem{Ikado:2006un}
K.~Ikado et~al., Phys. Rev. Lett. \textbf{97}, 251802 (2006),
  \texttt{hep-ex/0604018}

\bibitem{Goudzovski:2009me}
E.~Goudzovski (2009), \texttt{0908.3858}

\bibitem{Antonelli:2009ws}
M.~Antonelli et~al. (2009), \texttt{0907.5386}

\bibitem{Lueders:1992dq}
G.~Luders, Annals Phys. \textbf{2}, 1 (1957)

\bibitem{Pauli:1988xn}
W.~Pauli (1955), in *Enz, C. P. (ed.), Meyenn, K. V. (ed.): Wolfgang Pauli*
  459-479

\bibitem{Bell:1955sd}
J.~Bell, Proc. Roy. Soc. Lond. \textbf{A231}, 479 (1955)

\bibitem{Jost:1957zz}
R.~Jost, Helv. Phys. Acta \textbf{30}, 409 (1957)

\bibitem{Ambrosino:2006vr}
F.~Ambrosino et~al. (KLOE), Phys. Lett. \textbf{B642}, 315 (2006),
  \texttt{hep-ex/0607027}

\bibitem{DiDomenico:2009xw}
A.~Di~Domenico (2009), \texttt{0904.1976}

\bibitem{Bertlmann:2002wv}
R.A. Bertlmann, K.~Durstberger, B.C. Hiesmayr, Phys. Rev. \textbf{A68}, 012111
  (2003), \texttt{quant-ph/0209017}

\bibitem{Bertlmann:2006fn}
R.A. Bertlmann, W.~Grimus, B.C. Hiesmayr, Phys. Rev. \textbf{A73}, 054101
  (2006), \texttt{quant-ph/0602116}

\bibitem{Penrose:1998dg}
R.~Penrose, Phil. Trans. Roy. Soc. Lond. \textbf{A356}, 1927 (1998)

\bibitem{Mavromatos:2007xb}
N.E. Mavromatos, PoS \textbf{KAON}, 041 (2008), \texttt{0707.3422}

\bibitem{Ghirardi:1985mt}
G.C. Ghirardi, A.~Rimini, T.~Weber, Phys. Rev. \textbf{D34}, 470 (1986)

\bibitem{Ghirardi:1987nr}
G.C. Ghirardi, A.~Rimini, T.~Weber (1987), \texttt{Trieste ICTP Report
  IC-87-96}

\bibitem{Lindblad:1975ef}
G.~Lindblad, Commun. Math. Phys. \textbf{48}, 119 (1976)

\bibitem{Gorini:1975nb}
V.~Gorini, A.~Kossakowski, E.C.G. Sudarshan, J. Math. Phys. \textbf{17}, 821
  (1976)

\bibitem{Schrodinger:1935zz}
E.~Schrodinger, Naturwiss. \textbf{23}, 807 (1935)

\bibitem{Furry:1936xx}
W.H. Furry, Phys. Rev. \textbf{49}, 393 (1936)

\bibitem{Bertlmann:1996at}
R.A. Bertlmann, W.~Grimus, Phys. Lett. \textbf{B392}, 426 (1997),
  \texttt{hep-ph/9610301}

\bibitem{Bertlmann:1999np}
R.A. Bertlmann, W.~Grimus, B.C. Hiesmayr, Phys. Rev. \textbf{D60}, 114032
  (1999), \texttt{hep-ph/9902427}

\bibitem{Apostolakis:1997td}
A.~Apostolakis et~al. (CPLEAR), Phys. Lett. \textbf{B422}, 339 (1998)

\bibitem{Go:2007ww}
A.~Go et~al. (Belle), Phys. Rev. Lett. \textbf{99}, 131802 (2007),
  \texttt{quant-ph/0702267}

\bibitem{AmelinoCamelia:2008qg}
G.~Amelino-Camelia (2008), \texttt{0806.0339}

\bibitem{AmelinoCamelia:1997gz}
G.~Amelino-Camelia, J.R. Ellis, N.E. Mavromatos, D.V. Nanopoulos, S.~Sarkar,
  Nature \textbf{393}, 763 (1998), \texttt{astro-ph/9712103}

\bibitem{Gambini:1998it}
R.~Gambini, J.~Pullin, Phys. Rev. \textbf{D59}, 124021 (1999),
  \texttt{gr-qc/9809038}

\bibitem{Aloisio:2000cm}
R.~Aloisio, P.~Blasi, P.L. Ghia, A.F. Grillo, Phys. Rev. \textbf{D62}, 053010
  (2000), \texttt{astro-ph/0001258}

\bibitem{AmelinoCamelia:2000zs}
G.~Amelino-Camelia, T.~Piran, Phys. Rev. \textbf{D64}, 036005 (2001),
  \texttt{astro-ph/0008107}

\bibitem{AmelinoCamelia:2000mn}
G.~Amelino-Camelia, Int. J. Mod. Phys. \textbf{D11}, 35 (2002),
  \texttt{gr-qc/0012051}

\bibitem{Magueijo:2002am}
J.~Magueijo, L.~Smolin, Phys. Rev. \textbf{D67}, 044017 (2003),
  \texttt{gr-qc/0207085}

\bibitem{AmelinoCamelia:2003xp}
G.~Amelino-Camelia, L.~Smolin, A.~Starodubtsev, Class. Quant. Grav.
  \textbf{21}, 3095 (2004), \texttt{hep-th/0306134}

\bibitem{AmelinoCamelia:1997em}
G.~Amelino-Camelia, Mod. Phys. Lett. \textbf{A12}, 1387 (1997),
  \texttt{gr-qc/9706007}

\bibitem{AmelinoCamelia:1999pm}
G.~Amelino-Camelia, S.~Majid, Int. J. Mod. Phys. \textbf{A15}, 4301 (2000),
  \texttt{hep-th/9907110}

\bibitem{AmelinoCamelia:2007zzb}
G.~Amelino-Camelia, A.~Marciano, M.~Arzano (2007), in *Di Domenico, A. (ed.):
  Handbook of neutral kaon interferometry at a Phi-factory* 155-186

\bibitem{AmelinoCamelia:1998ax}
G.~Amelino-Camelia, Nature \textbf{398}, 216 (1999), \texttt{gr-qc/9808029}

\bibitem{Jacobson:2002ye}
T.~Jacobson, S.~Liberati, D.~Mattingly, Nature \textbf{424}, 1019 (2003),
  \texttt{astro-ph/0212190}

\bibitem{Jacob:2006gn}
U.~Jacob, T.~Piran, Nature Phys. \textbf{3}, 87 (2007), \texttt{hep-ph/0607145}

\bibitem{AmelinoCamelia:2009inprep}
G.~Amelino-Camelia, F.~Mercati (2010), in preparation

\bibitem{Bernabeu:2003ym}
J.~Bernabeu, N.E. Mavromatos, J.~Papavassiliou, Phys. Rev. Lett. \textbf{92},
  131601 (2004), \texttt{hep-ph/0310180}

\bibitem{Kostelecky:2008zz}
V.A. Kostelecky, ed., \emph{{CPT and Lorentz symmetry. Proceedings: 4th
  Meeting, Bloomington, USA, Aug 8-11}} (2007)

\bibitem{Greenberg:2002uu}
O.W. Greenberg, Phys. Rev. Lett. \textbf{89}, 231602 (2002),
  \texttt{hep-ph/0201258}

\bibitem{Greenberg:2003nv}
O.W. Greenberg, Found. Phys. \textbf{36}, 1535 (2006), \texttt{hep-ph/0309309}

\bibitem{Wheeler:1998vs}
J.A. Wheeler, K.~Ford (1998), {Geons, black holes, and quantum foam: A life in
  physic - New York, USA: Norton (1998) 380 p}

\bibitem{Wald:1980nm}
R.M. Wald, Phys. Rev. \textbf{D21}, 2742 (1980)

\bibitem{Ellis:1983jz}
J.R. Ellis, J.S. Hagelin, D.V. Nanopoulos, M.~Srednicki, Nucl. Phys.
  \textbf{B241}, 381 (1984)

\bibitem{Ellis:1995xd}
J.R. Ellis, J.L. Lopez, N.E. Mavromatos, D.V. Nanopoulos, Phys. Rev.
  \textbf{D53}, 3846 (1996), \texttt{hep-ph/9505340}

\bibitem{Benatti:1997rv}
F.~Benatti, R.~Floreanini, Nucl. Phys. \textbf{B488}, 335 (1997)

\bibitem{Huet:1994kr}
P.~Huet, M.E. Peskin, Nucl. Phys. \textbf{B434}, 3 (1995),
  \texttt{hep-ph/9403257}

\bibitem{Bernabeu:2005pm}
J.~Bernabeu, N.E. Mavromatos, J.~Papavassiliou, A.~Waldron-Lauda, Nucl. Phys.
  \textbf{B744}, 180 (2006), \texttt{hep-ph/0506025}

\bibitem{Bernabeu:2006av}
J.~Bernabeu, N.E. Mavromatos, S.~Sarkar, Phys. Rev. \textbf{D74}, 045014
  (2006), \texttt{hep-th/0606137}

\bibitem{Mavromatos:2008bz}
N.E. Mavromatos, S.~Sarkar, Phys. Rev. \textbf{D79}, 104015 (2009),
  \texttt{0812.3952}

\bibitem{Kostelecky:2003fs}
V.A. Kostelecky, Phys. Rev. \textbf{D69}, 105009 (2004),
  \texttt{hep-th/0312310}

\bibitem{Colladay:1998fq}
D.~Colladay, V.A. Kostelecky, Phys. Rev. \textbf{D58}, 116002 (1998),
  \texttt{hep-ph/9809521}

\bibitem{Colladay:1996iz}
D.~Colladay, V.A. Kostelecky, Phys. Rev. \textbf{D55}, 6760 (1997),
  \texttt{hep-ph/9703464}

\bibitem{Kostelecky:2000hz}
V.A. Kostelecky, R.~Potting, Phys. Rev. \textbf{D63}, 046007 (2001),
  \texttt{hep-th/0008252}

\bibitem{Kostelecky:1991ak}
V.A. Kostelecky, R.~Potting, Nucl. Phys. \textbf{B359}, 545 (1991)

\bibitem{Kostelecky:1988zi}
V.A. Kostelecky, S.~Samuel, Phys. Rev. \textbf{D39}, 683 (1989)

\bibitem{Alfaro:1999wd}
J.~Alfaro, H.A. Morales-Tecotl, L.F. Urrutia, Phys. Rev. Lett. \textbf{84},
  2318 (2000), \texttt{gr-qc/9909079}

\bibitem{Klinkhamer:2003ec}
F.R. Klinkhamer, C.~Rupp, Phys. Rev. \textbf{D70}, 045020 (2004),
  \texttt{hep-th/0312032}

\bibitem{Carroll:2001ws}
S.M. Carroll, J.A. Harvey, V.A. Kostelecky, C.D. Lane, T.~Okamoto, Phys. Rev.
  Lett. \textbf{87}, 141601 (2001), \texttt{hep-th/0105082}

\bibitem{Bertolami:2003qs}
O.~Bertolami, R.~Lehnert, R.~Potting, A.~Ribeiro, Phys. Rev. \textbf{D69},
  083513 (2004), \texttt{astro-ph/0310344}

\bibitem{ArkaniHamed:2004ar}
N.~Arkani-Hamed, H.C. Cheng, M.~Luty, J.~Thaler, JHEP \textbf{07}, 029 (2005),
  \texttt{hep-ph/0407034}

\bibitem{Kostelecky:2002ca}
V.A. Kostelecky, R.~Lehnert, M.J. Perry, Phys. Rev. \textbf{D68}, 123511
  (2003), \texttt{astro-ph/0212003}

\bibitem{Jackiw:1999yp}
R.~Jackiw, V.A. Kostelecky, Phys. Rev. Lett. \textbf{82}, 3572 (1999),
  \texttt{hep-ph/9901358}

\bibitem{Kostelecky:2001jc}
V.A. Kostelecky, C.D. Lane, A.G.M. Pickering, Phys. Rev. \textbf{D65}, 056006
  (2002), \texttt{hep-th/0111123}

\bibitem{Kostelecky:2000mm}
V.A. Kostelecky, R.~Lehnert, Phys. Rev. \textbf{D63}, 065008 (2001),
  \texttt{hep-th/0012060}

\bibitem{Lehnert:2004ri}
R.~Lehnert, J. Math. Phys. \textbf{45}, 3399 (2004), \texttt{hep-ph/0401084}

\bibitem{Lehnert:2003ue}
R.~Lehnert, Phys. Rev. \textbf{D68}, 085003 (2003), \texttt{gr-qc/0304013}

\bibitem{Altschul:2005mu}
B.~Altschul, V.A. Kostelecky, Phys. Lett. \textbf{B628}, 106 (2005),
  \texttt{hep-th/0509068}

\bibitem{Bluhm:2008yt}
R.~Bluhm, N.L. Gagne, R.~Potting, A.~Vrublevskis, Phys. Rev. \textbf{D77},
  125007 (2008), \texttt{0802.4071}

\bibitem{Bluhm:2004ep}
R.~Bluhm, V.A. Kostelecky, Phys. Rev. \textbf{D71}, 065008 (2005),
  \texttt{hep-th/0412320}

\bibitem{Kostelecky:2001mb}
V.A. Kostelecky, M.~Mewes, Phys. Rev. Lett. \textbf{87}, 251304 (2001),
  \texttt{hep-ph/0111026}

\bibitem{Xia:2007qs}
J.Q. Xia, H.~Li, X.l. Wang, X.m. Zhang, Astron. Astrophys. \textbf{483}, 715
  (2008), \texttt{0710.3325}

\bibitem{Dehmelt:1999jh}
H.~Dehmelt, R.~Mittleman, R.S. van Dyck, P.~Schwinberg, Phys. Rev. Lett.
  \textbf{83}, 4694 (1999), \texttt{hep-ph/9906262}

\bibitem{Humphrey:2001wm}
M.A. Humphrey et~al., Phys. Rev. \textbf{A68}, 063807 (2003),
  \texttt{physics/0103068}

\bibitem{Muller:2003zzc}
H.~Muller, S.~Herrmann, C.~Braxmaier, S.~Schiller, A.~Peters, Phys. Rev. Lett.
  \textbf{91}, 020401 (2003), \texttt{physics/0305117}

\bibitem{Kostelecky:1997mh}
V.A. Kostelecky, Phys. Rev. Lett. \textbf{80}, 1818 (1998),
  \texttt{hep-ph/9809572}

\bibitem{Nguyen:2001tg}
H.~Nguyen (2001), \texttt{hep-ex/0112046}

\bibitem{Link:2002fg}
J.M. Link et~al., Phys. Lett. \textbf{B556}, 7 (2003), \texttt{hep-ex/0208034}

\bibitem{Aubert:2007bp}
B.~Aubert et~al. (BaBar), Phys. Rev. Lett. \textbf{100}, 131802 (2008),
  \texttt{0711.2713}

\bibitem{Bennett:2007yc}
G.W. Bennett et~al. (Muon (g-2)), Phys. Rev. Lett. \textbf{100}, 091602 (2008),
  \texttt{0709.4670}

\bibitem{Hohensee:2009zk}
M.A. Hohensee, R.~Lehnert, D.F. Phillips, R.L. Walsworth, Phys. Rev. Lett.
  \textbf{102}, 170402 (2009), \texttt{0904.2031}

\bibitem{Battat:2007uh}
J.B.R. Battat, J.F. Chandler, C.W. Stubbs, Phys. Rev. Lett. \textbf{99}, 241103
  (2007), \texttt{0710.0702}

\bibitem{Kostelecky:2008ts}
V.A. Kostelecky, N.~Russell (2008), \texttt{0801.0287}

\bibitem{Kostelecky:2001ff}
V.A. Kostelecky, Phys. Rev. \textbf{D64}, 076001 (2001),
  \texttt{hep-ph/0104120}

\bibitem{Kostelecky:1999bm}
V.A. Kostelecky, Phys. Rev. \textbf{D61}, 016002 (2000),
  \texttt{hep-ph/9909554}

\bibitem{Kostelecky:1994rn}
V.A. Kostelecky, R.~Potting, Phys. Rev. \textbf{D51}, 3923 (1995),
  \texttt{hep-ph/9501341}

\bibitem{Hiesmayr:2007bt}
B.C. Hiesmayr, M.~Huber, Phys. Lett. \textbf{A372}, 3608 (2008),
  \texttt{0711.1368}

\bibitem{Greenberger:1988zz}
D.M. Greenberger, A.~Yasin, Phys. Lett. \textbf{A138}, 391 (1988)

\bibitem{Englert:1996zz}
B.G. Englert, Phys. Rev. Lett. \textbf{77}, 2154 (1996)

\bibitem{Bramon:2003aj}
A.~Bramon, G.~Garbarino, B.C. Hiesmayr, Phys. Rev. \textbf{A69}, 022112 (2004),
  \texttt{quant-ph/0311179}

\bibitem{Scully:1982zz}
M.O. Scully, K.~Dr\"uhl, Phys. Rev. A \textbf{25}(4), 2208 (1982)

\bibitem{Aharonov:2005dc}
Y.~Aharonov, M.S. Zubairy, Science \textbf{307}, 875 (2005)

\bibitem{Bramon:2003bj}
A.~Bramon, G.~Garbarino, B.~Hiesmayr, Phys. Rev. Lett. \textbf{92}, 020405
  (2004), \texttt{quant-ph/0306114}

\bibitem{Bramon:2004zp}
A.~Bramon, G.~Garbarino, B.C. Hiesmayr, Phys. Rev. \textbf{A69}, 062111 (2004),
  \texttt{quant-ph/0402212}

\bibitem{Bramon:2004pc}
A.~Bramon, G.~Garbarino, B.C. Hiesmayr (2004), \texttt{quant-ph/0404086}

\bibitem{Bramon:2003te}
A.~Bramon, G.~Garbarino, B.C. Hiesmayr (2003), \texttt{hep-ph/0311232}

\bibitem{Weinberg:1978kz}
S.~Weinberg, Physica \textbf{A96}, 327 (1979)

\bibitem{Gasser:1983yg}
J.~Gasser, H.~Leutwyler, Ann. Phys. \textbf{158}, 142 (1984)

\bibitem{D'Ambrosio:1996nm}
G.~D'Ambrosio, G.~Isidori, Int. J. Mod. Phys. \textbf{A13}, 1 (1998),
  \texttt{hep-ph/9611284}

\bibitem{deRafael:1995zv}
E.~de~Rafael (1995), \texttt{hep-ph/9502254}

\bibitem{Pich:1995bw}
A.~Pich, Rept. Prog. Phys. \textbf{58}, 563 (1995), \texttt{hep-ph/9502366}

\bibitem{Ecker:1994gg}
G.~Ecker, Prog. Part. Nucl. Phys. \textbf{35}, 1 (1995),
  \texttt{hep-ph/9501357}

\bibitem{D'Ambrosio:2003ef}
G.~D'Ambrosio, Mod. Phys. Lett. \textbf{A18}, 1273 (2003),
  \texttt{hep-ph/0305249}

\bibitem{Ecker:1992de}
G.~Ecker, J.~Kambor, D.~Wyler, Nucl. Phys. \textbf{B394}, 101 (1993)

\bibitem{D'Ambrosio:1997tb}
G.~D'Ambrosio, J.~Portoles, Nucl. Phys. \textbf{B533}, 494 (1998),
  \texttt{hep-ph/9711211}

\bibitem{Cabibbo:2005ez}
N.~Cabibbo, G.~Isidori, JHEP \textbf{03}, 021 (2005), \texttt{hep-ph/0502130}

\bibitem{D'Ambrosio:1994km}
G.~D'Ambrosio, G.~Isidori, A.~Pugliese, N.~Paver, Phys. Rev. \textbf{D50}, 5767
  (1994), \texttt{hep-ph/9403235}

\bibitem{Ambrosino:2006ek}
F.~Ambrosino et~al. (KLOE), JHEP \textbf{12}, 011 (2006),
  \texttt{hep-ex/0610034}

\bibitem{D'Ambrosio:1986ze}
G.~D'Ambrosio, D.~Espriu, Phys. Lett. \textbf{B175}, 237 (1986)

\bibitem{Buccella:1991ni}
F.~Buccella, G.~D'Ambrosio, M.~Miragliuolo, Nuovo Cim. \textbf{A104}, 777
  (1991)

\bibitem{Lai:2002sr}
A.~Lai et~al. (NA48), Phys. Lett. \textbf{B551}, 7 (2003),
  \texttt{hep-ex/0210053}

\bibitem{Buchalla:2003sj}
G.~Buchalla, G.~D'Ambrosio, G.~Isidori, Nucl. Phys. \textbf{B672}, 387 (2003),
  \texttt{hep-ph/0308008}

\bibitem{Lai:2003vc}
A.~Lai et~al. (NA48), Phys. Lett. \textbf{B578}, 276 (2004),
  \texttt{hep-ex/0309022}

\bibitem{Ecker:1987fm}
G.~Ecker, A.~Pich, E.~de~Rafael, Phys. Lett. \textbf{B189}, 363 (1987)

\bibitem{Ecker:1991ru}
G.~Ecker, A.~Pich, Nucl. Phys. \textbf{B366}, 189 (1991)

\bibitem{Angelopoulos:1997gu}
A.~Angelopoulos et~al. (CPLEAR), Phys. Lett. \textbf{B413}, 232 (1997)

\bibitem{Buras:2003dj}
A.J. Buras, R.~Fleischer, S.~Recksiegel, F.~Schwab, Phys. Rev. Lett.
  \textbf{92}, 101804 (2004), \texttt{hep-ph/0312259}

\bibitem{Ecker:1987qi}
G.~Ecker, A.~Pich, E.~de~Rafael, Nucl. Phys. \textbf{B291}, 692 (1987)

\bibitem{D'Ambrosio:1998yj}
G.~D'Ambrosio, G.~Ecker, G.~Isidori, J.~Portoles, JHEP \textbf{08}, 004 (1998),
  \texttt{hep-ph/9808289}

\bibitem{Appel:1999yq}
R.~Appel et~al. (E865), Phys. Rev. Lett. \textbf{83}, 4482 (1999),
  \texttt{hep-ex/9907045}

\bibitem{:2009pv}
J.R. Batley et~al. (NA48/2), Phys. Lett. \textbf{B677}, 246 (2009),
  \texttt{0903.3130}

\bibitem{Batley:2003mu}
J.R. Batley et~al. (NA48/1), Phys. Lett. \textbf{B576}, 43 (2003),
  \texttt{hep-ex/0309075}

\bibitem{Batley:2004wg}
J.R. Batley et~al. (NA48/1), Phys. Lett. \textbf{B599}, 197 (2004),
  \texttt{hep-ex/0409011}

\bibitem{Cappiello:2007rs}
L.~Cappiello, G.~D'Ambrosio, Phys. Rev. \textbf{D75}, 094014 (2007),
  \texttt{hep-ph/0702292}

\bibitem{D'Ambrosio:1992bf}
G.~D'Ambrosio, M.~Miragliuolo, P.~Santorelli (1992), \texttt{LNF-92/066(P)}

\bibitem{D'Ambrosio:1994du}
G.~D'Ambrosio, G.~Isidori, Z. Phys. \textbf{C65}, 649 (1995),
  \texttt{hep-ph/9408219}

\bibitem{Raggi:200739}
M.~Raggi et~al. (NA48/2), Nucl. Phys.B - Proceedings Supplements \textbf{167},
  39 (2007), proceedings of BEACH 2006

\bibitem{Lai:2003ad}
A.~Lai et~al. (NA48), Eur. Phys. J. \textbf{C30}, 33 (2003)

\bibitem{Bernard:2006gx}
V.~Bernard, U.G. Meissner, Ann. Rev. Nucl. Part. Sci. \textbf{57}, 33 (2007),
  \texttt{hep-ph/0611231}

\bibitem{Bijnens:2006zp}
J.~Bijnens, Prog. Part. Nucl. Phys. \textbf{58}, 521 (2007),
  \texttt{hep-ph/0604043}

\bibitem{Flynn:1988gy}
J.~Flynn, L.~Randall, Phys. Lett. \textbf{B216}, 221 (1989)

\bibitem{Bijnens:2005sj}
J.~Bijnens, Acta Phys. Slov. \textbf{56}, 305 (2006), \texttt{hep-ph/0511076}

\bibitem{Witten:1979vv}
E.~Witten, Nucl. Phys. \textbf{B156}, 269 (1979)

\bibitem{'tHooft:1973jz}
G.~'t~Hooft, Nucl. Phys. \textbf{B72}, 461 (1974)

\bibitem{Kaiser:2000gs}
R.~Kaiser, H.~Leutwyler, Eur. Phys. J. \textbf{C17}, 623 (2000),
  \texttt{hep-ph/0007101}

\bibitem{Fariborz:1999gr}
A.H. Fariborz, J.~Schechter, Phys. Rev. \textbf{D60}, 034002 (1999),
  \texttt{hep-ph/9902238}

\bibitem{AbdelRehim:2002an}
A.M. Abdel-Rehim, D.~Black, A.H. Fariborz, J.~Schechter, Phys. Rev.
  \textbf{D67}, 054001 (2003), \texttt{hep-ph/0210431}

\bibitem{Beisert:2002ad}
N.~Beisert, B.~Borasoy, Nucl. Phys. \textbf{A705}, 433 (2002),
  \texttt{hep-ph/0201289}

\bibitem{Fujiwara:1984mp}
T.~Fujiwara, T.~Kugo, H.~Terao, S.~Uehara, K.~Yamawaki, Prog. Theor. Phys.
  \textbf{73}, 926 (1985)

\bibitem{Ecker:1988te}
G.~Ecker, J.~Gasser, A.~Pich, E.~de~Rafael, Nucl. Phys. \textbf{B321}, 311
  (1989)

\bibitem{Anisovich:1996tx}
A.V. Anisovich, H.~Leutwyler, Phys. Lett. \textbf{B375}, 335 (1996),
  \texttt{hep-ph/9601237}

\bibitem{Kambor:1995yc}
J.~Kambor, C.~Wiesendanger, D.~Wyler, Nucl. Phys. \textbf{B465}, 215 (1996),
  \texttt{hep-ph/9509374}

\bibitem{Beisert:2003zd}
N.~Beisert, B.~Borasoy, Phys. Rev. \textbf{D67}, 074007 (2003),
  \texttt{hep-ph/0302062}

\bibitem{Borasoy:2005du}
B.~Borasoy, R.~Nissler, Eur. Phys. J. \textbf{A26}, 383 (2005),
  \texttt{hep-ph/0510384}

\bibitem{Borasoy:2004qj}
B.~Borasoy, R.~Nissler, Nucl. Phys. \textbf{A740}, 362 (2004),
  \texttt{hep-ph/0405039}

\bibitem{Borasoy:2003yb}
B.~Borasoy, R.~Nissler, Eur. Phys. J. \textbf{A19}, 367 (2004),
  \texttt{hep-ph/0309011}

\bibitem{Sutherland:1966zz}
D.G. Sutherland, Phys. Lett. \textbf{23}, 384 (1966)

\bibitem{Bell:1996mi}
J.S. Bell, D.G. Sutherland, Nucl. Phys. \textbf{B4}, 315 (1968)

\bibitem{Leutwyler:1996qg}
H.~Leutwyler, Phys. Lett. \textbf{B378}, 313 (1996), \texttt{hep-ph/9602366}

\bibitem{Tippens:2001fm}
W.B. Tippens et~al. (Crystal Ball), Phys. Rev. Lett. \textbf{87}, 192001 (2001)

\bibitem{Ambrosino:2007wi}
F.~Ambrosino et~al. (KLOE) (2007), \texttt{0707.4137}

\bibitem{:2007iy}
M.~Bashkanov et~al., Phys. Rev. \textbf{C76}, 048201 (2007), \texttt{0708.2014}

\bibitem{Adolph:2008vn}
C.~Adolph et~al. (WASA-at-COSY), Phys. Lett. \textbf{B677}, 24 (2009),
  \texttt{0811.2763}

\bibitem{Unverzagt:2008ny}
M.~Unverzagt et~al. (Crystal Ball at MAMI), Eur. Phys. J. \textbf{A39}, 169
  (2009), \texttt{0812.3324}

\bibitem{Prakhov:2008ff}
S.~Prakhov et~al. (Crystal Ball at MAMI), Phys. Rev. \textbf{C79}, 035204
  (2009), \texttt{0812.1999}

\bibitem{Bijnens:2007pr}
J.~Bijnens, K.~Ghorbani, JHEP \textbf{11}, 030 (2007), \texttt{0709.0230}

\bibitem{Ambrosino:2008ht}
F.~Ambrosino et~al. (KLOE), JHEP \textbf{05}, 006 (2008), \texttt{0801.2642}

\bibitem{Gross:1979ur}
D.J. Gross, S.B. Treiman, F.~Wilczek, Phys. Rev. \textbf{D19}, 2188 (1979)

\bibitem{Borasoy:2006uv}
B.~Borasoy, U.G. Meissner, R.~Nissler, Phys. Lett. \textbf{B643}, 41 (2006),
  \texttt{hep-ph/0609010}

\bibitem{Binon:1984fe}
F.G. Binon et~al. (Serpukhov-Brussels-Annecy(LAPP)), Phys. Lett. \textbf{B140},
  264 (1984)

\bibitem{Alde:1987jt}
D.~Alde et~al. (Serpukhov-Brussels-Los Alamos-Annecy(LAPP)), Z. Phys.
  \textbf{C36}, 603 (1987)

\bibitem{Blik:2008zz}
A.M. Blik et~al., Phys. Atom. Nucl. \textbf{71}, 2124 (2008)

\bibitem{:2008tb}
P.~Naik et~al. (CLEO), Phys. Rev. Lett. \textbf{102}, 061801 (2009),
  \texttt{0809.2587}

\bibitem{Benayoun:2003we}
M.~Benayoun, P.~David, L.~DelBuono, P.~Leruste, H.B. O'Connell, Eur. Phys. J.
  \textbf{C31}, 525 (2003), \texttt{nucl-th/0306078}

\bibitem{Holstein:2001bt}
B.R. Holstein, Phys. Scripta \textbf{T99}, 55 (2002), \texttt{hep-ph/0112150}

\bibitem{Gormley:1970qz}
M.~Gormley et~al., Phys. Rev. \textbf{D2}, 501 (1970)

\bibitem{Layter:1973ti}
J.G. Layter et~al., Phys. Rev. \textbf{D7}, 2565 (1973)

\bibitem{Borasoy:2007dw}
B.~Borasoy, R.~Nissler, Eur. Phys. J. \textbf{A33}, 95 (2007),
  \texttt{0705.0954}

\bibitem{Abele:1997yi}
A.~Abele et~al. (Crystal Barrel), Phys. Lett. \textbf{B402}, 195 (1997)

\bibitem{Acciarri:1997yx}
M.~Acciarri et~al. (L3), Phys. Lett. \textbf{B418}, 399 (1998)

\bibitem{Dzhelyadin:1980tj}
R.I. Dzhelyadin et~al., Phys. Lett. \textbf{B102}, 296 (1981)

\bibitem{:2009wb}
R.~Arnaldi et~al. (NA60), Phys. Lett. \textbf{B677}, 260 (2009),
  \texttt{0902.2547}

\bibitem{Terschluesen:2010ik}
C.~Terschluesen, S.~Leupold (2010), \texttt{1003.1030}

\bibitem{Achasov:2000ne}
M.N. Achasov et~al., Phys. Lett. \textbf{B504}, 275 (2001)

\bibitem{MeijerDrees:1992qb}
R.~Meijer~Drees et~al. (SINDRUM-I), Phys. Rev. \textbf{D45}, 1439 (1992)

\bibitem{Abouzaid:2008cd}
E.~Abouzaid et~al. (KTeV), Phys. Rev. Lett. \textbf{100}, 182001 (2008),
  \texttt{0802.2064}

\bibitem{Dzhelyadin:1980kh}
R.I. Dzhelyadin et~al., Phys. Lett. \textbf{B94}, 548 (1980)

\bibitem{Bijnens:1999jp}
J.~Bijnens, F.~Perrsson (1999), \texttt{hep-ph/0106130}

\bibitem{Ambrosino:2008cp}
F.~Ambrosino et~al. (KLOE), Phys. Lett. \textbf{B675}, 283 (2009),
  \texttt{0812.4830}

\bibitem{Akhmetshin:2000bw}
R.R. Akhmetshin et~al. (CMD-2), Phys. Lett. \textbf{B501}, 191 (2001),
  \texttt{hep-ex/0012039}

\bibitem{Landsberg:1986fd}
L.G. Landsberg, Phys. Rept. \textbf{128}, 301 (1985)

\bibitem{Gao:2002gq}
D.N. Gao, Mod. Phys. Lett. \textbf{A17}, 1583 (2002), \texttt{hep-ph/0202002}

\bibitem{He:2008zw}
X.G. He, J.~Tandean, G.~Valencia, JHEP \textbf{06}, 002 (2008),
  \texttt{0803.4330}

\bibitem{Ametller:1991dp}
L.~Ametller, J.~Bijnens, A.~Bramon, F.~Cornet, Phys. Lett. \textbf{B276}, 185
  (1992)

\bibitem{DiMicco:2005rv}
B.~Di~Micco et~al. (KLOE), Acta Phys. Slov. \textbf{56}, 403 (2006)

\bibitem{Ambrosino:2009sc}
F.~Ambrosino et~al. (KLOE), JHEP \textbf{07}, 105 (2009), \texttt{0906.3819}

\bibitem{Rosner:1982ey}
J.L. Rosner, Phys. Rev. \textbf{D27}, 1101 (1983)

\bibitem{Cheng:2008ss}
H.Y. Cheng, H.n. Li, K.F. Liu, Phys. Rev. \textbf{D79}, 014024 (2009),
  \texttt{0811.2577}

\bibitem{Ambrosino:2006gk}
F.~Ambrosino et~al. (KLOE), Phys. Lett. \textbf{B648}, 267 (2007),
  \texttt{hep-ex/0612029}

\bibitem{Escribano:2007cd}
R.~Escribano, J.~Nadal, JHEP \textbf{05}, 006 (2007), \texttt{hep-ph/0703187}

\bibitem{Jaffe:1976ig}
R.L. Jaffe, Phys. Rev. \textbf{D15}, 267 (1977)

\bibitem{Maiani:2004uc}
L.~Maiani, F.~Piccinini, A.D. Polosa, V.~Riquer, Phys. Rev. Lett. \textbf{93},
  212002 (2004), \texttt{hep-ph/0407017}

\bibitem{Hooft:2008we}
G.~'t~Hooft, G.~Isidori, L.~Maiani, A.D. Polosa, V.~Riquer, Phys. Lett.
  \textbf{B662}, 424 (2008), \texttt{0801.2288}

\bibitem{Giacosa:2006rg}
F.~Giacosa, Phys. Rev. \textbf{D74}, 014028 (2006), \texttt{hep-ph/0605191}

\bibitem{Isidori:2006we}
G.~Isidori, L.~Maiani, M.~Nicolaci, S.~Pacetti, JHEP \textbf{05}, 049 (2006),
  \texttt{hep-ph/0603241}

\bibitem{Black:2006mn}
D.~Black, M.~Harada, J.~Schechter, Phys. Rev. \textbf{D73}, 054017 (2006),
  \texttt{hep-ph/0601052}

\bibitem{Giacosa:2008st}
F.~Giacosa, G.~Pagliara, Nucl. Phys. \textbf{A812}, 125 (2008),
  \texttt{0804.1572}

\bibitem{Achasov:2005hm}
N.N. Achasov, A.V. Kiselev, Phys. Rev. \textbf{D73}, 054029 (2006),
  \texttt{hep-ph/0512047}

\bibitem{Bramon:2002iw}
A.~Bramon, R.~Escribano, J.L. Lucio~M, M.~Napsuciale, G.~Pancheri, Eur. Phys.
  J. \textbf{C26}, 253 (2002), \texttt{hep-ph/0204339}

\bibitem{Escribano:2003xa}
R.~Escribano, Nucl. Phys. Proc. Suppl. \textbf{126}, 204 (2004),
  \texttt{hep-ph/0307038}

\bibitem{Oller:2002na}
J.A. Oller, Nucl. Phys. \textbf{A714}, 161 (2003), \texttt{hep-ph/0205121}

\bibitem{Ivashyn:2009te}
S.~Ivashyn, A.~Korchin (2009), \texttt{0904.4823}

\bibitem{Heinz:2008cv}
A.~Heinz, S.~Struber, F.~Giacosa, D.H. Rischke, Phys. Rev. \textbf{D79}, 037502
  (2009), \texttt{0805.1134}

\bibitem{Ambrosino:2006hb}
F.~Ambrosino et~al. (KLOE), Eur. Phys. J. \textbf{C49}, 473 (2007),
  \texttt{hep-ex/0609009}

\bibitem{Ambrosino:2009py}
F.~Ambrosino et~al. (KLOE), Phys. Lett. \textbf{B681}, 5 (2009),
  \texttt{0904.2539}

\bibitem{Ambrosino:2005wk}
F.~Ambrosino et~al. (KLOE), Phys. Lett. \textbf{B634}, 148 (2006),
  \texttt{hep-ex/0511031}

\bibitem{Aloisio:2002vm}
A.~Aloisio et~al. (KLOE), Phys. Lett. \textbf{B541}, 45 (2002),
  \texttt{hep-ex/0206010}

\bibitem{Ambrosino:2009rg}
F.~Ambrosino et~al. (KLOE), Phys. Lett. \textbf{B679}, 10 (2009),
  \texttt{0903.4115}

\bibitem{Hanneke:2008tm}
D.~Hanneke, S.~Fogwell, G.~Gabrielse, Phys. Rev. Lett. \textbf{100}, 120801
  (2008), \texttt{0801.1134}

\bibitem{Jegerlehner:2003rx}
F.~Jegerlehner, Nucl. Phys. Proc. Suppl. \textbf{131}, 213 (2004),
  \texttt{hep-ph/0312372}

\bibitem{Jegerlehner:2006ju}
F.~Jegerlehner, Nucl. Phys. Proc. Suppl. \textbf{162}, 22 (2006),
  \texttt{hep-ph/0608329}

\bibitem{Jegerlehner:2008rs}
F.~Jegerlehner, Nucl. Phys. Proc. Suppl. \textbf{181-182}, 135 (2008),
  \texttt{0807.4206}

\bibitem{Marciano:1980pb}
W.J. Marciano, A.~Sirlin, Phys. Rev. \textbf{D22}, 2695 (1980)

\bibitem{Sirlin:1980nh}
A.~Sirlin, Phys. Rev. \textbf{D22}, 971 (1980)

\bibitem{Sirlin:1989uf}
A.~Sirlin, Phys. Lett. \textbf{B232}, 123 (1989)

\bibitem{Degrassi:1990tu}
G.~Degrassi, S.~Fanchiotti, A.~Sirlin, Nucl. Phys. \textbf{B351}, 49 (1991)

\bibitem{Ferroglia:2001cr}
A.~Ferroglia, G.~Ossola, A.~Sirlin, Phys. Lett. \textbf{B507}, 147 (2001),
  \texttt{hep-ph/0103001}

\bibitem{Veltman:1977kh}
M.J.G. Veltman, Nucl. Phys. \textbf{B123}, 89 (1977)

\bibitem{Abe:1995hr}
F.~Abe et~al. (CDF), Phys. Rev. Lett. \textbf{74}, 2626 (1995),
  \texttt{hep-ex/9503002}

\bibitem{Abachi:1995iq}
S.~Abachi et~al. (D0), Phys. Rev. Lett. \textbf{74}, 2632 (1995),
  \texttt{hep-ex/9503003}

\bibitem{Degrassi:1997iy}
G.~Degrassi, P.~Gambino, M.~Passera, A.~Sirlin, Phys. Lett. \textbf{B418}, 209
  (1998), \texttt{hep-ph/9708311}

\bibitem{Degrassi:1999jd}
G.~Degrassi, P.~Gambino, Nucl. Phys. \textbf{B567}, 3 (2000),
  \texttt{hep-ph/9905472}

\bibitem{Ferroglia:2002rg}
A.~Ferroglia, G.~Ossola, M.~Passera, A.~Sirlin, Phys. Rev. \textbf{D65}, 113002
  (2002), \texttt{hep-ph/0203224}

\bibitem{Awramik:2003rn}
M.~Awramik, M.~Czakon, A.~Freitas, G.~Weiglein, Phys. Rev. \textbf{D69}, 053006
  (2004), \texttt{hep-ph/0311148}

\bibitem{Awramik:2004ge}
M.~Awramik, M.~Czakon, A.~Freitas, G.~Weiglein, Phys. Rev. Lett. \textbf{93},
  201805 (2004), \texttt{hep-ph/0407317}

\bibitem{Awramik:2006uz}
M.~Awramik, M.~Czakon, A.~Freitas, JHEP \textbf{11}, 048 (2006),
  \texttt{hep-ph/0608099}

\bibitem{Burkhardt:2005se}
H.~Burkhardt, B.~Pietrzyk, Phys. Rev. \textbf{D72}, 057501 (2005),
  \texttt{hep-ph/0506323}

\bibitem{leplsd:2005ema}
S.~Schael et~al. (ALEPH), Phys. Rept. \textbf{427}, 257 (2006),
  \texttt{hep-ex/0509008}

\bibitem{LEPEWWG}
LEP-EW-WG ({Last checked 25 February 2010}), {Electroweak theory tests},
  \texttt{http://lepewwg.web.cern.ch}

\bibitem{Weiglein:2004hn}
G.~Weiglein et~al. (LHC/LC Study Group), Phys. Rept. \textbf{426}, 47 (2006),
  \texttt{hep-ph/0410364}

\bibitem{Steinhauser:1998rq}
M.~Steinhauser, Phys. Lett. \textbf{B429}, 158 (1998), \texttt{hep-ph/9803313}

\bibitem{Cabibbo:1961sz}
N.~Cabibbo, R.~Gatto, Phys. Rev. \textbf{124}, 1577 (1961)

\bibitem{Eidelman:1995ny}
S.~Eidelman, F.~Jegerlehner, Z. Phys. \textbf{C67}, 585 (1995),
  \texttt{hep-ph/9502298}

\bibitem{Burkhardt:1995tt}
H.~Burkhardt, B.~Pietrzyk, Phys. Lett. \textbf{B356}, 398 (1995)

\bibitem{Martin:1994we}
A.D. Martin, D.~Zeppenfeld, Phys. Lett. \textbf{B345}, 558 (1995),
  \texttt{hep-ph/9411377}

\bibitem{Swartz:1995hc}
M.L. Swartz, Phys. Rev. \textbf{D53}, 5268 (1996), \texttt{hep-ph/9509248}

\bibitem{Davier:1997vd}
M.~Davier, A.~Hoycker, Phys. Lett. \textbf{B419}, 419 (1998),
  \texttt{hep-ph/9801361}

\bibitem{Davier:1998si}
M.~Davier, A.~Hocker, Phys. Lett. \textbf{B435}, 427 (1998),
  \texttt{hep-ph/9805470}

\bibitem{Kuhn:1998ze}
J.H. Kuhn, M.~Steinhauser, Phys. Lett. \textbf{B437}, 425 (1998),
  \texttt{hep-ph/9802241}

\bibitem{Groote:1998pk}
S.~Groote, J.G. Korner, K.~Schilcher, N.F. Nasrallah, Phys. Lett.
  \textbf{B440}, 375 (1998), \texttt{hep-ph/9802374}

\bibitem{Erler:1998sy}
J.~Erler, Phys. Rev. \textbf{D59}, 054008 (1999), \texttt{hep-ph/9803453}

\bibitem{Martin:2000by}
A.D. Martin, J.~Outhwaite, M.G. Ryskin, Phys. Lett. \textbf{B492}, 69 (2000),
  \texttt{hep-ph/0008078}

\bibitem{Alemany:1997tn}
R.~Alemany, M.~Davier, A.~Hocker, Eur. Phys. J. \textbf{C2}, 123 (1998),
  \texttt{hep-ph/9703220}

\bibitem{Akhmetshin:2003zn}
R.R. Akhmetshin et~al. (CMD-2), Phys. Lett. \textbf{B578}, 285 (2004),
  \texttt{hep-ex/0308008}

\bibitem{Bai:2001ct}
J.Z. Bai et~al. (BES), Phys. Rev. Lett. \textbf{88}, 101802 (2002),
  \texttt{hep-ex/0102003}

\bibitem{Burkhardt:2001xp}
H.~Burkhardt, B.~Pietrzyk, Phys. Lett. \textbf{B513}, 46 (2001)

\bibitem{Jegerlehner:2001wq}
F.~Jegerlehner, J. Phys. \textbf{G29}, 101 (2003), \texttt{hep-ph/0104304}

\bibitem{Hagiwara:2003da}
K.~Hagiwara, A.D. Martin, D.~Nomura, T.~Teubner, Phys. Rev. \textbf{D69},
  093003 (2004), \texttt{hep-ph/0312250}

\bibitem{deTroconiz:2004tr}
J.F. de~Troconiz, F.J. Yndurain, Phys. Rev. \textbf{D71}, 073008 (2005),
  \texttt{hep-ph/0402285}

\bibitem{Hagiwara:2006jt}
K.~Hagiwara, A.D. Martin, D.~Nomura, T.~Teubner, Phys. Lett. \textbf{B649}, 173
  (2007), \texttt{hep-ph/0611102}

\bibitem{Aloisio:2004bu}
A.~Aloisio et~al. (KLOE), Phys. Lett. \textbf{B606}, 12 (2005),
  \texttt{hep-ex/0407048}

\bibitem{Akhmetshin:2006wh}
R.R. Akhmetshin et~al. (CMD-2), JETP Lett. \textbf{84}, 413 (2006),
  \texttt{hep-ex/0610016}

\bibitem{Akhmetshin:2006bx}
R.R. Akhmetshin et~al. (CMD-2), Phys. Lett. \textbf{B648}, 28 (2007),
  \texttt{hep-ex/0610021}

\bibitem{Aulchenko:2006na}
V.M. Aulchenko et~al. (CMD-2), JETP Lett. \textbf{82}, 743 (2005),
  \texttt{hep-ex/0603021}

\bibitem{Achasov:2006vp}
M.N. Achasov et~al., J. Exp. Theor. Phys. \textbf{103}, 380 (2006),
  \texttt{hep-ex/0605013}

\bibitem{Aubert:2007ur}
B.~Aubert et~al. (BaBar), Phys. Rev. \textbf{D76}, 012008 (2007),
  \texttt{0704.0630}

\bibitem{Aubert:2007ef}
B.~Aubert et~al. (BaBar), Phys. Rev. \textbf{D76}, 092005 (2007),
  \texttt{0708.2461}

\bibitem{Aubert:2007uf}
B.~Aubert et~al. (BaBar), Phys. Rev. \textbf{D76}, 092006 (2007),
  \texttt{0709.1988}

\bibitem{Jegerlehner:1999hg}
F.~Jegerlehner (1999), \texttt{hep-ph/9901386}

\bibitem{Eidelman:1998vc}
S.~Eidelman, F.~Jegerlehner, A.L. Kataev, O.~Veretin, Phys. Lett.
  \textbf{B454}, 369 (1999), \texttt{hep-ph/9812521}

\bibitem{Kuhn:2007vp}
J.H. Kuhn, M.~Steinhauser, C.~Sturm, Nucl. Phys. \textbf{B778}, 192 (2007),
  \texttt{hep-ph/0702103}

\bibitem{Brown:2000sj}
H.N. Brown et~al. (Muon (g-2)), Phys. Rev. \textbf{D62}, 091101 (2000),
  \texttt{hep-ex/0009029}

\bibitem{Brown:2001mga}
H.N. Brown et~al. (Muon g-2), Phys. Rev. Lett. \textbf{86}, 2227 (2001),
  \texttt{hep-ex/0102017}

\bibitem{Bennett:2002jb}
G.W. Bennett et~al. (Muon g-2), Phys. Rev. Lett. \textbf{89}, 101804 (2002),
  \texttt{hep-ex/0208001}

\bibitem{Bennett:2004pv}
G.W. Bennett et~al. (Muon g-2), Phys. Rev. Lett. \textbf{92}, 161802 (2004),
  \texttt{hep-ex/0401008}

\bibitem{Bennett:2006fi}
G.W. Bennett et~al. (Muon g-2), Phys. Rev. \textbf{D73}, 072003 (2006),
  \texttt{hep-ex/0602035}

\bibitem{Czarnecki:2001pv}
A.~Czarnecki, W.J. Marciano, Phys. Rev. \textbf{D64}, 013014 (2001),
  \texttt{hep-ph/0102122}

\bibitem{Stockinger:2006zn}
D.~Stockinger, J. Phys. \textbf{G34}, R45 (2007), \texttt{hep-ph/0609168}

\bibitem{Stockinger:2008zz}
D.~Stockinger, Nucl. Phys. Proc. Suppl. \textbf{181-182}, 32 (2008)

\bibitem{Roberts:2010cj}
B.L. Roberts (2010), \texttt{1001.2898}

\bibitem{Carey:2009zz}
R.M. Carey et~al. (2009), \texttt{FERMILAB-PROPOSAL-0989}

\bibitem{Kinoshita:2005zr}
T.~Kinoshita, M.~Nio, Phys. Rev. \textbf{D73}, 013003 (2006),
  \texttt{hep-ph/0507249}

\bibitem{Aoyama:2007dv}
T.~Aoyama, M.~Hayakawa, T.~Kinoshita, M.~Nio, Phys. Rev. Lett. \textbf{99},
  110406 (2007), \texttt{0706.3496}

\bibitem{Aoyama:2007mn}
T.~Aoyama, M.~Hayakawa, T.~Kinoshita, M.~Nio, Phys. Rev. \textbf{D77}, 053012
  (2008), \texttt{0712.2607}

\bibitem{Laporta:1992pa}
S.~Laporta, E.~Remiddi, Phys. Lett. \textbf{B301}, 440 (1993)

\bibitem{Laporta:1996mq}
S.~Laporta, E.~Remiddi, Phys. Lett. \textbf{B379}, 283 (1996),
  \texttt{hep-ph/9602417}

\bibitem{Passera:2006gc}
M.~Passera, Phys. Rev. \textbf{D75}, 013002 (2007), \texttt{hep-ph/0606174}

\bibitem{Kataev:2006yh}
A.L. Kataev, Phys. Rev. \textbf{D74}, 073011 (2006), \texttt{hep-ph/0608120}

\bibitem{Passera:2004bj}
M.~Passera, J. Phys. \textbf{G31}, R75 (2005), \texttt{hep-ph/0411168}

\bibitem{Czarnecki:2002nt}
A.~Czarnecki, W.J. Marciano, A.~Vainshtein, Phys. Rev. \textbf{D67}, 073006
  (2003), \texttt{hep-ph/0212229}

\bibitem{Czarnecki:1995sz}
A.~Czarnecki, B.~Krause, W.J. Marciano, Phys. Rev. Lett. \textbf{76}, 3267
  (1996), \texttt{hep-ph/9512369}

\bibitem{Czarnecki:1995wq}
A.~Czarnecki, B.~Krause, W.J. Marciano, Phys. Rev. \textbf{D52}, 2619 (1995),
  \texttt{hep-ph/9506256}

\bibitem{Gourdin:1969dm}
M.~Gourdin, E.~De~Rafael, Nucl. Phys. \textbf{B10}, 667 (1969)

\bibitem{Teubner:2008zz}
T.~Teubner, Nucl. Phys. Proc. Suppl. \textbf{181-182}, 20 (2008)

\bibitem{Ambrosino:2008en}
F.~Ambrosino et~al. (KLOE), Phys. Lett. \textbf{B670}, 285 (2009),
  \texttt{0809.3950}

\bibitem{Davier:2007ua}
M.~Davier, Nucl. Phys. Proc. Suppl. \textbf{169}, 288 (2007),
  \texttt{hep-ph/0701163}

\bibitem{Eidelman:2007zz}
S.~Eidelman, Acta Phys. Polon. \textbf{B38}, 3499 (2007)

\bibitem{Jegerlehner:2008zz}
F.~Jegerlehner, Nucl. Phys. Proc. Suppl. \textbf{181-182}, 26 (2008)

\bibitem{Jegerlehner:2009ry}
F.~Jegerlehner, A.~Nyffeler, Phys. Rept. \textbf{477}, 1 (2009),
  \texttt{0902.3360}

\bibitem{Aubert:2009fg}
B.~Aubert et~al. (BaBar), Phys. Rev. Lett. \textbf{103}, 231801 (2009),
  \texttt{0908.3589}

\bibitem{Davier:2009zi}
M.~Davier, A.~Hoecker, B.~Malaescu, C.Z. Yuan, Z.~Zhang (2009),
  \texttt{0908.4300}

\bibitem{Davier:2002dy}
M.~Davier, S.~Eidelman, A.~Hocker, Z.~Zhang, Eur. Phys. J. \textbf{C27}, 497
  (2003), \texttt{hep-ph/0208177}

\bibitem{Davier:2003pw}
M.~Davier, S.~Eidelman, A.~Hocker, Z.~Zhang, Eur. Phys. J. \textbf{C31}, 503
  (2003), \texttt{hep-ph/0308213}

\bibitem{Marciano:1988vm}
W.J. Marciano, A.~Sirlin, Phys. Rev. Lett. \textbf{61}, 1815 (1988)

\bibitem{Sirlin:1981ie}
A.~Sirlin, Nucl. Phys. \textbf{B196}, 83 (1982)

\bibitem{Cirigliano:2001er}
V.~Cirigliano, G.~Ecker, H.~Neufeld, Phys. Lett. \textbf{B513}, 361 (2001),
  \texttt{hep-ph/0104267}

\bibitem{Cirigliano:2002pv}
V.~Cirigliano, G.~Ecker, H.~Neufeld, JHEP \textbf{08}, 002 (2002),
  \texttt{hep-ph/0207310}

\bibitem{Davier:2009ag}
M.~Davier et~al., Eur. Phys. J. \textbf{C66}, 127 (2010), \texttt{0906.5443}

\bibitem{FloresBaez:2006gf}
F.~Flores-Baez, A.~Flores-Tlalpa, G.~Lopez~Castro, G.~Toledo~Sanchez, Phys.
  Rev. \textbf{D74}, 071301 (2006), \texttt{hep-ph/0608084}

\bibitem{FloresBaez:2007es}
F.V. Flores-Baez, G.L. Castro, G.~Toledo~Sanchez, Phys. Rev. \textbf{D76},
  096010 (2007), \texttt{0708.3256}

\bibitem{Benayoun:2007cu}
M.~Benayoun, P.~David, L.~DelBuono, O.~Leitner, H.B. O'Connell, Eur. Phys. J.
  \textbf{C55}, 199 (2008), \texttt{0711.4482}

\bibitem{Benayoun:2009im}
M.~Benayoun, P.~David, L.~DelBuono, O.~Leitner, Eur. Phys. J. \textbf{C65}, 211
  (2010), \texttt{0907.4047}

\bibitem{Benayoun:2009fz}
M.~Benayoun, P.~David, L.~DelBuono, O.~Leitner (2009), \texttt{0907.5603}

\bibitem{Krause:1996rf}
B.~Krause, Phys. Lett. \textbf{B390}, 392 (1997), \texttt{hep-ph/9607259}

\bibitem{Knecht:2001qf}
M.~Knecht, A.~Nyffeler, Phys. Rev. \textbf{D65}, 073034 (2002),
  \texttt{hep-ph/0111058}

\bibitem{Knecht:2001qg}
M.~Knecht, A.~Nyffeler, M.~Perrottet, E.~de~Rafael, Phys. Rev. Lett.
  \textbf{88}, 071802 (2002), \texttt{hep-ph/0111059}

\bibitem{Melnikov:2003xd}
K.~Melnikov, A.~Vainshtein, Phys. Rev. \textbf{D70}, 113006 (2004),
  \texttt{hep-ph/0312226}

\bibitem{Prades:2009tw}
J.~Prades, E.~de~Rafael, A.~Vainshtein (2009), \texttt{0901.0306}

\bibitem{Nyffeler:2009tw}
A.~Nyffeler, Phys. Rev. \textbf{D79}, 073012 (2009), \texttt{0901.1172}

\bibitem{Jegerlehner:2007xe}
F.~Jegerlehner, Acta Phys. Polon. \textbf{B38}, 3021 (2007),
  \texttt{hep-ph/0703125}

\bibitem{Miller:2007kk}
J.P. Miller, E.~de~Rafael, B.L. Roberts, Rept. Prog. Phys. \textbf{70}, 795
  (2007), \texttt{hep-ph/0703049}

\bibitem{Passera:2005mx}
M.~Passera, Nucl. Phys. Proc. Suppl. \textbf{155}, 365 (2006),
  \texttt{hep-ph/0509372}

\bibitem{Passera:2007fk}
M.~Passera, Nucl. Phys. Proc. Suppl. \textbf{169}, 213 (2007),
  \texttt{hep-ph/0702027}

\bibitem{Davier:2004gb}
M.~Davier, W.J. Marciano, Ann. Rev. Nucl. Part. Sci. \textbf{54}, 115 (2004)

\bibitem{Knecht:2003kc}
M.~Knecht, Lect. Notes Phys. \textbf{629}, 37 (2004), \texttt{hep-ph/0307239}

\bibitem{Prades:2009qp}
J.~Prades (2009), \texttt{0909.2546}

\bibitem{Passera:2008jk}
M.~Passera, W.J. Marciano, A.~Sirlin, Phys. Rev. \textbf{D78}, 013009 (2008),
  \texttt{0804.1142}

\bibitem{Passera:2010ev}
M.~Passera, W.J. Marciano, A.~Sirlin (2010), \texttt{1001.4528}

\bibitem{Aubert:2004kj}
B.~Aubert et~al. (BaBar), Phys. Rev. \textbf{D70}, 072004 (2004),
  \texttt{hep-ex/0408078}

\bibitem{Aubert:2005eg}
B.~Aubert et~al. (BaBar), Phys. Rev. \textbf{D71}, 052001 (2005),
  \texttt{hep-ex/0502025}

\bibitem{Denig:2008zz}
A.G. Denig, P.A. Lukin (BaBar), Nucl. Phys. Proc. Suppl. \textbf{181-182}, 111
  (2008)

\bibitem{Aubert:2006jq}
B.~Aubert et~al. (BaBar), Phys. Rev. \textbf{D73}, 052003 (2006),
  \texttt{hep-ex/0602006}

\bibitem{Aubert:2007ym}
B.~Aubert et~al. (BaBar), Phys. Rev. \textbf{D77}, 092002 (2008),
  \texttt{0710.4451}

\bibitem{Arbuzov:1998te}
A.B. Arbuzov, E.A. Kuraev, N.P. Merenkov, L.~Trentadue, JHEP \textbf{12}, 009
  (1998), \texttt{hep-ph/9804430}

\bibitem{Binner:1999bt}
S.~Binner, J.H. Kuhn, K.~Melnikov, Phys. Lett. \textbf{B459}, 279 (1999),
  \texttt{hep-ph/9902399}

\bibitem{Benayoun:1999hm}
M.~Benayoun, S.I. Eidelman, V.N. Ivanchenko, Z.K. Silagadze, Mod. Phys. Lett.
  \textbf{A14}, 2605 (1999), \texttt{hep-ph/9910523}

\bibitem{Rodrigo:2001jr}
G.~Rodrigo, A.~Gehrmann-De~Ridder, M.~Guilleaume, J.H. Kuhn, Eur. Phys. J.
  \textbf{C22}, 81 (2001), \texttt{hep-ph/0106132}

\bibitem{Kuhn:2002xg}
J.H. Kuhn, G.~Rodrigo, Eur. Phys. J. \textbf{C25}, 215 (2002),
  \texttt{hep-ph/0204283}

\bibitem{Rodrigo:2001kf}
G.~Rodrigo, H.~Czyz, J.H. Kuhn, M.~Szopa, Eur. Phys. J. \textbf{C24}, 71
  (2002), \texttt{hep-ph/0112184}

\bibitem{Czyz:2002np}
H.~Czyz, A.~Grzelinska, J.H. Kuhn, G.~Rodrigo, Eur. Phys. J. \textbf{C27}, 563
  (2003), \texttt{hep-ph/0212225}

\bibitem{Czyz:2003ue}
H.~Czyz, A.~Grzelinska, J.H. Kuhn, G.~Rodrigo, Eur. Phys. J. \textbf{C33}, 333
  (2004), \texttt{hep-ph/0308312}

\bibitem{Kluge:2008fb}
W.~Kluge, Nucl. Phys. Proc. Suppl. \textbf{181-182}, 280 (2008),
  \texttt{0805.4708}

\bibitem{Actis:2009gg}
S.~Actis et~al. (2009), \texttt{0912.0749}

\bibitem{Godfrey:1985xj}
S.~Godfrey, N.~Isgur, Phys. Rev. \textbf{D32}, 189 (1985)

\bibitem{Achasov:2003ir}
M.N. Achasov et~al., Phys. Rev. \textbf{D68}, 052006 (2003),
  \texttt{hep-ex/0305049}

\bibitem{Akhmetshin:2000wv}
R.R. Akhmetshin et~al. (CMD-2), Phys. Lett. \textbf{B489}, 125 (2000),
  \texttt{hep-ex/0009013}

\bibitem{Antonelli:1992jx}
A.~Antonelli et~al. (DM2), Z. Phys. \textbf{C56}, 15 (1992)

\bibitem{Bisello:1990du}
D.~Bisello et~al. (DM2), Nucl. Phys. Proc. Suppl. \textbf{21}, 111 (1991)

\bibitem{Akhmetshin:2004dy}
R.R. Akhmetshin et~al. (CMD-2), Phys. Lett. \textbf{B595}, 101 (2004),
  \texttt{hep-ex/0404019}

\bibitem{Achasov:2005rg}
M.N. Achasov et~al., J. Exp. Theor. Phys. \textbf{101}, 1053 (2005),
  \texttt{hep-ex/0506076}

\bibitem{Fujikawa:2008ma}
M.~Fujikawa et~al. (Belle), Phys. Rev. \textbf{D78}, 072006 (2008),
  \texttt{0805.3773}

\bibitem{Barnes:1996ff}
T.~Barnes, F.E. Close, P.R. Page, E.S. Swanson, Phys. Rev. \textbf{D55}, 4157
  (1997), \texttt{hep-ph/9609339}

\bibitem{Balazs:1998sb}
C.~Balazs, H.J. He, C.P. Yuan, Phys. Rev. \textbf{D60}, 114001 (1999),
  \texttt{hep-ph/9812263}

\bibitem{Bityukov:1986yd}
S.I. Bityukov et~al., Phys. Lett. \textbf{188B}, 383 (1987)

\bibitem{Frabetti:2003pw}
P.L. Frabetti et~al., Phys. Lett. \textbf{B578}, 290 (2004),
  \texttt{hep-ex/0310041}

\bibitem{Antonelli:1996xn}
A.~Antonelli et~al. (FENICE), Phys. Lett. \textbf{B365}, 427 (1996)

\bibitem{Castro:1994pi}
A.~Castro (DM2), Nuovo Cim. \textbf{A107}, 1807 (1994)

\bibitem{Ablikim:2006hp}
M.~Ablikim et~al. (BES), Phys. Rev. Lett. \textbf{97}, 142002 (2006),
  \texttt{hep-ex/0606047}

\bibitem{Mane:1982si}
F.~Mane et~al., Phys. Lett. \textbf{B112}, 178 (1982)

\bibitem{Shen:2009zze}
C.P. Shen et~al. (Belle), Phys. Rev. \textbf{D80}, 031101 (2009),
  \texttt{0808.0006}

\bibitem{Link:2002mp}
J.M. Link et~al. (FOCUS), Phys. Lett. \textbf{B545}, 50 (2002),
  \texttt{hep-ex/0208027}

\bibitem{Ablikim:2007yt}
M.~Ablikim et~al. (BES), Phys. Rev. Lett. \textbf{100}, 102003 (2008),
  \texttt{0712.1143}

\bibitem{Amendolia:1986wj}
S.R. Amendolia et~al. (NA7), Nucl. Phys. \textbf{B277}, 168 (1986)

\bibitem{Muller:2009pj}
S.E. Muller et~al. (KLOE) (2009), \texttt{0912.2205}

\bibitem{Blinov:2009zza}
V.E. Blinov et~al., Nucl. Instrum. Meth. \textbf{A598}, 23 (2009)

\bibitem{Bartalini:2005wx}
O.~Bartalini et~al. (GRAAL), Eur. Phys. J. \textbf{A26}, 399 (2005)

\bibitem{Gurzadyan:2004rx}
V.G. Gurzadyan et~al., Mod. Phys. Lett. \textbf{A20}, 19 (2005),
  \texttt{astro-ph/0410742}

\bibitem{jparcLoIamu}
M.~Aoki et~al. (2009), {J-PARC proposal},
  \texttt{http://j-parc.jp/jhf-np/Proposal\_e.html}

\bibitem{Babusci:2009sg}
D.~Babusci et~al. (KLOE-2) (2009), \texttt{0906.0875}

\bibitem{Brodsky:1970vk}
S.J. Brodsky, T.~Kinoshita, H.~Terazawa, Phys. Rev. Lett. \textbf{25}, 972
  (1970)

\bibitem{Brodsky:1971ud}
S.J. Brodsky, T.~Kinoshita, H.~Terazawa, Phys. Rev. \textbf{D4}, 1532 (1971)

\bibitem{Yang:1950rg}
C.N. Yang, Phys. Rev. \textbf{77}, 242 (1950)

\bibitem{Boyer:1990vu}
J.~Boyer et~al., Phys. Rev. \textbf{D42}, 1350 (1990)

\bibitem{Marsiske:1990hx}
H.~Marsiske et~al. (Crystal Ball), Phys. Rev. \textbf{D41}, 3324 (1990)

\bibitem{Oest:1990ki}
T.~Oest et~al. (JADE), Z. Phys. \textbf{C47}, 343 (1990)

\bibitem{Mori:2006jj}
T.~Mori et~al. (Belle), Phys. Rev. \textbf{D75}, 051101 (2007),
  \texttt{hep-ex/0610038}

\bibitem{Uehara:2008pf}
S.~Uehara et~al. (Belle), Phys. Rev. \textbf{D78}, 052004 (2008),
  \texttt{0810.0655}

\bibitem{Klempt:2007cp}
E.~Klempt, A.~Zaitsev, Phys. Rept. \textbf{454}, 1 (2007), \texttt{0708.4016}

\bibitem{Pennington:2007zy}
M.R. Pennington (2007), \texttt{0711.1435}

\bibitem{Barnes:1985cy}
T.~Barnes, Phys. Lett. \textbf{B165}, 434 (1985)

\bibitem{Barnes:1992sg}
T.~Barnes (1992), invited paper to Int. Workshop on Photon-Photon Collisions,
  La Jolla, CA, Mar 22-26, 1992

\bibitem{Achasov:1987ks}
N.N. Achasov, G.N. Shestakov, Z. Phys. \textbf{C41}, 309 (1988)

\bibitem{Achasov:2006cq}
N.N. Achasov, A.V. Kiselev, Phys. Rev. \textbf{D76}, 077501 (2007),
  \texttt{hep-ph/0606268}

\bibitem{Hanhart:2007wa}
C.~Hanhart, Y.S. Kalashnikova, A.E. Kudryavtsev, A.V. Nefediev, Phys. Rev.
  \textbf{D75}, 074015 (2007), \texttt{hep-ph/0701214}

\bibitem{Branz:2008ha}
T.~Branz, T.~Gutsche, V.E. Lyubovitskij, Phys. Rev. \textbf{D78}, 114004
  (2008), \texttt{0808.0705}

\bibitem{Volkov:2009pc}
M.K. Volkov, E.A. Kuraev, Y.M. Bystritskiy (2009), \texttt{0904.2484}

\bibitem{Mennessier:2008kk}
G.~Mennessier, S.~Narison, W.~Ochs, Phys. Lett. \textbf{B665}, 205 (2008),
  \texttt{0804.4452}

\bibitem{Mennessier:2008mc}
G.~Mennessier, S.~Narison, W.~Ochs, Nucl. Phys. Proc. Suppl. \textbf{181-182},
  238 (2008), \texttt{0806.4092}

\bibitem{Achasov:2007qr}
N.N. Achasov, G.N. Shestakov, Phys. Rev. \textbf{D77}, 074020 (2008),
  \texttt{0712.0885}

\bibitem{Achasov:2009ee}
N.N. Achasov, G.N. Shestakov (2009), \texttt{0905.2017}

\bibitem{Giacosa:2008xp}
F.~Giacosa, AIP Conf. Proc. \textbf{1030}, 153 (2008), \texttt{0804.3216}

\bibitem{Giacosa:2007bs}
F.~Giacosa, T.~Gutsche, V.E. Lyubovitskij, Phys. Rev. \textbf{D77}, 034007
  (2008), \texttt{0710.3403}

\bibitem{vanBeveren:2008st}
E.~van Beveren, F.~Kleefeld, G.~Rupp, M.D. Scadron, Phys. Rev. \textbf{D79},
  098501 (2009), \texttt{0811.2589}

\bibitem{Pennington:2008xd}
M.R. Pennington, T.~Mori, S.~Uehara, Y.~Watanabe, Eur. Phys. J. \textbf{C56}, 1
  (2008), \texttt{0803.3389}

\bibitem{Mori:2007bu}
T.~Mori et~al. (Belle), J. Phys. Soc. Jap. \textbf{76}, 074102 (2007),
  \texttt{0704.3538}

\bibitem{Morgan:1991zx}
D.~Morgan, M.R. Pennington, Phys. Lett. \textbf{B272}, 134 (1991)

\bibitem{Lyth:1971pm}
D.H. Lyth, Nucl. Phys. \textbf{B30}, 195 (1971)

\bibitem{Mennessier:1982fk}
G.~Mennessier, Z. Phys. \textbf{C16}, 241 (1983)

\bibitem{Mennessier:1986cp}
G.~Mennessier, T.N. Truong, Phys. Lett. \textbf{B177}, 195 (1986)

\bibitem{Pennington:1992vw}
M.R. Pennington (1992), \texttt{Durham Report DTP-92-32}

\bibitem{Pennington:1995fv}
M.R. Pennington (1995), \texttt{Durham Report DTP-95-18}

\bibitem{Boglione:1998rw}
M.~Boglione, M.R. Pennington, Eur. Phys. J. \textbf{C9}, 11 (1999),
  \texttt{hep-ph/9812258}

\bibitem{Behrend:1992hy}
H.J. Behrend et~al. (CELLO), Z. Phys. \textbf{C56}, 381 (1992)

\bibitem{Bienlein:1992en}
J.K. Bienlein (Crystal Ball) (1992), {Crystal Ball contributions to 9th Int.
  Workshop on Photon- Photon Collisions, La Jolla, CA, Mar 22-26, 1992}

\bibitem{:2009cf}
S.~Uehara et~al. (Belle), Phys. Rev. \textbf{D80}, 032001 (2009),
  \texttt{0906.1464}

\bibitem{Pennington:2006dg}
M.R. Pennington, Phys. Rev. Lett. \textbf{97}, 011601 (2006)

\bibitem{Oller:2007sh}
J.A. Oller, L.~Roca, C.~Schat, Phys. Lett. \textbf{B659}, 201 (2008),
  \texttt{0708.1659}

\bibitem{Bernabeu:2008wt}
J.~Bernabeu, J.~Prades, Phys. Rev. Lett. \textbf{100}, 241804 (2008),
  \texttt{0802.1830}

\bibitem{Pennington:2007eg}
M.R. Pennington, Prog. Theor. Phys. Suppl. \textbf{168}, 143 (2007),
  \texttt{hep-ph/0703256}

\bibitem{Pennington:2007yt}
M.R. Pennington, Mod. Phys. Lett. \textbf{A22}, 1439 (2007), \texttt{0705.3314}

\bibitem{Nguyen:2006sr}
F.~Nguyen, F.~Piccinini, A.D. Polosa, Eur. Phys. J. \textbf{C47}, 65 (2006),
  \texttt{hep-ph/0602205}

\bibitem{Bdsim:MC}
S.~Malton et~al. (2008), {BdSim MonteCarlo},
  \texttt{https://www.pp.rhul.ac.uk/twiki/bin/view/JAI/BdSim}

\bibitem{Prades:2009pn}
J.~Prades (2009), \texttt{0905.3164}

\bibitem{Kou:1999tt}
E.~Kou, Phys. Rev. \textbf{D63}, 054027 (2001), \texttt{hep-ph/9908214}

\bibitem{Courau:1984ia}
A.~Courau (1984), \texttt{SLAC-PUB-3363}

\bibitem{Alexander:1993rz}
G.~Alexander et~al., Nuovo Cim. \textbf{A107}, 837 (1994)

\bibitem{Bellucci:1994id}
S.~Bellucci, A.~Courau, S.~Ong (1994), \texttt{LNF-94/078(P)}

\bibitem{Ong:1999gp}
S.~Ong (1999), \texttt{hep-ex/9902030}

\bibitem{Uehara:1996di}
S.~Uehara (1996), \texttt{KEK-REPORT-96-11}

\bibitem{Gronberg:1997fj}
J.~Gronberg et~al. (CLEO), Phys. Rev. \textbf{D57}, 33 (1998),
  \texttt{hep-ex/9707031}

\bibitem{Aubert:2009mc}
B.~Aubert et~al. (BaBar), Phys. Rev. \textbf{D80}, 052002 (2009),
  \texttt{0905.4778}

\bibitem{Budnev:1974de}
V.M. Budnev, I.F. Ginzburg, G.V. Meledin, V.G. Serbo, Phys. Rept. \textbf{15},
  181 (1975)

\bibitem{Czyz:2003gb}
H.~Czyz, E.~Nowak, Acta Phys. Polon. \textbf{B34}, 5231 (2003),
  \texttt{hep-ph/0310335}

\bibitem{Czyz:2005ab}
H.~Czyz, E.~Nowak-Kubat, Acta Phys. Polon. \textbf{B36}, 3425 (2005),
  \texttt{hep-ph/0510287}

\bibitem{Czyz:2006dm}
H.~Czyz, E.~Nowak-Kubat, Phys. Lett. \textbf{B634}, 493 (2006),
  \texttt{hep-ph/0601169}

\bibitem{Schuler:1997ex}
G.A. Schuler, Comput. Phys. Commun. \textbf{108}, 279 (1998),
  \texttt{hep-ph/9710506}

\bibitem{Berends:2001ta}
F.A. Berends, R.~van Gulik, Comput. Phys. Commun. \textbf{144}, 82 (2002),
  \texttt{hep-ph/0109195}

\bibitem{Czyz:inprep}
H.~Czyz, S.~Ivashyn (2010), {in preparation}

\bibitem{:2008en}
F.~Ambrosino et~al. (KLOE), Phys. Lett. \textbf{B670}, 285 (2009),
  \texttt{0809.3950}

\bibitem{Hayakawa:1995ps}
M.~Hayakawa, T.~Kinoshita, A.I. Sanda, Phys. Rev. Lett. \textbf{75}, 790
  (1995), \texttt{hep-ph/9503463}

\bibitem{Hayakawa:1996ki}
M.~Hayakawa, T.~Kinoshita, A.I. Sanda, Phys. Rev. \textbf{D54}, 3137 (1996),
  \texttt{hep-ph/9601310}

\bibitem{Hayakawa:1997rq}
M.~Hayakawa, T.~Kinoshita, Phys. Rev. \textbf{D57}, 465 (1998),
  \texttt{hep-ph/9708227}

\bibitem{Bijnens:1995xf}
J.~Bijnens, E.~Pallante, J.~Prades, Nucl. Phys. \textbf{B474}, 379 (1996),
  \texttt{hep-ph/9511388}

\bibitem{Bijnens:1995cc}
J.~Bijnens, E.~Pallante, J.~Prades, Phys. Rev. Lett. \textbf{75}, 1447 (1995),
  \texttt{hep-ph/9505251}

\bibitem{Bijnens:2001cq}
J.~Bijnens, E.~Pallante, J.~Prades, Nucl. Phys. \textbf{B626}, 410 (2002),
  \texttt{hep-ph/0112255}

\bibitem{Prades:2008zz}
J.~Prades, Nucl. Phys. Proc. Suppl. \textbf{181-182}, 15 (2008),
  \texttt{0806.2250}

\bibitem{Bijnens:2007pz}
J.~Bijnens, J.~Prades, Mod. Phys. Lett. \textbf{A22}, 767 (2007),
  \texttt{hep-ph/0702170}

\bibitem{Bijnens:2007fp}
J.~Bijnens, J.~Prades, Acta Phys. Polon. \textbf{B38}, 2819 (2007),
  \texttt{hep-ph/0701240}

\bibitem{DeRafael:2008iu}
E.~De~Rafael, Nucl. Phys. Proc. Suppl. \textbf{186}, 211 (2009),
  \texttt{0809.3085}

\bibitem{Hertzog:2007hz}
D.W. Hertzog, J.P. Miller, E.~de~Rafael, B.~Lee~Roberts, D.~Stockinger (2007),
  \texttt{0705.4617}

\bibitem{Jegerlehner:2008zzb}
F.~Jegerlehner, Lect. Notes Phys. \textbf{745}, 9 (2008)

\bibitem{deRafael:1993za}
E.~de~Rafael, Phys. Lett. \textbf{B322}, 239 (1994), \texttt{hep-ph/9311316}

\bibitem{Bijnens:2003rc}
J.~Bijnens, E.~Gamiz, E.~Lipartia, J.~Prades, JHEP \textbf{04}, 055 (2003),
  \texttt{hep-ph/0304222}

\bibitem{Dorokhov:2008pw}
A.E. Dorokhov, W.~Broniowski, Phys. Rev. \textbf{D78}, 073011 (2008),
  \texttt{0805.0760}

\bibitem{RamseyMusolf:2002cy}
M.J. Ramsey-Musolf, M.B. Wise, Phys. Rev. Lett. \textbf{89}, 041601 (2002),
  \texttt{hep-ph/0201297}

\bibitem{Kampf:2005tz}
K.~Kampf, M.~Knecht, J.~Novotny, Eur. Phys. J. \textbf{C46}, 191 (2006),
  \texttt{hep-ph/0510021}

\bibitem{Kampf:2009tk}
K.~Kampf, B.~Moussallam, Phys. Rev. \textbf{D79}, 076005 (2009),
  \texttt{0901.4688}

\bibitem{Kampf:2009kg}
K.~Kampf (2009), \texttt{0905.0585}

\bibitem{Dorokhov:2009dg}
A.E. Dorokhov (2009), \texttt{0905.4577}

\bibitem{Dorokhov:2009jd}
A.E. Dorokhov (2009), \texttt{0912.5278}

\bibitem{Behrend:1990sr}
H.J. Behrend et~al. (CELLO), Z. Phys. \textbf{C49}, 401 (1991)

\bibitem{Jean:2003ci}
P.~Jean et~al., Astron. Astrophys. \textbf{407}, L55 (2003),
  \texttt{astro-ph/0309484}

\bibitem{Adriani:2008zr}
O.~Adriani et~al. (PAMELA), Nature \textbf{458}, 607 (2009), \texttt{0810.4995}

\bibitem{Chang:2008zzr}
J.~Chang et~al., Nature \textbf{456}, 362 (2008)

\bibitem{Abdo:2009zk}
A.A. Abdo et~al. (Fermi LAT), Phys. Rev. Lett. \textbf{102}, 181101 (2009),
  \texttt{0905.0025}

\bibitem{Collaboration:2008aaa}
F.~Aharonian et~al. (H.E.S.S.), Phys. Rev. Lett. \textbf{101}, 261104 (2008),
  \texttt{0811.3894}

\bibitem{Aharonian:2009ah}
F.~Aharonian et~al. (H.E.S.S), Astron. Astrophys. \textbf{508}, 561 (2009),
  \texttt{0905.0105}

\bibitem{Bernabei:2005hj}
R.~Bernabei et~al., Int. J. Mod. Phys. \textbf{D13}, 2127 (2004),
  \texttt{astro-ph/0501412}

\bibitem{Bernabei:2008yi}
R.~Bernabei et~al. (DAMA), Eur. Phys. J. \textbf{C56}, 333 (2008),
  \texttt{0804.2741}

\bibitem{Pospelov:2007mp}
M.~Pospelov, A.~Ritz, M.B. Voloshin, Phys. Lett. \textbf{B662}, 53 (2008),
  \texttt{0711.4866}

\bibitem{ArkaniHamed:2008qn}
N.~Arkani-Hamed, D.P. Finkbeiner, T.R. Slatyer, N.~Weiner, Phys. Rev.
  \textbf{D79}, 015014 (2009), \texttt{0810.0713}

\bibitem{Alves:2009nf}
D.S.M. Alves, S.R. Behbahani, P.~Schuster, J.G. Wacker (2009),
  \texttt{0903.3945}

\bibitem{Pospelov:2008jd}
M.~Pospelov, A.~Ritz, Phys. Lett. \textbf{B671}, 391 (2009), \texttt{0810.1502}

\bibitem{Hisano:2003ec}
J.~Hisano, S.~Matsumoto, M.M. Nojiri, Phys. Rev. Lett. \textbf{92}, 031303
  (2004), \texttt{hep-ph/0307216}

\bibitem{Cirelli:2008pk}
M.~Cirelli, M.~Kadastik, M.~Raidal, A.~Strumia, Nucl. Phys. \textbf{B813}, 1
  (2009), \texttt{0809.2409}

\bibitem{MarchRussell:2008yu}
J.~March-Russell, S.M. West, D.~Cumberbatch, D.~Hooper, JHEP \textbf{07}, 058
  (2008), \texttt{0801.3440}

\bibitem{Cholis:2008wq}
I.~Cholis, G.~Dobler, D.P. Finkbeiner, L.~Goodenough, N.~Weiner, Phys. Rev.
  \textbf{D80}, 123518 (2009), \texttt{0811.3641}

\bibitem{Cholis:2008qq}
I.~Cholis, D.P. Finkbeiner, L.~Goodenough, N.~Weiner, JCAP \textbf{0912}, 007
  (2009), \texttt{0810.5344}

\bibitem{ArkaniHamed:2008qp}
N.~Arkani-Hamed, N.~Weiner, JHEP \textbf{12}, 104 (2008), \texttt{0810.0714}

\bibitem{Fayet:1980ad}
P.~Fayet, Phys. Lett. \textbf{B95}, 285 (1980)

\bibitem{Fayet:1980rr}
P.~Fayet, Nucl. Phys. \textbf{B187}, 184 (1981)

\bibitem{Holdom:1986eq}
B.~Holdom, Phys. Lett. \textbf{B178}, 65 (1986)

\bibitem{TuckerSmith:2001hy}
D.~Tucker-Smith, N.~Weiner, Phys. Rev. \textbf{D64}, 043502 (2001),
  \texttt{hep-ph/0101138}

\bibitem{Chang:2008gd}
S.~Chang, G.D. Kribs, D.~Tucker-Smith, N.~Weiner, Phys. Rev. \textbf{D79},
  043513 (2009), \texttt{0807.2250}

\bibitem{Batell:2009yf}
B.~Batell, M.~Pospelov, A.~Ritz, Phys. Rev. \textbf{D79}, 115008 (2009),
  \texttt{0903.0363}

\bibitem{Essig:2009nc}
R.~Essig, P.~Schuster, N.~Toro, Phys. Rev. \textbf{D80}, 015003 (2009),
  \texttt{0903.3941}

\bibitem{Reece:2009un}
M.~Reece, L.T. Wang, JHEP \textbf{07}, 051 (2009), \texttt{0904.1743}

\bibitem{Bossi:2009uw}
F.~Bossi (2009), \texttt{0904.3815}

\bibitem{Borodatchenkova:2005ct}
N.~Borodatchenkova, D.~Choudhury, M.~Drees, Phys. Rev. Lett. \textbf{96},
  141802 (2006), \texttt{hep-ph/0510147}

\bibitem{Yin:2009mc}
P.F. Yin, J.~Liu, S.h. Zhu, Phys. Lett. \textbf{B679}, 362 (2009),
  \texttt{0904.4644}

\bibitem{Bjorken:2009mm}
J.D. Bjorken, R.~Essig, P.~Schuster, N.~Toro, Phys. Rev. \textbf{D80}, 075018
  (2009), \texttt{0906.0580}

\bibitem{Batell:2009di}
B.~Batell, M.~Pospelov, A.~Ritz (2009), \texttt{0906.5614}

\bibitem{Essig:2010xa}
R.~Essig, P.~Schuster, N.~Toro, B.~Wojtsekhowski (2010), \texttt{1001.2557}

\bibitem{Freytsis:2009bh}
M.~Freytsis, G.~Ovanesyan, J.~Thaler, JHEP \textbf{01}, 111 (2010),
  \texttt{0909.2862}

\bibitem{Strassler:2006im}
M.J. Strassler, K.M. Zurek, Phys. Lett. \textbf{B651}, 374 (2007),
  \texttt{hep-ph/0604261}

\bibitem{Schuster:2009au}
P.~Schuster, N.~Toro, I.~Yavin, Phys. Rev. \textbf{D81}, 016002 (2010),
  \texttt{0910.1602}

\bibitem{Strassler:2006qa}
M.J. Strassler (2006), \texttt{hep-ph/0607160}

\bibitem{Han:2007ae}
T.~Han, Z.~Si, K.M. Zurek, M.J. Strassler, JHEP \textbf{07}, 008 (2008),
  \texttt{0712.2041}

\bibitem{Strassler:2008bv}
M.J. Strassler (2008), \texttt{0801.0629}

\bibitem{Baumgart:2009tn}
M.~Baumgart, C.~Cheung, J.T. Ruderman, L.T. Wang, I.~Yavin, JHEP \textbf{04},
  014 (2009), \texttt{0901.0283}

\bibitem{Gunion:2005rw}
J.F. Gunion, D.~Hooper, B.~McElrath, Phys. Rev. \textbf{D73}, 015011 (2006),
  \texttt{hep-ph/0509024}

\bibitem{McElrath:2005bp}
B.~McElrath, Phys. Rev. \textbf{D72}, 103508 (2005), \texttt{hep-ph/0506151}

\bibitem{Dermisek:2006py}
R.~Dermisek, J.F. Gunion, B.~McElrath, Phys. Rev. \textbf{D76}, 051105 (2007),
  \texttt{hep-ph/0612031}

\bibitem{Galloway:2008yh}
J.~Galloway, B.~McElrath, J.~McRaven, Phys. Lett. \textbf{B670}, 363 (2009),
  \texttt{0807.2657}

\bibitem{Pospelov:2008zw}
M.~Pospelov, Phys. Rev. \textbf{D80}, 095002 (2009), \texttt{0811.1030}

\bibitem{Boehm:2003hm}
C.~Boehm, P.~Fayet, Nucl. Phys. \textbf{B683}, 219 (2004),
  \texttt{hep-ph/0305261}

\bibitem{Aubert:2009pw}
B.~Aubert et~al. (BaBar) (2009), \texttt{0908.2821}

\end{thebibliography}

\end{document}
